%% file: thesis.tex
\documentclass[pagesize,twoside,BCOR2mm]{scrreprt}
\usepackage{amsfonts}
\usepackage[]{graphicx}
\usepackage[ansinew]{inputenc}
\usepackage[french,british]{babel}
\usepackage[T1]{fontenc}
\usepackage{textcomp}
\setcapindent{1em}
\usepackage[pdftex]{hyperref}

\newcommand{\chem}[1]{$\mathsf{#1}$}
\newcommand{\chemr}[1]{$\mathrm{#1}$}

\begin{document}

\selectlanguage{british}

\include{leader}
\include{frabstract}
\include{introduction}
\include{literature}
\include{theory}
\include{experiment}
\include{results}
\include{conclusion}
\include{appendix}

\bibliography{falk11}
\bibliographystyle{plain}

\end{document}

%% file: leader.tex
\titlehead{
\includegraphics[width=7cm]{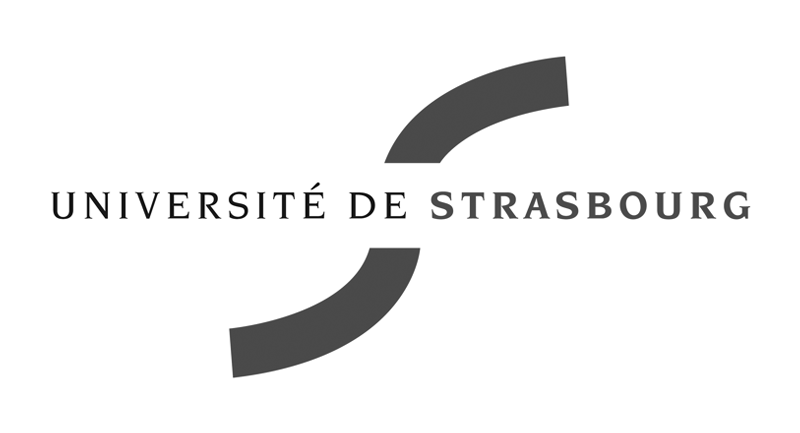}
}
%\titlehead{Institut National de la Santé et de la Recherche Medicale (INSERM)\\ Unit 977 \\ ``Biomaterials and Tissue Engineering''\\ 11, rue Humann \\ 67085 Strasbourg \\ France}
\subject{Dissertation submitted to obtain the Doctor of Philosophy degree in Physics by the University of Strasbourg}
\title{Melanin Made by Dopamine Oxidation: Thin Films and Interactions with Polyelectrolyte Multilayers}
%\subtitle{}
\author{Candidate: Falk Bernsmann}
\date{Publicly defended on 12 July 2010}
\publishers{Jury:\\
Vincent Ball, Strasbourg\\
Claudine Filiatre, Besan\c{c}on \\
Mir Wais Hosseini, Strasbourg\\
Marie-Hélène Metz-Boutigue, Strasbourg\\
Helmut M\"ohwald, Potsdam \\
Marie-France Vallat, Mulhouse
}
%\uppertitleback{}
\lowertitleback{Front cover: Photograph of a blue peacock (File ``Pfau\_imponierend.jpg'' from the free media database Wikimedia Commons [\href{http://commons.wikimedia.org}{http://commons.wikimedia.org}] published under the Creative Commons Attribution ShareAlike 3.0 License [\href{http://creativecommons.org/licenses/by-sa/3.0}{http://creativecommons.org/licenses/by-sa/3.0}]. Author of the image: BS Thurner Hof). The colour of the feathers is a combined effect of pigmentation and diffraction at melanin-rich, periodic nanostructures \cite{kinoshita:2005} \cite{yoshioka:2002} \cite{zi:2003}.}
%\dedication{}
\maketitle

\section*{Acknowledgements}
I would like to thank the following people for their valuable help during the preparation of this work:
\begin{itemize}
	\item Jean-Claude Voegel, director of the laboratory ``Biomatériaux et Ingénierie Tissulaire'' (unit 977 of Inserm, the French National Institute for Health and Medical Research), for receiving me in his team and supporting my research project.
	\item Vincent Ball (u 977) for proposing an interesting research topic, for help in editing journal articles and this text and for numerous fruitful discussions.
	\item Claudine Filiatre (Laboratoire Univers, Transport, Interfaces, Nanostructures, Atmosphère et Environnement, Molécules, Besan\c{c}on), Mir Wais Hosseini (Laboratoire de Chimie de Coordination Organique, Strasbourg), Marie-Hélène Metz-Boutigue (Laboratoire de Biologie de la Communication Cellulaire, Strasbourg), Helmut Möhwald (Max Planck Institut für Kolloid- und Grenzflächenforschung (MPIKG), Potsdam) and Marie-France Vallat (Institut des Sciences de Matériaux, Mulhouse) for evaluating my work.
	\item Cosette Betscha (u 977) for accompanying me during my first days in the laboratory and for several ellipsometry and UV--visible spectroscopy experiments.
	\item Christian Ringwald (u 977) for always offering a helping hand, for measuring $\zeta$-potentials and for preparing many dopamine-melanin deposits.
	\item Bernhard Senger (u 977) for numerically fitting fluorescence recovery after photobleaching data close to ``real-time'' and for many interesting discussions.
	\item Alae el Haitami (u 977) for an introduction to cyclic voltammetry experiments.
	\item Ludovic Richert (u 977) for help with atomic force microscopy.
	\item Philippe Lavalle (u 977) and Jérôme Mutterer (Institut de Biologie Moléculaire des Plantes, Strasbourg) for an introduction to confocal laser scanning microscopy and for help in preparation of the experiments.
	\item Hajare Mjahed (u 977) for good coordination of our confocal microscopy sessions and for salsa dancing.
	\item Jesus Raya (Institut de Chimie, Strasbourg) for performing and explaining nuclear magnetic resonance spectroscopy.
	\item Arnaud Ponche (Institut de Chimie des Surfaces et Interfaces, Mulhouse) for x-ray photoelectron spectroscopy measurements.
	\item Jérome Combet (Institut Charles Sadron (ICS), Strasbourg) for small angle x-ray scattering experiments.
		\item Fouzia Boulmedais (ICS) for help with the electrochemical quartz crystal microbalance experiments (and especially with the subsequent cleaning procedure).
	\item Francine Valenga (ICS) for acquiring UV-visible spectra when our own spectrometer failed at the most unsuitable moment.
	\item Pascal Marie (ICS) for allowing me to measure contact angles using his equipment.
	\item Dayang Wang and Wei Li (MPIKG) for interesting days at their institute. 
	\item All members of Inserm unit 977 for the warm welcome and the pleasant atmosphere.
	\item Tax payers of Alsace region for financial support.
\end{itemize}

\tableofcontents
%\listoffigures

%% file: frabstract.tex
\chapter{French abstract}
\begin{otherlanguage*}{french}
\section{Introduction}
Le travail de recherche présenté dans cette thèse est situé dans le domaine interdisciplinaire des biomatériaux, à l'interface entre la physique, la chimie et la biologie. Les sciences des biomatériaux étudient les propriétés chimiques et physiques de matériaux d'origine biologique, par exemple l'os, la nacre ou les pieds adhésifs de moules, pour comprendre la relation entre la structure et les propriétés de ces matériaux. On essaye de mimer les propriétés souvent surprenantes de matériaux biologiques dans beaucoup de domaines d'ingénierie et notamment dans l'ingénierie biomédicale pour concevoir entre autres des tissus artificiels ou des systèmes de libération de médicaments. En général, les performances de matériaux biomimétiques restent largement inférieures à celles de leurs modèles naturels. Par exemple au niveau de l'efficience de production, la biocompatibilité ou la biodégradabilité il reste encore beaucoup de progrès à faire. De plus, l'augmentation de l'espérance de vie dans les pays occidentaux accroit le besoin de matériaux de prothèses ayant une durée de vie comparable à celle du patient. En outre les matériaux implantés ne doivent pas introduire de réaction immunitaire trop aigüe après l'intervention chirurgicale faute de quoi la prothèse risque d'être rejetée. On voit sur ce simple exemple des prothèses que la mise au point d'un tel dispositif biomédical nécessite une recherche poussée à la fois dans le domaine mécanique et dans celui de la chimie de surface.

Ce travail a été réalisé au sein de l'unité mixte de recherche (UMR) 977 \guillemotleft ~Biomatériaux et Ingénierie Tissulaire \guillemotright \ de la Faculté de Chirurgie Dentaire de l'Université de Strasbourg et de l'Institut National de la Santé et de la Recherche Médicale (Inserm). Le sujet principal de l'UMR 977 est la modification de surfaces artificielles, par exemple d'implants trachéaux ou dentaires, afin de mieux les intégrer dans le corps humain. Pour ce faire, la majorité des projets dans ce laboratoire consiste à fonctionaliser la surface des matériaux utilisés par un film multicouche de polyélectrolytes (Sous-chapitre \ref{sec:literaturelbl}). Un tel film est construit par le dépôt alternatif de polymères (naturels ou synthétiques) chargés positivement ou négativement sur une surface de géométrie quelconque \cite{decher:1997}. On obtient ainsi des films d'une épaisseur de quelques nm à quelques $\mu$m et de propriétés physicochimiques (par exemple l'hydrophilie ou la dureté) contrôlées \cite{voegel:2003}. Il est possible d'incorporer des composés actifs comme des enzymes, des médicaments voire des cellules entières dans des multicouches de polyélectrolytes puis de les libérer de façon contrôlée.

Malgré ses possibilités multiples à l'échelle du laboratoire, la méthode de recouvrement de surfaces par multicouches de polyélectrolytes n'a pas trouvé d'applications industrielles parce qu'il s'agit d'une mé\-thode multi-étapes nécessitant beaucoup de temps et un équipement sophistiqué d'automation. Par conséquent il faut trouver des méthodes pour former des recouvrements de surfaces fonctionnels de façon simple en une seule étape. Une méthode possible est la formation de mélanine synthétique à partir de la dopamine proposée par Lee et collaborateurs \cite{lee:2007.2}, qui va être étudiée en détail dans cette thèse.

La mélanine (Sous-chapitre \ref{sec:melanin}) est un pigment biologique, qu'on trouve dans beaucoup d'animaux et de plantes, notamment dans la peau humaine où elle sert comme photo-protecteur. Elle fait aussi partie du système immunitaire inné \cite{mackintosh:2001}, et elle joue probablement un rôle dans le développement de maladies neurodégénératives comme la maladie de Parkinson \cite{kienzl:1999} \cite{linert:2000}.

La mélanine est constituée de molécules de 5,6-di\-hydroxy\-indole (DHI) et de l'acide carboxylique correspondant, le 5,6-di\-hydroxy\-indole-2-carboxylate (DHICA, Figure \ref{fig:dopamine}) \cite{ito:1986}. La structure macromoléculaire de la mélanine est l'objet d'une controverse: deux modèles différents ayant été proposés et soumis à la validation par l'expérience. Le premier modèle \cite{clark:1990} décrit la mélanine comme \emph{polymère hétérogène} d'unités de DHI et de DHICA liés aléatoirement par leurs positions 2, 3, 4 ou 7 \cite{peter:1989} \cite{pezzella:2007} \cite{reinheimer:1999} (voir figure \ref{fig:dopamine} pour la numérotation). Le deuxième modèle décrit la mélanine comme agrégat non-covalent \emph{d'unités fondamentales} formées d'empilements de trois à cinq feuillets oligomériques de quatre à huit molécules de DHI ou de DHICA. Ce modèle a été déduit d'expériences de diffraction de rayons x \cite{cheng:1994}, et il est supporté par d'autres expériences \cite{clancy:2001} \cite{nofsinger:2000} \cite{pezzella:1997} \cite{zajac:1994} et des simulations \cite{kaxiras:2006} \cite{stark:2003} \cite{stark:2005} sans qu'il y ait à ce jour une \emph{preuve directe} de l'existence de ces unités fondamentales. Les deux modèles peuvent expliquer les propriétés macroscopiques de la mélanine. Récemment Watt et collaborateurs \cite{watt:2009} ont observé  pour la première fois de manière directe une structure en couches dans de la mélanine naturelle et synthétique.

La mélanine possède des propriétés physicochimiques très particulières pour une molécule biologique: son spectre d'absorption est monotone de l'ultraviolet à l'infrarouge \cite{meredith:2006.1}, et elle convertit de façon efficace l'énergie absorbée en chaleur \cite{forest:1998}. Ceci ouvre la voie pour des applications en photo-protection, photo-détection ou photo-thermie. La mélanine est capable de capturer \cite{hong:2007} et de réduire \cite{lee:2007.2} des ions métalliques ce qui permet de créer des (nano)particules métalliques qui sont utilisées par exemple en catalyse chimique. En outre la mélanine possède une conductivité électrique qui dépend fortement de son hydratation allant de $10^{-11}$ S/m sous vide à $10^{-3}$ S/m à $100 \%$ d'humidité relative \cite{jastrzebska:1995}. En plus la conductivité dépend aussi de l'illumination de la mélanine \cite{ligonzo:2009} \cite{subianto:2005}.

Les cathécholamines telles que l'adrénaline, la noradrénaline et la dopamine (2-(3,4-di\-hydroxy\-phényl)\-éthyl\-amine) jouent un rôle important comme neurotransmetteurs (\cite{patrick:2003}, chapitre 16), et on étudie leurs mécanismes de réactions électrochimiques depuis longtemps \cite{hawley:1967}. La 3,4-di\-hydroxy-L-phényl\-alanine (DOPA, Figure \ref{fig:dopamine}), une forme hydroxylée de l'acide aminé L-phényl\-alanine, est un précurseur dans la synthèse naturelle de la dopamine (\cite{patrick:2003}, chapitre 16.4). Plus récemment il a été observé que la DOPA est aussi un constituent important des protéines trouvés dans les pieds adhésifs des moules. Le rôle de la DOPA dans la forte adhésion de moules sur toute sorte de substrat a été confirmé par des expériences de microscopie à force atomique \cite{lee:2006}. Depuis, des nombreuses méthodes de modifications de surfaces se sont basées sur les propriétés adhésives de la DOPA (Sous-chapitre \ref{sec:dopamineapplications}).

En 2007 Lee et collaborateurs ont communiqué une méthode simple pour recouvrir des surfaces de métaux, d'oxydes et de polymères en une seule étape \cite{lee:2007.2}. La méthode consiste à immerger un objet dans une solution de dopamine en présence de 	tampon Tris (10 mmol/L tris(hydroxy\-méthyl) amino\-méthane, pH 8.5). Sous ces conditions la dopamine est oxydée pour former de la mélanine synthétique en solution et à la surface de l'objet immergé. De cette manière on obtient des films de mélanine d'une épaisseur allant jusqu'à 50 nm après 25 h d'immersion. Les dépôts de mélanine ont été utilisés comme plate-forme pour des modifications de surfaces secondaires \cite{lee:2007.2}. Il était par exemple possible d'immobiliser des peptides avec une orientation déterminée par le pH lors de l'immobilisation \cite{lee:2009}. Dans ce dernier article les auteurs ont proposé un mécanisme de liaison covalente entre des groupements catéchols de la mélanine et des groupements amines du peptide pour expliquer leurs résultats.

La méthode développée par Lee et collaborateurs a été utilisée pour créer des films de polymères codés destinés à l'identification électrochimique \cite{zhang:2009}; pour préparer des capsules creuses de mélanine afin d'encapsuler des médicaments \cite{postma:2009} \cite{yu:2009}; pour modifier la perméabilité de membranes à utiliser dans des piles à combustible \cite{wang:2009} \cite{xi:2009}; pour fonctionnaliser la surface de nanotubes de carbone \cite{fei:2008} et pour préparer des surfaces imprimées par molécules dans le but de détecter par exemple des protéines \cite{zhou:2009} (Sous-chapitre \ref{sec:literaturelee}). Néanmoins toutes les publications citées traitent d'applications de dépôts de mélanine sans s'occuper du mécanisme de dépôt et des propriétés physicochimiques des films de mélanine.

Le but du travail présenté ici est donc d'étudier en détail la méthode de dépôt de la mélanine proposée par Lee et collaborateurs \cite{lee:2007.2} et de développer d'autres méthodes sur cette base. La formation de mélanine par oxydation de la dopamine est aussi observée au sein de multicouches de polyélectrolytes. Ces films biocompatibles de poly(L-lysine) (PLL) et de hyaluronate (HA) sont renforcés par la mélanine et peuvent être détachés de leurs supports comme membranes autosupportées.

\newpage
\section{Résultats}
\subsection{Formation de la mélanine en solution}
Le sous-chapitre \ref{sec:resultsdopamineinsolution} décrit l'étude de l'oxydation spontanée de la dopamine en solution en présence de 50 mmol/L Tris à pH 8.5. Le précipité noir qui se forme est identifié comme de la mélanine par spectroscopie de résonance magnétique nucléaire du carbone (\chem{^{13}C}) à l'état solide (Figures \ref{fig:nmrmelanin}, \ref{fig:nmrdopamine}). La vitesse de la réaction augmente avec la concentration initiale d'hydrochlorure de dopamine et atteint un plateau pour des concentrations supérieures à 1 g/L (Figure \ref{fig:dopaminepolymerisationrate}). La mélanine obtenue est capable de lier des amines avec une capacité de $(1.5 \pm 0.2) \cdot 10^{-3}$ moles de groupements amines par gramme de mélanine (Figure \ref{fig:aminobindingcap}). Probablement les groupements amines sont liés de façon covalente à des groupements catéchols présents à la surface de la mélanine comme cela a été proposé par d'autres groupes \cite{lee:2009} (Figure \ref{fig:pteamelanin}). Par conséquent la mélanine peut servir comme substrat pour immobiliser des biomolécules de façon contrôlée.

En accord avec l'observation trouvée dans la littérature \cite{bothma:2008}, il est possible de rédisperser le précipité de mélanine dans une solution fortement basique. Des spectres UV--visible (Figure \ref{fig:uvvismelaninph12}) couplés à des expériences de dialyse (Figure \ref{fig:dialysismelanin}) indiquent que les solutions obtenues contiennent des grands agrégats de mélanine ayant un spectre monotone. En plus, la solution contient des grains plus petits engendrant des maxima d'absorbance dans l'UV. Des images de microscopie électronique en transmission (Figure \ref{fig:temmelaningrain}) montrent une structure hiérarchique de la mélanine compatible avec le modèle structural en oligomères empilés. La densité et l'indice de réfraction de la mélanine rédispersée valent respectivement $(1.2 \pm 0.1)$ g/mL et $(1.73 \pm 0.05) + (0.027 \pm 0.002)i$ à une longueur d'onde de 589 nm. Les valeurs de l'indice de réfraction sont proches de celles trouvées dans la littérature pour de la mélanine naturelle ($N=2.0 + 0.01i$, \cite{yoshioka:2002} \cite{zi:2003}).

Les grains obtenus par redispersion du précipité de mélanine sont utilisés pour cons\-truire des films multicouches avec le polycation poly(chlorure de di\-allyl\-di\-méthyl\-ammo\-nium) (Figures \ref{fig:ellipsopdadmacmelanin}, \ref{fig:qcmpdadmacmelanin}). Contrairement aux solutions de mélanine, les films multicouches possèdent un spectre UV--visible monotone (Figure \ref{fig:uvvispdadmacmelanin}), ce qui indique l'adsorption préférentielle de grands agrégats de mélanine sur le film multicouche. La morphologie de surface des films multicouches est composée de plaquettes ayant une extension latérale comprise entre 100 nm et 500 nm (Figure \ref{fig:sfmpdadmacmelanin}). Une comparaison avec la morphologie de surface d'autres mélanines synthétiques et naturelles trouvée dans la littérature scientifique \cite{clancy:2001} \cite{moses:2006} \cite{nofsinger:2000} et dans ce travail (Figure \ref{fig:compmelaninsfm}) mène à la conclusion que la morphologie observée est une propriété typique de la mélanine.

Dans ce contexte il serait intéressant d'étudier si les grains de mélanine en solution ont aussi une forme anisotrope. Des expériences de diffusion dynamique de la lumière pour déterminer leur rayon hydrodynamique pourraient répondre à cette question. En effet Gallas et collaborateurs ont utilisé des expériences de diffusion de rayons x à petits angles et de neutrons pour conclure que la mélanine faite à partir de tyrosine forme des feuillets dans une solution légèrement basique \cite{gallas:1999}. 

\subsection{Méthodes pour déposer des films minces de mélanine}
Dans le sous-chapitre \ref{sec:resdepostitionmethods} le matériau, qui se forme à la surface d'un objet immergé dans une solution d'hydrochlorure de dopamine (2 g/L dans 50 mmol/L Tris, pH 8.5), est identifié comme de la mélanine par son spectre UV-vis monotone (Figure \ref{fig:compmelaninuvvis}) et sa composition chimique déduite de spectres de photoélectrons à rayons x (XPS, Figures \ref{fig:xps32x15min}, \ref{fig:highresxps}). En outre quatre méthodes différentes pour former des films minces de mélanine par oxydation de la dopamine sont développées et comparées. Les méthodes diffèrent principalement dans la façon par laquelle la dopamine est oxydée pour initier la formation de la mélanine. On peut utiliser par exemple de l'oxygène comme molécule oxydante \cite{lee:2007.2} \cite{postma:2009}. Dans ce cas on immerge l'objet à recouvrir dans des solutions multiples de dopamine fraichement préparées et non-aérées (méthode A) ou dans une seule solution aérée de façon permanente (méthode B). La méthode C emploie des ions cuivriques (\chem{Cu(II)}) comme molécules oxydantes. En utilisant une électrode pour oxyder de la dopamine \cite{li:2006.1} il est possible de limiter la formation de mélanine à l'interface électrode-solution (méthode D).

Toutes les méthodes présentées mènent à la formation de films homogènes de mélanine d'épaisseur contrôlée (Figure \ref{fig:compmelaningrowth}). Les films sont hydrophiles, une propriété utile pour leur utilisation possible comme substrat afin d'immobiliser des molécules biologiques. Des images de microscopie à force atomique révèlent que tous les films possèdent une morphologie de surface similaire (Figure \ref{fig:compmelaninsfm}) comme discuté auparavant. Au delà d'une certaine épaisseur, les films de mélanine deviennent imperméables aux ions de ferrocyanure (\chem{Fe(CN)_6^{4-}}, Figure \ref{fig:compmelaninvolta}). En plus il a été confirmé pour une méthode de dépôt, que la mélanine est plus perméable à des sondes neutres et cationiques qu'à une sonde anionique (Figure \ref{fig:voltaprobeairedmelanin}). Cette permsélectivité ouvre la voie vers des applications de détection électrochimique de la dopamine \cite{li:2006.1} \cite{rubianes:2001}.

Concernant leurs différences, les méthodes de dépôt de mélanine ont les avantages et les inconvénients suivants:
\begin{description}
	\item[La méthode A] est applicable pour n'importe quel substrat \cite{lee:2007.2}, facile à mettre en place et mène a une croissance linéaire et \guillemotleft ~infinie \guillemotright \ de l'épaisseur du film, mais ce processus multi-étapes est fastidieux et consomme beaucoup de dopamine.  
	\item[La méthode B] est aussi applicable pour n'importe quel substrat \cite{lee:2007.2}. Elle consiste en une seule étape pour des épaisseurs de mélanine allant jusqu'à 40 nm et peut être répétée pour obtenir des épaisseurs plus importantes, mais cette méthode est difficile à mettre en place dans des petites cellules de mesure. 
	\item[La méthode C] peut être utilisée dans un environnement acide et anaérobie, et se réalise en une seule étape, mais il faut un réactif supplémentaire, le sulfate de cuivre.
	\item[La méthode D] se réalise aussi en une seule étape, et la formation non-contrôlée de la mélanine en solution est évitée, mais cette méthode nécessite un substrat conducteur et l'épaisseur du film est limitée à 45 nm.
\end{description}

Si on remplace le tampon Tris par un tampon phosphate dans les solutions de dopamine, la croissance des films de mélanine devient plus lent mais n'est plus limitée à une épaisseur limite d'environ 40 nm (Figure \ref{fig:ellipsophosphatemelanin}). Cependant les propriétés physiques des dépôts de mélanine restent inchangées et il n'est pas possible de détecter une incorporation de Tris ou de phosphate dans les dépôts par XPS (Figure \ref{fig:xpscomposition}). 

\subsection{Dépôt de mélanine par immersion dans des solutions multiples de dopamine}
Dans le sous-chapitre \ref{sec:resnx15min} le dépôt de films de mélanine par immersions dans des solutions multiples de dopamine selon la méthode A est étudiée plus en détail. Des expériences en présence de méthanol suggèrent qu'une molécule radicalaire comme la dopamine semiquinone initie la formation de la mélanine à l'interface solution--substrat (Figure \ref{fig:ellipsomethanol}). L'atténuation du signal du support dans les spectres XPS indique que la déshydratation de films de mélanine dans ultravide induit une forte diminution de leurs épaisseurs (Figure \ref{fig:thicknessellipsoxps}). Ce tassement est en accord avec les observations d'autres groupes que des propriétés physiques de la mélanine dépendent fortement de son état d'hydratation. Par exemple Jastrzebska \cite{jastrzebska:1995} et Subianto \cite{subianto:2005} ont observés un effet de l'hydratation sur la conductivité électrique et la masse de la mélanine respectivement. Les films de mélanine préparés selon la méthode A sont stables dans une gamme de pH de 1 à 11, mais ils sont rapidement dissouts à pH 13 (Figure \ref{fig:melaninph}). Grâce à cette dernière observation il est possible de nettoyer facilement des surfaces recouvertes de mélanine dans une solution fortement basique.

Pour la première fois le potentiel $\zeta$ d'un dépôt de mélanine a été mesuré. Sa valeur est de -40 mV à pH 8.5 pour des dépôts faits par au moins trois immersions du support dans des solutions de dopamine (Figure \ref{fig:zetamelanin}). Le potentiel $\zeta$ augmente si le pH de la solution de mesure descend, et il atteint des valeurs positives à un pH inférieur à 4.5 (Figure \ref{fig:zetatitration}). La charge variable de la mélanine peut être expliquée par la protonation successive de groupements catéchols, quinone imines et quinones lors de la diminution de pH \cite{szpoganicz:2002}. Ces groupements se trouvent probablement dans un environnement caractérisé par un désordre chimique \cite{cheng:1994} \cite{dischia:2009} \cite{meredith:2006.1} \cite{tran:2006}. Dans ces conditions, chaque groupement acido-basique est caractérisé par une distribution relativement large de la valeur de sa constante de dissociation ce qui empêche l'apparition de \guillemotleft ~marches \guillemotright \ individuelles dans le graphe du potentiel $\zeta$ en fonction de pH.

La charge de surface négative de la mélanine peut expliquer en partie l'adsorption des protéines lysozyme, myoglobine et $\alpha$-lactalbumine par des interactions électrostatiques (Table \ref{tab:proteinmasses}). Cependant des expériences de désorption utilisant du dodecylsulfate de sodium ainsi que des expériences d'adsorption à taux de sel élevé (Figure \ref{fig:qcmproteins}) suggèrent qu'une deuxième interaction plus forte existe. Basé sur la quantification de sites de liaison de groupements amines sur des grains de mélanine (Sous-chapitre \ref{sec:resultsaminebinding}) et le schéma de réaction proposé par d'autres groupes \cite{lee:2009} \cite{merrit:1996}, des liaisons covalentes sont supposées se former entre des groupements amines des protéines et des groupements catéchols de la mélanine. Grâce à ce mécanisme de liaison, des films de mélanine pourraient servir comme substrat fonctionnel pour l'immobilisation contrôlée de biomolécules, par exemple d'enzymes afin de construire des capteurs.

\subsection{Mélanine dans des films de polyélectrolytes}
Le sous-chapitre \ref{sec:resdopaminepllha} décrit une nouvelle approche pour renforcer des films de polyélectrolytes en utilisant de la mélanine. Le renforcement de films biocompatibles de poly(L-lysine) (PLL) et de hyaluronate (HA) est d'un grand intérêt parce qu'il améliore fortement l'adhésion et la prolifération cellulaire  \cite{engler:2004} \cite{richert:2004}. Des expériences de spectroscopie UV--visible (Figure \ref{fig:dopamineincorporation}) et infrarouge (Figure \ref{fig:ftirpllhadopamine}) confirment que la mélanine peut se former dans des films \chemr{(PLL-HA)_n}. D'après des observations en microscopie à force atomique (Figure \ref{fig:roughness_pllha30_dopa}) et en microscopie confocale à balayage laseré (Figure \ref{fig:confocalline}), des films de polyélectrolytes restent homogènes et relativement lisses lors de l'incorporation de la mélanine. La technique du rétablissement de fluorescence après photoblanchiment a été utilisée pour montrer que l'exposition de films \chemr{(PLL-HA)_{30}} à des solutions de dopamine de concentrations croissantes (dans 50 mmol/L Tris à pH 8.5) mène à une diminution progressive de la mobilité latérale des chaines PLL dans les films.

L'illustration la plus convaincante, que la mélanine renforce des films de polyélectrolytes, est la formation de membranes autosupportées pour un temps de contact élevé \mbox{($\geq 10$ h)} entre des films \chemr{(PLL-HA)_{30}} et des solutions de dopamine (Figure \ref{fig:membranepreparation}). En plus ces membranes d'une taille latérale macroscopique et d'une épaisseur de quelques $\mu$m sont préparées sous des conditions relativement douces (dans 0.1 mol/L d'acide hydrochlorique). Ceci est un avantage important pour des applications biomédicales, par exemple l'ingénierie tissulaire. D'autres méthodes pour préparer des membranes à base de polyélectrolytes nécessitent souvent la dissolution d'un support sacrificiel dans de l'acétone \cite{mamedov:2000} ou dans de l'acide hydrofluorique \cite{mamedov:2002} \cite{podsiadlo:2007} \cite{tang:2003}. A ce jour ils n'existe que peu de publications sur des membranes autosupportées de polyélectrolytes obtenues en solution aqueuse basique \cite{lavalle:2005}, voire neutre \cite{ono:2006}.

Ce travail n'apporte pas de preuve directe d'une réticulation chimique des chaines de PLL par la mélanine. Néanmoins les indications suivantes mènent à la conclusion que des films \chemr{(PLL-HA)_{n}} sont renforcés par des liaisons covalentes entre des chaines de PLL et la mélanine: la mélanine est capable de lier de façon covalente la 2-(2-pyridine\-di\-thiol)\-éthyl\-amine par des liaisons catéchol--amine (Sous-chapitre \ref{sec:resultsaminebinding}), qui jouent probablement aussi un rôle dans l'adsorption de protéines (\cite{lee:2009} \cite{merrit:1996}, Sous-chapitre \ref{sec:resultsproteinonmelanin}). Des expériences de spectroscopie infrarouge et de microscopie confocale indiquent que l'épaisseur de films \chemr{(PLL-HA)_{n}} diminue lors de l'incorporation de la mélanine. Le nombre de chaines de PLL libres dans les films diminue, un comportement difficile à expliquer si la mélanine ne formait que des obstacles physiques à la diffusion de la PLL. Et finalement il n'est pas possible d'obtenir des membranes autosupportées d'un film de polyélectrolytes sans amines primaires (Sous-chapitre \ref{sec:respdadmapaadopamine}). Pour prouver définitivement l'existence d'une réticulation chimique il faudrait des expériences supplémentaires, par exemple de la spectroscopie de résonance magnétique nucléaire.

\newpage
\section{Conclusion}
\subsection{Résumé}
Basé sur la publication de Lee et collaborateurs \cite{lee:2007.2}, l'oxydation spontanée de la dopamine en solution légèrement basique a été étudiée, et le produit de la réaction a été identifié comme de la mélanine. La capacité de la mélanine de lier des groupements amines de façon covalente a été confirmée par la quantification des sites de liaison correspondants sur des agrégats de mélanine. En outre il est possible de rédisperser des agrégats de mélanine dans des solutions fortement basiques. Les grains de mélanine ainsi obtenus ont été utilisés pour construire des films multicouches avec le poly\-(di\-allyl\-di\-methyl\-ammonium) (PDADMA) impliquant une adsorption préférentielle de grains plus grands aux films multicouches.

Après l'étude de la formation de mélanine en solution, des méthodes différentes d'oxydation de la dopamine pour former des films de mélanine à l'interface solide--liquide ont été développées. Toutes les méthodes mènent à la formation de films continus de mélanine ayant des morphologies de surface très similaires. Par ailleurs ces morphologies, consistant en agrégats fortement anisotropes, sont aussi trouvées sur les films \chemr{(PDADMA-m\acute{e}lanine)_n} étudiés auparavant et sur des échantillons de mélanine synthétique et naturelle examinés par d'autres groupes. Par conséquent la formation d'agrégats en forme de plaquettes semble être une propriété intrinsèque de la mélanine. Les films de mélanine deviennent imperméables à des sondes électrochimiques à partir d'une épaisseur de l'ordre de 10 nm. Dans ce contexte une plus grande perméabilité à des sondes chargées positivement ou neutres qu'à des sondes négatives a été confirmée pour une méthode de préparation.

L'adsorption de protéines à des revêtements de mélanine a été expliquée par une combinaison d'interactions électrostatiques et fortes, très probablement covalentes. Pour arriver à cette explication le potentiel $\zeta$ de dépôts de mélanine a été mesuré en fonction du pH.

Finalement la formation de la mélanine par oxydation de la dopamine dans des films multicouches de poly(L-lysine) (PLL) et de hyaluronate (HA) a été étudiée. Il a été observé que la mélanine est capable de remplir des films \chemr{(PLL-HA)_n} de manière homogène en ne modifiant que légèrement la morphologie des films. Par contre la mobilité des chaines de PLL dans les films est fortement diminuée en présence de mélanine. En plus les composés polyélectrolyte--mélanine ainsi obtenus peuvent être détachés de leurs substrats comme membranes autosupportées préparées par une méthode biomimétique sous conditions relativement douces. De nombreuses indications mènent à la conclusion que le renforcement observé des films \chemr{(PLL-HA)_n} est due à une réticulation chimique des chaines de PLL par la mélanine.

\subsection{Questions ouvertes}
Les questions les plus importantes évoquées dans ce travail auxquelles il faudrait répondre par des études ultérieures sont:

Quel est le mécanisme qui fait que des films de mélanine préparés par voie électrochimique sont plus perméables aux ions de ferrocyanure que les films préparés par d'autres méthodes? Aucune différence morphologique n'a été détectée en microscopie à force atomique. La réponse pourrait être trouvée dans l'utilisation de techniques d'investigation sensibles à des échelles de longueur du nanomètre, par exemple la diffusion des rayons x aux petits angles.

Pourquoi le sulfate de cuivre est-il capable d'induire la formation de mélanine en milieu acide contrairement aux autres agents d'oxydation examinés? Et pourquoi y a-t-il des maxima intenses d'absorbance dans le spectre UV des films de mélanine contenant un taux très faible de cuivre? Probablement les maxima ne sont pas liés directement au cuivre mais à des changements de la structure de la mélanine induits par le cuivre ou le milieu acide pendant la formation de la mélanine.

Comment le choix d'une molécule de tampon particulière influence-t-il la croissance de films de mélanine malgré le fait que les molécules de tampon ne sont pas détectées dans les films? Une analyse de la structure moléculaire de la mélanine, par exemple par la diffraction de rayons x ou la spectroscopie de résonance magnétique nucléaire, pourrait donner une réponse à cette question. 

Est-il possible de \guillemotleft ~figer \guillemotright \ la formation de la mélanine pour étudier des produits intermédiaires? Un élément de réponse est peut être donné par la modification de sites de liaison de la dopamine en utilisant des sucres comme proposé dans \cite{pezzella:2009}.

\subsection{Perspectives}
Le travail commencé par cette thèse pourrait être continué par exemple par les projets suivants:

Des membranes composites polyélectrolytes--mélanine pourraient être fonctionnalisées en utilisant des molécules biologiques pour former des feuilles de culture cellulaire ou des pansements bioactifs. 

Puisque les films de mélanine étudiés sont facilement déposés sur une grande variété de supports et possèdent des sites chimiques actifs, ils pourraient être utilisés comme substrat polyvalent pour des fonctionnalisations secondaires. Le potentiel de réduction de la mélanine peut être utilisé par exemple pour synthétiser des particules d'argent afin de créer des revêtements bactéricides.

Les couleurs iridescentes des plumes du paon sont dues à la diffraction de nanostructures périodiques riches en mélanine \cite{kinoshita:2005} \cite{yoshioka:2002} \cite{zi:2003}. Des cristaux photoniques fait par un arrangement périodique de grains de mélanine et d'un matériau d'indice de réfraction différent, par exemple des billes de silice dans un cristal colloïdal \cite{norris:2004}, pourraient mimer le comportement diffractif des plumes du paon. En plus de leur beauté, les cristaux obtenus pourraient trouver des applications dans la transmission ou le traitement optique de données.

L'incorporation de la mélanine dans un matériau de fort coefficient d'expansion thermique mènerait à un composite possédant un indice de réfraction dépendant de la température. Ceci pourrait être utile pour construire de thermomètres ou bolomètres grâce à l'absorption à large bande de la mélanine.
\end{otherlanguage*}{french}

%% file: introduction.tex
\chapter{Introduction}
The present thesis is situated in the field of biomaterial science. In this field, the chemical and physical properties of materials of biological origin like mussel feet, nacre or bone are investigated to understand the relationship between their structure and function. The obtained knowledge can be used to create bioinspired or biomimetic materials in all fields of engineering science and especially in biomedical engineering to design for example artificial soft tissue, implants or drug delivery systems. Due to the immense variety of materials found in nature with properties often superior to man-made materials, biomaterial science has an important influence on technological development.

The laboratory ``Biomatériaux et Ingénierie Tissulaire'' of Inserm (Institut National de la Sant\'e et de le Recherche M\'edicale), where this thesis was prepared, is focused on surface modifications for biomedical applications using most often the technique of layer-by-layer (LbL) deposition of polyelectrolytes. Multiple studies have been carried out to characterise the growth of LbL films, to functionalise them by incorporation of biologically active molecules and to control their physical properties. %In particular it was shown that cell adhesion on LbL films critically depends on the films' mechanical stiffness.
In spite of their versatility, LbL coatings have not found any industrial applications, because as a multi-step procedure their application is very slow and needs sophisticated automation equipment. Hence there is a need to find simple one-step procedures to obtain functional coatings of controlled thickness.

One possible method is the deposition of a polymer made by spontaneous oxidation of dopamine as first described by Lee and others \cite{lee:2007.2}. A further examination of this method in the present thesis will reveal that the deposits are made of melanin, a material with several interesting properties: 
\begin{itemize}
\item Monotonous absorption from the ultraviolet to the infrared and efficient conversion of electromagnetic radiation into heat make melanin a candidate for photodetection, photoprotection or photothermal applications.
\item The ability to capture and reduce metal cations can be used to build metal particles, for example for chemical catalysis.
\item Melanin plays a role in innate immunity and in neurodegenerative diseases.
\item Despite huge efforts, the macromolecular structure of melanin remains unclear.
\item Melanin films can easily be formed on virtually any kind of substrate by a biocompatible method.
\item These films can be used as a versatile platform for further functionalisation or to build melanin capsules.
\end{itemize}

Different protocols will be established to build melanin thin films by dopamine oxidation, and the properties of the obtained films will be examined. Furthermore melanin will also be formed in poly(L-lysine)-hyaluronate LbL films leading to an important enhancement of the films' strength by a biocompatible method.

%% file: literature.tex
\chapter{Literature overview}
\section{Catecholamine-containing coatings}
\label{sec:dopamineapplications}
Catecholamines like adrenaline, noradrenaline and dopamine (2-(3,4-di\-hydroxy\-phe\-nyl)\-ethyl\-amine) play an important role as neurotransmitters (\cite{patrick:2003}, chapter 16) and their electrochemical reaction pathways are studied for a long time \cite{hawley:1967}. 3,4-dihydroxy-L-phenylalanine (DOPA), an hydroxylated form of the amino acid L-phenyl\-alanine, is a precursor molecule in natural dopamine synthesis (\cite{patrick:2003}, chapter 16.4). It was found more recently to be also an important constituent of the proteins found near the plaque-substrate interface of mussels \cite{waite:2001}. The role of DOPA in the strong adhesion of mussels to virtually any kind of substrate was confirmed by scanning force microscopy measurements showing strong and reversible adhesion between DOPA molecules and titanium as well as silanised silicon surfaces \cite{lee:2006}.

The adhesion properties of DOPA were used by Statz and others \cite{statz:2005} to create antifouling coatings using a peptide containing DOPA, L-lysine and N-methoxyethyl glycine. The modified surfaces resisted to protein adsorption from serum and to adhesion of 3T3 fibroblasts.

Podsiadlo and others used the layer-by-layer (LbL) deposition method to produce a high-strength nanocomposite of clay platelets and a DOPA-polymer \cite{podsiadlo:2007}. In this material the DOPA-polymer serves as glue to connect the hard but brittle clay platelets. The authors showed that the mechanical properties of the composite material were comparable to natural nacre and lamellar bones.

%As another application of the strong interaction of dopamine with metal oxide surfaces it has been demonstrated that the adsorbed catechol can be used to initiate atom transfer radical polymerisation of poly(methyl methacrylate) (PMMA) from the surface of titanium dioxide nanoparticles \cite{fan:2006}.

An ``Electrochemical quartz crystal microbalance study on growth and property of the polymer deposit at gold electrodes during oxidation of dopamine in aqueous solutions'' was published by Li and others \cite{li:2006.1}. They proposed a mechanism of sequential dop\-amine oxidations leading to the growth of a melanin-like polymer on the working electrode during cyclic voltammetry. The polymer was selectively permeable to dopamine compared to ascorbic acid. Since dopamine and ascorbic acid have similar redox properties, a permselective coating discriminating the two molecules might be useful for electrochemical dopamine sensing. The concept of modifying electrodes for dopamine-sensing with melanin-like films had been introduced previously by Rubianes and Rivas \cite{rubianes:2001}. Li and others succeeded furthermore in incorporating active anti-(human immunoglobulin G) (IgG) during electrochemical dopamine polymer deposition \cite{he:2005}. The films showed higher reactivity against human IgG than anti-(human IgG) in poly(pyrrole) films of comparable thickness. 

Lee and others combined the adhesion strategies of geckos and mussels to create a reversible wet/dry adhesive \cite{lee:2007.1}. They prepared a nanostructured poly(methyl methacrylate) (PMMA) pad mimicking the structure of a gecko's foot and coated it with a polymer of high DOPA content inspired by mussel foot proteins. Scanning force microscopy showed strong and reversible adhesion forces between the silicon nitride cantilever of the microscope and the adhesive pad in air as well as in water. The DOPA-polymer coating enhanced the adhesion of the nanostructured PMMA in dry and even more in wet conditions.

The same authors modified poly(ethyleneimine) (PEI) and hyaluronate (HA) with DOPA to obtain stronger adhesion in layer-by-layer assembly \cite{lee:2008}. They showed that the modified polyelectrolytes allowed for deposition of multilayers on polymeric surfaces that are difficult to functionalise with unmodified polyelectrolytes. Furthermore the redox activity of the catechol groups in the LbL film was used for in-situ reduction of silver nitrate to silver nanoparticles to give the film bactericidal properties.

\subsection{Coatings made by spontaneous dopamine oxidation}
\label{sec:literaturelee}
The present research project was initiated by an article by Lee and others, reporting a simple one-step coating method for various metal, metal oxide and polymer surfaces \cite{lee:2007.2}. The method consists in putting the material of interest into a solution of 2 g/L of dopamine in tris\-(hydroxy\-methyl)\-amino\-methane (Tris) buffer (10 mmol/L, pH 8.5). Under these conditions, dopamine is oxidised to form synthetic melanin (Section \ref{sec:melanin}) in solution and on the surface of the immersed material yielding a film of up to 50 nm in thickness within 25 h. In the following, melanin coatings made this way by spontaneous dopamine oxidation will be called \emph{dopamine-melanin} coatings. Lee and others  used these coatings to induce different secondary reactions: The metal binding ability of dopamine was used to deposit adherent and uniform metal coatings. A monolayer of alkanethiol was spontaneously formed by simple immersion. Grafting methoxy-poly(ethylene oxide) compounds led to strongly reduced protein adhesion and fibroblast attachment, whereas the melanin coating itself did not influence cell attachment. Specific biomolecular interactions were induced by grafting hyaluronic acid (HA) to melanin surfaces. The bioactivity of HA was shown by the specific adhesion of human megakaryocytic M07e cells. Later, the same group showed the possibility to conjugate a biomolecule with histidine and lysine functionalities on dopamine-melanin coatings with a pH-controlled orientation \cite{lee:2009}.

The group of Zhang used similar dopamine-melanin films with incorporated semiconductor nanocrystals to create encoded polymer films for electrochemical identification \cite{zhang:2009}. By combining polymer spots containing varying concentrations of different nanocrystals, multiple voltammetric signal sequences could be generated. Since dopamine-melanin firmly adheres to various substrates, it might be used to create identification tags on various products for example for counterfeit protection. Nevertheless a simpler, non-destructive readout method remains to be found for possible industrial applications.

Postma and others \cite{postma:2009} used dopamine oxidation to build hollow dopamine-melanin capsules. Therefore dopamine-melanin was deposited on silica beads that were subsequently dissolved in hydrofluoric acid. The wall thickness of the obtained capsules could be controlled by the reaction time in dopamine solutions. Since the authors showed that the capsules had no toxic effect on LIM 1215 colon cancer cells, they predict a possible use as drug or gene carriers. Yu and colleagues further examined the build-up of dopamine-melanin capsules with varying diameters and observed surprising unidirectional permeability to the fluorophore rhodamine 3G \cite{yu:2009}. If unidirectional permeability is confirmed for other molecules, it would be an important property for possible drug-delivery applications.

Xi and others \cite{xi:2009} used hydrophobic polymer membranes instead of rigid supports for the deposition of dopamine- and DOPA-based coatings by spontaneous oxidation. The coatings rendered the membranes more hydrophilic and increased the water flux through the membranes. Furthermore the optimum pH for the dopamine- or DOPA-melanin deposition was found to be 8.5, and the coatings were stable in a hot water bath for up to 36 days. The properties of the coated membranes make them candidates for separation membranes in fuel cells.

In the same research area, Wang and others proposed to coat Nafion membranes with dopamine-melanin for their use as proton conductors in direct methanol fuel cells (DMFC) \cite{wang:2009}. They showed that the coating reduced the diffusion of methanol through the membranes, while the proton conductivity remained nearly constant. Therefore the dopamine-melanin coating might enhance the performance of Nafion membranes in DMFC.

Films obtained by spontaneous oxidation of dopamine can also be deposited on carbon nanotubes (CNT) \cite{fei:2008}. In this case, most of the dopamine-melanin formed on the nanotubes and not in solution. The authors explained this by a strong affinity between the carbon nanotube sidewalls and aromatic rings contained in dopamine-melanin. The dopamine-melanin coating might be used as a versatile platform for surface modifications on CNT, as exemplified by the reduction of gold(III) chloride (\chem{HAuCl_4}) to gold nanoparticles on melanin-coated CNT.

Another application of dopamine-melanin deposits are molecularly imprinted surfaces. Therefore dopamine-melanin is deposited on a surface in presence of a template molecule, for example a protein. Upon removal of the template, specific binding sites remain at the modified surface, that can be used for sensing or purification applications. This concept was described for example in \cite{zhou:2009} to prepare imprinted dopamine-melanin coatings on iron(II,III) oxide (\chem{Fe_3O_4}) nanoparticles for specific protein binding. After binding to the nanoparticles, the proteins could be magnetically separated. %due to the particles' superparamagnetism. 

\section{Melanin}
\label{sec:melanin}
Melanin is an important biomolecule found abundantly in nature. Its most well-known role in the skin is protection against ultraviolet (UV) radiation. Furthermore, melanin forms an important part of the innate immune system \cite{mackintosh:2001}. This is due to the antimicrobial activity of intermediate molecules in melanin synthesis and to melanin's ability to bind micro-organisms and toxins \cite{boman:1987}. In humans there are two types of melanin, the brown-black eumelanin and the yellow-reddish pheomelanin, both of them present in varying quantities in skin, hair and eyes. Neuromelanin found in the substantia nigra of the brain has been proposed to be a copolymer of eumelanin and pheomelanin \cite{odh:1994}.

\begin{figure}
	\centering
		\includegraphics[width=\textwidth]{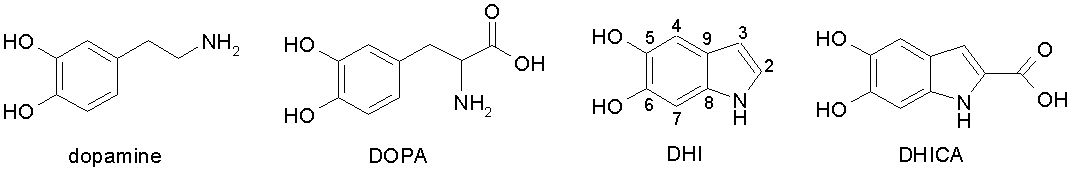}
	\caption{Molecular structures of dopamine, 3,4-di\-hydroxy-L-phenyl\-alanine (DOPA), 5,6-di\-hydroxy\-indole (DHI) and 5,6-di\-hydroxy\-indole carboxylic acid (DHICA)}
	\label{fig:dopamine}
\end{figure}

In vivo, eumelanins are produced from tyrosine that is first enzymatically converted to 3,4-dihydroxy-L-phenylalanine (DOPA) and then in several steps to 5,6-dihydroxyindole carboxylic acid (DHICA) (Figure \ref{fig:dopamine}), which finally polymerises to form melanin in a process that is not well understood yet \cite{clark:1990} \cite{ito:2003}. In the synthesis of pheomelanin, sulfur-containing cysteine binds to DOPA before further oxidation and polymerisation steps \cite{ito:2003}. Synthetic melanins are commonly produced by spontaneous oxidation of dopamine \cite{herlinger:1995} \cite{peter:1989}, DOPA \cite{peter:1989} or 5,6-dihydroxyindole (DHI) \cite{herve:1994} or using the enzyme tyrosinase to obtain melanins from the same precursors \cite{clark:1990}.

It is generally accepted that eumelanins are macromolecules of DHI and DHICA in proportions depending on the origin and preparation method of the eumelanin \cite{ito:1986}. Furthermore, it has been shown that eumelanin contains reduced (catecholic) as well as oxidised (quinonoid) substructures \cite{pezzella:2009}.

On the contrary, there is no consensus yet on the macromolecular structure, which is explained by two different models. The first one describes melanin as an \emph{extended heteropolymer} consisting of DHI and DHICA units \cite{clark:1990} to explain its broadband absorption spectrum. Nuclear magnetic resonance experiments showed that the units are bound randomly at their 2, 3, 4 or 7 positions \cite{peter:1989} \cite{pezzella:2007} \cite{reinheimer:1999} (see figure \ref{fig:dopamine} for the numbering). The other model describes melanin as non-covalent aggregates of \emph{fundamental units}. The fundamental unit is a stack of three to five oligomeric sheets of four to eight DHI or DHICA units, which are stacked by $\pi -\pi$ interactions like graphene sheets with an inter-sheet-distance of 0.35 nm. This model was deduced from x-ray scattering experiments \cite{cheng:1994} and supported by scanning tunnelling microscopy \cite{zajac:1994}, matrix-assisted laser desorption ionisation mass spectrometry \cite{pezzella:1997}, scanning electron microscopy \cite{nofsinger:2000} and scanning force microscopy \cite{clancy:2001}. Density functional theory (DFT) calculations were used on oligomers of DHI \cite{stark:2003} and on stacks of up to three of these oligomers in different oxidation states \cite{stark:2005} to reproduce the UV--visible absorption spectrum of eumelanin. Another group \cite{kaxiras:2006} used DFT on a tetramer of DHI molecules in different oxidation states to reproduce the UV--visible absorption spectrum of eumelanin and its x-ray scattering data as well as its metal-binding capability. Despite numerous publications supporting the stacked sheet model, there is no \emph{direct proof} of the existence of fundamental melanin units yet. Both models can explain the macroscopic properties of melanin. Recently Watt and others presented first direct evidence of an onion-like stacked sheet structure in synthetic and natural eumelanins by high resolution transmission electron microscopy \cite{watt:2009}.

In contrast to most organic chromophores, melanin has a monotonic absorption spectrum over the whole visible and UV region, which can be described as a function of wavelength by a simple exponential function \cite{meredith:2006.1}. Most of the absorbed energy is dissipated as heat within one nanosecond after excitation in the UVA and visible region \cite{forest:1998}. This efficient energy conversion by melanin is important for photoprotection. Scattering, which could explain the monotonic spectrum, contributes only to a small extent ($<6$~\%) to the attenuation of light by melanin solutions \cite{riesz:2006}. The absorption spectrum was explained by an organic semiconductor model with a fundamental band gap of 1.7~e\/V in the near infrared region \cite{albuquerque:2006}.

The electron delocalisation induced by the indolic structure of melanin suggests that it may be electrically conducting. Consequently Jastrzebska and others investigated the electrical conductivity of synthetic DOPA-melanin \cite{jastrzebska:1995}. They found that it depends strongly on the hydration of the sample with values reaching from $10^{-11}$ S/m in vacuum to $10^{-3}$ S/m at $100 \%$ relative humidity. Furthermore the conductivity increases with increasing temperature as in a semiconductor \cite{abbas:2009} \cite{jastrzebska:1995}. In \cite{goncalves:2006}, the authors additionally described an influence of the protonation of melanin on its conductivity. Actually the question whether electrons or protons are responsible for the electrical conductivity of melanin has not been answered yet \cite{meredith:2006.2}. In addition to dark conductivity melanin films also show photoconductivity \cite{ligonzo:2009} \cite{subianto:2005}.

An important feature of melanin is its ability to bind many metal ions. In vivo it can act as a reservoir by reversibly binding calcium(II) and zinc(II) ions, or as a sink for potentially dangerous reactive metal species like iron(II) or copper(I) by irreversibly binding them. Reference \cite{hong:2007} gives an overview over the current understanding of the binding sites, capacity, affinity and biological significance of metals in melanin. An alteration of the iron-binding capability of neuromelanin is possibly involved in the development of Parkinson's disease \cite{kienzl:1999} \cite{linert:2000}.

In some microorganisms melanin plays an important role in radioprotection. Dadachova and others examined the protection of fungi against $\gamma$-radiation by melanin \cite{dadachova:2007}. In the examined fungi, melanin forms multilayered shells of approximately 100 nm in thickness and of the same shape as the fungal cells. The protective effect of melanin is explained as a combination of efficient Compton scattering and quenching of free electrons and radicals. The high efficiency of the Compton scattering is due to the high number of $\pi$-electrons and the form of the melanin shells. In the authors' opinion melanin is an interesting candidate to create new lightweight radioprotective materials.

The ability of melanin to chemisorb radioactive metals might be used in radioprotective applications too. Howell and others showed that different synthetic melanins are able to chemisorb radioactive isotopes of actinium (\chem{^{225}Ac}), bismuth (\chem{^{213}Bi}) and indium (\chem{^{111}In}) \cite{howell:2008}. Thus melanin might be used to eliminate these elements from a contaminated body. 

\section{Layer-by-layer films of polyelectrolytes}
\label{sec:literaturelbl}
Layer-by-layer (LbL) assembly of polyelectrolyte films is a method to build tailored molecular assemblies on a vast variety of supports, which was made popular in the early 1990s by the group of Decher \cite{decher:1997}. Electrostatic attraction of oppositely charged polyelectrolytes is the driving force for the build-up. Consequently the support material has to posses an electric charge to allow LbL deposition. This is true for most metals, silicones and glasses carrying a net negative charge in solution due to surface oxidation and hydrolysis \cite{tang:2006.1}. %Instead of pairs of polycations and polyanions one can also use hydrogen-bond donor and acceptor pairs to build multilayers \cite{lutkenhaus:2005}, \cite{ono:2006}.

To build a film, a charged support is alternately brought in contact with aqueous solutions of a polycation and a polyanion. Since each molecule carries more than one charge, there can be charge overcompensation at each deposition step. This leads to reversal of the surface charge limiting on the one hand the adsorption of a given polyelectrolyte and allowing on the other hand the adsorption of the next polyelectrolyte of opposite charge. Hence one can obtain films of quasi unlimited thickness by repeating the adsorption cycle. After n deposition cycles the film is denoted \chemr{(polycation-polyanion)_n}. Charge overcompensation was shown for example in \cite{ladam:2000} by reversal of the $\zeta$-potential of the multilayer after each deposition step in a multilayer of poly\-(styrene\-sulfonate) (PSS) and poly\-(allyl\-amine hydro\-chloride) (PAH). Usually one or more washing steps are effectuated after each adsorption step to remove loosely bound polyelectrolytes and to avoid contamination of the following polyelectrolyte solution. 

The support can be brought in contact with polyelectrolyte solutions either by dipping it into the solutions, by spraying the solutions on the surface or by spin coating. Due to its simplicity, the dipping method is most abundantly used, although it is very slow (typically 20 h are needed to perform 30 deposition cycles). Izquierdo and others \cite{izquierdo:2005} showed, that the build-up can be significantly faster by spraying compared to dipping while both methods allow the use of a large variety of support sizes and topologies. The drawback of the spraying method is its very high consumption of polyelectrolyte solutions. Spin coating \cite{seo:2008} is also a rapid method and offers a low solution consumption but it can only be used on flat substrates.

%Polyelectrolytes are used instead of small molecules because good adhesion of a layer on a support requires a large number of ionic bonds. Furthermore polymers can bridge over defects of the underlying support so that film properties such as surface roughness mainly depend on the employed polyelectrolytes and not on the support \cite{decher:1997}. Experiments of x-ray and neutron reflectometry carried out on films of PSS and PAH have confirmed the layered structure of these multilayers showing additionally a large overlap of adjacent layers \cite{decher:1997}.

There are basically two regimes of layer-by-layer film growth. The thickness of the film growths either linearly or exponentially with the number of deposited polyelectrolyte layers. In the case of linear growth, a polyelectrolyte interacts only with the last deposited polyelectrolyte layer during its deposition, and in every deposition step the same amount of polyelectrolyte is adsorbed. One of the most studied systems of this kind are \chemr{(PSS-PAH)_n} multilayers \cite{voegel:2003}. For exponential growth, at least one of the constituents has to be able to freely diffuse in and out of the film during its build-up. Picart and others proposed the following mechanism for the case that only the polycation can freely diffuse \cite{picart:2001}:
\begin{enumerate}
\item When the multilayer is brought in contact with the solution of polycations after the deposition of a polyanion layer, the polycations form a layer at the surface creating a positive charge excess and diffuse at the same time into the film forming a reservoir of free polycations.
\item In the following washing step only a part of the free polycations leaves the film because of the barrier of positive charges at the surface.
\item When polyanions adsorb at the surface in the next deposition step, the barrier of positive charges disappears. Thus, the free polycations can diffuse to the surface of the film and form complexes with further polyanions. These complexes are an integral part of the polyelectrolyte film.
\end{enumerate}
Since the thickness of the new layer is proportional to the number of free polycations, and this number is in a first approximation proportional to the thickness of the film, the mechanism leads to an exponential growth. For films of poly(L-lysine) (PLL) and hyaluronate (HA), an extensively studied exponentially growing film, it was shown by confocal laser scanning microscopy that PLL diffuses in and out of the film during its build-up \cite{picart:2002}. Furthermore PLL is able to diffuse in the film plane \cite{richert:2004}. Due to the diffusion of their components, exponentially growing LbL films do not possess a layered structure. More recently it was observed that initially exponentially growing films can change to linear growth after a certain number of deposition steps. Porcel and others \cite{porcel:2006} showed that this transition takes place after about twelve deposition cycles for the \chemr{(PLL-HA)_n} films. They explained the transition by a progressive restructuring of the interior of the film that hinders the diffusion of PLL over part of the film.

\subsection{Free-standing membranes}
The mechanical properties of LbL films are often not suitable for the envisioned applications. Native \chemr{(PLL-HA)_n} films for example are very soft and behave like a viscous liquid \cite{francius:2006}. If these films shall be used in tissue engineering, they have to be stiffened, because cells usually prefer hard substrates for adhesion \cite{discher:2005} \cite{pelham:1997}. This can be done by chemical crosslinking \cite{francius:2006} or by capping with a stiff \chemr{(PSS-PAH)_m} film \cite{francius:2007} leading to a better cell adhesion \cite{engler:2004} \cite{richert:2004}.

At a sufficient degree of stiffening, layer-by-layer assemblies can be separated from their support to obtain free-standing membranes. These membranes might find applications for example as separation membranes, sensors or micromechanical devices or in tissue engineering \cite{ono:2006}. Over the last decade, different approaches have been established to obtain free-standing LbL assemblies.

Mamedov and Kotov \cite{mamedov:2000} deposited a film of magnetite nanoparticles and poly\-(di\-allyl\-di\-methyl\-ammonium) (PDADMA) on cellulose acetate supports, which were subsequently dissolved in acetone to liberate a magnetic polyelectrolyte-nanoparticle membrane. They observed furthermore that the addition of clay platelets greatly enhanced the mechanical stability of the membrane. The same authors also prepared multilayer composites from single-wall carbon nanotubes (SWNT) with poly\-(acrylic acid) (PAA) and PEI \cite{mamedov:2002} with thermal and chemical crosslinking of the components. These multilayers were separated from their silicon supports by immersion in hydrofluoric acid (\chem{HF}). They showed high ultimate tensile stress approaching the values of hard ceramics. The authors concluded that the SWNT as well as the clay platelets in the previous work acted as ``molecular armour'' reinforcing the polyelectrolyte films.

Two further articles about nacre-like materials obtained by separation from their supports with \chem{HF} should be mentioned. Nacre is a very strong natural material characterized be a layered structure of organic and inorganic components. This structure is mimicked by layer-by-layer assembly of polyelectrolytes and clay particles. In \cite{tang:2003}, multilayers of PDADMA and clay platelets were examined. The clay platelets formed parallel layers in the composites, and the ultimate tensile stress as well as the modulus of elasticity were close to the ones of natural nacre. In \cite{podsiadlo:2007}, the concept was further improved by replacing PDADMA with a poly(ethyleneoxide) (PEO) derivative modified with lysine and 3,4-di\-hydroxy-L-phenyl\-alanine (DOPA). This modification was inspired by adhesive proteins found in mussel feet (Section \ref{sec:dopamineapplications}) and served to increase the adherence between the organic and the inorganic components of the membrane. Indeed when crosslinking the DOPA groups with ferric ions (\chem{Fe^{3+}}), the ultimate tensile strength of the membranes could be increased by a factor of two compared to \cite{tang:2003}. Despite the impressive mechanical properties of the aforementioned composites, their preparation by immersion in \chem{HF} or acetone is not suitable for biomedical applications and implies ecological problems. Thus additional methods for membrane preparation had to be found.

Consequently Lavalle and others presented a way to separate a chemically crosslinked film of PLL and HA from a glass substrate by immersion in an aqueous solution at pH 13 \cite{lavalle:2005}. The authors successfully immobilised the enzyme alkaline phosphatase in an active state on the membranes and envisioned applications of the membranes as bioactive patches carrying specific drugs. However residues of the chemical crosslinking agents might corrupt the biocompatibility of \chemr{(PLL-HA)_n} membranes.  

A non-aqueous method to obtain free-standing polyelectrolyte membranes was proposed by Lutkenhaus and others \cite{lutkenhaus:2005}. They prepared hydrogen bonded multilayers of PEO and PAA on supports of poly(tetrafluorethylene). A film of 100 bilayers could be peeled from the support with tweezers without destroying it. This method should exclude possible alterations of the film during separation from its support that might occur using other separation methods.

Ono and Decher described a possibility to obtain ultrathin self-standing multilayer membranes at physiological conditions \cite{ono:2006}. First they prepared a two compartment film consisting of an electrostatically bonded \chemr{(PAH-PSS)_n} part on top of a hydrogen-bonded \chemr{(PAA-PEO)_m} part in acidic medium (pH = 2). When the pH was increased to 7, the sacrificial hydrogen-bonded part dissolved to liberate the electrostatically bonded part. Despite its thickness of less than 200 nm, the obtained membrane was strong enough to be handled with tweezers.

%% file: theory.tex
\chapter{Major characterisation techniques}
\section{Scanning force microscopy}
Scanning force microscopy (SFM) is a scanning probe microscopy method. It allows imaging of the topography of non conducting samples in vacuum, in air and in liquids with a typical lateral resolution of some nanometres and a height resolution of less than 0.1 nm \cite{binnig:1986}. Since it is possible to achieve a resolution of atomic scale, this method is also called atomic force microscopy (AFM).

\begin{figure}
	\centering
		\includegraphics[width=100mm]{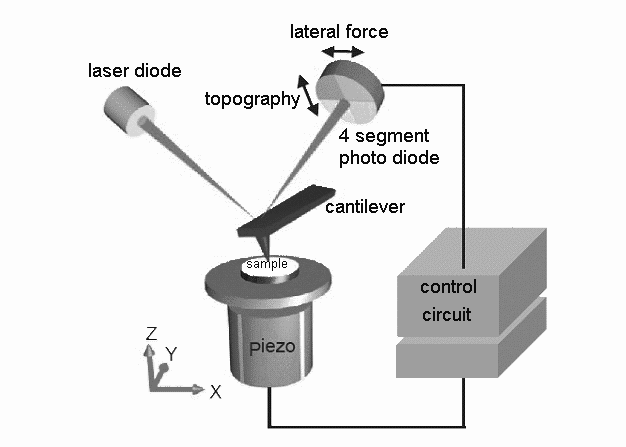}
	\caption[Layout of a scanning force microscope]{Layout of a scanning force microscope (modified from \cite{schmitt:2006})}
	\label{fig:AFMSchema}
\end{figure}
Figure \ref{fig:AFMSchema} shows the layout of a scanning force microscope. The very fine tip of a cantilever serves as probe to explore the surface of a sample placed on a stage, which can be moved in three dimensions by piezoelectric actuators. A laser spot is reflected from the cantilever onto the centre of a four segment photo diode. By measuring the light intensities on the four segments and calculating the differences between the intensity arriving on the upper and lower half and between the right and left half of the photo diode, movements of the cantilever can be detected. If the cantilever is bent, the first difference is non-zero; if it is twisted, the second difference is non-zero.

In \emph{contact mode} the sample is approached to the cantilever until its tip feels the repulsive force caused by the sample surface. This force causes an upwards bending of the cantilever and can thus be detected via the photo diode. There are two possibilities to scan the sample surface:

In \emph{constant height mode} the z position of the sample is held constant while scanning in x and y directions. Here x, y and z are the axes of a Cartesian coordinate system with the z axis perpendicular to the sample surface and the x and y axes in the sample surface plane (see figure \ref{fig:AFMSchema}). Any height differences of the sample surface cause a change in the force acting on the cantilever tip which is recorded to calculate an image of the sample surface. Due to the limited flexibility of the cantilever, this mode can only be used on very flat surfaces.

In \emph{constant force mode} the z position of the sample is modified by a control circuit to maintain the force acting on the cantilever tip constant while scanning. This way the distance between cantilever tip and sample surface remains constant. Hence the variations of the position of the z piezoelectric drive correspond directly to the height variations on the sample surface and can be used to create an image. This mode can be used on flat as well as on rough surfaces, but it is slower than the constant height mode. 

Another possibility for imaging the sample surface is the \emph{dynamic mode} (also called tapping mode or AC mode). In this case the cantilever is set to oscillation by a piezoelectric element at a frequency near its resonance frequency. The amplitude of the cantilever oscillation is detected via the photo diode. When the cantilever approaches the sample surface and forces act, the amplitude changes. Similar to the constant force mode, the z position of the sample is modified while scanning in the x and y directions to maintain the amplitude change constant. Hence  the position variations of the z piezoelectric drive can be used to create an image of the surface topography. Since in dynamic mode the force acting on the sample surface is smaller than in contact mode, the former is especially useful for imaging soft and delicate samples. The advantages of the contact mode lie in its higher lateral resolution \cite{schmitt:2006} and easier operation.

\section{Quartz crystal microbalance}
\label{sec:theoryqcm}
In a quartz crystal microbalance (QCM) the piezoelectric effect is used to monitor the influence of an adsorbed mass on the oscillations of a quartz crystal.

\begin{figure}
	\centering
		\includegraphics[width=12cm]{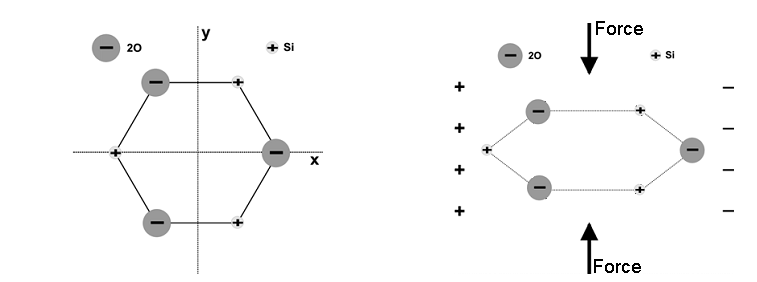}
	\caption[Illustration of the piezoelectric effect]{Simplified crystal structure of quartz (left) and illustration of the piezoelectric effect (right) (modified from \cite{mondon:2002}). Silicon atoms are represented by positive charges, pairs of oxygen atoms by negative charges.}
	\label{fig:piezoeffect}
\end{figure}

A mechanical deformation creates electric charges at the surface of a piezoelectric material (direct piezoelectric effect). Inversely, the application of electric charges at the surface leads to a mechanical deformation (inverse piezoelectric effect). To show piezoelectricity, a material has to possess a polar axis with a dipole moment that is compensated in the rest state. In the case of the simplified crystal structure of quartz illustrated in figure \ref{fig:piezoeffect}, the x-axis is the polar axis. When a force is applied along the y-axis, electrical polarisation occurs along the x-axis leading to the appearance of charges at the corresponding surfaces of the crystal. This transverse piezoelectric effect is reversible: Application of an electric field along the x-axis leads to mechanical deformation of the crystal.

To excite the quartz crystal in a QCM and to monitor its oscillations, gold electrodes are deposited on the top and bottom surface of a cylindrical quartz slide. At resonance the quartz slide oscillates in a transversal standing shear wave with maximum amplitude at its top and bottom surface and zero amplitude in the middle of the crystal. In this case the thickness of the crystal $d$, the wavelength $\lambda$, the frequency $f_\nu$ and the propagation speed $v$ of the shear wave and the overtone number $\nu$ are related by
\begin{eqnarray}
d&=&\frac{\nu \lambda}{2}=\frac{\nu v}{2f_\nu} \\
\Rightarrow f_\nu&=&\nu \frac{v}{2d} = \nu f_1. \label{eqn:frequency}
\end{eqnarray}
$f_1$ is the ground frequency of the crystal. If an adsorbed layer is thin compared to the crystal, rigid, tightly bound to the crystal surface and completely covering it, its adsorption can be approximated as an increase in the thickness of the crystal \cite{sauerbrey:1959}. In a linear approximation this leads to the following change in frequency:
\begin{equation}
\frac{\Delta f_\nu}{f_\nu}=-\frac{\Delta d}{d} \label{eqn:frequencychange}
\end{equation}
The increase in thickness $\Delta d$ corresponds to an added mass $\Delta m_Q=\rho A \Delta d$ of quartz. $A$ and $\rho$ are the surface area and mass density of the crystal. For a thin deposit the added mass of quartz is taken equal to the mass $\Delta m$ of any adsorbed material. Inserting this and equation \ref{eqn:frequency} into equation \ref{eqn:frequencychange} leads to:
\begin{eqnarray}
\Delta f_\nu &=& -\frac{\Delta m_Q f_\nu}{\rho A d} = -\frac{2 \Delta m f_\nu^2}{\rho A \nu v} = -\frac{2 \Delta m f_1^2 \nu}{\rho A v} \\
\Rightarrow \frac{\Delta m}{A} &=& -\frac{\rho v}{2 f_1^2}\frac{\Delta f_\nu}{\nu}=-C\frac{\Delta f_\nu}{\nu}. \label{eqn:sauerbrey}
\end{eqnarray}
This is the Sauerbrey equation that relates the normalized frequency change $\Delta f_\nu /\nu$ to the adsorbed mass per unit area $\Delta m /A$ via the constant $C$ describing material properties of the quartz crystal. The Sauerbrey equation leads to a good approximation of the adsorbed mass, if the reduced frequency changes are equal for different overtones \cite{sauerbrey:1959}. In the general case, if the adsorbed deposit does not fulfil the assumptions made above, a more complex model taking into account the viscoelastic properties of the deposit \cite{voinova:1999} has to be used to calculate the adsorbed mass.

\section{Ellipsometry}
Ellipsometry is a method of surface characterisation first described by Paul Drude in 1887 \cite{drude:1887}. A light wave of known polarisation is reflected at a surface and its polarisation after reflection is measured to deduce interaction parameters that cause a difference between the initial and reflected polarisation state. These parameters are usually the complex refractive index $N$ of the reflecting material or the thickness $d_f$ and the refractive index of a thin surface layer. A sample is typically probed with a laser beam of some millimetres in diameter assuming that the surface is homogeneous over this area, but there are also imaging ellipsometers with spatial resolution in the micrometer range. 

\subsection{Theory of ellipsometry}
A polarised quasi-monochromatic wave of circular frequency $\omega$ propagating along the z-axis with the speed $c$ in an isotropic,homogeneous medium of complex refractive index $N$ can be characterised by its electric field vector $\vec E$:
\begin{equation}
\vec E = (E_x \vec e_x + E_y \vec e_y) \exp \left[i \omega \left( \frac{Nz}{c}-t \right) \right].
\end{equation}
In this equation $\vec e_x$ and $\vec e_y$ denote the Cartesian basis vectors in x- and y-direction, $i$ the imaginary unit, $t$ the time, and $c$ the vacuum speed of light. $E_x$ and $E_y$ are complex constants whose ratio defines the polarisation state. For linear polarisation $E_x/E_y$ is a real constant, for circular polarisation $E_x/E_y = \pm i$. The real and imaginary part of the complex index of refraction are 
\begin{equation}
N = n + i k.
\end{equation}
Instead of the index of refraction the dielectric function
\begin{equation}
\epsilon=\epsilon_1 + i \epsilon_2 = N^2
\end{equation}
can also be used to describe the material properties. The components of dielectric function and refractive index are linked by
\begin{eqnarray}
\epsilon_1 &=& n^2 - k^2 \\
\epsilon_2 &=& 2 n k \\
n &=& \sqrt{\frac{\epsilon_1 + \sqrt{\epsilon_1^2+\epsilon_2^2}}{2}} \\ 
k &=& \sqrt{\frac{- \epsilon_1 + \sqrt{\epsilon_1^2+\epsilon_2^2}}{2}}.
\end{eqnarray}

\begin{figure}
	\centering
		\includegraphics[width=9cm]{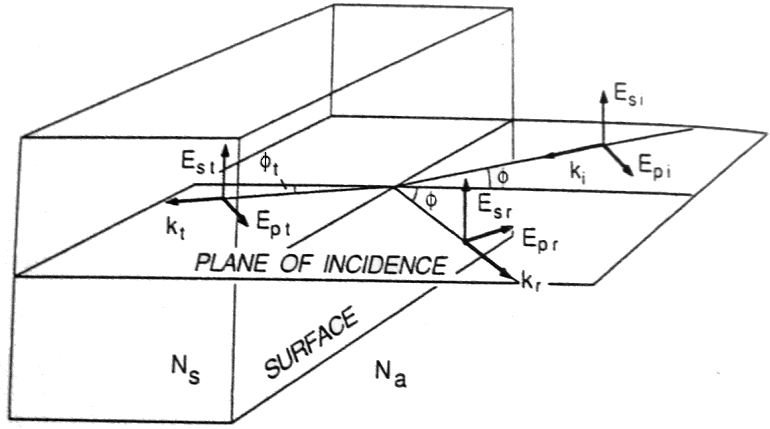}
	\caption[Electric field vectors in ellipsometry]{Electric field vectors E resolved into their components parallel (p) and perpendicular (s) to the plane of incidence. The indices i, t and r correspond to incident, transmitted and reflected light at the surface between two media of refractive indices \chemr{N_a} and \chemr{N_s}. \chemr{k_i}, \chemr{k_t} and \chemr{k_r} are the corresponding propagation vectors and $\phi$ and $\phi_t$ are the angles of incidence and transmission (from \cite{collins:1993}).}
	\label{fig:ellipsometry}
\end{figure}
To study the reflection of a polarised wave at one or more interfaces, the electric field vectors are decomposed into their components in the plane of incidence $E_p$ and perpendicular to it $E_s$. The plane of incidence is defined as the plane containing the incident, transmitted and reflected propagation vectors as visualised in figure \ref{fig:ellipsometry}. The complex amplitude reflection coefficients $r_p$ and $r_s$ for the two independent directions are defined as ratio of reflected to incident electric field amplitude in the respective direction:
\begin{eqnarray}
r_p&=&|r_p|\exp(i\delta _p) = \frac{E_{rp}}{E_{ip}} \\
r_s&=&|r_s|\exp(i\delta _s) = \frac{E_{rs}}{E_{is}}.
\end{eqnarray}
By measuring the incident and reflected polarisation one can obtain the complex amplitude reflection ratio $\rho$ defined as
\begin{equation}
\rho = \frac{r_p}{r_s} = \frac{E_{rp} E_{is}}{E_{ip}E_{rs}} = \frac{|r_p|}{|r_s|} \exp(i(\delta _p -\delta _s))= \tan(\psi) \exp(i\Delta).
\label{eq:ellipsometry}
\end{equation}
$\psi$ and $\Delta$ are the \emph{ellipsometry angles} that are usually used to express the interaction of the incident wave with the reflecting system.

In many experimental set ups the incident wave is linearly polarised at 45° with respect to the plane of incidence leading to $E_{is}/E_{ip} = 1$. In this case equation \ref{eq:ellipsometry} simplifies to
\begin{equation}
\rho = \frac{E_{rp}}{E_{rs}} = \frac{|E_{rp}|}{|E_{rs}|} \exp(i(\delta _{rp} -\delta _{rs}))= \tan(\psi) \exp(i\Delta).
\end{equation}
In this configuration $\tan(\psi)$ is the amplitude ratio of the components of the reflected electric field vector in the plane of incidence and perpendicular to it and $\Delta$ is the phase shift between the two components. This way the ellipsometry angles define the polarisation state of the reflected wave.

In this work ellipsometry is used to measure the thickness $d_f$ of surface films in a two-interface system with three media: ambient air, the film and the substrate. The dielectric properties of the silicon substrate are known. This case leads to the complex amplitude reflection coefficients $r_p$ and $r_s$ given by \cite{collins:1993}:
\begin{eqnarray}
r_p&=&\frac{r_{p,af}+r_{p,fs}Z}{1+r_{p,af}r_{p,fs}Z} \label{eq:rp}\\
r_s&=&\frac{r_{s,af}+r_{s,fs}Z}{1+r_{s,af}r_{s,fs}Z} \label{eq:rs}
\end{eqnarray}
$$
  \mathrm{with} \ Z=\exp\left(\frac{4 \pi i d_f}{\lambda}\sqrt{\epsilon_f-\epsilon_a \sin^2(\phi)}\right). 
%\\ \rho &=& \frac{r_p}{r_s} = \tan(\psi) \exp(i\Delta)
$$
Herein $\lambda$ is the wavelength of the reflected light, $\epsilon_f$ and $\epsilon_a$ are the dielectric function of the film as well as the ambient medium, and $r_{j,af}$ and $r_{j,fs}$ ($j = p\ \mathrm{or}\ s$) are the reflection coefficients at the ambient-film and film-substrate interface \cite{collins:1993}:
\begin{eqnarray}
r_{p,af}&=&\frac{\epsilon_f N_{a\bot}-\epsilon_a N_{f\bot}}{\epsilon_f N_{a\bot}+\epsilon_a N_{f\bot}} \\
r_{s,af}&=&\frac{N_{a\bot}-N_{f\bot}}{N_{a\bot}+N_{f\bot}} \label{eq:rsaf}
\end{eqnarray}
$$
 \mathrm{with} \ N_{a\bot}=N_a\cos(\phi) \ \mathrm{and} \
N_{f\bot}=N_f\cos(\phi_t)=\sqrt{\epsilon_f-\epsilon_a \sin^2(\phi)}.
$$
$N_a$ and $N_f$ are the indices of refraction of the ambient medium and the film, and $\phi$ and $\phi_t$ are the angles of incidence and transmission (figure \ref{fig:ellipsometry}). Analogous equations hold for $r_{j,fs}$.

A measurement determines the ellipsometric angles that are according to equations \ref{eq:rp} to \ref{eq:rsaf} a function of $d_f,\epsilon_a, \epsilon_f, \epsilon_s, \lambda$ and $\phi$:
\begin{equation}
\tan(\psi)\exp(i\Delta)=\frac{r_p}{r_s} =f(d_f,\epsilon_a, \epsilon_f, \epsilon_s, \lambda, \phi) \label{eq:rho}.
\end{equation}
The dielectric functions of the ambient air $\epsilon_a$ and of the substrate $\epsilon_s$ as well as the wavelength $\lambda$ and the angle of incidence $\phi$ are known. Thus the problem of ellipsometry consists in determining three film parameters, the two components of the dielectric function $\epsilon_{f1,2}$ and the thickness $d_f$, from the two ellipsometric angles. This is only possible if multiple measurements for example at different wavelengths or angles of incidence are performed. Even for a transparent film ($k_f=\epsilon_{f2}=0$), equation \ref{eq:rho} cannot be inverted to provide a unique thickness from a single measurement. Equations \ref{eq:rp} and \ref{eq:rs} show that with $d_f$, all
\begin{equation}
\left\{d_f + \frac{m\lambda}{\sqrt{\epsilon_f-\epsilon_a \sin^2(\phi)}}\ \mathrm{with}\ m \in \mathbb{Z}\right\}
\label{eq:periodicity}
\end{equation}
are also solutions. This ambiguity is often solved by measuring at two different wavelengths to obtain by numerical inversion of (\ref{eq:rho}) two sets of possible thickness values. The one common element of the two sets is the actual thickness of the examined film.

\subsection{Rotating-element ellipsometers}
\begin{figure}
	\centering
		\includegraphics[width=100mm]{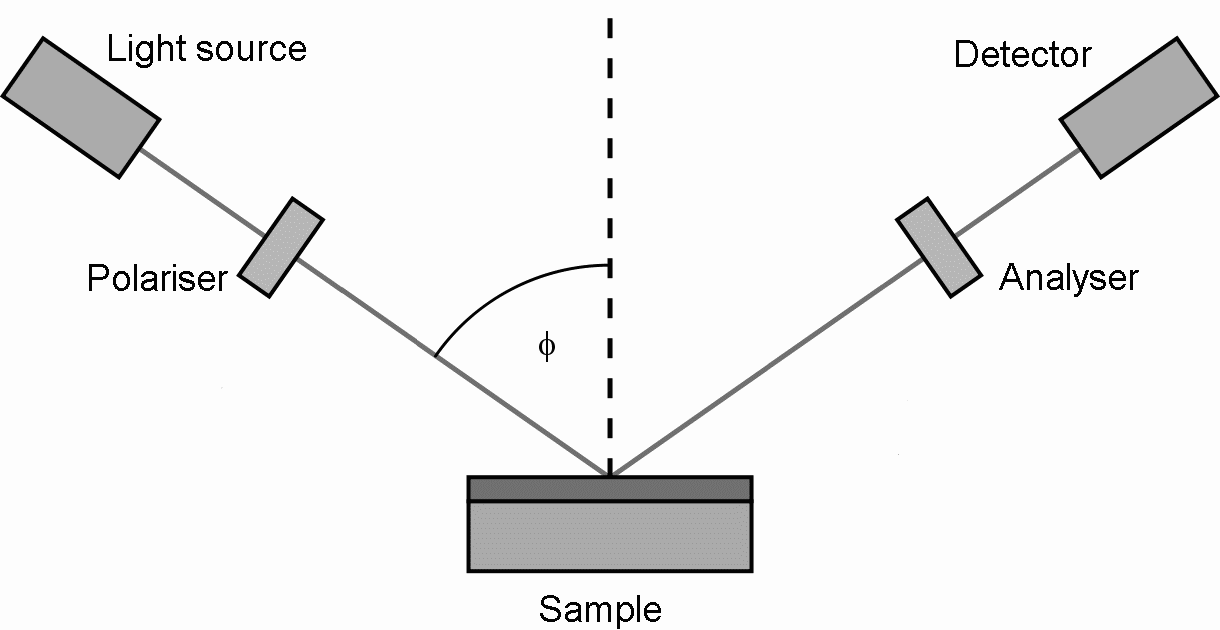}
	\caption{Principal elements of an ellipsometer}
	\label{fig:ellipsometer}
\end{figure}
The principal elements of an ellipsometer shown in figure \ref{fig:ellipsometer}
are a monochromatic light source, a polariser, the sample, an analyser and a detector. The optical axes of the source and detector sides lie in the plane of incidence and intersect at the sample surface. In a rotating-element ellipsometer the analyser or polariser is rotated at a frequency $\omega$. The time-dependent intensity at the detector is according to \cite{collins:1993}
\begin{equation}
I(t)=I_0[1+\alpha \cos(2(\omega t + \gamma)) + \beta \sin(2(\omega t + \gamma))].
\end{equation}
$\gamma$ is a phase angle determined by calibration. The coefficients $\alpha$ and $\beta$ are extracted by Fourier analysis of the detector signal. For a rotating analyser instrument they are related to the ellipsometer angles by \cite{collins:1993}
\begin{eqnarray}
\tan(\psi)&=&\sqrt{\frac{1+\alpha}{1-\alpha}}\tan(P) \\
\Delta&=&\arccos\left(\frac{\beta}{\sqrt{1-\alpha^2}}\right).
\end{eqnarray}
$P$ is the angle measured in a counter-clockwise sense looking into the light beam between the plane of incidence and the transmission axis of the polariser. For the ellipsometer employed in this work it is $P=45$\textdegree \ leading to $\tan(P)=1$.

\section{Cyclic voltammetry}
\begin{figure}
	\centering
		\includegraphics[width=4cm]{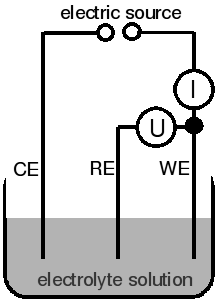}
	\caption[Setup for cyclic voltammetry]{Three electrode setup for cyclic voltammetry containing a counter electrode (CE), a reference electrode (RE), a working electrode (WE) a voltmeter (U) and an amperemeter (I)}
	\label{fig:voltammetry}
\end{figure}
Cyclic voltammetry is an electrochemical method to characterise the redox properties of analytes in an electrolyte solution. In a three electrode setup as shown in figure \ref{fig:voltammetry}, the potential between the working electrode and the reference electrode is varied periodically while the current between the working electrode and the counter electrode is measured. Depending on the potential, a chemical species in the electrolyte solution can be transformed from its reduced form \chem{Red} to its oxidised form \chem{Ox^+} and vice versa:
\begin{equation}
\mathsf{Red} \rightleftharpoons \mathsf{Ox^+} + e^-
\end{equation}
The resulting current versus potential graphs are called cyclic voltammograms. An example, monitoring the oxidation of hexacyanoferrate(II) (\chem{Fe(CN)_6^{4-}}) to hexacyanoferrate(III) \chem{Fe(CN)_6^{3-}}, is shown in figure \ref{fig:voltaaeratedmelanin}. As the potential is raised from negative to positive values, the absolute value of the current rises when the potential approaches the oxidation potential of \chem{Fe(CN)_6^{4-}}. A further increase of the potential leads to a decrease in current, because \chem{Fe(CN)_6^{4-}} is depleted in the region close to the working electrode. When the potential is decreased again, the current shows another peak indicating the reduction of \chem{Fe(CN)_6^{3-}}. If the process is reversible, the voltammograms of subsequent cycles coincide. The redox potential is the arithmetic mean of the potentials of the oxidation (anodic) and the reduction (cathodic) peaks. The peak current depends on the diffusion coefficient of the examined species close to the working electrode. High diffusion coefficients lead to high peak currents and vice versa. It also strongly depends on the potential sweep rate and the surface area of the working electrode.

The working electrode is usually made of glassy carbon or gold. It is enclosed in an insulating material so that only a disk of some millimetres in diameter at its end is exposed to the solution. The counter electrode should offer a good conductibility and it must not react with the solution. Thus platinum is a common material. The reference electrode has to have a constant electrode potential. Therefore a  redox system with constant concentrations of the participating species is used. For example in a silver chloride reference electrode the redox reaction is
\begin{equation}
\mathsf{Ag + Cl^-} \rightleftharpoons \mathsf{AgCl} + e^-.
\end{equation}
A silver electrode is coated with silver chloride and kept in a saturated potassium chloride solution. A porous plug assures the contact with the environmental solution. The potential of a silver chloride electrode in 3.5 mol/L potassium chloride versus a standard hydrogen electrode is + 0.205 V at a temperature of 25 \textdegree C \cite{sawyer:1995}.

\section{Confocal laser scanning microscopy}
\begin{figure}
	\centering
		\includegraphics[width=\textwidth]{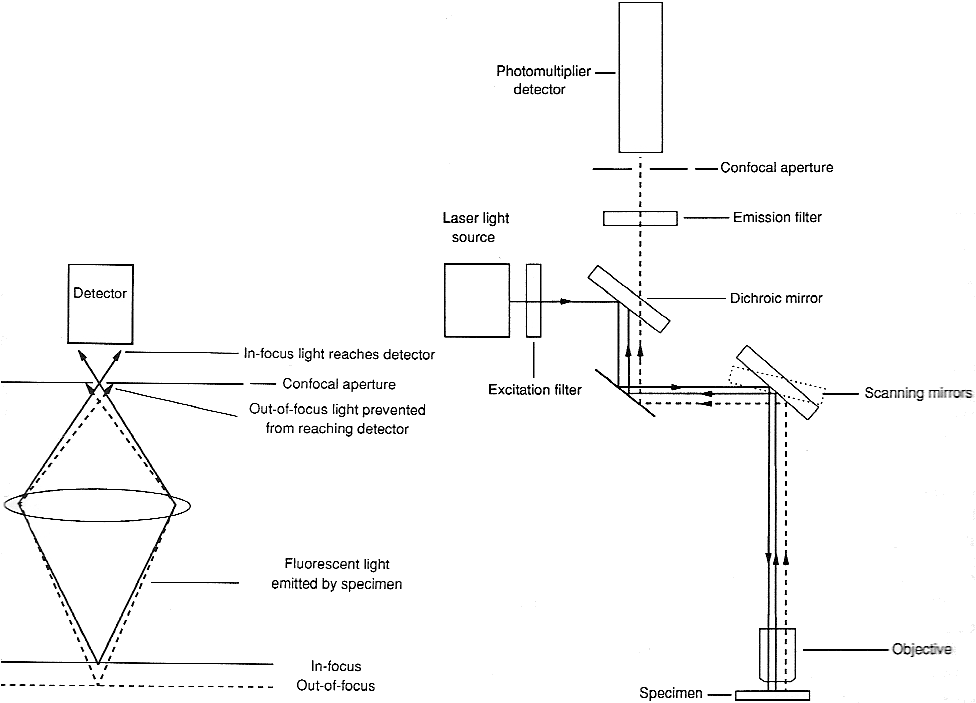}
	\caption[Principle and layout of a confocal microscope]{Principle of confocal microscopy (left) \cite{rawlins:1992} and layout of a confocal laser scanning microscope (right)\cite{rawlins:1992}. In the left image, light emitted from the focal plane is represented by solid lines, light emitted from an out-of-focus plane by dashed lines. In the right image, excitation light is represented by solid lines and fluorescence light by dashed lines.}
	\label{fig:confocal-CLSM}
\end{figure}

Confocal microscopy is usually used on fluorescent samples and allows for optical sectioning of specimens of up to $100 \; \mu m$ thickness \cite{rawlins:1992}. In confocal laser scanning microscopy (CLSM) a laser serves as light source and the excitation wavelength is selected by an excitation filter (Figure \ref{fig:confocal-CLSM}). The laser beam is focused at a certain depth in the sample and it is scanned in the sample plane by a set of scanning mirrors. Light emitted by the fluorescent molecules returns through the objective and leaves the path of the exciting light through a dichroic mirror, which reflects the exciting light but lets pass the fluorescence light due to its longer wavelength. Afterwards, wavelengths not originating from the fluorescent molecules are blocked by an emission filter. In front of the detector the light passes the confocal aperture. This is a very small aperture, also called \emph{pinhole}, that is optically at the same point as the focal plane of the scanning laser beam. As shown in figure \ref{fig:confocal-CLSM}, only light emitted from the focal plane can pass the confocal aperture. The detector is usually a photo multiplier tube and its output signal is digitised for further treatment with a computer. By scanning the laser beam in the sample plane, an image is created point by point. A stack of images at different depths within the sample can be acquired and combined numerically to create a three-dimensional representation of the specimen.

\subsection{Optical image formation}
The image quality of a CLSM depends on two main factors: optical image formation and signal processing. The point spread function (PSF) describes the limitations of optical image formation. It is the image of an ideal point object and most of its energy is concentrated in a rotational ellipsoid with respect to the optical axis in the ideal deflection limited case (no optical aberrations, homogeneous illumination). The lateral and axial resolution are given by the full widths at half maximum (${FWHM}$) of the PSF in lateral (perpendicular to the optical axis) and axial (along the optical axis) direction. This way the resolution is defined as the minimum distance two point objects have to have to be separated in the image. 
%The total point spread function of a CLSM is the product of PSF of the illumination pathway $PSF_{ill}$ and the one of the detection pathway $PSF_{det}$. 

Depending on the size of the pinhole, there are two different regimes. If the diameter of the pinhole is larger than one Airy unit ($AU$) one speaks of \emph{geometric optic confocality}. If the diameter is smaller than one Airy unit one speaks of \emph{wave-optic confocality}. One Airy unit corresponds to the diameter of the Airy disk, the diffraction image of a point source created by the microscope objective, and it is defined as
\begin{equation}
1 AU = \frac{1.22 \bar{\lambda}}{N\!A}
\end{equation}
$$
\mathrm{with} \ \bar{\lambda} = \frac{\sqrt{2} \lambda_{em} \lambda_{exc}}{\sqrt{\lambda_{em}^2 + \lambda_{exc}^2}} \ \mathrm{and} \ N\!A=n\sin(\theta).
$$
$\bar{\lambda}$ is a mean wavelength calculated from the excitation wavelength $\lambda_{exc}$ and emission wavelength $\lambda_{em}$, $n$ is the refractive index of the medium between objective and sample, $N\!A$ and $\theta$ are the numerical aperture and the half opening angle of the microscope objective.

In the case of geometric optic confocality, diffraction effects at the pinhole are negligible and the resolution is limited by the size of the scanning laser spot in the sample. The axial and lateral resolution are according to \cite{wilhelm}
\begin{eqnarray}
{FWHM}_{axial}&=&0.88\frac{\lambda_{exc}}{n-\sqrt{n^2-N\!A^2}} \label{eq:geometricaxialresolution}\\
{FWHM}_{lateral}&=&0.51\frac{\lambda_{exc}}{N\!A} \label{eq:geometriclateralresolution}.
\end{eqnarray}
These formulae are the same as for conventional microscopy, except for the fact that the resolution depends on the excitation wavelength instead of the emission wavelength. This leads to a gain in resolution by the factor $\lambda_{exc}/\lambda_{em}$.

In the case of wave-optic confocality, diffraction effects at the pinhole have to be taken into account leading to an influence of excitation \emph{and} emission wavelength on the resolution. Thus equations \ref{eq:geometricaxialresolution} and \ref{eq:geometriclateralresolution} are transformed for the limit of a pinhole diameter of $0 AU$ into \cite{wilhelm}
\begin{eqnarray}
{FWHM}_{axial}&=&0.64\frac{\bar{\lambda}}{n-\sqrt{n^2-N\!A^2}} \label{eq:waveaxialresolution}\\
{FWHM}_{lateral}&=&0.37\frac{\bar{\lambda}}{N\!A} \label{eq:wavelateralresolution}. \\
&=&0.30 AU
\end{eqnarray}

\begin{figure}
	\centering
		\includegraphics[width=85mm]{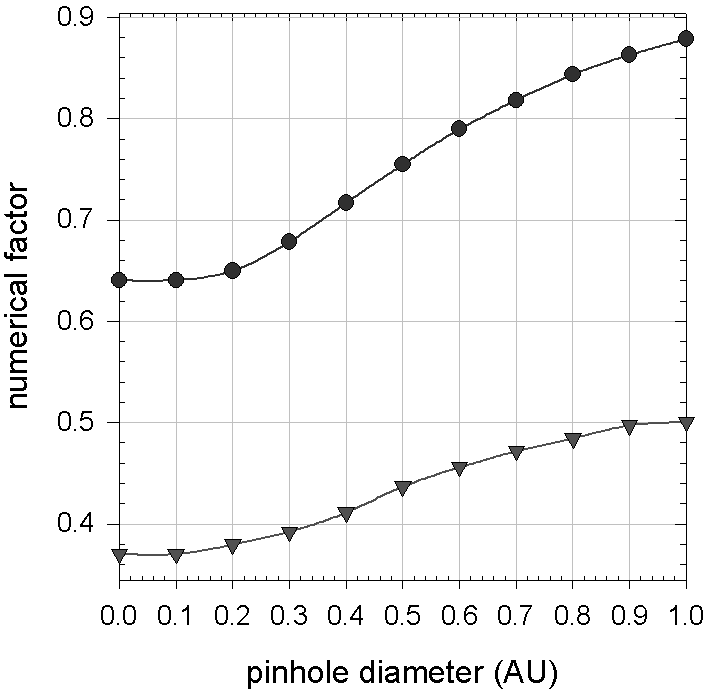}
	\caption[Numerical factors for wave-optic confocal resolution]{Numerical factors for equations \ref{eq:waveaxialresolution} (circles) and \ref{eq:wavelateralresolution} (triangles) for pinhole diameters from $0 AU$ to $1 AU$. Values are taken from \cite{wilhelm}.}
	\label{fig:numericalfactors}
\end{figure}

Since there would not be any intensity arriving at the detector, the pinhole diameter $0 AU$ cannot be used in practice. Nevertheless equations \ref{eq:waveaxialresolution} and \ref{eq:wavelateralresolution} remain good approximations for diameters up to $1 AU$ provided the numerical factors are changed as represented in figure \ref{fig:numericalfactors}.

\subsection{Image processing}
\label{sec:clsmimageprocessing}
Information on the sample is obtained from the electrical output signal of the detector, usually a photomultiplier tube. This continuous signal has a variable intensity with time $I(t)$. Time $t$ and position $x$ on the sample are linked by the scanning speed $v_{scan}=x/t$. The continuous signal is transformed into a discrete series of points (pixels) with an analog to digital (A/D) converter by sampling it at fixed time intervals. To detect as much of the emitted light as possible, the signal intensity at one pixel is the integrated output signal between two adjacent sample points. The integration time can be changed by changing the scanning speed.

The optimum pixel spacing to sample a periodic signal is given by the \emph{Nyquist theorem} as half the period of the signal. Thus for optimum sampling of an image characterised by the ${FWHM}$ of a point object, the pixel spacing $d_{opt}$ has to be
\begin{equation}
d_{opt}=0.5 {FWHM}.
\end{equation}
This holds for the lateral pixel spacing in one single image as well as for the axial distance between adjacent images in case of optical sectioning. If the pixel spacing is larger than $d_{opt}$, part of the information from the sample is lost. On the other hand a too small pixel spacing enlarges the amount of data without adding information. Furthermore the pixel integration time gets smaller leading to a smaller signal to noise ratio.

\subsection{Noise}
The main kinds of statistical noise in a CLSM are detector noise, laser noise and shot noise of the emitted light from the sample. With modern photomultiplier tubes, detector noise is usually negligible. The relation between the other two types of noise depends on the signal intensity. Laser noise dominates for strong signals when a sample is observed in reflection. It is caused by statistically fluctuating occupations of excited states in the laser medium.

For the weak signals observed in fluorescence imaging, shot noise is dominating. If the number of detected photons per pixel $p$ is below $1000$ \cite{wilhelm}, the photons cannot be described as a continuous current any more. Instead single photons are detected in a way governed by Poisson statistics. Consequently structures in the sample close to the optical resolution limit can be resolved or not depending on the noise pattern. In other words, the resolution of the microscope becomes a resolution probability strongly depending on the pinhole diameter. Since the number of detected photons tends to zero for very small pinhole diameters, the signal to noise ratio $S/N$ given by Poisson statistics ($S/N=\sqrt{p}$) also tends to zero leading to a vanishing resolution probability. For large pinhole diameters the resolution probability is also reduced, because the optical resolution is diminished. The best resolution probability is achieved for most fluorescence applications with a pinhole diameter of about $1 AU$.

The resolution probability can be enhanced by longer pixel integration times or by summation over several scans of the same image. In both cases the number of detected photons per pixel is increased. According to Poisson statistics an increase by a factor $f$ leads to an increase of the signal to noise ratio by a factor $\sqrt{f}$. The drawback of these methods are longer imaging times leading to stronger bleaching of the sample. 

%% file: experiment.tex
\chapter{Materials and methods}
\section{Materials}
\textbf{Solvents:}
\begin{itemize}
	\item Deuterium oxide (atom fraction of deuterium: 99 \% , Sigma-Aldrich, St. Louis, Missouri, USA, ref. 435767)
	\item Dimethylsulfoxide (DMSO, SdS, Peypin, France)
	\item Ultrapure water with a resistivity of 182 k$\Omega$m purified in a Milli Q Plus water purification system (Millipore, Billerica, Massachusetts, USA)
\end{itemize}
\textbf{Electrolytes:}
\begin{itemize}
	\item Copper(II) sulphate (\chem{CuSO_4}, molar mass $M=159.6$g/mol, Sigma-Aldrich, ref. 61230)
	\item di-Potassium hydrogenphosphate trihydrate (\chem{K_2HPO_4 * 3 H_2O}, molar mass $M=228.2$ g/mol, Merck, Darmstadt, Germany, ref. 5099)
	\item Sodium chloride (\chem{NaCl}, $M = 58.4$ g/mol, VWR International, Lutterworth, United Kingdom, ref. 27810.295)
	\item Sodium nitrate (\chem{NaNO_3}, $M = 85.0$ g/mol, Sigma-Aldrich, ref. S5506)
	%\item Calcium chloride dihydrate(\chem{CaCl_2 * 2(H_2O)}, $M = 147.0$ g/mol, VWR International, ref. 22317.297)
	%\item Calcium nitrate tetrahydrate(\chem{Ca(NO_3)_2 * 4(H_2O)}, $M = 236.2$ g/mol, Sigma-Aldrich, ref. C5676)
	\item Tris(hydroxymethyl)aminomethane (Tris, $M = 121.1$ g/mol, acid dissociation con\-stant $pK_a=8.1$, Sigma-Aldrich, ref. T1503)
	%\item Sodium acetate trihydrate (\chem{NaCH_3COO * 3(H_2O)}, $M = 136.1$ g/mol, Sigma-Aldrich, ref. 23650-0)
\end{itemize}
\textbf{Polyanions:}
\begin{itemize}
	%\item Poly(sodium 4-styrenesulfonate) (PSS, $M = 7 \cdot 10^5$ g/mol, Sigma-Aldrich, ref. 24,305-1)
	\item Poly(acrylic acid sodium salt) (PAA, $M\approx 3\cdot~10^4$~g/mol, weight fraction in water: 40~\%, Sigma-Aldrich, ref. 41,604-5)
	\item Sodium hyaluronate (HA, $M = 4.2 \cdot 10^5$ g/mol, LifecoreBiomedical, Chaska, Minnesota, USA, ref. 80190)
\end{itemize}
\textbf{Polycations:}
\begin{itemize}
	\item Poly(diallyldimethyl ammonium chloride) (PDADMA, $M=(1 - 2)\cdot 10^5$~g/mol, weight fraction in water: 20~\%, Sigma-Aldrich, ref. 409014)
	\item Poly(ethyleneimine) (PEI, $M = 7.5 \cdot 10^5$ g/mol, Sigma-Aldrich, ref. P3143)
	%\item Poly(allylamine hydrochloride) (PAH, Sigma-Aldrich, ref. 28322-3)
	\item Poly(L-lysine hydrobromide)(PLL, $M = 4.6 \cdot 10^4$ g/mol, Sigma-Aldrich, ref. P2636)
	%\item Poly-L-lysine (PLL, $m_{m,vis} = 6.8 \cdot 10^4$ g/mol, Sigma-Aldrich, ref. P2636)
\end{itemize}
\begin{figure}
	\centering
		\includegraphics[width=90mm]{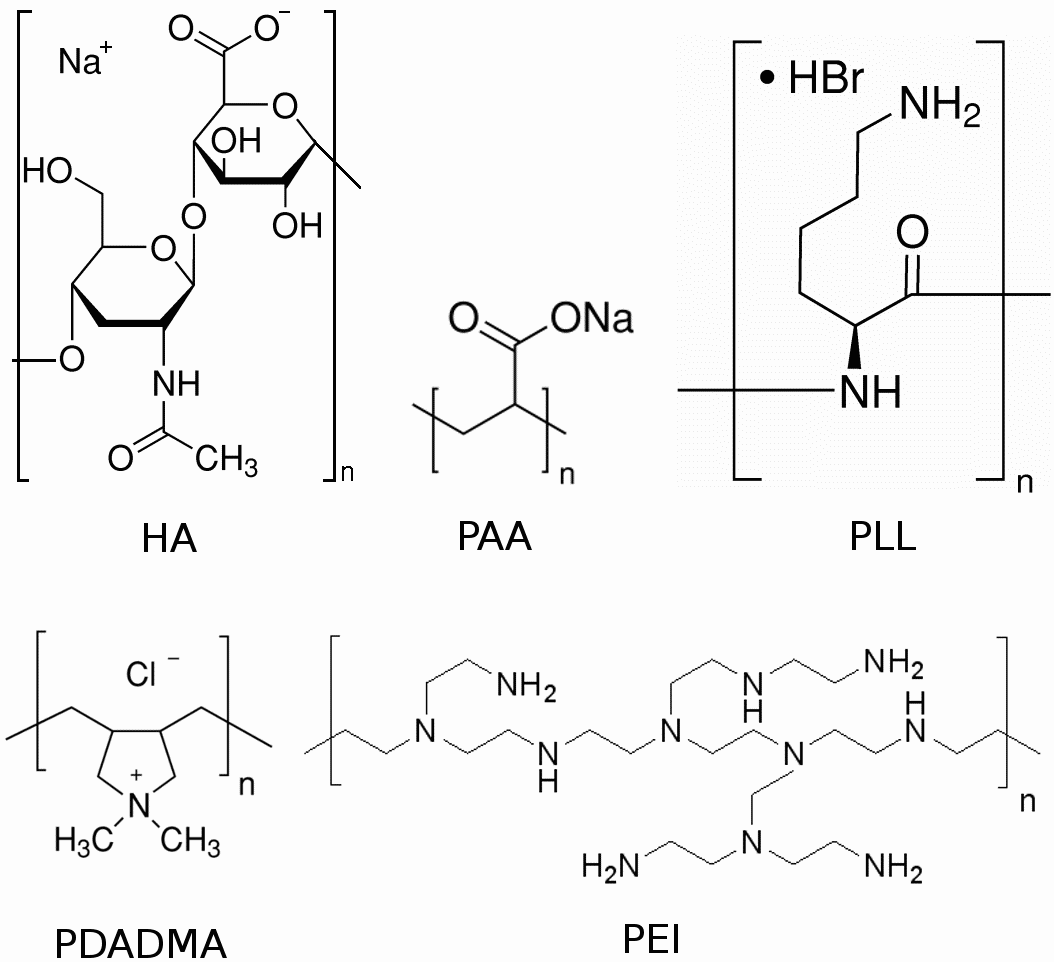}
	\caption[Monomer structures of employed polyelectrolytes]{Monomer structures of sodium hyaluronate (HA), poly(acrylic acid sodium salt) (PAA), poly(L-lysine hydrobromide) (PLL), poly(diallyldimethyl ammonium chloride) (PDADMA) and poly(ethylene imine) (PEI)}
	\label{fig:polyelectrolytes}
\end{figure}
\textbf{Proteins:}
\begin{itemize}
	\item $\alpha$-Lactalbulmine from bovine milk ($M=14.2$ kg/mol, isoelectric point $pI=4.5$, Sigma-Aldrich, ref. L5385)
	%\item Alkaline phosphatase from porcine intestinal mucosa (activity: 1.1 U/mg, Sigma-Aldrich, ref. P-4002)
	\item Lysozyme from chicken egg white ($M=14.3$ kg/mol, $pI=11.4$, Sigma-Aldrich, ref. L6876)
	\item Myoglobine from horse heart ($M=17.0$ kg/mol, $pI=7.2$, Sigma-Aldrich, ref. M1882) 
\end{itemize}
\textbf{Electrochemical and fluorescent probes:}
\begin{itemize}
	\item Ferrocenemethanol (\chem{C_{11}H_{12}FeO}, $M=216.1$ g/mol, Sigma-Aldrich, ref. 335061)
	\item Fluorescein isothiocyanate (FITC, $M = 398.4$ g/mol)
	\item Hexaamineruthenium(II) chloride (\chem{Ru(NH_3)_6Cl_2}, $M=274.2$ g/mol, Sig\-ma-Al\-drich, ref. 303690)
	\item Potassium hexacyanoferrate(II) trihydrate (\chem{K_4(CN)_6Fe * 3 H_2O}, $M=422.4$ g/mol, Sig\-ma-Ald\-rich, ref. P9387)	
	%\item Para-nitrophenylphosphate disodium salt hexahydrate (pNPP, Sigma-Aldrich, ref. N9389)	
	%\item Rhodamine B isothiocyanate (RhoB, $M = 536.1$ g/mol, excitation maximum: 540 nm, emission maximum: 573 nm, Sigma-Aldrich, ref. 83692)
\end{itemize}
\textbf{Other chemical reagents:}
\begin{itemize}
	%\item Nitric acid solution (\chem{HNO_3})
	%\item Titanium(IV) bis(ammonium lactato) dihydroxide solution(TBALDH,\chem{C_6H_{18}N_2O_8Ti}, $M = 294.1$ g/mol, Sigma-Aldrich, ref. 388165)
	%\item Sodium silicate solution (\chem{Na_2Si_3O_7}, $M = 242.2$ g/mol, Sigma-Aldrich, ref. 13729)
	\item Dopamine hydrochloride (3-hydroxytyramine hydrochloride, \chem{C_8H_{11}NO_2 \cdot HCl}, $M = 189.6$ g/mol, Sigma-Aldrich, ref. H8502)
	%\item Aniline hydrochloride (\chem{C_6H_5NH_2 \cdot HCl},$M = 129.59$ g/mol, Sigma-Aldrich, ref. 10414)
	%\item Ammonium peroxodisulphate (\chem{(NH_4)_2S_2O_8}, $M = 228.2$ g/mol, Sigma-Aldrich, ref. 09920)
	\item 2-mercaptoethanol (\chem{C_2H_6OS}, pure (14.2 mol/L), arviresco, Solon, Ohio, USA, ref. 0482)
	\item 2-(2-pyridyldithio)ethylamine hydrochloride (PTEAHCl, $M = 208.8$~g/mol, Laboratoire de Chimie Bioorganique, Faculté de Pharmacologie, Illkirch, France)
		\item Sodium hydroxide (\chem{NaOH}, Sigma-Aldrich, ref. S-5881)
\end{itemize}
\textbf{Supports for polyelectrolyte or melanin deposition:}
\begin{itemize}
	\item Microscope cover glasses (Fisher Bioblock Scientific, Illkirch, France)
	\item Quartz crystals (silicon dioxide-covered, Q Sense, Göteborg, Sweden, ref. QSX 303)
	\item Quartz crystals (gold-covered, Q Sense, ref. QSX 301)
	\item Quartz slides (Fisher Bioblock Scientific)
	\item Silicon wafers (polished, phosphorus-doped, orientation (100), Siltronix, Ar\-champs, France, ref. 12765)
	\item Zinc plates (size: 2.0 x 2.0 x 0.1 cm$^3$, Rheinzink, Datteln, Germany)
\end{itemize}
\textbf{Cleaning reagents:}
\begin{itemize}
	\item Hellmanex II solution (Hellma, M\"ullheim, Germany)
	\item Hydrochloric acid solution (\chem{HCl}, 37 \%, Sigma-Aldrich, ref. 258148)
	\item Sodium dodecylsulphate (SDS, Euromedex, Souffelweyersheim, France, ref. 1012-C)
	%\item Aqueous hydrogen peroxide solution (30 \% \chem{H_2O_2}, Sigma-Aldrich, ref. H1009)
	%\item Sulfuric acid (95 \% \chem{H_2SO_4}, Prolabo, ref. 20700.298)
\end{itemize} 

\section{General remarks}
This document is typeset using the open source program pdf\LaTeX in the {Mik\TeX} distribution (version 2.8, Christian Schenk, \href{http://miktex.org}{http://miktex.org}) with KOMA-Script (version 3.05, Markus Kohm, \href{http://komascript.de}{http://komascript.de}) and the editor \TeX nicCenter (version 1.0, Sven Wiegand, \href{http://www.texniccenter.org}{www.texniccenter.org}). Graphs are created with the software SigmaPlot (version 10.0.0.54, Systat Software Inc., San Jose, California, USA, \href{http://www.sigmaplot.com}{www.sigmaplot.com}).

The pH of solutions is measured with a pH-meter HI8417 (Hanna Instruments, Woonsocket, Rhode Island, USA) and adjusted by addition of hydrochloric acid solution (37~\%) or sodium hydroxide solution (1~mol/L) if not otherwise stated.

Supports for polyelectrolyte or melanin deposition are cleaned by the following procedure before use: 15 minutes immersion in Hellmanex (volume fraction in water: 2 \%) or SDS (0.01 mol/L in water) solutions at about 70 °C, rinsing with water, further 15 minutes immersion in hydrochloric acid solutions (0.1 mol/L in water) at about 70 °C, rinsing with water and drying by a nitrogen stream.

Experiments are carried out at room temperature (between 15 °C and 30 °C due to the poor thermal isolation of the laboratory building).

\section{Melanin deposition from dopamine solutions}
\label{sec:expdepositionmethods}
\begin{figure}
	\centering
		\includegraphics[width=11cm]{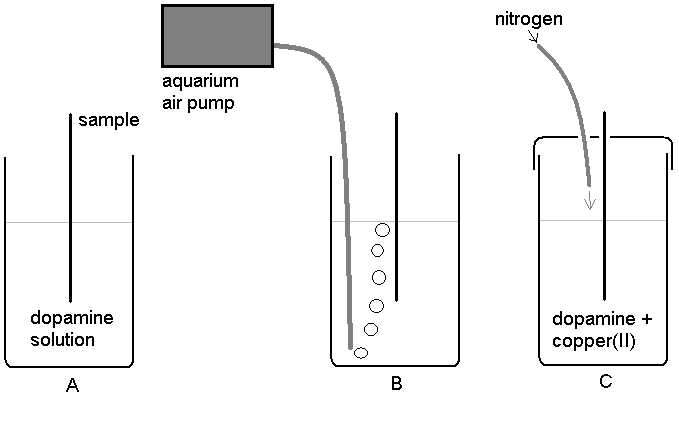}
	\caption[Illustration of melanin deposition methods]{Illustration of melanin deposition methods by oxidation of dopamine in solution. Method A: without aeration. Method B: with aeration. Method C: using copper as oxidant}
	\label{fig:melanindepositionsketch}
\end{figure}
Different methods are developed to build melanin deposits from aqueous dopamine solutions. The methods described in the following list, differ mainly in the way melanin formation is initiated by the oxidation of dopamine. Oxygen (Methods A and B), copper(II) ions (Method C) or the working electrode of an electrochemical cell (Method D) serve as oxidising agents (Figure \ref{fig:melanindepositionsketch}):
\begin{description}
\item[Method A:] Substrates are successively immersed for $n$ x 15 min or $n$ x 5 min in freshly prepared dopamine hydrochloride solutions (2 g/L in 50 mmol/L Tris at pH 8.5) in contact with ambient air without stirring.
\item[Method B:] Same as A but a substrate is immersed in only one single solution continuously aerated with an aquarium pump (RENA Air 50, Mars Fishcare, Metz, France).
\item[Method C:] Same as B but 30 mmol/L of copper sulphate (\chem{CuSO_4}) are added to the dopamine solutions that are not aerated but deoxygenated before and kept under nitrogen during the experiment. Due to the addition of copper sulphate the pH of the solutions decreases to about 4.5 and is not readjusted.
\item[Method D:] The working electrode in an electrochemical cell serves as substrate. The potential (vs. \chem{Ag/AgCl}) is cycled between -0.4 V and 0.3 V at a speed of 0.01 V/s in a deoxygenated dopamine hydrochloride solution (0.5 g/L in 10 mmol/L Tris with 150 mmol/L \chem{NaNO_3}, pH 7.5) kept under nitrogen for the whole deposition time. Under these conditions no melanin is formed in the bulk of the solution.
\end{description}

The adsorption substrates are silicon dioxide- or gold-covered quartz crystals for quartz crystal microbalance with dissipation (QCM-D) experiments, glass cover slips (24 x 60 mm$^2$) for streaming potential measurements, quartz slides for UV-visible spectroscopy, vitreous carbon electrodes for cyclic voltamperometry and silicon wafers cut into rectangles of 4 x 1 \chemr{cm^2} for ellipsometry, contact angle measurements and x-ray photoelectron spectroscopy (XPS).

\begin{figure}
	\centering
		\includegraphics[width=\textwidth]{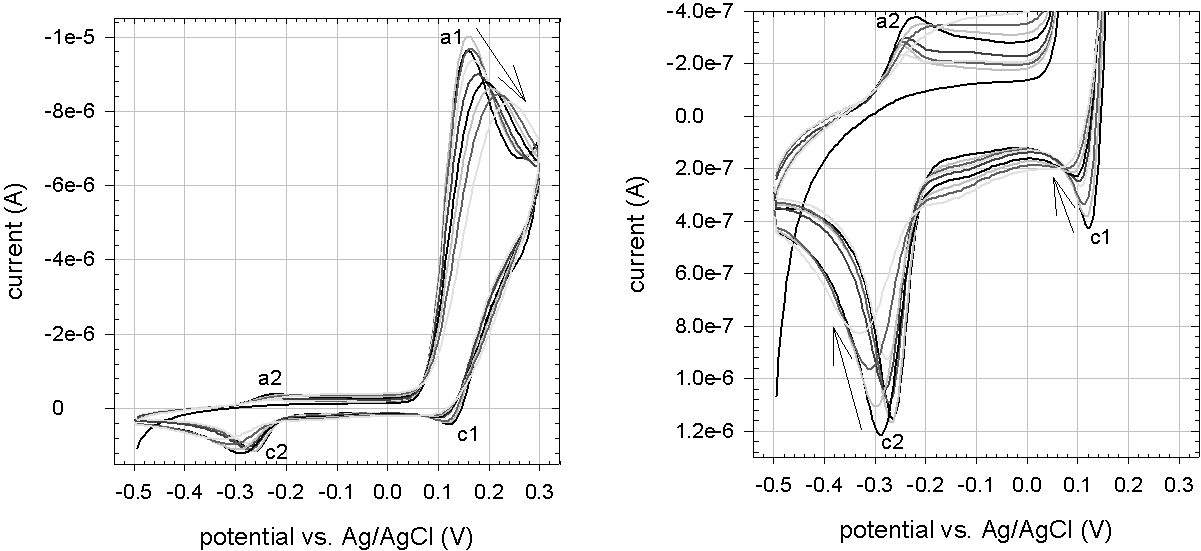}
	\caption[Cyclic voltammograms during dopamine-melanin deposition]{Cyclic voltammograms during dopamine-melanin deposition by method D. The graphs represent the \chemr{1^{st}, 5^{th}, 10^{th},\ldots,40^{th}} cycle of one experiment. Peaks are labelled as described in the text. Their evolution with increasing number of cycles is indicated by arrows. The right graph is a magnified detail of the left one.}
	\label{fig:voltamelanindeposition}
\end{figure}
\begin{figure}
	\centering
		\includegraphics[width=120mm]{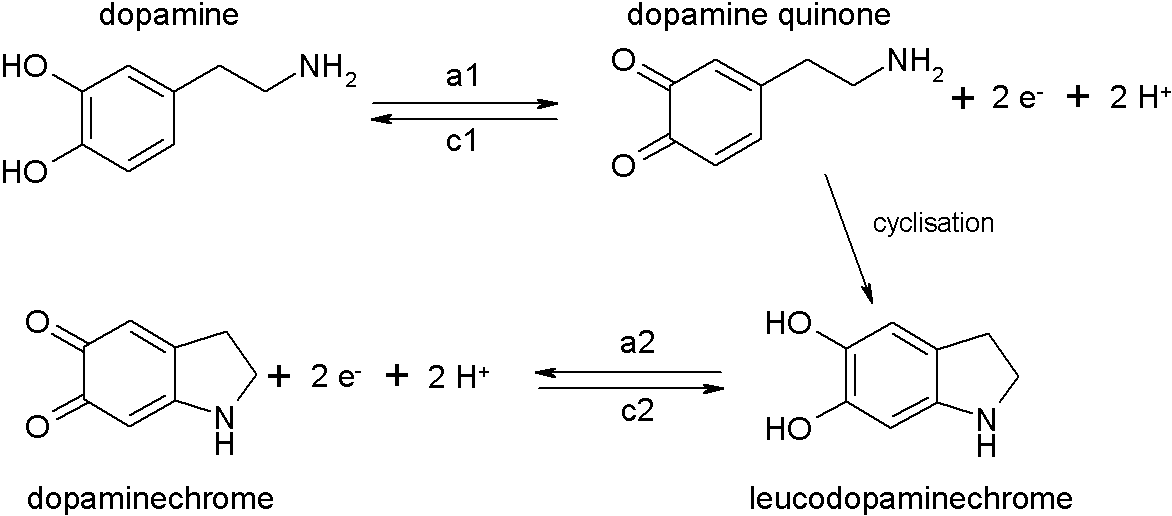}
	\caption[First steps of electrochemical dopamine-melanin formation]{First steps of electrochemical dopamine-melanin formation according to \cite{li:2006.1}. a1 and a2 designate electrochemical oxidations, c1 and c2 the corresponding reductions.}
	\label{fig:dopamineoxidations}
\end{figure}
Figure \ref{fig:voltamelanindeposition} shows the development of the cyclic voltammograms during dopamine-melanin deposition by method D. In the first cycle, one strong oxidation peak (anodic peak a1 at a potential of 0.16 V versus \chem{Ag/AgCl}) and two weak reduction peaks (cathodic peaks c1 at 0.12 V, c2 at -0.29 V) are visible. In the second cycle, a second very weak oxidation peak appears (a2 at -0.25 V). According to Li and others \cite{li:2006.1} peak a1 corresponds to the oxidation of dopamine to dopamine quinone and peak a2 to the oxidation of leucodopaminechrome to dopaminechrome during the first steps of dopamine-melanin formation as depicted in figure \ref{fig:dopamineoxidations}. Peaks c1 and c2 are caused by the corresponding reduction reactions. Peak a2 appears only from the second cycle on, because leucodopaminechrome has to be formed by internal cyclisation of dopamine quinone at the end of the first half-cycle before its oxidation. In fact the potential sweep rate has to be chosen low enough to allow for cyclisation to take place between the formation of dopamine quinone at peak a1 and its reduction at peak c1 \cite{li:2006.1}. The peak currents decrease when the number of voltamperometric cycles increases. This decrease indicates that a compact layer of melanin forms at the surface of the working electrode impeding the access of dopamine and its oxidation products to the electrode.

\section{Build-up of polyelectrolyte films}
\subsection{Poly(L-lysine) and hyaluronate}
These films are built on cover glasses of 12 mm in diameter for confocal microscopy and scanning force microscopy. Poly(L-lysine hydrobromide) (PLL) and sodium hyaluronate (HA) are dissolved at 1 g/L in buffer solutions containing 50 mmol/L Tris at a pH of 8.5. Since the acid dissociation constants $pK_a$ of the amine group in PLL and the carboxyl group in HA are 9.4 and 3.1 \cite{burke:2003}, PLL is charged positively and HA negatively at pH 8.5. The opposed electric charges lead to attractive electrostatic interactions between the two components. An automated dipping robot (DR3, Riegler and Kirstein GmbH, Berlin, Germany) immerses the cover glasses alternately for 8 min in solutions of PLL and HA. Between the deposition steps the sample is rinsed for 40 s and twice for 5 min in pure buffer solution. By repeating the deposition cycle (1 PLL deposition \emph{and} 1 HA deposition) $n$ times a film denoted \chemr{(PLL-HA)_n} is obtained.

For ellipsometric measurements, the films are manually built on silicon substrates. The substrates are alternately dipped for 5 min in polyelectrolyte solutions and twice for 2.5 min in pure buffer solution. Since the adsorption of PLL and HA is usually completed within less than 5 min \cite{picart:2001}, the reduced dipping time compared to the automated procedure should not influence the film growth. The build-up is interrupted after the deposition of 6, 12, 18 and 21 double layers to dry the samples with a nitrogen stream and measure their thickness by ellipsometry.

For UV-visible spectroscopy experiments, the films are built on quartz slides the same way as for ellipsometry but without intermediate drying steps.

To incorporate melanin in \chemr{(PLL-HA)_n} films, they are put in contact with dopamine solutions in Tris buffer (50 mmol/L, pH 8.5), which have been prepared just before. The samples are maintained in a vertical orientation to avoid the sedimentation of melanin aggregates, which appear in the dopamine solutions after a few hours of reaction at pH 8.5. The presence of such aggregates on the surface of the films might lead to a considerable increase in film roughness.

To prepare free standing membranes, melanin is incorporated in \chemr{(PLL-HA)_{n}} films deposited on glass slides (size: 4 x 2 cm$^2$) Then the samples are rinsed with Tris buffer and water. They are blown dry under a stream of nitrogen and the edges of the glass slide are cut with a razor blade. This procedure allows for the diffusion of the liquid to be used for film detachment between the film and the support. Afterwards the samples are put in contact with hydrochloric acid solutions at increasing concentrations: 0.001~mol/L, 0.01 mol/L and 0.1 mol/L. If a composite membrane detaches, it is deposited on paper and air dried.

\subsection{PDADMA and PAA}
To built these films, 10 g/L poly\-(di\-methyl\-di\-allyl\-ammonium chloride) (PDADMA) and 5 g/L poly\-(acrylic acid sodium salt) (PAA) are dissolved in water. Glass slides are immersed using the automated dipping robot for 8 min in polyelectrolyte solutions and rinsed for 40 s and twice for 5 min in water. After 30 cycles a film denoted \chemr{(PDADMA-PAA)_{30}} is obtained.

\subsection{PDADMA and melanin particles}
Dopamine hydrochloride at a concentration of 2 g/L is allowed to react with ambient oxygen for two hours in Tris buffer (50 mmol/L) at pH 8.5. Then the solution is titrated to pH 13 by addition of concentrated sodium hydroxide solution and back to pH 12 with concentrated hydrochloric acid. The solution is kept at pH 12 for at least 24 hours before further using it. PDADMA is dissolved at 1 g/L in Tris buffer (50 mmol/L) at pH 8.5 and titrated to pH 12. Supports for film build-up are silicon slides for ellipsometry, quartz slides for UV-visible spectroscopy and cover glasses for scanning force microscopy. Silicon and quartz slides are manually dipped for 5 min each in PDADMA and melanin solutions. The adsorption steps are separated by 5 min of rinsing in water. The cover glasses are coated using the automated dipping robot. In this case the adsorption steps last for 8 min and are separated by three successive rinsing steps for 40 s and twice 5 min in water. After $n$ adsorption cycles the obtained films are denoted \chemr{(PDADMA-melanin)_n}.   

\section{Fluorescence labelling}
\label{sec:fluorescencelabelling}
Poly(L-lysine hydrobromide) (PLL) is labelled with fluorescein isothiocyanate (FITC) for confocal microscopy. Therefore approximately 1 mg of FITC is dissolved in 1 mL of di\-methyl\-sulf\-oxide (DMSO). PLL is dissolved at a concentration of 1 g/L in 10 mL to 15 mL of Tris buffer (50 mmol/L,pH 8.5). The molar masses of FITC (389.4 g/mol) and PLL (45.8 kg/mol) are used to calculate the amounts of these molecules in the prepared solutions. Then the volume of FITC solution necessary to obtain a ratio of about two molecules of FITC per PLL chain is added to the PLL solution.

After 1 h of reaction in the dark, unbound FITC is removed from the solution by dialysing it twice for several hours in the dark against 500 mL of pure buffer solution. The cellulose membrane (Spectra/Por 7, Spectrum Laboratories, Rancho Dominguez, California, USA, ref. 132119) used for dialysis has a molar mass cut-off at 10 kg/mol. Before the end of the dialysis, the absorbance of the buffer solution at wavelengths between 450 nm and 550 nm is measured to verify the absence of FITC molecules. These would show a strong absorbance at approximately 490 nm.

The \chemr{PLL_{FITC}} solution is stored at a temperature of $-20$ \textdegree C  until used. To introduce the labelled PLL into polyelectrolyte films, multilayers built with unlabelled PLL of the same molecular mass as the labelled one are exposed for five minutes to a \chemr{PLL_{FITC}} solution and rinsed with buffer solution afterwards.

\section{UV--visible spectroscopy}
UV--visible absorbance spectra are acquired with a mc$^2$ spectrophotometer (Safas, Mo\-na\-co). Spectra of buffer solutions containing a molecule of interest are acquired in quartz cuvettes (10 mm light path, Hellma, M\"ullheim, Germany, ref. 6040-UV) or poly(styrene) cuvettes (10 mm light path, Brand GmbH \& Co. KG, Wertheim, Germany, ref. Plastibrand 7590 15) with the corresponding spectra of pure buffer solutions as baseline. For the spectra of deposits on quartz slides, the spectra of the same quartz slides before coating are taken as baseline.

\section{Nuclear magnetic resonance spectroscopy}
Dopamine hydrochloride is dissolved at 2 g/L in Tris buffer (50 mmol/L, pH 8.5) and oxidation followed by melanin formation is allowed for 48 h in a vessel in continuous contact with ambient oxygen. Then the solution is centrifuged during 5 min at a centripetal acceleration of 380 g (BRi4 multifunction centrifuge, Thermo Electron Corporation, Waltham, Massachusetts, USA) and the black-brown powder from the pellet is washed with water and redispersed before further centrifugation during 10 min at 380~g. This purification step is repeated once more and the final powder is dried in an oven at 70 °C during 6 h. At this temperature melanin should not be destroyed, because it does not decomposes below 200 °C according to thermogravimetric analyses \cite{fei:2008} \cite{jaber:2010}.

The dopamine-melanin powder is characterised by solid state carbon-13 cross-pola\-ri\-sa\-tion magic angle spinning nuclear magnetic resonance (\chem{^{13}C} CP/MAS NMR) spectroscopy. Experiments are performed using an AVANCE 500 MHz wide bore spectrometer (Bru\-ker, Wissembourg, France) operating at a frequency of 125.7 MHz for \chem{^{13}C}. A triple resonance MAS probe with spinners of 3.2 mm external diameter is used in order to acquire spectra at a spinning rate of 20 kHz to completely remove spinning sidebands. To get undistorted \chem{^{13}C} lineshapes, a spin echo ($\tau - \pi - \tau$) was added prior to the acquisition of the data points for the regular CP/MAS experiments. All experiments were carried out with a proton pulse of $3.12\: \mu$s, a spin-lock field of 80 kHz for protons and 60 kHz for carbon, a decoupling field of 100 kHz and a recycle time of 5 s. A contact time of 1.2 ms was used to get the full spectrum, while a cross-polarisation time of $35\: \mu$s allowed to polarise only protonated carbons. The total echo time was kept identical in all spectra and equal to two rotation periods ($\tau  = 50\: \mu$s). For the non-quaternary carbon suppression (NQS) experiment the dephasing time was set to $44\: \mu$s.

\section{Scanning force microscopy}
\label{sec:experimentsfm}
Scanning force microscopy (SFM) topographies of \chemr{(PLL-HA)_{30}} films deposited on 12 mm cover glasses are acquired in presence of Tris buffer solution (50 mmol/L, pH 8.5) in dynamic mode using a Nanoscope IV (Veeco, Santa Barbara, California, USA) microscope. The employed cantilevers (model NP 10, Veeco) have a nominative spring constant of 0.06 N/m and are terminated with a silicon nitride tip with a nominative radius of curvature of 20 nm. For determination of the film thickness, the sample is scratched with a syringe needle previously cleaned with ethanol, and the height difference between the scratched and the unscratched area is used as a measure of the thickness.

SFM in contact mode is performed on dopamine-melanin deposits after rinsing with water and drying under a stream of nitrogen. Images are acquired in air at a scanning frequency of 2 Hz using model MSCT (Veeco, spring constant: 0.01 N/m, tip radius: 10 nm) cantilevers.

Images are acquired using the operating software Nanoscope (version 6.13r1, Veeco), which also serves to calculate the arithmetic roughness $R_a$ and the root mean square roughness $R_q$ of the surfaces defined by
\begin{eqnarray}
R_a&=&\frac{\sum_{j=1}^n{|z_j-\bar z|}}{n}, \\
R_q&=&\sqrt{\frac{\sum_{j=1}^n{(z_j-\bar z)^2}}{n}}.
\end{eqnarray}
$z_j$ is the height at one pixel, $\bar z$ the mean height and $n$ the number of pixels in the examined area. In images of scratched regions a plane is fit to the scratched area and subtracted from the whole image. Then sections perpendicular to the scratch are obtained with the Nanoscope software to determine the film height.

Images of unscratched regions are flattened with the open source software Gwyddion (version 2.19, Petr Klapetek, \href{http://www.gwyddion.net}{www.gwyddion.net}). Therefore a second order surface is fit to and subtracted from the height image. Furthermore the height scale is adjusted to obtain a good visibility of the surface topography.

\section{Determination of the refractive index of melanin}
\label{sec:expmelaninrefractiveindex}
Solutions of dopamine hydrochloride are prepared at a concentration of 10 g/L in Tris buffer (50 mmol/L) at pH 8.5. After a given time of contact with ambient oxygen under continuous stirring, the pH of the solutions is adjusted to 13.0 by addition of concentrated sodium hydroxide and fifteen minutes later to 12.0 by addition of concentrated hydrochloric acid. Then the solutions are kept in a closed vessel for at least 24 hours before performing measurements on them.

The real part $n$ of the solutions' refractive index $N = n + ik$ is measured at a wavelength of 589 nm as a function of the initial dopamine hydrochloride concentration with a RFM340 refractometer (Bellingham and Stanley Ltd., Tunbridge Wells, United Kingdom). Therefore the solutions are diluted up to ten times in Tris buffer (50 mmol/L, pH 12.0). To determine the imaginary part $k$ of the solutions' refractive index their absorbance $A$ at wavelengths of 589 nm and 633 nm is measured in poly(styrene) cuvettes. The solutions are diluted 10 to 100 times in Tris buffer (50 mmol/L, pH 12.0) and the absorbance of an empty cuvette is taken as baseline.

The measured absorbance $A$ is transformed to $k$ by the following calculations. $A$ is defined as
\begin{equation}
A=\log \left( \frac{I_0}{I} \right)
\end{equation}
with the initial intensity $I_0$ and the detected intensity $I$. The Lambert-Beer law links these variables to the extinction coefficient $\epsilon$ and the concentration $C$ of the absorbing species as well as the optical path length $d$:
\begin{eqnarray}
I&=&I_0 \exp(-\epsilon C d) \label{eq:lambert-beer}\\
\Rightarrow \epsilon C &=& \frac{\log(I_0/I)}{d\log(e)} = \frac{A}{d\log(e)}
\label{eq:extinctioncoefficient}
\end{eqnarray}
An electromagnetic wave of angular frequency $\omega$ and wavevector $\vec{k}$ with
\begin{equation}
|\vec{k}|=\frac{\omega}{c}(n+ik)
\end{equation}
propagating a distance $d$ in the time $t$ can be described by its electric field vector $\vec E$:
\begin{equation}
\vec E = \vec E_0 \exp[i(|\vec k|d-\omega t)]=\vec E_0 \exp\left[i\omega \left( \frac{nd}{c}-t\right) \right] \exp\left(-\frac{\omega k}{c}d \right).
\end{equation}
$c$ denotes the vacuum speed of light. Since the intensity is proportional to the square of the absolute value of the electric field vector it follows
\begin{equation}
I=I_0 \exp\left(-\frac{2\omega k}{c}d\right) = I_0 \exp\left(-\frac{4\pi k}{\lambda}d \right).
\end{equation}
$\lambda$ is the wavelength of the absorbed light. Comparison with the Lambert-Beer law (Equation \ref{eq:lambert-beer}) leads to
\begin{eqnarray}
\frac{4\pi k}{\lambda}&=& \epsilon C \\
\Rightarrow k &=& \frac{\epsilon C \lambda}{4 \pi} = \frac{A \lambda}{4\pi \log(e) d}.
\label{eq:imaginaryrefractiveindex}
\end{eqnarray}

\begin{figure}
	\centering
		\includegraphics[width=\textwidth]{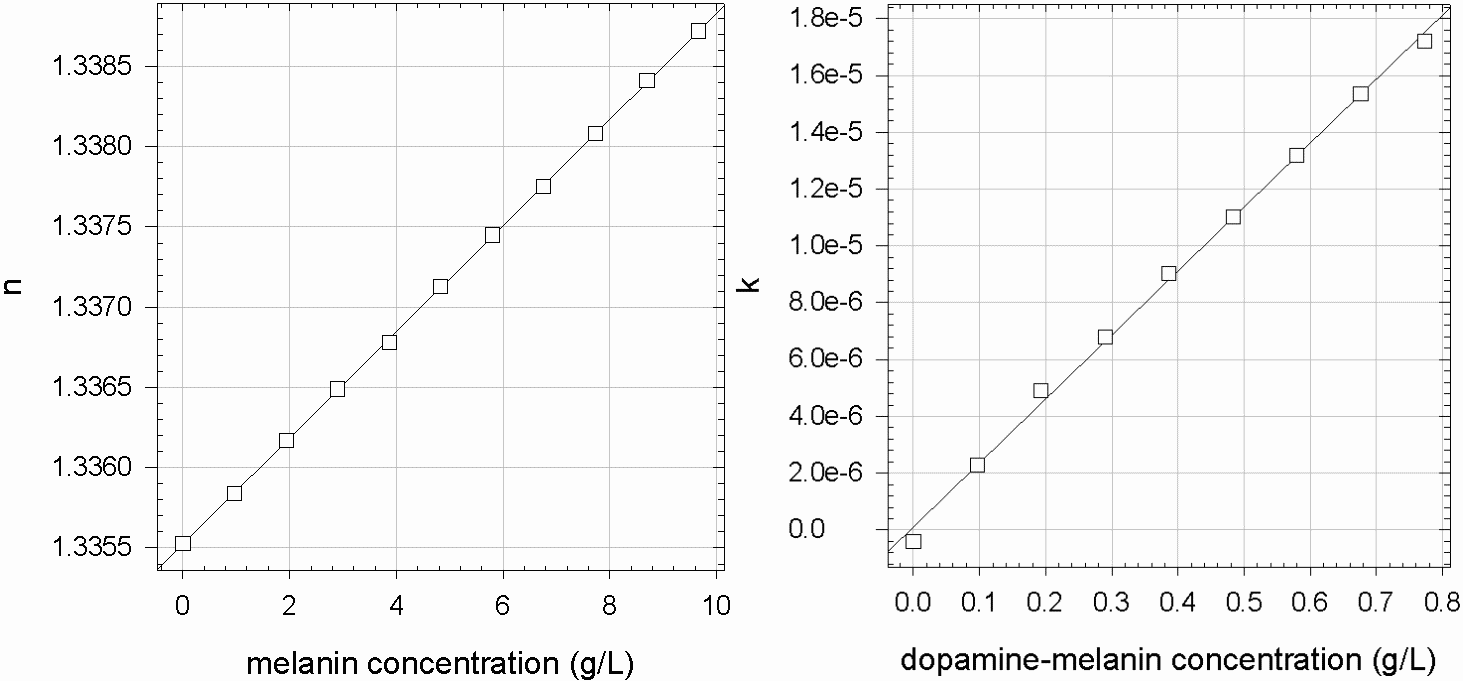}
	\caption[Refractive index of dopamine-melanin solutions versus concentration]{Real part $n$ (left) and imaginary part $k$ (right) of the refractive index of dopamine-melanin solutions  at a wavelength of 589 nm versus concentration with linear regressions. The solutions are obtained from a dopamine hydrochloride solution at 10 g/L allowed to react for 2 h at pH 8.5, titrated to pH 13, then to pH 12.0. For the measurements they are diluted in 50 mmol/L Tris buffer at pH 12.0.}
	\label{fig:refractiveindexmelanin}
\end{figure}

To calculate the refractive index of melanin $N_m$ from the one measured for the solution $N$ and the one of the solvent $N_s$, the following relationship is used:
\begin{equation}
N=\theta N_m + (1-\theta )N_s
\end{equation}
$\theta$ is the volume fraction of melanin in the solution calculated as its concentration $C$ divided by its density $\rho$. Thus
\begin{eqnarray}
N &=& N_s + \frac{(N_m-N_s)}{\rho}C = N_s + \frac{dn}{dC} C \ // \ \mathrm{{for}\  {small}}\ C\\ 
\Rightarrow N_m &=& \rho \frac{dn}{dC}+N_S
\end {eqnarray}
can be used to calculate the refractive index of melanin from the linear concentration dependence of the solution's refractive index. Figure \ref{fig:refractiveindexmelanin} shows that for the employed concentrations the linear relationship is very well fulfilled.

The density of melanin is measured by dissolving 1.607 g of dopamine hydrochloride in 5.0 mL of an aqueous sodium hydroxide solution at 0.1 mol/L. After one day of reaction the volume of the black solution is measured. The density of the formed melanin is given by the ratio of the initial mass to the difference in volume. Therefore it is supposed that the mass of melanin is identical to the initial dopamine hydrochloride mass and evaporation of the solvent is negligible.

\section{Quantification of amine binding sites on melanin}
\label{sec:aminebinding}
Lee and others proposed a possible mechanism for amines to covalently bind to catechol groups present at the surface of dopamine-melanin deposits \cite{lee:2009}. Based on this mechanism, the compound 2-(2-pyridyldithio)ethylamine (PTEA) will be used to quantify the amine binding sites on dopamine-melanin aggregates.

Melanin aggregates are formed in solutions of 2 g/L dopamine hydrochloride in Tris buffer (50 mmol/L, pH 8.5) continuously aerated with an aquarium pump (Rena AIR 50) for 20 h to 24 h. The obtained suspensions are centrifuged for ten minutes at a centripetal acceleration of 380 g in a BRi4 multifunction centrifuge. Then the precipitates are washed twice with water followed by centrifugation steps before being dried in an oven at 90 °C. Weighed aliquots of the dried powder (between 10 mg and 30 mg) are resuspended in 50 mL of Tris buffer (50 mmol/L, pH 8.5) containing a given amount of PTEA (between $10^{-4}$ and $10^{-2}$ mol per gram of melanin).

\begin{figure}
	\centering
		\includegraphics[width=130mm]{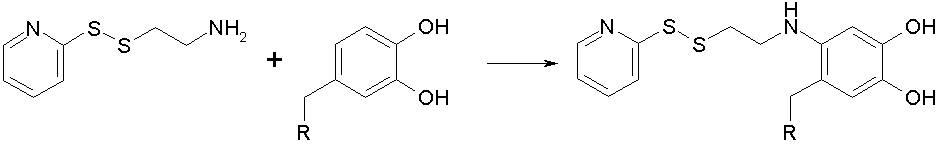}
	\caption[Binding mechanism of PTEA to melanin]{Proposed binding mechanism of 2-(2-pyridyldithio)ethylamine to catechols on dopamine-melanin grains according to \cite{lee:2009}}
	\label{fig:pteamelanin}
\end{figure}
According to \cite{lee:2009} PTEA should bind to catechols on dopamine-melanin grains as shown in figure \ref{fig:pteamelanin}. Since the reaction takes place at a pH of 8.5 and the negative logarithm of the first dissociation constant of catechol groups in dihydroxyindole-melanin is $pK_a = 9.4$ \cite{szpoganicz:2002}, most of the catechol groups are in their fully hydrogenated form.
%The primary amine of PTEA is protonated at pH 8.5 ($pK_a=9.6$, calculated using Marvin Sketch version 5.3.1, ChemAxon Ltd., Budapest, Hungary, available free of charge at \href{http://www.chemaxon.com/marvin/sketch}{www.chemaxon.com/marvin/sketch})

\begin{figure}
	\centering
		\includegraphics[width=85mm]{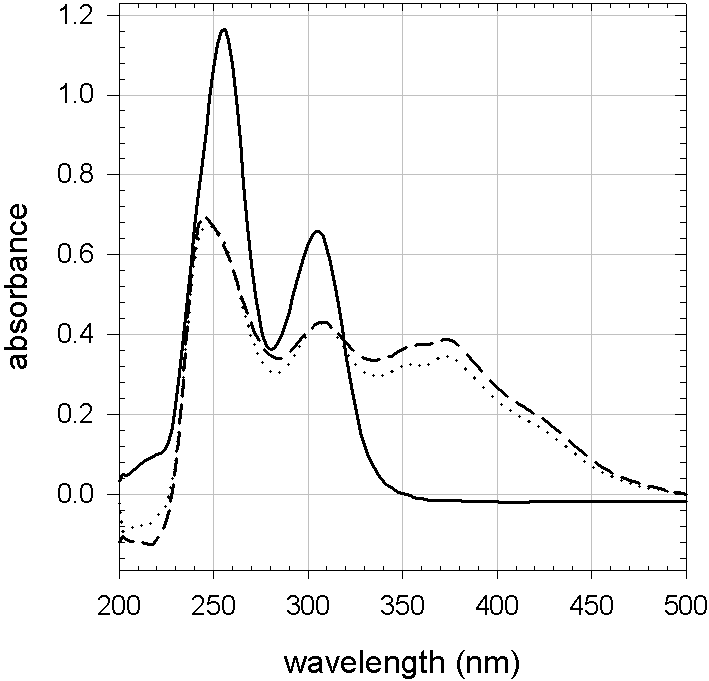}
	\caption[UV-visible spectra of PTEA]{UV-visible spectra of $2.6 \cdot 10^{-4}$ mol/L PTEA in Tris buffer (50 mmol/L, pH 8.5) before (full line) and after 1 h (dashed line) and 2 h (dotted line) of contact with melanin grains. The latter two spectra are corrected for the background caused by melanin remaining in solution.}
	\label{fig:uvvisptea}
\end{figure}

After the addition of melanin to PTEA solutions, the obtained suspensions are continuously shaken in closed centrifuge tubes during 2 h. In the first experiment the suspension is centrifuged (10 min, 380 g) after one and two hours and 0.5 mL of the supernatant are taken for UV--visible spectroscopy in quartz cuvettes. PTEA shows two absorbance peaks at the wavelengths 256 nm and 304 nm. Their intensities after one hour of reaction are smaller than the ones measured for the PTEA solution before melanin addition, but the intensities do not change between 1 h and 2 h (figure \ref{fig:uvvisptea}). Thus the PTEA binding reaction to melanin is completed within less than one hour.

After two hours of reaction with PTEA, the melanin suspensions are centrifuged (10 min, 380 g). The precipitates are washed twice with water followed by centrifugation steps before resuspension in 50 mL of Tris buffer (50 mmol/L, pH 8.5). This procedure removes all unbound PTEA as confirmed by the absence of PTEA-related peaks in the UV--visible spectra of the supernatants after the first washing step. 

\begin{figure}
	\centering
		\includegraphics[width=130mm]{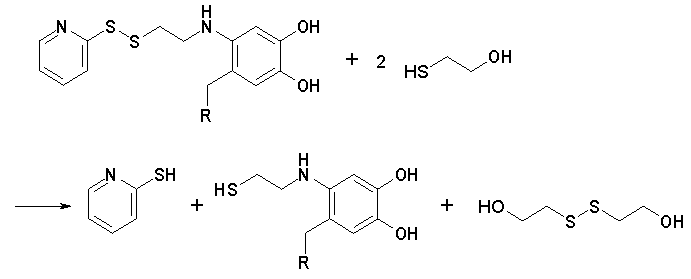}
	\caption[Cleavage of PTEA by 2-mercaptoethanol]{Cleavage of melanin-bound PTEA by 2-mercaptoethanol to liberate pyri\-dine-2-thi\-one.}
	\label{fig:2mercaptoethanol}
\end{figure}
The necessary amount of 2-mercaptoethanol to obtain a 20-fold excess compared to the initial number of PTEA molecules is added to the melanin suspensions after reaction with PTEA. This is done under a fume hood due to the \emph{toxicity of 2-mercaptoethanol}. Mercaptoethanol cleaves the disulfur bond in melanin-bound PTEA molecules to liberate pyridine-2-thione as depicted in figure \ref{fig:2mercaptoethanol}. Pyridine-2-thione has an absorbance peak at the wavelength of 354 nm and the liberated amount is equal to the amount of accessible amine-binding sites. Therefore its absorbance can be used to quantify the amine-binding sites of melanin as proposed in \cite{carlsson:1978} for amine-binding sites of proteins.

\begin{figure}
	\centering
		\includegraphics[width=85mm]{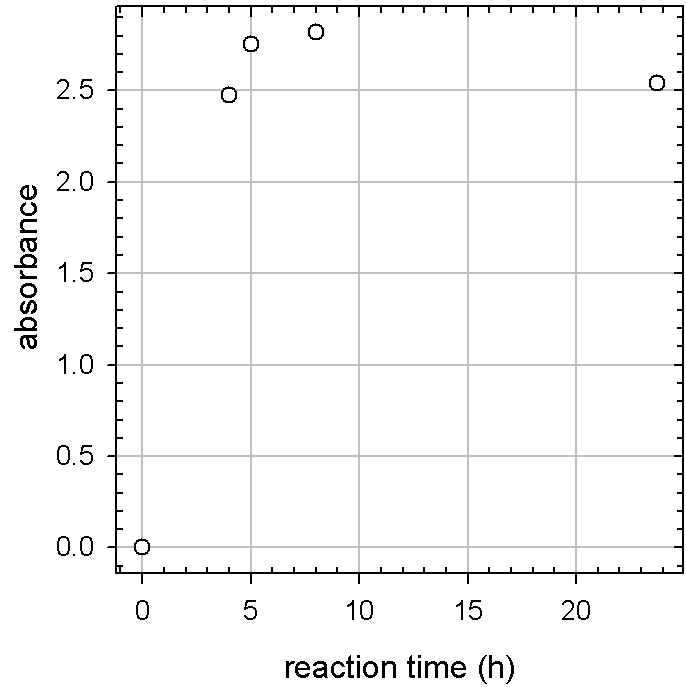}
	\caption[Pyridine-2-thione liberation from melanin-PTEA]{Absorbance of pyridine-2-thione at 354 nm versus reaction time of PTEA bound to 24.3 mg melanin with $2.4 \cdot 10 ^{-3}$ mol 2-mercaptoethanol in 50 mL Tris buffer (50 mmol/L, pH 8.5)}
	\label{fig:abspyridinethione}
\end{figure}

The melanin-PTEA suspensions containing 2-mercaptoethanol are shaken in closed centrifuge tubes protected from light. At given times the suspensions are centrifuged (10 min, 380 g) and aliquots of 0.5 mL of the supernatant are taken for UV--visible spectroscopy. The absorbance at 354 nm is immediately measured for each aliquot to find the maximal absorbance corresponding to the liberation of all pyridine-2-thione molecules from the melanin aggregates. An exemplary plot of the absorbance versus liberation time is shown in figure \ref{fig:abspyridinethione}. The number $n$ of pyridine-2-thione molecules can be calculated from the absorbance $A$, the volume of the suspension $V=50$ mL, the light path of the spectrophotometer cuvette $d=1$ cm, and the extinction coefficient $\epsilon$ using the Lambert-Beer law (Equation \ref{eq:lambert-beer}):
\begin{eqnarray}
\epsilon C &=&\frac{A}{d\log(e)} \ // \ \mathrm{concentration} \ C=\frac{n}{V} \\
\Rightarrow n&=&\frac{AV}{\epsilon \log(e) d}
\end{eqnarray}
This number will be transformed to an amine binding capacity of melanin by dividing it by the mass of melanin used in the experiment.  

\begin{figure}
	\centering
		\includegraphics[width=85mm]{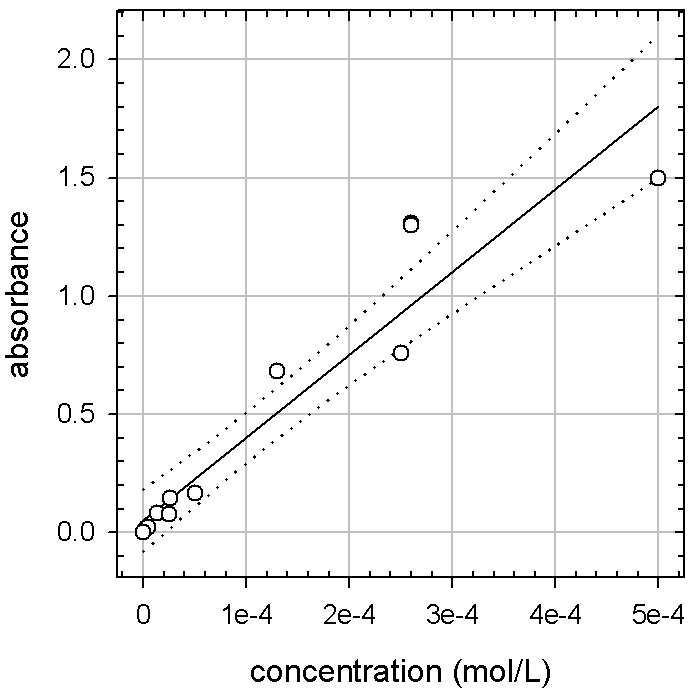}
	\caption[Absorbance of pyridine-2-thione versus concentration]{Absorbance of pyridine-2-thione solutions at 354 nm versus concentration in Tris buffer (50 mmol/L, pH 8.5) in two independent experiments. The solid line corresponds to a linear fit with slope $(3.5 \pm 0.3) \cdot 10^3$ L/mol. The dotted lines delimit the 95 \% confidence interval.}
	\label{fig:dosagepyridinethione}
\end{figure}
The extinction coefficient $\epsilon$ of pyridine-2-thione at a wavelength of 354 nm is determined as follows: A stock solution of $5.0 \cdot 10^{-4}$ mol/L or $2.6 \cdot 10^{-4}$ mol/L PTEA in Tris buffer (50 mmol/L, pH 8.5) is treated with an amount of 2-mercaptoethanol 20-fold larger than the amount of PTEA present in the solution. The UV--visible spectra of this solution stabilize within less than ten minutes after the addition of 2-mercaptoethanol indicating that at this time all PTEA molecules have been cleaved to release pyridine-2-thione. The absorbance at a wavelength of 354 nm is plotted in figure \ref{fig:dosagepyridinethione} as a function of pyridine-2-thione concentration for different dilutions in Tris buffer. The slope of the linear regression $A/C$ and the light path of the spectrophotometer cuvette $d=1$ cm are used to calculate $\epsilon \log(e)$ from the Lambert-Beer law (Equation \ref{eq:lambert-beer}):
\begin{equation}
\epsilon \log(e)=\frac{A}{dC}=(3.5 \pm 0.3)\cdot 10^3 \mathrm{\frac{L}{mol \: cm}}
\end{equation}

\section{Quartz crystal microbalance}
\label{sec:experimentqcm}
Quartz crystal microbalance with dissipation (QCM-D, model D300, Q-Sense, Gö\-te\-borg, Sweden) is used to follow the deposition of melanin from dopamine solutions by method A in situ. 0.5 mL of dopamine solution are injected through the flow cell (about 0.1 mL in volume) above the quartz crystal and regularly replaced by injection of fresh dopamine solutions.

The oscillations of the quartz crystal are excited close to its resonance frequency (about 5 MHz). The exciting signal is stopped every 3 s, and the decay of the crystal's oscillation is followed as a function of time at the fundamental frequency as well as at the third, fifth and seventh overtone. From these decay curves, the operating software of the QCM-D (QSoft301) calculates the oscillation frequency and the dissipation, which measures the energy loss of the crystal in contact with the deposit. The change in the resonance frequency divided by the overtone number with respect to the frequency in contact with pure buffer solution is calculated by the software QTools (version 1.2.2.35, Q-Sense). It will be called reduced frequency change $\Delta f_\nu /\nu$, where $\nu$ is the overtone number ($\nu \in \{1; 3; 5; 7\}$). For ideally thin and rigid films, the reduced frequency changes for all overtones overlap and the adsorbed mass per unit area of the surface can be calculated using the Sauerbrey equation \ref{eqn:sauerbrey} as explained in section \ref{sec:theoryqcm}. The  employed crystals are characterised by the following relationship between the reduced frequency change and the adsorbed mass per surface area $\Gamma$:
\begin{equation}
\Gamma = \frac{\Delta m}{A} = -16.6 \frac{\Delta f_\nu}{\nu} \mathrm{\frac{ng}{cm^2 Hz}}.
\label{eqn:sauerbreynumerical}
\end{equation}

Furthermore $\Gamma$ and the density $\rho$ of the deposit can be used to calculate its thickness
\begin{equation}
d=\frac{\Gamma}{\rho}.
\label{eqn:qcmthickness}
\end{equation}

The quartz crystal microbalance also serves to monitor the adsorption of proteins on melanin films. After the last melanin deposition step and five minutes of rinsing with buffer solution, 0.5 mL of protein solution (1 g/L) in Tris buffer (50 mmol/L, pH 8.5) are injected through the measuring chamber. After 20 min the protein solution is replaced by pure buffer solution for 10 min, then for another 10 min to 15 min by sodium dodecylsulphate (SDS, 10 mmol/L in water) and finally by pure buffer solution. During the whole procedure, the adsorption and desorption of proteins is monitored via the oscillation frequency changes of the quartz crystal.

Melanin deposition experiments using the electrochemical method (Method D, Section \ref{sec:expdepositionmethods}) are carried out in the electrochemical module (QEM 401, Q-Sense) of another QCM-D (model E4, Q-Sense). Here, a gold-covered quartz crystal serves as working electrode and the counter electrode is a platinum plate parallel to the crystal surface. The reference electrode is a \chem{Ag/AgCl} electrode. The adsorbed mass is followed as described above for the D300 QCM-D during electrochemical melanin deposition.

\section{Ellipsometry}
\label{sec:expellipsometry}
Ellipsometry measurements are performed at a wavelength of 632.8 nm and an angle of incidence of 70\textdegree \ with a rotating analyser ellipsometer (model PZ 2000, HORIBA Jobin Yvon, Longjumeau, France). The angle of the polariser transmission axis is 45\textdegree \ with respect to the plane of incidence. All measurements are taken on nitrogen-dried samples in ambient air. The thickness values are given as the average ($\pm$ one standard deviation) over 5 to 10 measurements taken along the major axis of the rectangular samples.

Before the deposition of polyelectrolyte multilayers (PEM), measurements are taken on the cleaned silicon substrates. The operating software of the ellipsometer (PQ Diamond, version 1.4b) calculates the thickness of the silicon oxide layer on the substrate from the measured ellipsometric angles using the refractive index 1.465 of silicon oxide and assuming the layer to be homogeneous and isotropic. After PEM deposition, the total thickness of the PEM and the oxide layer is determined the same way. The thickness of the PEM is obtained by subtracting the thickness of the oxide layer from the measured total thickness. It is a good approximation to fix the refractive index of the PEM in the dry state at 1.465 because fully hydrated $\mathrm{(PLL-HA)_n}$ films have a refractive index between 1.42 and 1.43 at 632.8 nm \cite{picart:2002}.

Single-wavelength measurements lead to a set of possible thickness values with a given periodicity and not to one unique value (Equation \ref{eq:periodicity}). This ambiguity is solved either by following the thickness during film build-up at thickness intervals much smaller than the periodicity (281.6 nm for the experimental conditions) or by comparing the thickness to direct measurements by scanning force microscopy (SFM).

Contrary to the employed polyelectrolytes, melanin strongly absorbs light in the visible range and has thus a non-zero imaginary part of the refractive index. As a first approximation the thickness is calculated anyway with an imaginary part of 0. Consequently the obtained thickness values are only rough estimations.

In a second more refined analysis, the thickness of dopamine-melanin is calculated using a two-layer model. The first layer consists of silicon oxide, and its thickness is measured as described above before the deposition of dopamine-melanin. Then the thickness and refractive index of this layer are fixed in the operating software of the ellipsometer. The refractive index of dopamine-melanin is independently determined from absorbance and refraction measurements in solution (Section \ref{sec:melaninrefractiveindex}) as well as absorbance measurements of dopamine-melanin deposits (Table \ref{tab:compmelaninabs}). Using this refractive index, the operating software can model the dopamine-melanin deposit as a second layer on top of the silicon oxide layer to calculate its thickness.

\section{X-ray photoelectron spectroscopy}
XPS is carried out using a SES 200-2 X-ray photoelectron spectrometer (Gammadata Scienta, Uppsala, Sweden) under ultra-high vacuum (pressure: $10^{-11}$ Pa). The monochromatized \chem{Al}K source (photon energy: 1486.6 e\/V) is operated at a current of 30~mA and a voltage of 14~kV. Spectra are acquired at a take-off angle (TOA) of 90\textdegree \ or 30\textdegree \ between the sample surface and photoemission direction. The samples are outgassed in several vacuum chambers with isolated pumping systems and pressure control before transfer to the analysis chamber. This procedure leads to removal of all degassing species and to complete dehydration of the melanin deposits. The samples are analysed without any further cleaning process and a possible carbon contamination might remain on the surface. During acquisition, the pass energy is set to 500 e\/V for the survey spectrum and to 100 e\/V for high-resolution spectra.

Classical Scofield sensitivity factors are used for peak fitting procedures with the CASA\-XPS software: \chem{C1s: 1.00, O1s: 2.83, N1s: 1.80}. All line shapes used in peak fitting procedures are a mix of 30 \% Gaussian and 70 \% Lorentzian. To limit errors on background determination due to low signal to noise ratio for the \chem{N1s} high-resolution spectra, the two limit points of the Shirley-type background are averaged over 21 experimental points. All components on high-resolution spectra are referenced to the \chem{CH_x} aromatic component of melanin at 284.6 e\/V.

Based on the attenuation of the silicon 2p signal, it is possible to evaluate the mean thickness $d$ of the dopamine-melanin deposit under ultra-high vacuum conditions using the following equation:
\begin{eqnarray}
I_t&=&I_0 \exp \left( \frac{-d}{\lambda \sin (\theta)} \right) \\
\Leftrightarrow d &=& \lambda \sin(\theta) \ln\left(\frac{I_0}{I_t} \right)
\label{eqn:xpsthickness}
\end{eqnarray}
$I_0$ is the \chem{Si2p} peak intensity of a pristine silicon slide, $I_t$ the corresponding intensity after melanin deposition, $\theta$ the take-off angle and $\lambda$ the inelastic mean free path of \chem{Si2p} photoelectrons in dopamine-melanin, which is estimated to be 3.0 nm \cite{tanuma:1993}. Using this value, one can calculate the probing depth of the technique by the relation 
\begin{equation}
L_{XPS}=3\lambda \sin(\theta)
\label{eqn:xpsprobingdepth}
\end{equation}
giving a value of 9 nm for a take-off angle of 90\textdegree \ (maximum probing depth) and 4.5 nm  at a take-off angle of 30\textdegree \ (for more surface sensitive analysis). The intensity emitted by atoms located at the limit of the probing depth is attenuated by more than 95 \% by the overlying deposit.

\section{Contact angles}
The static contact angles of water with dopamine-melanin deposits on silicon slides are measured in a Digidrop device (model ASE, GBX, Bourg de Peage, France). There\-fore four to six drops of 6 $\mu$L of water are automatically placed on the sample. The operating software WinDrop++ allows calculating the contact angle between a drop and the sample from an image of the drop taken by a digital camera.

\section{Cyclic voltamperometry}
The permeability of dopamine-melanin deposits to ferrocene\-methanol, hexa\-cyano\-ferrate and hexa\-amine\-ruthenium is determined by cyclic voltamperometry (CV) in a conventional three electrode set-up (model CHI 604B, CH Instruments, Austin, Texas, USA). The reference and counter electrodes are a \chem{Ag/AgCl} electrode (CHI 111) and a platinum wire (CHI 115). The working electrodes are made of amorphous carbon (CHI 104). They are polished on $\gamma$-alumina powder with a particle diameter of 50 nm (Buehler, Lake Bluff, Illinois, USA, ref. 40-6325-008) using an Escil polisher (Chassieu, France). Three successive polishing steps of two minutes are performed, separated by rinsing with water. Then the electrodes are sonicated twice during 3 min in a Transonic TI-H-50 sonicator (Laval Lab, Laval, Canada) at a frequency of 130 kHz in water.

The capacitive and the faradaic currents are measured by cycling the potential versus \chem{Ag/AgCl} between -0.1~V and 0.65~V at a scanning rate of 0.05 V/s. Solutions are deoxygenated before each measurement by nitrogen bubbling for at least 5 min. CV is first performed on the pristine electrode. The capacitive current is measured in a buffer solution containing 10 mmol/L of Tris and 0.15 mol/L of sodium nitrate (\chem{NaNO_3}) at pH 7.5. Sodium nitrate is added to the solutions, because CV experiments need to be performed in the presence of a high supporting electrolyte concentration to avoid a contribution of ion migration to the measured currents. The faradaic current is measured in the same buffer solution in the presence of 1 mmol/L of the electrochemical probe. The experiment is only continued if the potential difference between the oxidation and reduction peak is lower than 80 mV. Theoretically it should be of 59 mV for a reversible one-electron process taking place at a temperature of 298 K \cite{bard:2001}.

The working electrode is coated with dopamine-melanin by one of the methods described in section \ref{sec:expdepositionmethods} and rinsed with a buffer solution containing Tris (10 mmol/L) and sodium nitrate (150 mmol/L, pH 7.5) before performing CV as on the pristine electrode. According to Lee and others \cite{lee:2007.2}, the deposition of dopamine-melanin should follow the same mechanism on amorphous carbon as on silicon oxide.

\section{Streaming potential}
The streaming potentials of dopamine-melanin films deposited on glass slides are measured with a ZetaCAD device (CAD Instrumentation, Les Essarts le Roi, France). Two glass slides are mounted parallel to each other in the poly\-(methyl meth\-acrylate) (PMMA) sample holder separated by a poly\-(tetra\-fluor\-ethylene) (PTFE) spacer. For all measurements Tris buffer with a concentration of 5 mmol/L is circulated between the samples. The streaming potential is measured on the same substrate five times before and five times after coating with dopamine-melanin, and the reported values are given as averages with their standard deviation. Two kinds of experiments are done:
\begin{enumerate}
\item	Experiments aimed to define the number of immersion steps required to reach a steady value of the  $\zeta$-potential indicating that a dopamine-melanin film entirely covers the glass substrate. These experiments are performed at a constant pH of 8.5, the same pH as during dopamine-melanin deposition.
\item	pH titration experiments aimed to measure the $\zeta$-potential of a dopamine-melanin film as a function of pH and to correlate its behaviour to the film's chemical composition.
\end{enumerate}

The $\zeta$-potential is calculated from the streaming potential $\Delta E / \Delta P$ using the Smoluchowski relationship \cite{hunter:1988}:
\begin{equation}
\zeta = \frac{\eta \lambda}{\epsilon \epsilon_0}\frac{\Delta E}{\Delta P}.
\end{equation}
$\zeta$ is the  $\zeta$-potential, $\eta$ the solution viscosity, $\lambda$ the solution conductivity and $\epsilon \epsilon_0$ the dielectric permittivity of water. The potential difference $\Delta E$ is measured between two \chem{Ag/AgCl} reference electrodes located at both ends of the measurement cell. The pressure difference $\Delta P$ between the ends of the measurement cell is varied using compressed air by increments of 2 kPa between $-25$ kPa and $25$ kPa. Since the viscosity and the dielectric permittivity are temperature-dependant, their values are regularly calculated using the measured temperature of the buffer solution. The solution conductivity is directly measured in situ.

By neglecting the potential decrease occurring in the Stern layer and any contribution of specific ion adsorption\footnote{This would not be valid in the presence of transition metal cations that are specifically bound by melanin \cite{hong:2007}.}, the surface potential is identified with the $\zeta$-potential to estimate the surface charge density $\sigma$. Therefore the Grahame equation \cite{butt:2006}
\begin{equation}
\sigma \approx \sqrt{8 C_0 \epsilon \epsilon_0 k T} \, \sinh \! \left(\frac{e \zeta}{2kT}\right)
\label{eq:grahame}
\end{equation}
is used with the concentration of the supporting electrolyte $C_0$ (in m$^{-3}$), the Boltzmann constant $k$ ($1.38 \cdot 10^{-23}$J/K), the temperature $T$ ($298$ K), the dielectric constant of water $\epsilon$ ($78.4$ at $T = 298$ K), the permittivity of vacuum $\epsilon_0$ ($8.85\cdot 10^{-12} \mathrm{C^2 m^{-2} N^{-1}}$) and the elementary charge $e$ ($1.60\cdot 10^{-19}$C).  

\section{Infrared spectroscopy}
\chemr{PEI-(HA-PLL)_n} films are built on a zinc selenide (\chem{ZnSe}) crystal from deuterium oxide (\chem{D_2O}) solutions containing Tris buffer (50 mmol/L at pH 8.9 to account for the 0.4 pH units difference between \chem{H_2O} and \chem{D_2O}) and polyelectrolytes at a concentration of 1~g/L. In contrast to the other experiments, the films used for Fourier transform infrared spectroscopy in the attenuated total reflection mode (ATR-FTIR) are initiated with a layer of poly(ethyleneimine) (PEI), which is necessary as an anchoring layer on the zinc selenide crystal. Deuterium oxide is used instead of water for all solutions, because water shows a strong absorbance in the investigated wavenumber region.

After each polyelectrolyte adsorption from flowing solutions atop the trapezoidal zinc selenide crystal during 5 min, the polyelectrolyte solution is  replaced by Tris buffer and the infrared spectrum of the film is acquired by accumulating 512 scans at 2 cm$^{-1}$ spectral resolution in an Equinox 55 spectrometer (Bruker, Wissembourg, France). The detector is a liquid nitrogen cooled mercury cadmium telluride (MCT) detector. The transmitted intensity $I$ is compared to the one transmitted by the pristine crystal $I_0$ to calculate the absorbance $A$ of the film by
\begin{equation}
A=\log \left( \frac{I_0}{I} \right).
\end{equation}
When the film build-up is complete, the Tris buffer solution is replaced by a dopamine hydrochloride solution (2 g/L) in Tris buffer to follow the incorporation of dopamine and melanin in the film. The \chemr{PEI-(HA-PLL)_n} film is built to the level at which no further increase in absorbance can be observed upon deposition of additional layer pairs. Thus the thickness of the film is significantly higher than the penetration depth of the evanescent wave from the zinc selenide crystal into the polyelectrolyte film. There is no significant increase in the absorbance of the amide I band (between 1600~cm$^{-1}$ and 1700~cm$^{-1}$) of PLL between n = 9 and n = 12 (Figure \ref{fig:ftirpllhadopamine} A). Hence a \chemr{PEI-(HA-PLL)_{12}} film is used to investigate the incorporation of dopamine and melanin. 

\section{Confocal laser scanning microscopy}
To study the mobility of poly-L-lysine (PLL) chains labelled with fluorescein iso\-thio\-cyanate (\chemr{PLL_{FITC}}) in films of PLL and hyaluronate (HA), the fluorescence recovery after photobleaching (FRAP) technique is used. A LSM 510 inverted confocal  laser scanning microscope (CLSM) (Zeiss, Oberkochen, Germany) with a 25 mW argon ion laser for imaging and bleaching is employed for the FRAP experiments. Cover glasses supporting the polyelectrolyte films are placed in a home made sample holder in Tris buffer solution (50 mmol/L, pH 8.5). The FITC molecules are excited at a wavelength of 488 nm and the emitted fluorescence is collected at wavelengths between 505 nm and 530 nm. The employed immersion objective (Plan Neofluar, Zeiss) has a 40 x magnification and a numerical aperture of 1.3.

\label{sec:confocallinescan}
Before the actual FRAP experiments, some stacks of line scans in the sample plane with a length of $230.3 \; \mu$m are acquired at a resolution of 512 pixels using a confocal aperture of 1 Airy unit. The distance of the line scans normal to the sample plane are calculated by the operating software of the microscope to obtain an optimal resolution following the rules established in section \ref{sec:clsmimageprocessing}. The scans of one stack are combined to obtain virtual sections normal to the sample plane that are used to check whether the films present a homogeneous thickness.

In the FRAP experiment a square area in the sample plane of $(230.3 \; \mu m)^2$ is imaged using a raster of 512 by 512 pixels. To obtain a depth of field comparable to the film thickness, the confocal aperture is opened to 5 Airy units. A disk of 111 pixels ($50.0 \; \mu m$) diameter in the centre of the imaged area is bleached by scanning it up to 300 times with the full laser power. Then a time series of images is acquired with the laser power reduced to 2 \% to minimize further bleaching of the sample. For the first 10 min an image is taken every 2~min, from 10~min to 45~min the time interval between two images is 5~min, then it is 15~min up to a total time of 2~h. Further images are taken every 30~min up to a total time between 3~h and 4~h after bleaching. Longer observation times are not used since the sample position in the microscope shifts by some micrometers during the employed times and larger shifts would make exploitation of the images impossible.

\label{sec:expfrap}
The obtained image series is analysed with the open source software ImageJ (version 1.42q, Wayne Rasband, National Institute of Health, USA, \href{http://rsb.info.nih.gov/ij/}{http://rsb.info.nih.gov/ij/}). The mean fluorescence intensity in the bleached region is measured and divided by the mean intensity in the region between the circle and the square marked in figure \ref{fig:frapregions} using the macro frap.txt (see appendix \ref{sec:fraptxt}). The choice of this reference region minimises the influence of the bleached region and possible artefacts at the image borders on the reference intensity. The inevitable further bleaching of the sample during image acquisition should not influence the relative intensity $I_r$ obtained this way.
\begin{figure}
	\centering
		\includegraphics[width=70mm]{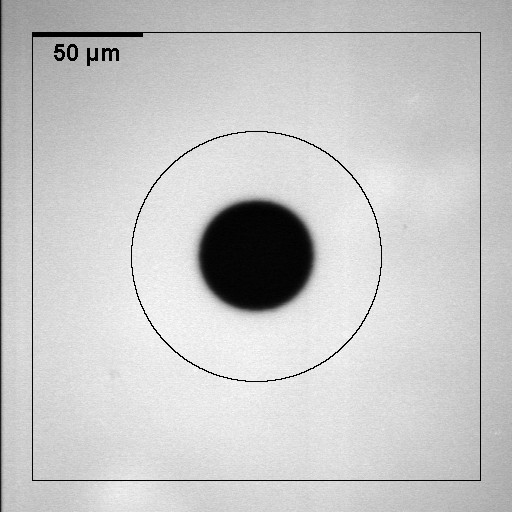}
	\caption[Confocal microscopy image of \chemr{(PLL-HA)_{30}-PLL_{FITC}}]{Confocal laser scanning microscopy image of a \chemr{(PLL-HA)_{30}-PLL_{FITC}} film. The bleached disk in the centre has a diameter of 111 pixels ($50.0 \; \mu$m), the circle (diameter: 250 pixels, $112.5 \; \mu$m) and the square (edge length: 448 pixels, $201.5 \; \mu$m) delimit the region used to measure the reference intensity for normalizing the fluorescence intensity in the bleached disk.}
	\label{fig:frapregions}
\end{figure}

The results are represented as plots of the relative fluorescence intensity in the bleached region versus the square root of the time since the end of bleaching. In this representation the speed of fluorescence recovery corresponds to the slope of the curves. To interpret these curves, the following assumptions are made:
\begin{itemize}
	\item The measured fluorescence intensity is proportional to the concentration of unbleached fluorescent molecules.
	\item The recovery of fluorescence in the bleached region is due to the diffusion of unbleached \chemr{PLL_{FITC}} molecules from outside into this region.
	\item The diffusion is two-dimensional and isotropic.
	\item There are two populations of FITC-labelled molecules, a mobile one with a constant diffusion coefficient $D$ and an immobile one. 
	\item At the end of bleaching, the bleached region is a disk of uniform concentration of unbleached molecules.
\end{itemize}
With these assumptions the speed of fluorescence recovery depends on the fraction $p$ of mobile \chemr{PLL_{FITC}} molecules, their diffusion coefficient $D$, the residual fluorescence intensity $\alpha$ right after bleaching and the radius $a$ of the bleached disk. By fitting the theoretical function 
\begin{equation}
I_r(p,\alpha,\tau)=\alpha+p(1-\alpha)\exp\left( -\frac{2}{\tau} \right) \left[ \mathbf{I_0}\left( \frac{2}{\tau} \right) + \mathbf{I_1} \left( \frac{2}{\tau} \right) \right] \ \mathrm{with} \ \tau = \frac{4 D t}{a^2} \label{eq:fraptheory}
\end{equation}
derived in \cite{picart:2005} to the experimental values, one obtains the parameters $D$, $p$ and $\alpha$. $I_r$ is the relative mean intensity in the bleached area at the time $t$ after bleaching, $\mathbf{I_\nu}(x)$ are the modified Bessel functions of first kind and order $\nu$.

%% file: results.tex
\chapter{Results}

\input{resultsdopamineinsolution}

\input{resultsdepositionmethods}

\input{resultsnx5min}

\input{resultsdopamineinpllha}

%% file: resultsdopamineinsolution.tex
\section{Formation of dopamine-melanin in solution}
\label{sec:resultsdopamineinsolution}
\subsection{Absorbance measurements}
When dopamine is dissolved in an alkaline buffer solution containing 50 mmol/L tris\-(hydroxy\-methyl)\-amino\-methane (Tris) at pH 8.5 in contact with ambient air, the solution immediately changes its colour from colourless via orange to dark brown. After some hours a black precipitate appears at the bottom of the reaction vessel. This behaviour, which is typical of the formation of melanin by spontaneous oxidation of dopamine, was described several times in the literature \cite{barreto:2001} \cite{herlinger:1995}. The product of the reaction will be called dopamine-melanin to remind its origin.

\begin{figure}
	\centering
		\includegraphics[width=85mm]{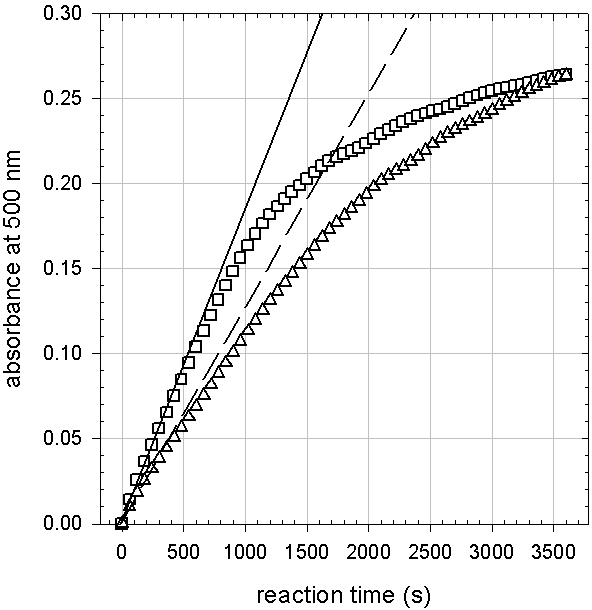}
	\caption[Absorbance of dopamine solutions versus reaction time]{Absorbance at 500 nm of dopamine hydrochloride solutions at 1 g/L (squares) and 0.5 g/L (triangles) in Tris buffer (50 mmol/L, pH 8.5) versus reaction time. Full and dashed lines: linear fits to the data sets over the first 300 s.}
	\label{fig:absorbancedopaminesolution}
\end{figure}
\begin{figure}
	\centering
		\includegraphics[width=85mm]{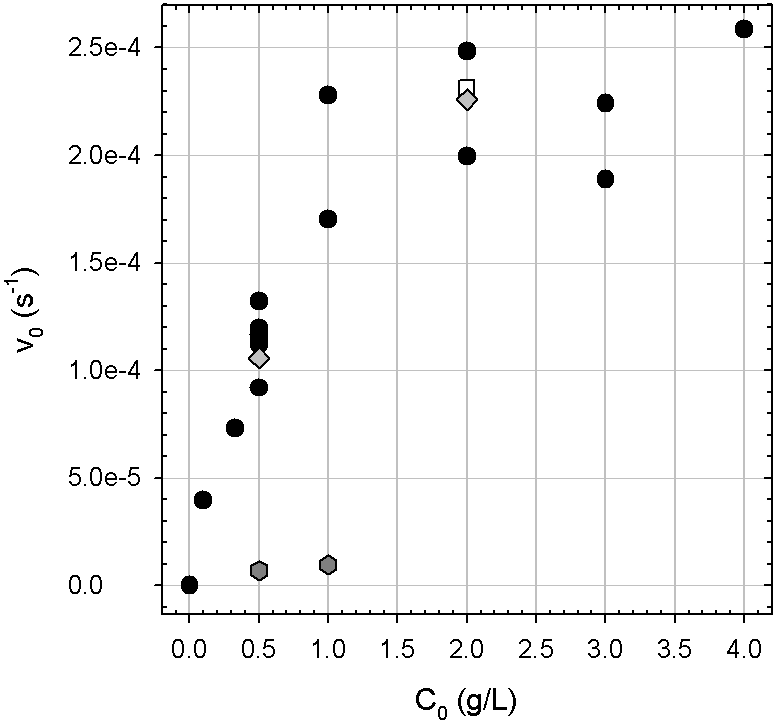}
	\caption[Initial reaction rate of dopamine versus concentration]{Initial reaction rate \chem{v_0} of dopamine hydrochloride versus its initial concentration \chem{C_0} in Tris buffer (50 mmol/L, pH 8.5) (circles). Similar experiments are performed in additional presence of 1 g/L HA (squares) or 1 g/L PLL (diamonds), or after deaerating the buffer by argon bubbling (hexagons).}
	\label{fig:dopaminepolymerisationrate}
\end{figure}

During the first minutes the absorbance of the dopamine solutions measured at a wavelength of 500 nm increases linearly with time (Figure \ref{fig:absorbancedopaminesolution}). The slope \chem{v_0} of the linear regression of the absorbance versus time curves during the first five minutes is used to express the initial reaction rate of dopamine. For dopamine hydrochloride concentrations below 1 g/L the initial reaction rate increases linearly with the initial dopamine concentration while for higher dopamine hydrochloride concentrations a plateau is reached (Figure \ref{fig:dopaminepolymerisationrate}). The following experiments are performed at an initial dopamine hydrochloride concentration of 2 g/L, if not otherwise mentioned, to work at the maximum reaction speed.

\begin{figure}
	\centering
		\includegraphics[width=120mm]{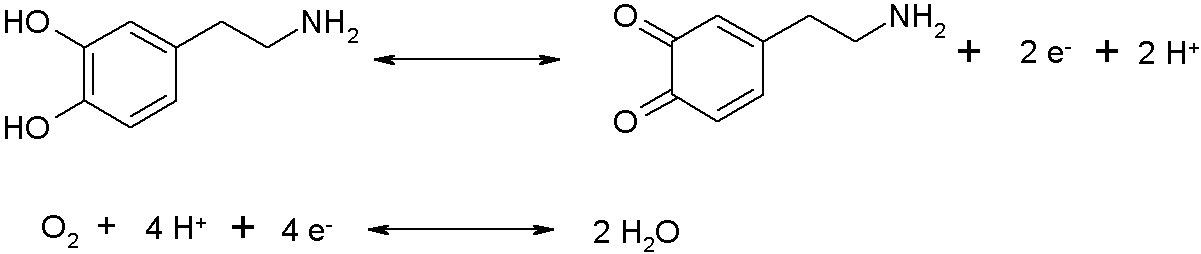}
	\caption[Oxiadation of dopamine to dopamine quinone]{Oxiadation of dopamine to dopamine quinone by reduction of oxygen to water}
	\label{fig:dopamineoxidation}
\end{figure}

The linear regime is consistent with a reaction of apparent first order with respect to dopamine. Electrochemical investigations have identified the oxidation of dopamine to dopamine quinone as the first step leading to melanin formation \cite{li:2006.1} (Figure: \ref{fig:dopamineoxidation}). Dissolved oxygen is the only species susceptible to oxide dopamine in pure water. Therefore the reaction is probably limited by the depletion of oxygen at high dopamine concentrations leading to a plateau value of the reaction rate. This assumption is supported by a strongly decreased initial reaction rate in deaerated solutions (Figure \ref{fig:dopaminepolymerisationrate}).

\begin{figure}
	\centering
		\includegraphics[width=85mm]{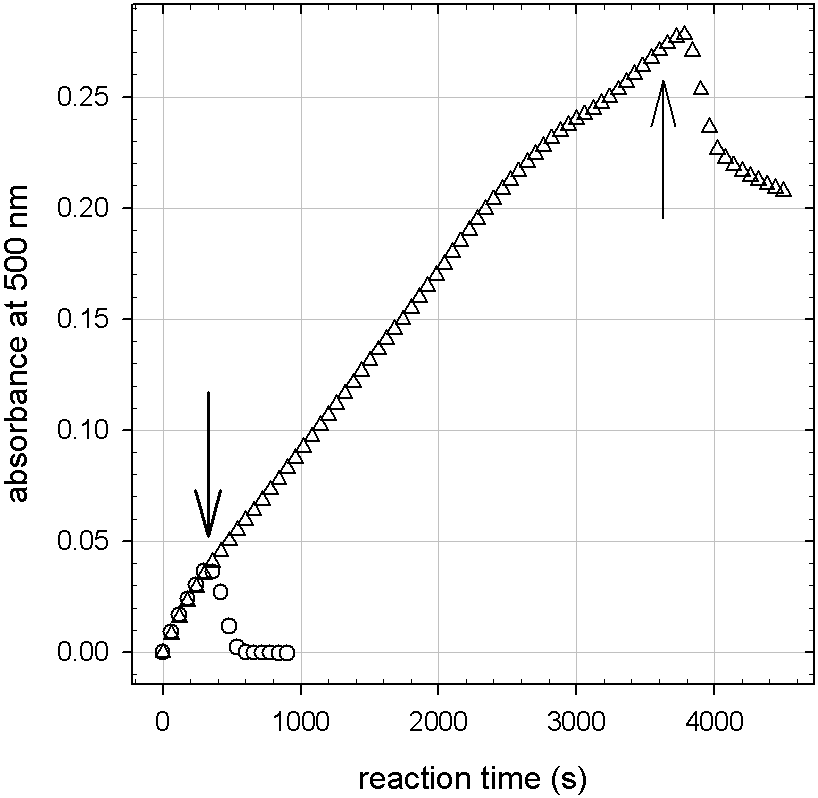}
	\caption[Influence of acidification on the kinetics of dopamine transformation]{Influence of a pH change from 8.5 to 3.5 on the reaction kinetics of dopamine hydrochloride solutions (0.5 g/L in 50 mmol/L Tris buffer) followed by the absorbance at a 500 nm. Arrows correspond to addition of 50 $\mu$L of concentrated hydrochloric acid to the solution (1.4 mL) at two different times (5 min for circles, 60 min for triangles).}
	\label{fig:polymerisationwithph}
\end{figure}

If the initial step of the reaction is oxidation of dopamine as depicted in figure \ref{fig:dopamineoxidation}, this step should be reversed by a decrease in pH according to Le Châtelier’s principle. This is the case as shown in figure \ref{fig:polymerisationwithph}. When the dopamine solution is acidified to pH 3.5 after five minutes of reaction, the brownish colour of the solution totally disappears and the absorbance at 500 nm decreases to zero. However, when the reaction is allowed to proceed for a longer time, the relative decrease in absorbance after addition of hydrochloric acid is lower. Thus after a longer reaction time dopamine quinone has been consumed, probably by cyclisation to leucodopaminechrome \cite{li:2006.1}, which will undergo further reactions to form dopamine-melanin.

\subsection{Nuclear magnetic resonance}
\begin{figure}
	\centering
		\includegraphics[width=110mm]{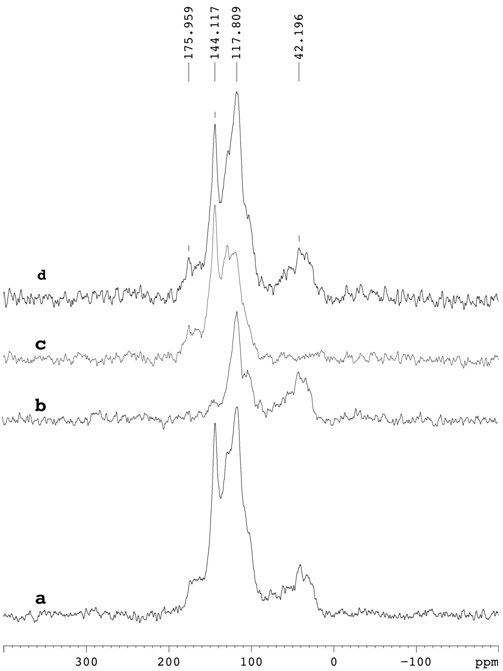}
	\caption[\chem{^{13}C} NMR spectra of dopamine-melanin]{Solid state \chem{^{13}C} NMR spectra of dopamine-melanin. All spectra obtained at 20 kHz spinning rate. (a) Regular spectrum with 1.2 ms contact-time showing all \chem{^{13}C} resonances. (b) Short contact-time ($35\: \mu$s) spectrum exhibiting only protonated carbons (aliphatics and aromatics). (c) Non-quarternary carbon suppression with a $43\: \mu$s dephasing delay for quaternary carbon detection. The sum (d) of (b) and (c) mimics well the regular spectrum shape (a).}
	\label{fig:nmrmelanin}
\end{figure}

The polymerisation of dopamine is commonly used to produce synthetic melanin \cite{herlinger:1995} \cite{peter:1989}. Solid-state carbon-13 nuclear magnetic resonance (\chem{^{13}C} NMR) spectra (figure \ref{fig:nmrmelanin}) of the precipitate from dopamine solutions at an initial dopamine hydrochloride concentration of 2 g/L are measured after 48 hours of reaction at pH 8.5. They allow identifying the reaction product as melanin, because the spectra closely resemble the ones published by Peter and F\"orster \cite{peter:1989} for dopamine-melanin.

\begin{figure}
	\centering
		\includegraphics[width=\textwidth]{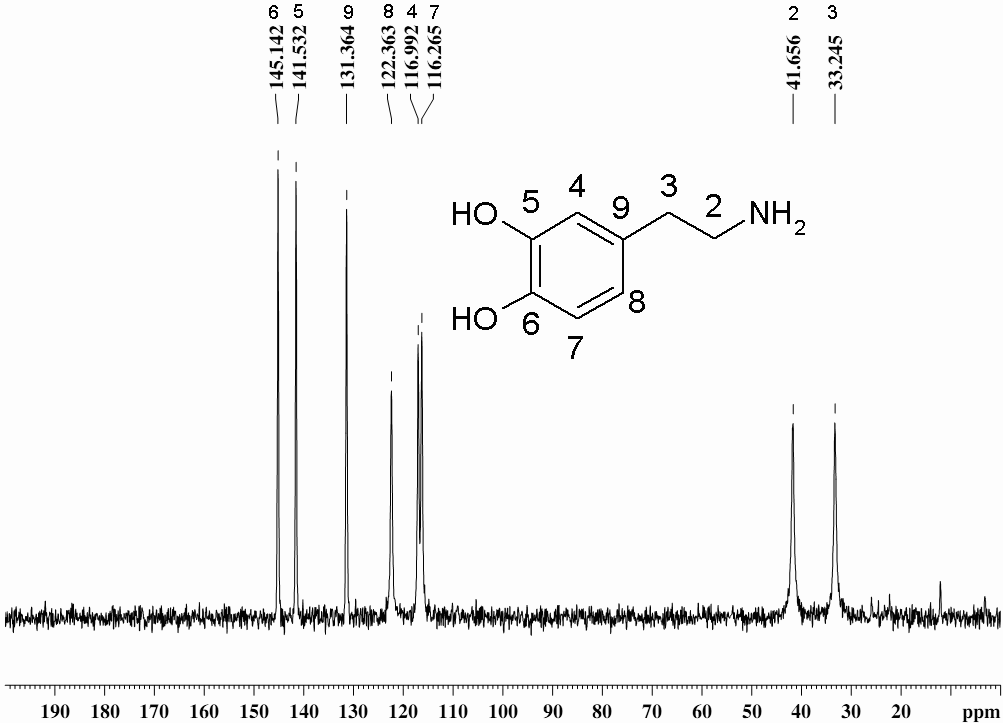}
	\caption[\chem{^{13}C} NMR spectrum of dopamine]{Solid-state \chem{^{13}C} NMR spectrum of dopamine obtained at 20 kHz spinning rate with 1.2 ms contact-time showing all \chem{^{13}C} resonances.}
	\label{fig:nmrdopamine}
\end{figure}

\label{sec:nmrresults}
To assign the peaks in the NMR spectra to the carbon atoms present in dopamine-melanin, a NMR spectrum of dopamine powder (Figure \ref{fig:nmrdopamine}) is acquired with the same parameters as the melanin spectra (Figure \ref{fig:nmrmelanin}). In figure \ref{fig:nmrdopamine} the peaks and the carbon atoms in the dopamine structure are numbered according to the assignment in \cite{adhyaru:2003} and \cite{peter:1989}. Thus the following protonated carbons are recognised in dopamine-melanin (figure \ref{fig:nmrmelanin} b): C4, C7 and C8 of the benzene ring with a chemical shift $\delta$ around 118 ppm and the side-chain carbons C2 and C3 at $\delta$ between 30 ppm and 40 ppm. The signals of C2 and C3 are weaker than in the dopamine spectrum but remain visible showing that only part of the dopamine molecules has undergone cyclisation to form an indole structure during the dopamine-melanin preparation. The non-quarterny suppression spectrum (Figure \ref{fig:nmrmelanin} c) contains the signals of unprotonated carbons: C5 and C6 at the hydroxy sites with $\delta \approx 144$ ppm and C9 with $\delta \approx 130$ ppm. Pyrrolic carbons from dopamine molecules after cyclization to indoles can contribute to the signal at $\delta \approx 120$ ppm \cite{jaber:2010} and the peak at $\delta \approx 176$ ppm is usually assigned to carboxyl groups that might be present in the sample as impurities. The appearance of indole structures is consistent with the melanin formation mechanism described in section \ref{sec:melanin}.

\subsection{Amine binding capacity of melanin}
\label{sec:resultsaminebinding}
\begin{figure}
	\centering
		\includegraphics[width=85mm]{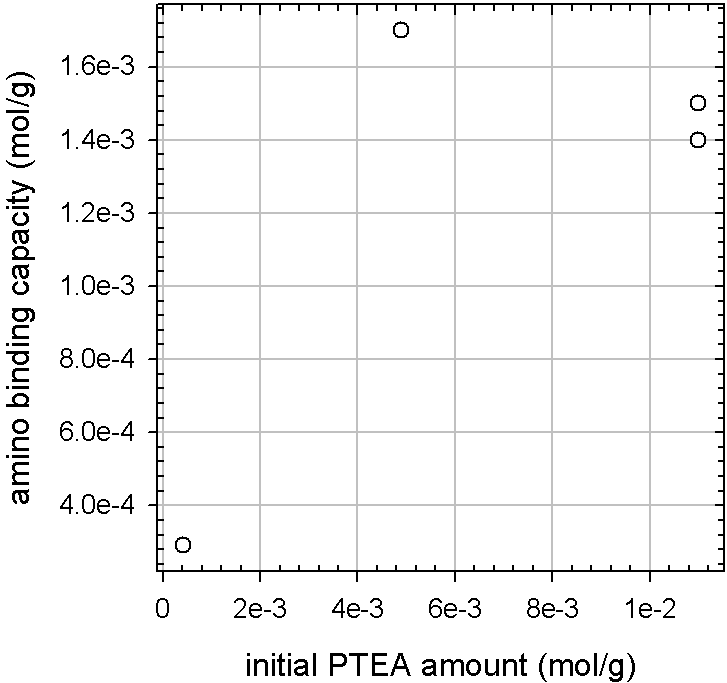}
	\caption[Amine binding capacity of melanin grains]{Amine binding capacity of dopamine-melanin powder (in moles of amine groups per gram of melanin) versus employed amount of 2-(2-pyridine\-di\-thiol)\-ethyl\-amine (in moles of PTEA per gram of melanin)}
	\label{fig:aminobindingcap}
\end{figure}

Lee and others have proposed that biomolecules can be immobilised by covalent binding of amine groups to catechol groups of melanin \cite{lee:2009}. To confirm the binding of amine groups to dopamine-melanin, the binding capacity of melanin powder obtained by precipitation from dopamine-melanin solutions for 2-(2-pyridinedithiol)ethylamine (PTEA) is quantified as described in section \ref{sec:aminebinding}. Figure \ref{fig:aminobindingcap} shows the amine binding capacity of melanin (in moles of amine groups per gram of melanin) as a function of the employed amount of PTEA (in moles of PTEA per gram of melanin). For amounts of PTEA higher than $4 \cdot 10^{-3}$ mol/g the calculated amine binding capacity is constant showing that all accessible sites are occupied by PTEA molecules. Using the three experiments in this regime, a mean binding capacity of $(1.5 \pm 0.2) \cdot 10^{-3}$ mol/g is calculated. The fact that the PTEA molecules are not removed in the rinsing steps strongly indicates that they are bound covalently to the dopamine-melanin powder.

\subsection{Dopamine-melanin grains}
\label{sec:resultsmelaningrains}
\subsubsection{UV--visible spectroscopy}
When a solution of dopamine hydrochloride in Tris buffer at pH 8.5 is allowed to react with ambient oxygen, a black precipitate made of dopamine-melanin forms. This precipitate can be redispersed to a black solution, which does not sediment, by titrating it to pH 13 and then back to pH 12. This is in line with the observation of Bothma and others that melanin is soluble in strongly alkaline media \cite{bothma:2008}. The solubilisation in basic conditions is most probably due to deprotonation of phenolic and carboxylic moieties carried by dopamine-melanin aggregates \cite{ito:1986}.

\begin{figure}
	\centering
		\includegraphics[width=85mm]{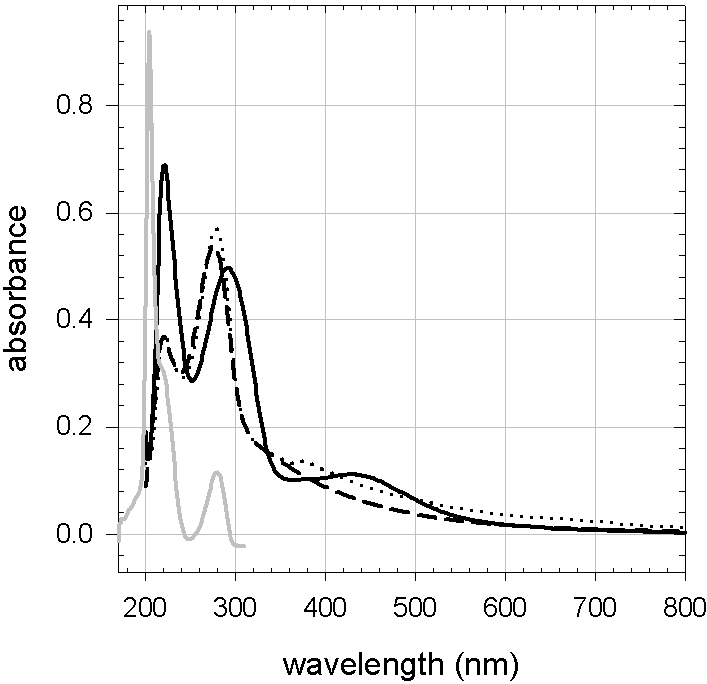}
	\caption[UV--visible spectra of melanin grains at pH 12]{UV--visible spectra of a dopamine-melanin solution 0.5 h (full black line), 24 h (dotted line) and 168 h (dashed line) after titration to pH 12. To prepare the solution, 2 g/L dopamine hydrochloride in  Tris buffer (50 mmol/L) is kept at pH 8.5 for 2 h, then the solution is titrated to pH 13 and back to pH 12. For spectrum acquisition samples are diluted 1/100 in Tris buffer at pH 12. The grey line is the spectrum of 0.01 g/L dopamine in 0.15 mol/L \chem{NaCl} at pH 5.9.}
	\label{fig:uvvismelaninph12}
\end{figure}

The evolution of a solution containing small melanin grains is followed by UV--visible spectroscopy as a function of time after titration to pH 12 (Figure \ref{fig:uvvismelaninph12}). The spectra stabilise after about 24 h and they do not correspond to the monotonic spectra of dopamine-melanin grown on quartz slides (Figure \ref{fig:compmelaninuvvis}). In addition to a monotonic background characteristic of melanin, there are two absorbance peaks around wavelengths of 220~nm and 280~nm. Their positions are close to the absorbance maxima of dopamine solutions under conditions where no melanin formation occurs (in presence of 0.15 mol/L \chem{NaCl} at pH 5.9). Others have also observed such peaks on low molar mass fractions of melanin \cite{simon:2000}. A third broad peak around 440 nm is only visible in the beginning and disappears after less than 2 h. The spectra can be explained by a different electronic structure of the assumed small melanin grains compared to the extended aggregates formed by dopamine-melanin growth on solid substrates. %In the following, solutions of dopamine-melanin grains are left for at least 24 h at pH 12, before further experiments are done.

\begin{figure}
	\centering
		\includegraphics[width=85mm]{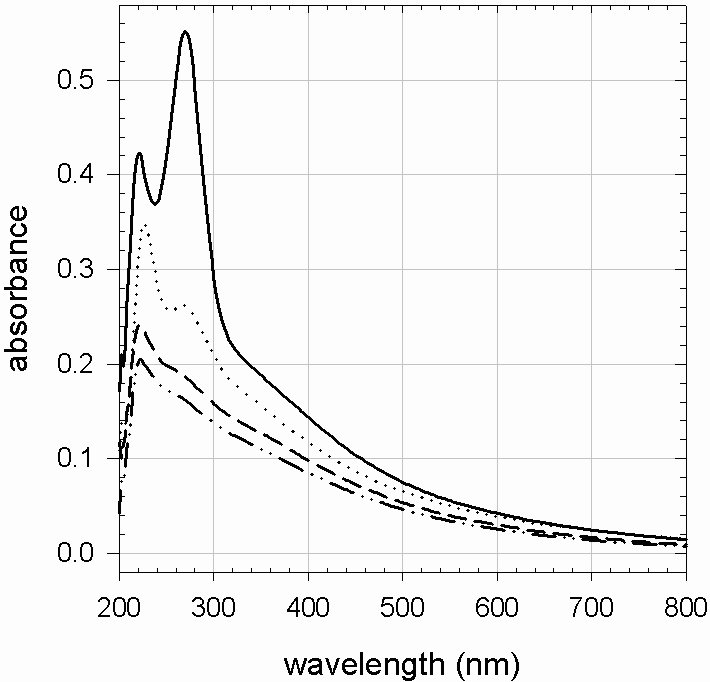}
	\caption[UV--visible monitoring of melanin grain dialysis]{UV--visible spectra of a melanin grain solution (5 mL with 50 mmol/L Tris at pH 12) after 0 h (full line), 5 h (dotted line), 23 h (dashed line) and 47 h (dash-dotted line) of dialysis against 495 mL Tris buffer (membrane with MWCO at $2\cdot 10^3$). For spectrum acquisition samples are diluted 1/100 in Tris buffer.}
	\label{fig:dialysismelanin}
\end{figure}

To have a first idea of the dopamine-melanin grains' size the solutions are dialysed against Tris buffer (50 mmol/L) at pH 12 through cellulose membranes with a relative molecular mass cut-off (MWCO) at $2\cdot 10^3$ or $5\cdot 10^4$ (Spectra/Por 7, Spectrum Laboratories, Rancho Dominguez, California, USA, ref. 132108 or 132130). Therefore melanin grains are prepared by reaction of 2 g/L dopamine hydrochloride in Tris buffer (50 mmol/L) for 2 h at pH 8.5. Then the pH is titrated to 13 and after 15 min to 12, and the dialysis is started after at least 3 d at pH 12. UV--visible spectra of a melanin grain solution at different dialysis times using the membrane with smaller MWCO are shown in figure \ref{fig:dialysismelanin}. The two absorbance peaks vanish at different speeds and the spectrum approaches the monotonic spectrum of a dopamine-melanin deposit (Figure \ref{fig:compmelaninuvvis}). This evolution is also seen using the other dialysis membrane. From these results one may conclude that the melanin grain solution contains at least three different species: two with a molar mass below 2 kg/mol and one with a molar mass above 50 kg/mol. Knowing the molar mass of dopamine (153 g/mol), these thresholds correspond to 13 and 327 monomers. Interestingly in the model of melanin structure as an aggregate of small oligomers first proposed by Cheng \cite{cheng:1994} and Zajac \cite{zajac:1994} (Section \ref{sec:melanin}), the \emph{fundamental unit} is composed of three to five sheets of four to eight indole units. Thus the number of monomers in one fundamental units is between 12 an 40, and the small entities of roughly 13 monomers or less, which cross the dialysis membranes, might be single fundamental units.

\subsubsection{Transmission electron microscopy}
\begin{figure}
	\centering
		\includegraphics[width=\textwidth]{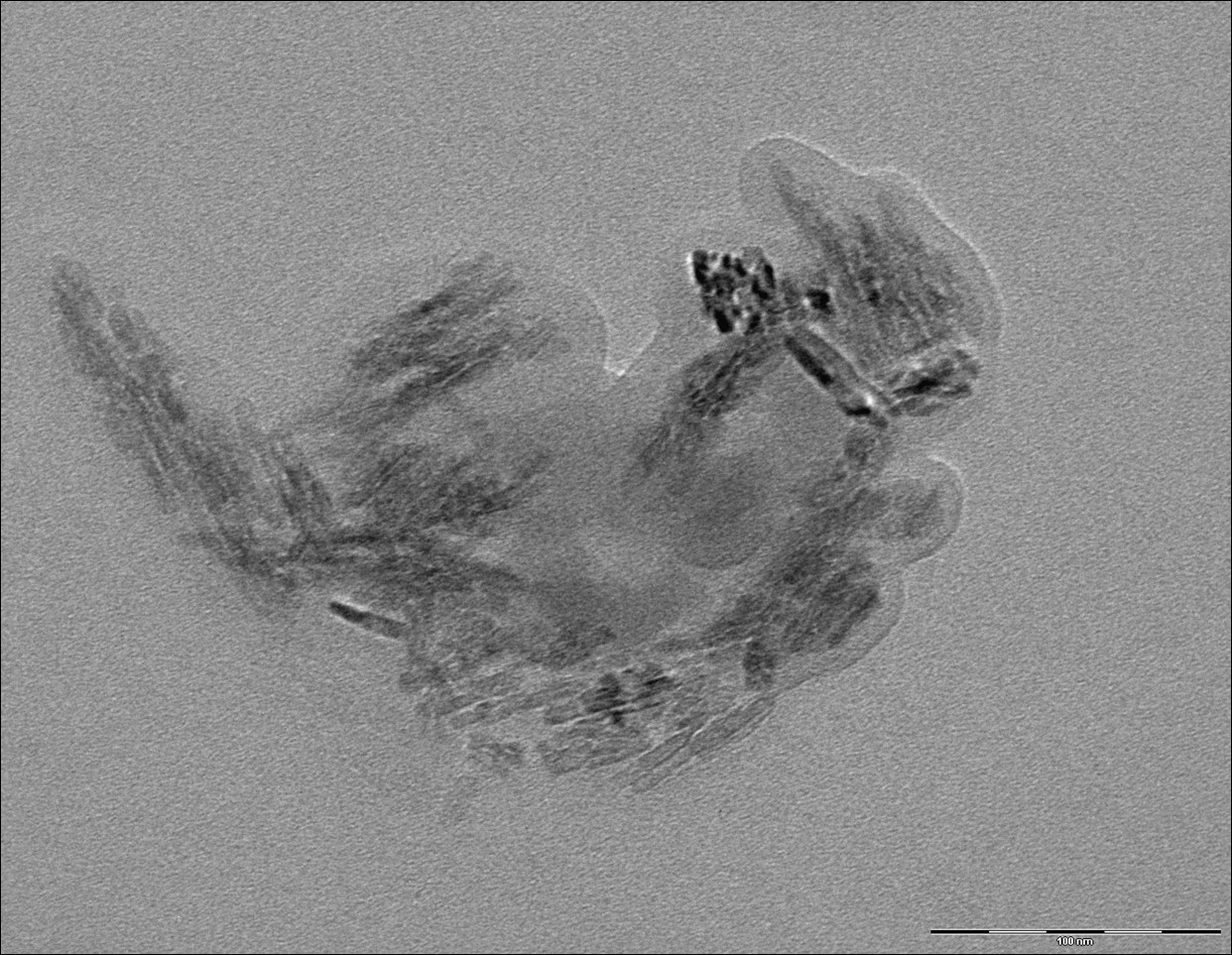}
	\caption[TEM image of a melanin grain]{TEM image of a melanin grain obtained from the reaction of 2 g/L dopamine hydrochloride at pH 8.5 for 24 h under agitation. Scale bar: 100 nm}
	\label{fig:temmelaningrain}
\end{figure}
Melanin grains are obtained by the reaction of 2 g/L dopamine hydrochloride at pH 8.5 in 50 mmol/L Tris for 24 h under agitation in contact with ambient air. Then the pH is titrated to 13 and subsequently to 12. The melanin solution is cast 24 h later on poly(vinyl) formal coated microscopy grids and dried under nitrogen. Transmission electron microscopy (TEM) images are acquired in ultra-high vacuum at 200 kV acceleration voltage and a resolution of 0.18 nm with a TOPCON 002B microscope.

A typical TEM image of a melanin grain of irregular shape and a lateral size of roughly 200 nm by 300 nm is shown in figure \ref{fig:temmelaningrain}. Within the grain, rod-like substructures of 10~nm in diameter and 20~nm to 60~nm in length are visible. Such a hierarchical structure was also observed by Clancy and Simon in SFM images of eumelanin from sepia officinalis \cite{clancy:2001}. The authors described aggregates of about 150~nm in size composed of particles with a diameter of 20~nm and filaments of 5~nm width and 35~nm length. These data can be explained by a melanin structure as aggregates of \emph{fundamental units}. The larger structures would consist of fundamental units held together by non-covalent interactions. Nevertheless the imaging data alone cannot prove the proposed model.   

\subsubsection{Density and refractive index}
\label{sec:melaninrefractiveindex}
Using the method described in section \ref{sec:expmelaninrefractiveindex} the density of the dopamine-melanin grains is calculated as ($1.2 \pm 0.1$) g/mL in agreement with theoretical (1.3 g/mL \cite{kaxiras:2006}) and experimental (1.27 g/mL \cite{cheng:1994} or 1.4 g/mL \cite{gallas:1999}) values published for synthetic melanin.

The refractive index of dopamine-melanin grains depends only weakly on the polymerisation time at pH 8.5 for time intervals from 0.5 to 48 h. For two hours of polymerisation the real and imaginary part of the refractive index are calculated as $n = 1.73 \pm 0.05$ and $k = 0.027 \pm 0.002$ at a wavelength of 589 nm. At 632.8 nm the imaginary part is $k = 0.022 \pm 0.002$. These values are close to the ones estimated in \cite{yoshioka:2002} and \cite{zi:2003} ($N=2.0 + 0.01i$) for natural melanin in peacock feathers by simulation of the reflection spectrum of the feathers.

\subsubsection{PDADMA-melanin multilayers}
Dopamine-melanin grains can be used to build a multilayer film with the polycation poly(diallyldimethylammonium) (PDADMA). Therefore melanin grains are prepared by oxidation of 2 g/L dopamine hydrochloride for 2 h at pH 8.5 in 50 mmol/L Tris buffer. Then the pH is titrated to 13.0 and after 15 min to 12.0. After ageing of the obtained solutions for up to seven days at pH 12, a support is dipped alternately in the solutions of melanin grains and in solutions of PDADMA (1 g/L in 50 mmol/L Tris buffer, pH 12.0) with intermediate rinsing in water to grow a film denoted \chemr{(PDADMA-melanin)_n} after $n$ deposition cycles.

\begin{figure}
	\centering
		\includegraphics[width=\textwidth]{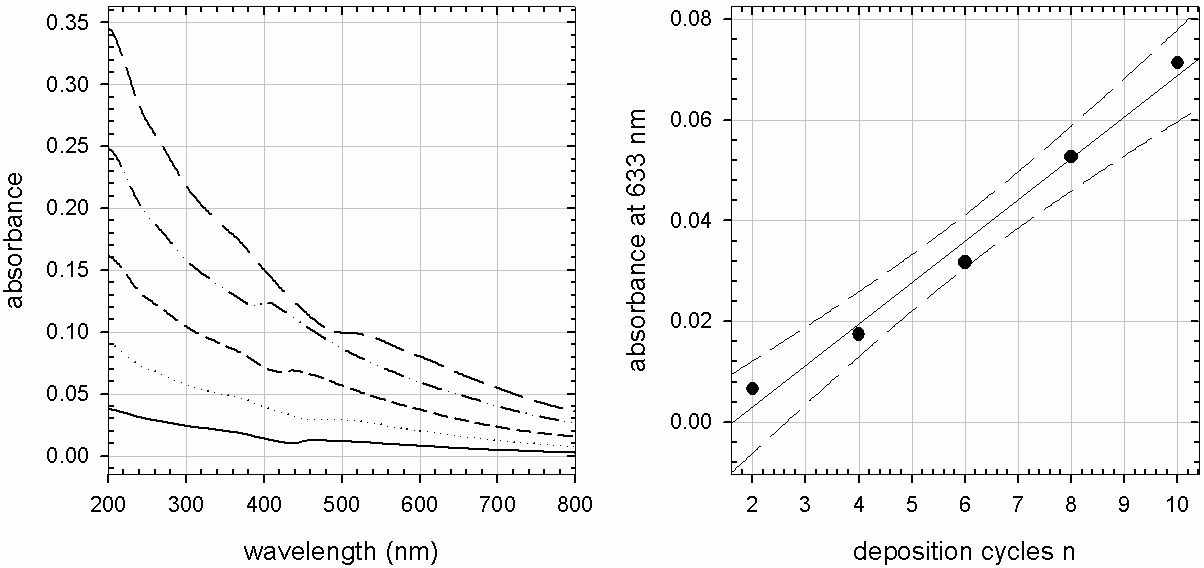}
	\caption[UV--visible spectroscopy of \chemr{(PDADMA-melanin)_n}]{UV--visible spectroscopy of a \chemr{(PDADMA-melanin)_n} film built on  a quartz slide using a melanin solution aged for 3.5 days at pH 12. Left: Spectra after 2 (full line), 4 (dotted line), 6 (short-dashed line), 8 (dash-dotted line) and 10 (long-dashed line) deposition cycles. Right: Absorbance at 633 nm of \chemr{(PDADMA-melanin)_n} versus n with a linear regression (full line) and its 95 \% confidence intervals (dashed line).}
	\label{fig:uvvispdadmacmelanin}
\end{figure}

Ellipsometry measurements assuming a refractive index of 1.465 of the film, a value commonly used for polyelectrolyte multilayers \cite{picart:2002}, are used to calculated the film thickness in the dry state. The choice of an entirely real refractive index is certainly an oversimplification for a material absorbing at the employed wavelength of 633 nm (Figure \ref{fig:uvvispdadmacmelanin}), but it is justified by the confirmation of the calculated thickness by SFM measurements (Figure \ref{fig:sfmpdadmacmelaninscratch}).

\begin{figure}
	\centering
		\includegraphics[width=\textwidth]{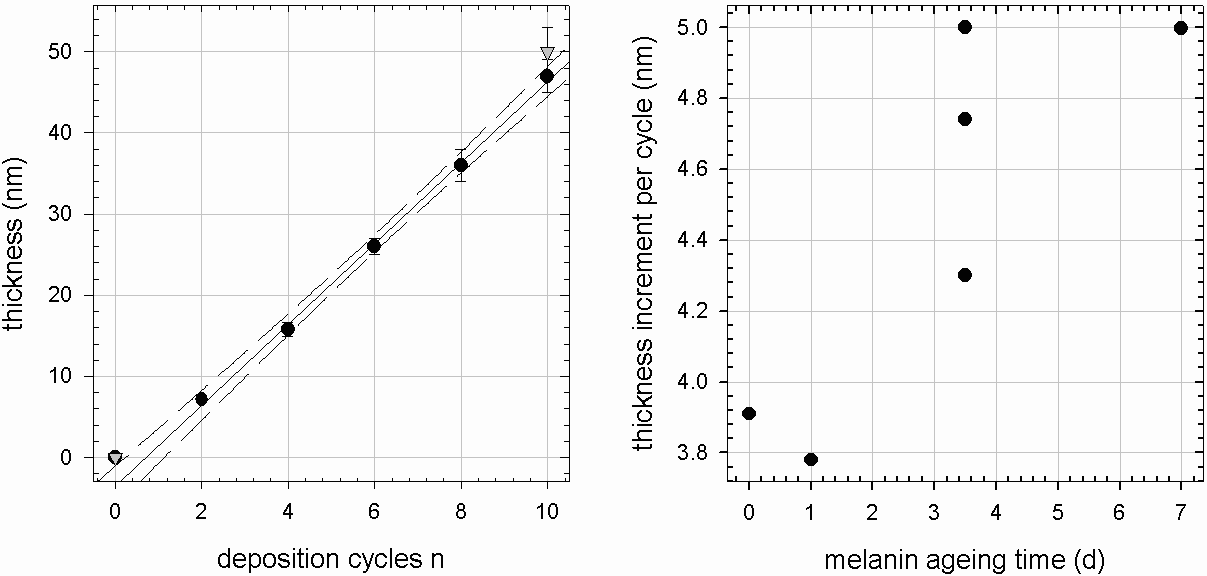}
	\caption[Ellipsometry of \chemr{(PDADMA-melanin)_n}]{Ellipsometry of \chemr{(PDADMA-melanin)_n} films built on silicon slides. Left: Thickness measured (assumed refractive index 1.465) every second cycle (circles) or only in the end (triangles) using a melanin solution aged for 3.5 days at pH 12 for film build-up. The straight line is a linear regression to the data represented by circles with its 95 \% confidence intervals delimited by dashed lines. Right: Thickness increment per deposition cycle (obtained by linear regressions) versus ageing time at pH 12 of the employed melanin solutions.}
	\label{fig:ellipsopdadmacmelanin}
\end{figure}

Figure \ref{fig:ellipsopdadmacmelanin} shows a linear increase of the thickness of a \chemr{(PDADMA-melanin)_n} film with $n$. This indicates that the melanin grains are negatively charged at pH 12 allowing them to form complexes with the positively charged PDADMA. There is no thickness increase upon repeated immersion of a substrate in melanin grain solutions. In figure \ref{fig:ellipsopdadmacmelanin} two types of experiments are represented: Either the sample is dried after every second deposition cycle to measure the film thickness or it is dried only once at the end. There is no visible influence of intermediate drying steps on the film growth. The thickness increment per deposition cycle, obtained by linear regression to the data, is about 4 nm to 5 nm and increases slightly with increased ageing time at pH 12 of the employed melanin solution. Furthermore the stability of \chemr{(PDADMA-melanin)_{10}} films in water or in Tris buffer (50 mmol/L, pH 8.5) at ambient temperature is followed by ellipsometry. It is found that in both cases the thickness does not change for at least 40 days.

\begin{figure}
	\centering
		\includegraphics[width=85mm]{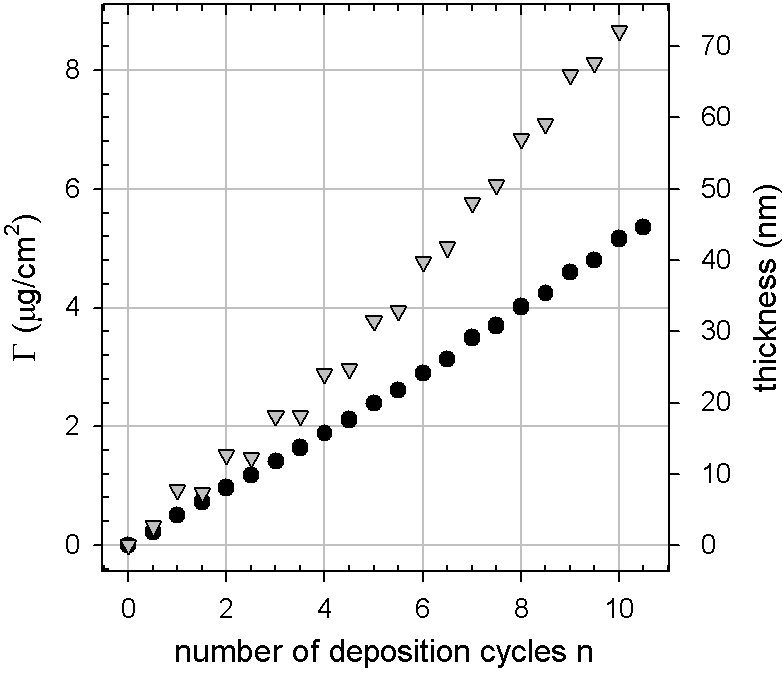}
	\caption[\chemr{(PDADMA-melanin)_n} deposition followed by QCM]{Mass per surface area $\Gamma$ and thickness of \chemr{(PDADMA-melanin)_n} films calculated using the Sauerbrey approximation from frequency changes at the third overtone measured by QCM at the end of the rinsing following each deposition step. The employed melanin solution was aged at pH 12 for 5 (triangles) or 6 (circles) days before film deposition.}
	\label{fig:qcmpdadmacmelanin}
\end{figure}

The build-up of \chemr{(PDADMA-melanin)_n} films is also followed by quartz crystal microbalance with dissipation (QCM-D). The dissipation of the deposits is small ($< 10^{-5}$) and the reduced frequency changes at different overtones overlap. Therefore the Sauerbrey approximation (Equation \ref{eqn:sauerbreynumerical}) is used to calculate the mass per surface area $\Gamma$ and the thickness of the deposit assuming a density of 1.2 g/cm$^3$ (Equation \ref{eqn:qcmthickness}). The values represented in figure \ref{fig:qcmpdadmacmelanin} are recorded at the third overtone (frequency $\approx 15$~MHz) at the end of the rinsing of the measurement cell following each deposition step. The calculated thickness after ten bilayers, which varies between 40~nm and 70~nm for different experiments, lies in the same range as the thickness obtained by ellipsometry measurements (Figure \ref{fig:ellipsopdadmacmelanin}). In one QCM experiment the thickness increment is much larger in melanin deposition steps than in PDADMA deposition steps, but this behaviour is not seen in the second experiment.

The UV--visible spectra of a \chemr{(PDADMA-melanin)_n} film built on a quartz slide using a melanin solution aged for 3.5 days at pH 12 are shown in figure \ref{fig:uvvispdadmacmelanin}. Just like the spectra of dopamine-melanin films grown on quartz (Figure \ref{fig:compmelaninuvvis}) and spectra of synthetic melanin found in the literature \cite{bothma:2008} \cite{diaz:2005} \cite{meredith:2006.1}, they are monotonic without of any peaks. This is in stark contrast to the spectra of melanin grain solutions (Figure \ref{fig:uvvismelaninph12}) showing two distinct peaks in the UV region. The difference may be explained by agglomeration of melanin particles in the \chemr{(PDADMA-melanin)_n} film or by preferential adsorption of larger entities from the melanin particle solution.

The absorbance of \chemr{(PDADMA-melanin)_n} increases linearly with the number of deposition cycles allowing for calculation of the film's extinction coefficient $\epsilon$ at a wavelength of 633 nm by equation \ref{eq:extinctioncoefficient}. Therefore the absorbance $A$ and the thickness $d$ are replaced by the absorbance increment ($(8.2 \pm 0.6) \cdot 10^{-3}$, Figure \ref{fig:uvvispdadmacmelanin}) and the thickness increment ($(4.7 \pm 0.4)$ nm, Figure \ref{fig:ellipsopdadmacmelanin}) per deposition cycle to obtain $\epsilon=(4.0 \pm 0.5) \cdot 10^{-6}$~m$^{-1}$. This value is determined for \chemr{(PDADMA-melanin)_n} films built with melanin solutions aged for 3.5 days at pH 12 and might be different for other ageing times. The extinction coefficient is close to the ones calculated for dopamine-melanin films obtained by different deposition methods (Table \ref{tab:compmelaninabs}).

\begin{figure}
	\centering
		\includegraphics[width=85mm]{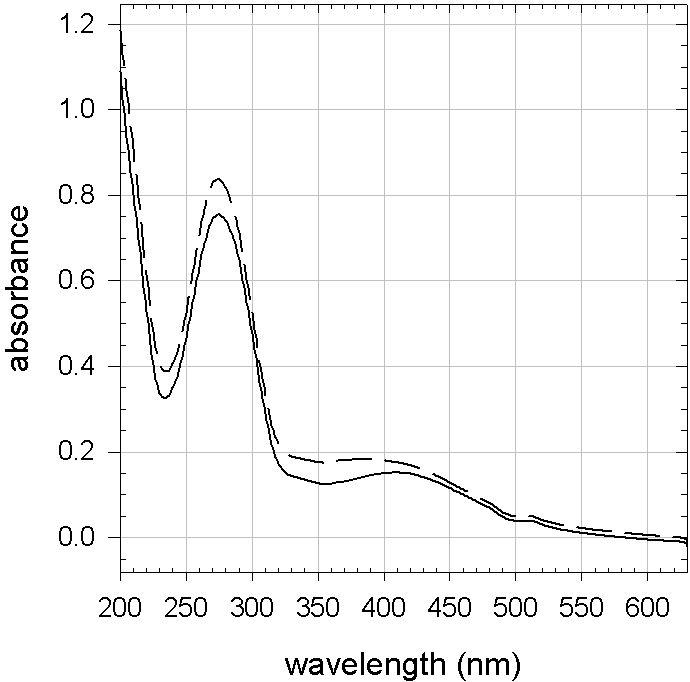}
	\caption[UV--visible spectra of melanin solutions with PDADMA]{UV--visible spectra of 2 g/L dopamine-melanin diluted 1/100 in 50 mmol/L Tris at pH 12 containing (dashed line) or not (solid line) 1 g/L PDADMA}
	\label{fig:uvvispdadmacmelaninsolution}
\end{figure}
The following experiment checks the influence of PDADMA on melanin grains in solution: 2 g/L of dopamine in Tris buffer (50 mmol/L) are allowed to react for 2 h at pH 8.5, then the pH is titrated to 13.0 and back to 12.0 before the obtained melanin solution is diluted 1/100 in pure Tris buffer at pH 12.0 or in the same buffer solution containing additionally 1 g/L PDADMA. The spectra of these solutions acquired immediately after dilution (Figure \ref{fig:uvvispdadmacmelaninsolution}) show the characteristic absorbance peaks of small dopamine-melanin grains (Figures \ref{fig:uvvismelaninph12}, \ref{fig:dialysismelanin}). Furthermore there is no visible influence of PDADMA on the spectra, excluding that PDADMA causes aggregation of dopamine-melanin grains in solution. Therefore dopamine-melanin grains probably do not aggregate at the solution-support interface upon contact with PDADMA either. Instead there is a preferential incorporation of larger melanin aggregates into the \chemr{(PDADMA-melanin)_n} films explaining their monotonic adsorption spectra.

\begin{figure}
	\centering
		\includegraphics[width=85mm]{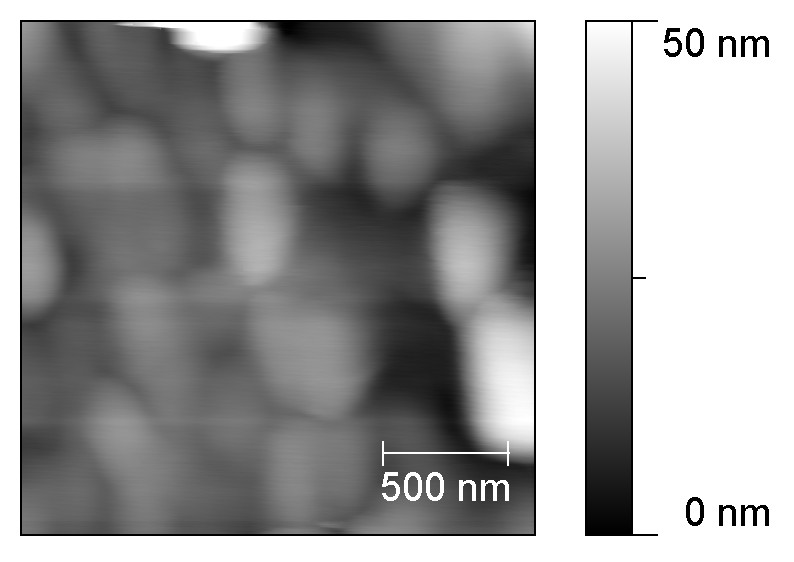}
	\caption[SFM image of \chemr{(PDADMA-melanin)_{10}}]{SFM height image of a \chemr{(PDADMA-melanin)_{10}} film taken at ambient humidity in contact mode. The film is built on a glass slide dipped alternately in 1 g/L PDADMA and 2 g/L melanin in Tris buffer (50 mmol/L, pH 12).}
	\label{fig:sfmpdadmacmelanin}
\end{figure}

\begin{figure}
	\centering
		\includegraphics[width=\textwidth]{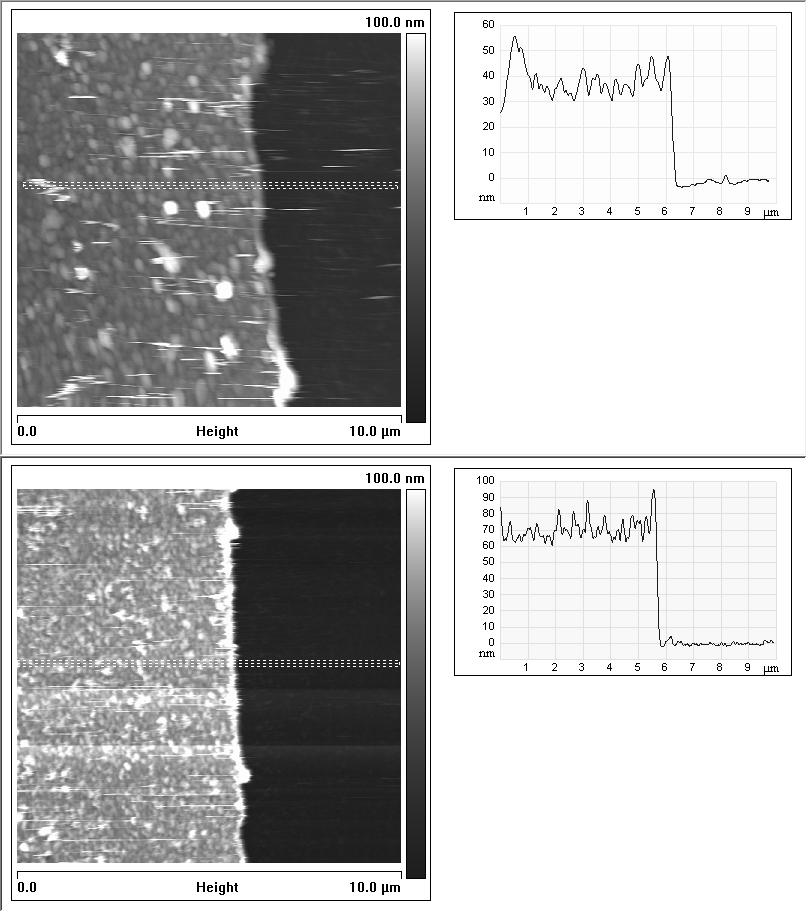}
	\caption[SFM height profiles of \chemr{(PDADMA-melanin)_{10}}]{SFM height images of the same \chemr{(PDADMA-melanin)_{10}} film as in figure \ref{fig:sfmpdadmacmelanin} after needle scratching to reveal its thickness in air (top) and under water (bottom). The panels on the right are height profiles of the dashed rectangles in the images on the left.}
	\label{fig:sfmpdadmacmelaninscratch}
\end{figure}

\chemr{(PDADMA-melanin)_{10}} films are built on glass slides using melanin solutions aged for 3 days at pH 12 to observe their surface morphology with scanning force microscopy (SFM). The images taken at ambient humidity (Figure \ref{fig:sfmpdadmacmelanin}) show particles of 100 nm to 500 nm in lateral size, while the total thickness of the film is only about 40 nm (Figure \ref{fig:sfmpdadmacmelaninscratch}). Thus the shape of the particles has to be anisotropic like platelets lying flat on the support surface. These structures resemble closely the ones observed on dopamine-melanin films grown directly on silicon oxide (Figure \ref{fig:compmelaninsfm}) as well as the particles found by others in melanin from glycera jaws \cite{moses:2006} or from sepia officinalis \cite{clancy:2001} \cite{nofsinger:2000}. Therefore the formation of the observed platelet-like structures seems to be an intrinsic property of melanin.

\begin{figure}
	\centering
		\includegraphics[width=\textwidth]{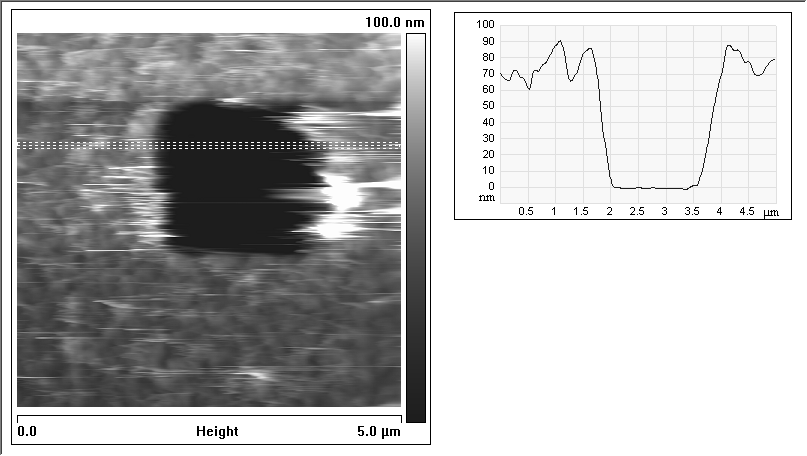}
	\caption[SFM image of \chemr{(PDADMA-melanin)_{10}} after tip scratching]{SFM height image in water of the same \chemr{(PDADMA-melanin)_{10}} film as in figure \ref{fig:sfmpdadmacmelanin} after scanning an area of $(2\mu m)^2$ several times at high deflection. The panel on the right is a height profile of the dashed rectangle in the image on the left.}
	\label{fig:sfmpdadmacmelaninhole}
\end{figure}

Furthermore the films are scratched with a syringe needle to reveal  their thickness at ambient humidity as well as under water (Figure \ref{fig:sfmpdadmacmelaninscratch}). The dry thickness is close to the one measured by ellipsometry (Figure \ref{fig:ellipsopdadmacmelanin}). Complete hydration of the films leads to an important swelling to nearly twice the initial thickness. Another important observation is that the films become very soft under water allowing their complete removal by repeated scanning with the SFM tip (Figure \ref{fig:sfmpdadmacmelaninhole}). According to \cite{moses:2006} melanin particles from sepia officinalis are very hard with an elastic modulus between 1 GPa and 7 GPa depending on their hydration. Thus the dopamine-melanin particles probably do not become soft upon hydration of the \chemr{(PDADMA-melanin)_{10}} films, but the cohesion between the particles and the PDADMA molecules weakens.

\subsection{Conclusion}
This section presents an investigation of the spontaneous oxidation of dopamine in alkaline solution. It is shown that oxygen is necessary to initiate the reaction leading to a black precipitate that is identified as dopamine-melanin by carbon-13 solid state nuclear magnetic resonance spectroscopy. The reaction speed increases with the initial dopamine hydrochloride concentration in solution and reaches a plateau for concentrations above 1~g/L. The obtained dopamine-melanin can bind $(1.5 \pm 0.2) \cdot 10^{-3}$ moles of amine groups per gram of melanin powder. The amine groups are probably bound covalently to catechol groups present at the surface of melanin as proposed in the literature. Thus melanin can be used as a platform to immobilize various biomolecules.

It is possible to redisperse the dopamine-melanin precipitate in strongly alkaline solutions. UV--visible spectroscopy and dialysis experiments indicate that these solutions contain large dopamine-melanin aggregates presenting a peakless absorbance spectrum typical of melanin, but also smaller grains having distinct absorbance peaks in the UV range. Transmission electron microscopy images show a hierarchical structure of melanin compatible with the stacked oligomer model of melanin structure. The density and the refractive index of the redispersed dopamine-melanin are determined as $(1.2 \pm 0.1)$~g/mL and as $(1.73 \pm 0.05) + (0.027 \pm 0.002)i$ at a wavelength of 589 nm.

The grains obtained by redispersion of the dopamine-melanin precipitate are used to build a multilayer with the polycation poly(diallyldimethylammonium chloride). In contrast to melanin solutions, the multilayers present a peakless UV--visible absorbance spectrum indicating preferential adsorption of larger melanin aggregates to the multilayer. The surface morphology of the multilayers observed by SFM is composed of platelets of 100 nm to 500 nm in lateral size. Comparison with the surface morphology of natural and synthetic melanins found in this work and in the literature lead to the conclusion that the formation of such a surface morphology is typical of melanin.

An interesting subject for further studies would be to find out whether the melanin grains in solution also present an anisotropic shape like the aggregates deposited on a support. This question might be answered by dynamic light scattering experiments to determine the hydrodynamic radius of the melanin grains. Indeed Gallas and others derived from small angle x-ray and neutron scattering that tyrosine-melanin is present as sheet-like particle in alkaline aqueous solution \cite{gallas:1999}.

%% file: resultsdepositionmethods.tex
\section{Comparison of dopamine-melanin deposition methods}
\label{sec:resdepostitionmethods}
\subsection{First observations}
\label{sec:resfirstobservations}
Lee and others have shown that an organic deposit grows on the surface of various materials when they are immersed in alkaline dopamine solutions (2 g/L) in the presence of 10 mmol/L Tris buffer \cite{lee:2007.2}. They supposed that the deposit, which reaches a plateau thickness of approximately 50 nm after 24 h of immersion, is composed of  melanin. In this section the deposit that grows on materials immersed in the dopamine solutions investigated in section \ref{sec:resultsdopamineinsolution} will be identified as dopamine-melanin. Furthermore different methods to obtain dopamine-melanin deposits will be compared.

\begin{figure}
	\centering
		\includegraphics[width=85mm]{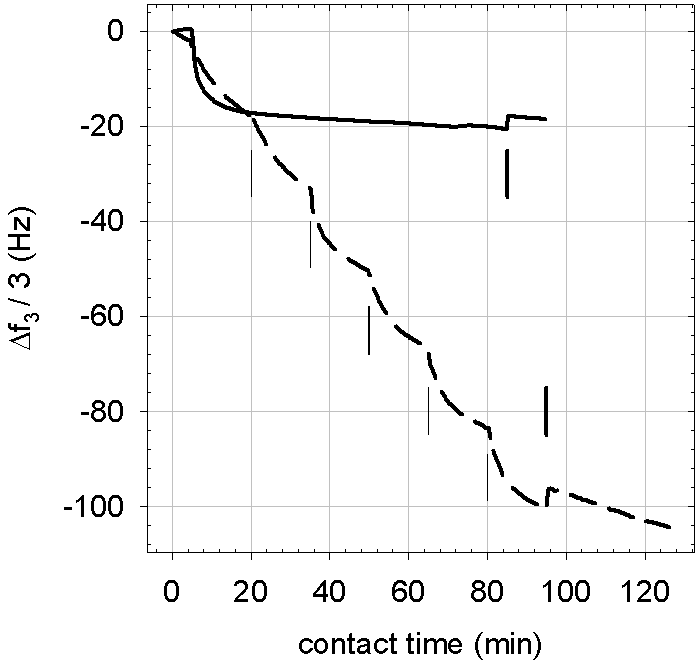}
	\caption[Melanin deposition monitored by QCM-D]{QCM-D experiments displaying the reduced change in resonance frequency at the third overtone (\chem{\Delta f_3/3}) of a quartz crystal put in contact with dopamine hydrochloride solutions at 2 g/L in Tris buffer (50 mmol/L, pH 8.5) as a function of time: full line: one single injection of dopamine solution, dashed line: 6 successive injections of freshly prepared dopamine solutions. A thin vertical line labels each new injection. The thick vertical lines correspond to the beginning of the final buffer rinse.}
	\label{fig:qcmmelanin}
\end{figure}

The deposition kinetics of melanin from freshly prepared dopamine hydrochloride solutions at 2 g/L is followed in situ using QCM-D: After an initial fast decrease of the reduced frequency of the quartz crystal, it reaches a steady state value within 15 min to 20 min (Figure \ref{fig:qcmmelanin}). When considering the different overtones, it appears that the reduced frequency shifts overlap within the limit of the experimental error of a few Hz. In addition the dissipation changes are small suggesting that the obtained deposits are thin and rigid and allow using the Sauerbrey relationship to calculate the deposited mass per unit area (Equation \ref{eqn:sauerbreynumerical}) as well as the thickness (Equation \ref{eqn:qcmthickness}). Using a density of 1.2 g/mL for dopamine-melanin (Section \ref{sec:melaninrefractiveindex}) a plateau thickness of 2.7 nm is calculated. In the literature the plateau is reached much later close to 50 nm after 24 h (\cite{lee:2007.2}, on silicon wafers) or near 20 nm after 6 h (\cite{postma:2009}, on silica spheres) under \emph{a priori} similar experimental conditions. The origin of this striking difference will be explained later on. For now it is only observed that the plateau can be surmounted by consecutive dopamine injections leading to a linear evolution of the reduced frequency shift with the number of injections (Figure \ref{fig:qcmmelanin}).

To check the proportionality between the number of injections of freshly prepared dopamine solution and the melanin thickness, ellipsometry experiments are performed. The silicon support is rinsed with Tris buffer and with water after each deposition step before drying it under a stream of nitrogen. Then the melanin thickness is measured before the next identical deposition step. It is found that the thickness is extremely variable from point to point along the main axis of the silicon slide suggesting that the drying-rehydration process strongly alters melanin deposition.

\begin{figure}
	\centering
		\includegraphics[width=85mm]{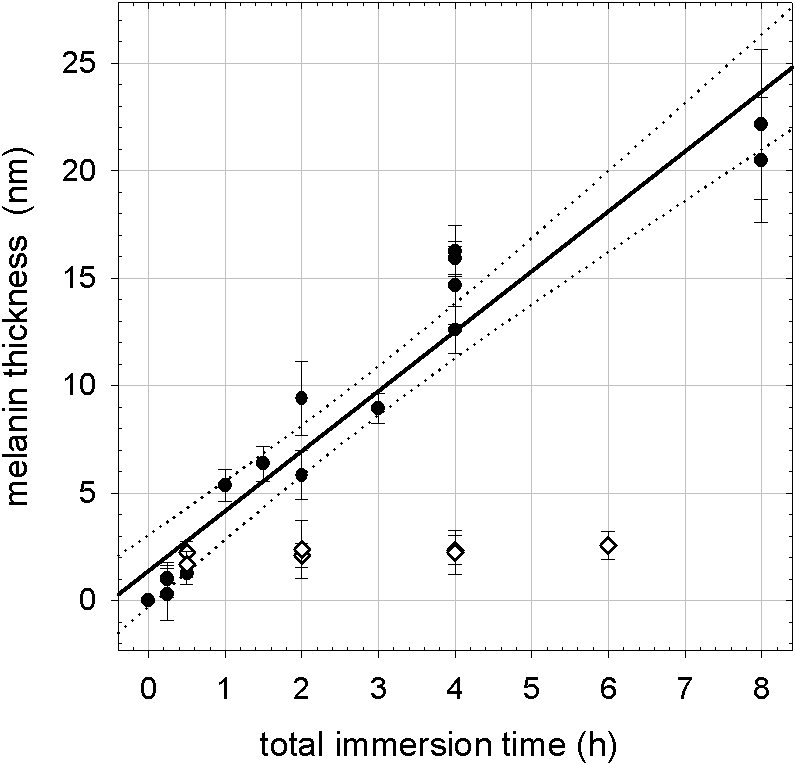}
	\caption[Optical thickness of melanin versus deposition time]{Thickness of melanin layers obtained by immersing silicon substrates in dopamine hydrochloride solutions at 2 g/L (measured by ellipsometry assuming a refractive index of 1.465) as a function of the total immersion time. Diamonds: immersion in one single dopamine solution. Circles: immersion in new dopamine solutions every 15 min. The full line corresponds to a linear regression to the data and the dotted lines represent the limits of the 95 \% confidence interval. Each point corresponds to an independent experiment.}
	\label{fig:ellipsomelanin}
\end{figure}

Hence the thickness is determined after $n$ immersions of the slides in freshly prepared dopamine solutions without intermediate rinsing and drying. Based on the QCM-D experiments the length of these immersion steps is chosen to be 15 min. Under these conditions the melanin thickness increases in proportion to the number of immersion steps in fresh dopamine solutions, and hence proportionally to the total reaction time, $t = n \cdot 15$ min (Figure \ref{fig:ellipsomelanin}). If the refractive index of dopamine-melanin ($1.73 + 0.022i$ see section \ref{sec:melaninrefractiveindex}) is used to calculate the thickness of the deposit, the obtained values are lower than those obtained by fixing the refractive index to 1.465 by about 15 \%, but the general trend is not modified. When the immersion time is reached by contact with one single dopamine solution, the thickness reaches a plateau value of about 2 nm after less than 30 min in agreement with the QCM-D results.

The saturation in the adsorption kinetics may be either due to saturation of the surface or to depletion of the reactive species in solution. The second assumption is supported by the fact that putting the silicon slide in fresh dopamine solutions allows for continued melanin growth. The following experiment provides further explanations: A dopamine hydrochloride solution (2 g/L) in Tris buffer (50 mmol/L, pH 8.5) is allowed to react during 4.5 h and then a silicon slide is put in this solution during 2 h. Ellipsometry detects no film deposition by this treatment. Hence the dopamine-melanin formed in solution does not adhere to the support in agreement with the observation by Lee and others \cite{lee:2007.2}. The deposition of melanin is probably a process initiated by the interaction of monomers or small oligomers with the substrate.

\begin{figure}
	\centering
		\includegraphics[width=85mm]{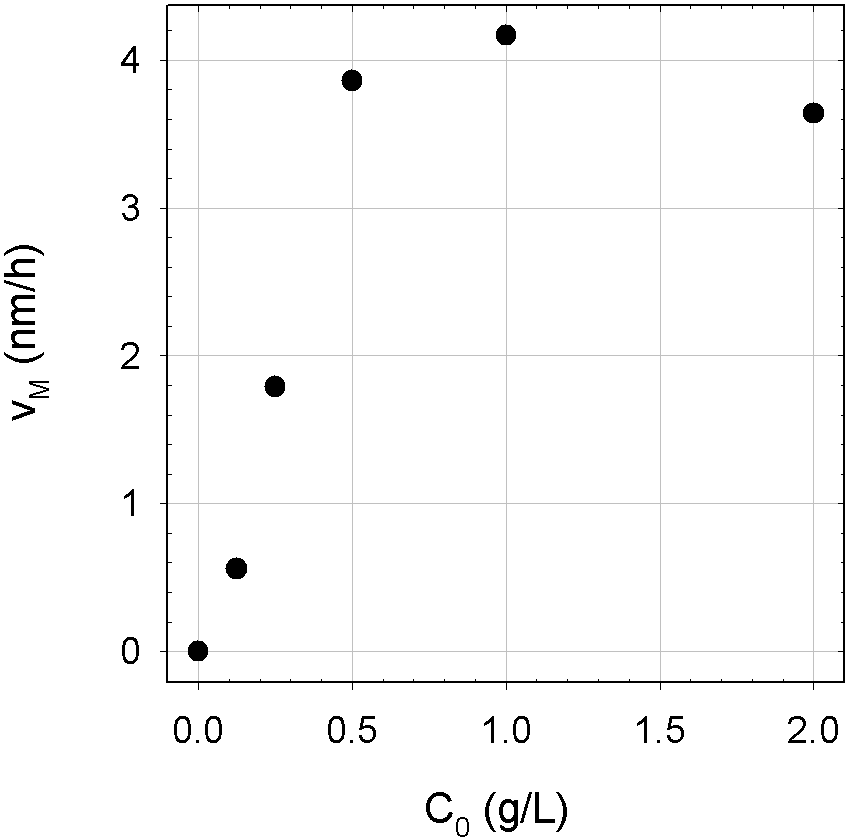}
	\caption[Growth rate of melanin versus dopamine concentration]{Growth rate \chem{v_M} of melanin on silicon slides versus initial dopamine hydrochloride concentration \chem{C_0} in Tris buffer (50 mmol/L, pH 8.5) for immersion cycles of 15 min. Each point originates from a linear regression as in figure \ref{fig:ellipsomelanin}.}
	\label{fig:melaningrowthrate}
\end{figure}

In additional experiments silicon slides are immersed for 15 min in multiple solutions of varying dopamine hydrochloride concentration. In all cases the melanin thickness increases linearly with the total immersion time. The slope of the thickness versus immersion time is plotted as a function of the initial dopamine hydrochloride concentration in figure \ref{fig:melaningrowthrate}. This quantity will be called the melanin growth rate. It appears that above an initial concentration of 0.5 g/L the growth rate reaches a plateau at about 4 nm/h. The growth rate at the plateau can be increased when the immersion cycle duration is decreased to 5 min instead of 15 min: 7 nm/h instead of 3.6 nm/h at an initial concentration of 2 g/L. This shows that thicker films can be obtained by regularly providing fresh dopamine. Ideally one should deposit melanin from a permanent flux of freshly prepared dopamine, for instance using a mixing chamber wherein the pH of the dopamine solution is increased to 8.5 just before reaching the adsorption substrate.

\subsection{Growth regimes}
In the following, four methods to obtain melanin deposits from aqueous dopamine solutions will be compared, which differ mainly in the way the formation of melanin is initiated by oxidation of dopamine. This can be done using dissolved oxygen as described in section \ref{sec:resultsdopamineinsolution} and in \cite{lee:2007.2} or \cite{postma:2009} (Methods A and B). Instead of dissolved oxygen one can also use other oxidants like \chem{Cu(II)} ions (Method C). Using an electrode to oxidise dopamine as described in \cite{li:2006.1} (Method D) localizes the melanin formation at the support-solution interface.

\begin{figure}
	\centering
		\includegraphics[width=\textwidth]{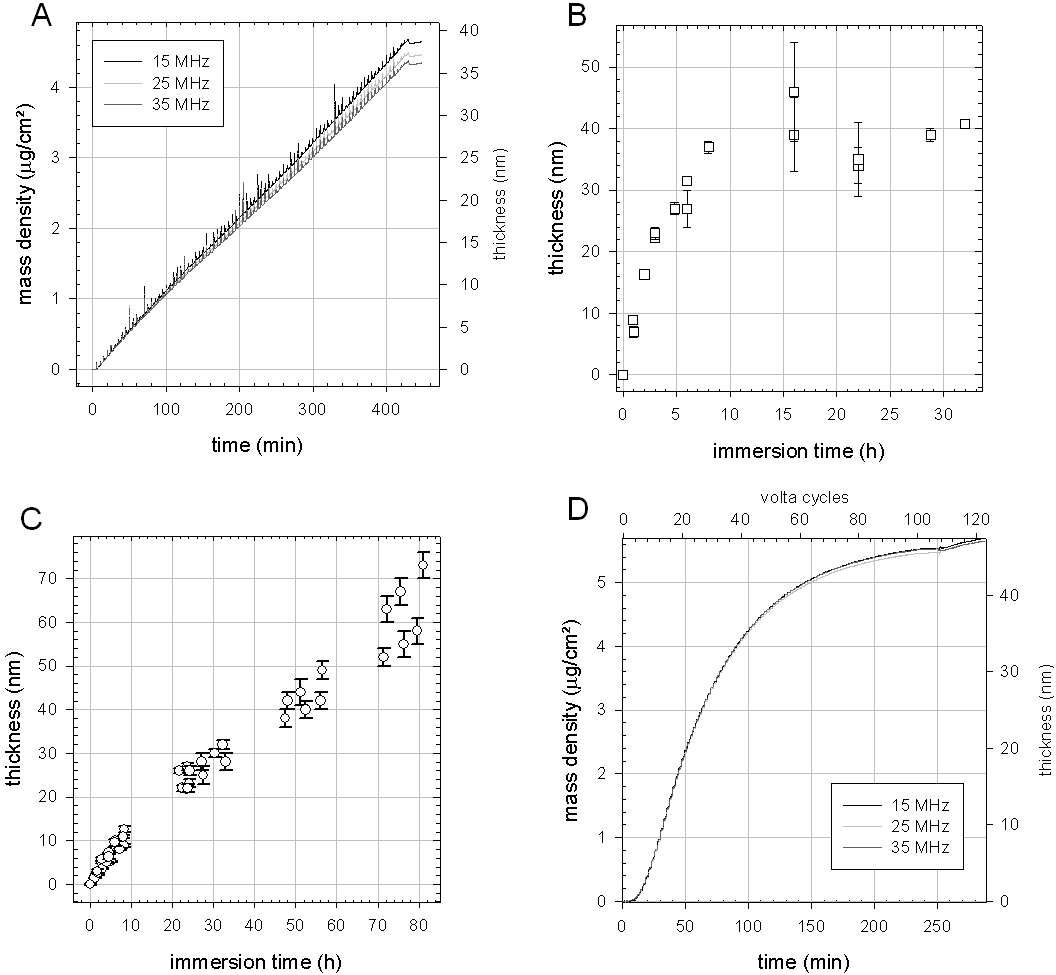}
	\caption[Growth of melanin deposits using different deposition methods]{Thickness of melanin deposits obtained by methods A, B, C or D versus contact time with dopamine solutions. The thickness is obtained from QCM measurements at different frequencies indicated in the figure using the Sauerbrey equation and a melanin density of 1.2 g/cm$^3$ (A,D) or from ellipsometry measurements assuming a refractive index of $N=1.8+0.14i$ (B,C).}
	\label{fig:compmelaningrowth}
\end{figure}
The growth of dopamine-melanin deposits obtained by the different methods is monitored by ellipsometry assuming a refractive index of $N=1.8+0.14i$ (Section \ref{sec:melaninrefractiveindex} and Table \ref{tab:compmelaninabs}) and quartz crystal microbalance (QCM). In QCM experiments the reduced frequency changes at different overtones overlap allowing for the use of the Sauerbrey approximation to calculate the mass (Equation \ref{eqn:sauerbreynumerical}) and thickness (Equation \ref{eqn:qcmthickness}) of melanin deposits using a melanin density of 1.2 g/cm$^3$ (Section \ref{sec:melaninrefractiveindex}). Method A leads to a linear growth of the melanin film with the deposition time for at least 85 injections of dopamine solutions lasting 5 min each (Figure \ref{fig:compmelaningrowth} A). The spikes superposed to the linear growth curves are artefacts caused by the injection of solutions into the measurement chamber of the QCM. The horizontal part at the end of the curves corresponds to a final rinsing of the QCM chamber with pure buffer solution.

In an aerated dopamine solutions the melanin thickness first grows linearly with the deposition time up to 25 nm within 4 h (Figure \ref{fig:compmelaningrowth} B). Afterwards the thickness approaches a plateau at 40 nm probably due to a depletion of dopamine in the solutions. This evolution is in line with the observations of Lee \cite{lee:2007.2} and Postma \cite{postma:2009} mentioned earlier. Unfortunately Lee and his colleagues do not mention in their article whether they aerated their dopamine solutions, but most probably they did because without aeration dopamine-melanin deposits do not surmount a thickness of 2 nm (Figures \ref{fig:qcmmelanin}, \ref{fig:ellipsomelanin}) in one step. When samples prepared by method B are stored in water at ambient temperature the dopamine-melanin thickness is stable for at least 56 days.

In presence of copper sulphate and without oxygen the dopamine-melanin thickness also grows linearly with the immersion time in a dopamine solution (Figure \ref{fig:compmelaningrowth} C). The initial growth speed of 1.5 nm/h is smaller then for the previously described methods (A: 5.5 nm/h, B: 7 nm/h) but the growth continues in a nearly linear way up to at least 60 to 70 nm within 80 hours of immersion in \emph{one single solution}. The formation of dopamine-melanin with copper sulphate takes place at a pH of 4.5 whereas it is completely inhibited in an acidic medium when oxygen is used as oxidant. According to Hawley \cite{hawley:1967} melanin formation does not take place at low pH because the protonation of the amine group in dopamine quinone (negative decadic logarithm of the dissociation constant $pK_a \approx 9$) precludes its cyclisation, which is a necessary step towards melanin formation \cite{li:2006.1}. Nevertheless others \cite{herlinger:1995} observe, like we do, that metal ions can initiate melanin formation in acidic solution by an up to now unexplained mechanism. In both methods B and C intermediate rinsing and drying steps do not change the growth of the dopamine-melanin thickness. The thickness of samples prepared by method C remains stable for at least one week of storage in water.

Using cyclic voltamperometry to initiate dopamine-melanin formation, leads initially to a faster growth of the melanin thickness than the other methods (25 nm/h, Figure \ref{fig:compmelaningrowth} D). During this stage one can distinguish the individual voltamperometry cycles as steps in the plotted curves. After 70 minutes of deposition time, corresponding to 30 voltamperometry cycles, the growth slows down and the melanin thickness reaches a plateau at 45 nm after 4 h corresponding to 100 voltamperometry cycles. Replacing the dopamine solution by a fresh one does not lead to further melanin deposition. This shows that the electrochemical melanin growth is not limited by depletion of dopamine but by the deposition of an insulating melanin film on the working electrode, which prevents the further oxidation of dopamine necessary for melanin growth. Electrochemical permeability measurements presented in section \ref{sec:compmelaninvolta} will support this explanation.

\begin{figure}
	\centering
		\includegraphics[width=\textwidth]{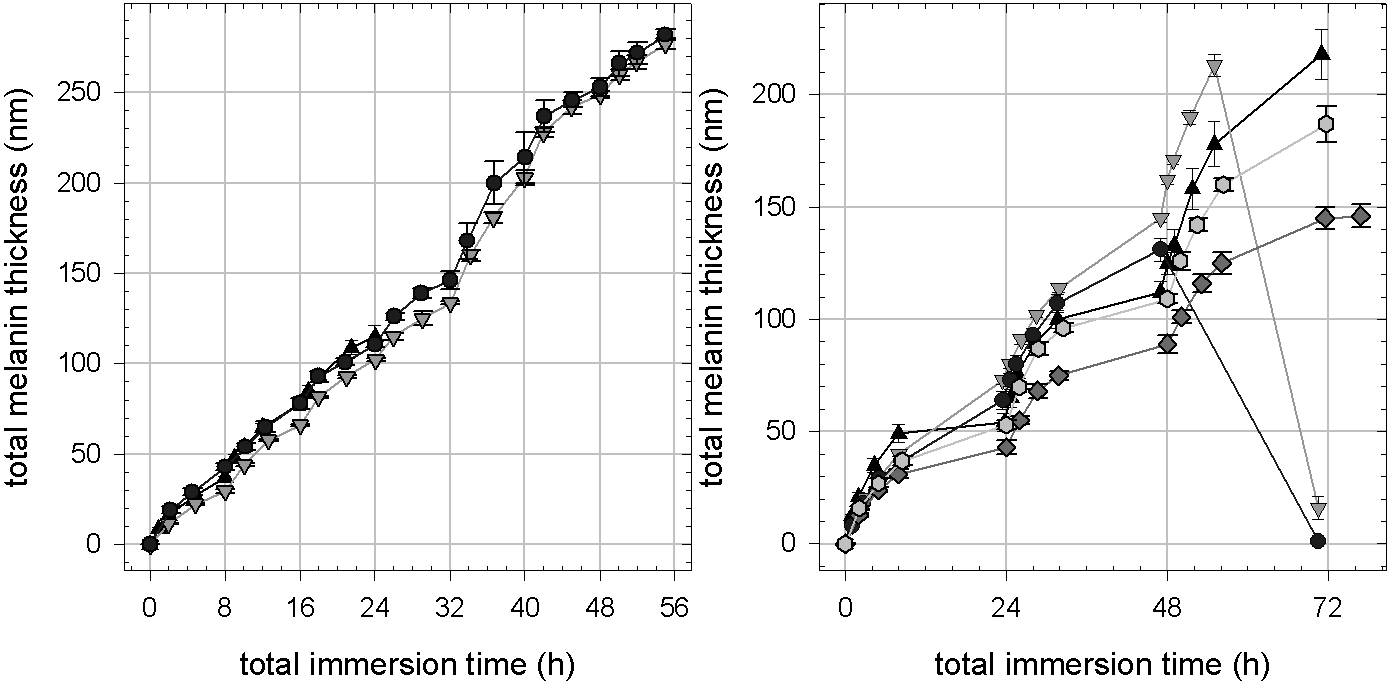}
	\caption[Growth of melanin deposits from multiple aerated dopamine solutions]{Ellipsometric thickness assuming a refractive index of $N=1.8+0.14i$ of melanin deposits obtained by immersion in aerated dopamine solutions (2 g/L in Tris 50 mmol/L, pH 8.5) replaced every 8 h (left) or 24 h (right). Different symbols represent independent experiments.}
	\label{fig:ellipsodopamineonmelanin}
\end{figure}
For method B it is possible to repeat the deposition steps several times as for method A to obtain thicker deposits. Figure \ref{fig:ellipsodopamineonmelanin} shows the melanin thickness measured by ellipsometry for experiments where the aerated dopamine solutions are replaced every 8~h. This way a regular growth of about 40 nm per deposition step is obtained for at least seven steps. In the represented experiments the samples are rinsed with water, dried under a stream of nitrogen and stored in the dry state over night between deposition steps. If the samples are kept in buffer solution between deposition steps, the dopamine-melanin deposit dissolves during the second or third step. Thus intermediate drying induces changes in the melanin film that are necessary for continued melanin deposition. Longer deposition steps (24 h instead of 8 h) sometimes also lead to the dissolution of the melanin deposit during the build-up and induce a poor reproducibility (Figure \ref{fig:ellipsodopamineonmelanin}). According to Lee and others dopamine-melanin adheres more firmly to organic than to inorganic supports \cite{lee:2007.2}. Maybe an organic anchoring layer as proposed by Ou and others \cite{ou:2009} would enhance the adherence of dopamine-melanin to silicon in the presented experiments.

\subsection{Melanin identification by UV--visible spectroscopy and XPS}
\begin{figure}
	\centering
		\includegraphics[width=\textwidth]{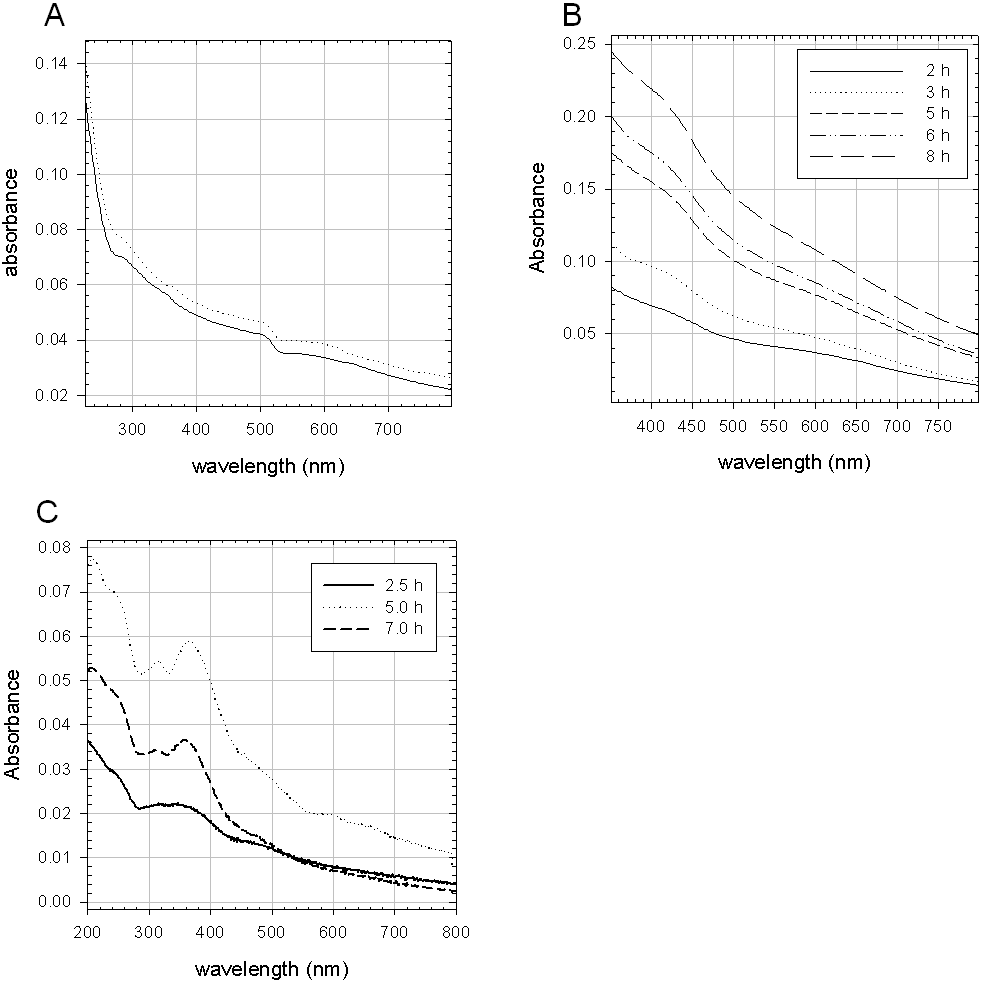}
	\caption[UV--visible spectra of melanin films]{UV--visible spectra of melanin films deposited on both sides of quartz slides. A: 8 x 15 min immersions in dopamine of two different slides. B: Immersion for different times indicated in the figure in aerated dopamine solutions. C: Immersion for different times indicated in the figure in copper-containing dopamine solutions}
	\label{fig:compmelaninuvvis}
\end{figure}
The UV--visible absorbance of dopamine-melanin deposits on quartz slides obtained by methods A and B (Figure \ref{fig:compmelaninuvvis}A, B) decreases monotonously with increasing wavelength as in published spectra of synthetic melanin \cite{bothma:2008} \cite{diaz:2005} \cite{meredith:2006.1}. The small differences between the spectra presented here and the spectra published in the literature may originate from the fact that the absorbance spectra in this work are measured in the transmission mode without using an integrating sphere to account for scattering losses. There are also some features in the spectra of samples prepared by method A, particularly around 280 nm and 520 nm, which may be due to the presence smaller melanin aggregates than the ones making up the film. Others observed similar features for partially polymerised melanin \cite{tran:2006} and low molar mass fractions of melanin \cite{simon:2000}. 
%The validity of the assumption that we observe partially polymerised species is confirmed by cyclic voltammetry: when capacitive measurements are done on melanin films, there is a small oxidation current at around 0.1 V vs. Ag/AgCl that progressively decreases when successive scans are done and completely vanishes after the fourth oxidation-reduction cycle.

The spectrum of dopamine-melanin built in presence of copper ions shows a shoulder at 250 nm and two peaks at 315 nm and 345 nm in addition to the monotonous melanin background (Figure \ref{fig:compmelaninuvvis} C). These spectral features might be due to copper-containing catechol or quinone imine complexes. Such complexes were described in \cite{szpoganicz:2002}, but unfortunately the publication does not contain any absorbance data of the species that is expected to predominate under acidic conditions. Gallas and others examined the solution structure of copper(II) induced melanin aggregates by small angle x-ray and neutron scattering but they did not examine the UV--visible spectra of the obtained aggregates \cite{gallas:1999}.

\begin{figure}
	\centering
		\includegraphics[width=\textwidth]{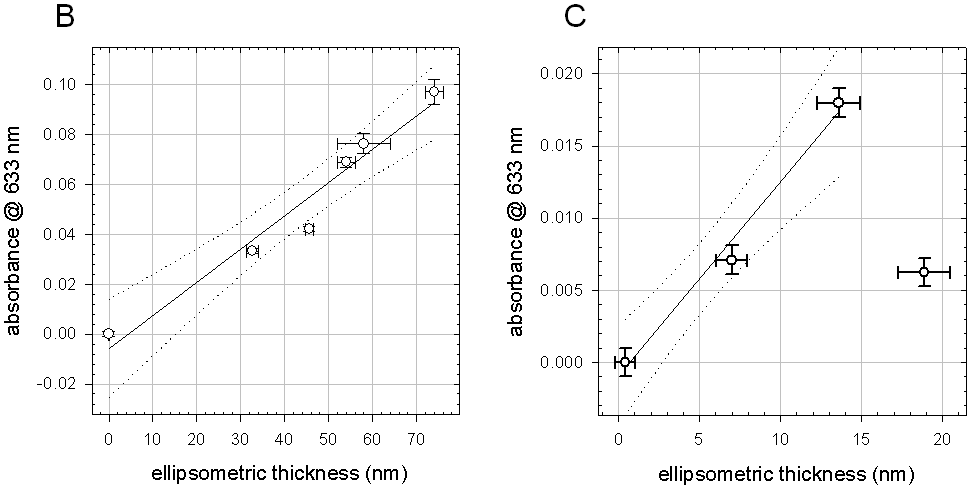}
	\caption[Absorbance of dopamine-melanin deposits versus thickness]{Absorbance at 633 nm of dopamine-melanin deposits obtained by method B and C versus their thickness. Full lines correspond to linear regressions to the data. Dotted lines delimit the 95 \% confidence intervals. In plot C the last point is not taken into account because the substrate was scratched in the measurement region.}
	\label{fig:compmelaninabs}
\end{figure}
For methods B and C the absorbance $A$ of the dopamine-melanin deposits at a wavelength of 633 nm is plotted as a function of the deposits' total thickness $d$ (Figure \ref{fig:compmelaninabs}) derived from ellipsometry measurements (Figure \ref{fig:compmelaningrowth}). The linear regressions of $A$ versus $d$ are used to calculate the extinction coefficient $\epsilon$ and the imaginary part $k$ of the refractive index of the melanin deposits using equations \ref{eq:extinctioncoefficient} and \ref{eq:imaginaryrefractiveindex} with a concentration $C=1$. Table \ref{tab:compmelaninabs} summarizes the results that are nearly identical for deposition methods B and C and close to values found in the literature for synthetic melanin films spin cast from ammonia solutions \cite{bothma:2008} or sprayed from solutions in dimethylsulfoxide and methanol \cite{abbas:2009}. Only the extinction coefficient of the deposits prepared by method A is higher. This could indicate a denser packing of dopamine-melanin in these films, but the observed difference in extinction coefficient should not be over-interpreted. Since measurements were performed at only one thickness ($(7 \pm 1)$ nm) compared to 3 to 6 thickness values examined for the other methods, the value of the extinction coefficient for method A is probably not as reliable as the others.
\begin{table}
	\centering
		\begin{tabular}{l|l|l}
		Method & $\epsilon$ ($10^6$ m$^{-1})$ & $k$ \\ \hline
		A: 8 x 15 min & $5.6 \pm 0.9$ & $0.28 \pm 0.05$ \\
		B: aerated dopamine & $2.8 \pm 0.2$ & $0.14 \pm 0.01$ \\
		B with phosphate buffer & $3.5 \pm 0.5$ & $0.17 \pm 0.02$ \\
		C: with \chem{CuSO_4}& $3.0 \pm 0.2$ & $0.15 \pm 0.01$ \\
		spin cast \cite{bothma:2008} & $2.0$ & $0.10$ \\
		sprayed \cite{abbas:2009}& $2.7$ & $0.14$	
		\end{tabular}
	\caption{Extinction coefficient $\epsilon$ and imaginary part $k$ of the refractive index at 633~nm of dopamine-melanin deposits obtained by different methods.}
	\label{tab:compmelaninabs}
\end{table}

\begin{figure}
	\centering
		\includegraphics[width=85mm]{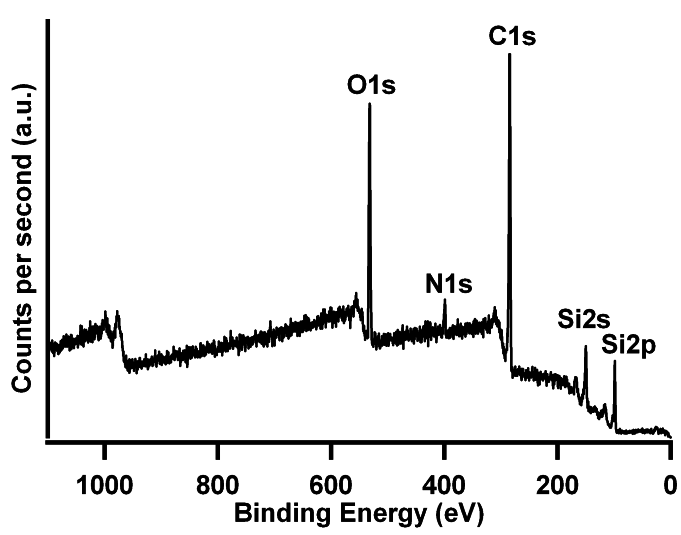}
	\caption[XPS of dopamine-melanin]{XPS at 90° take-off angle of a melanin deposit obtained by method A (32 immersions of 15 min in dopamine solutions). Peaks are labelled with the corresponding orbitals.}
	\label{fig:xps32x15min}
\end{figure}
\begin{figure}
	\centering
		\includegraphics[width=85mm]{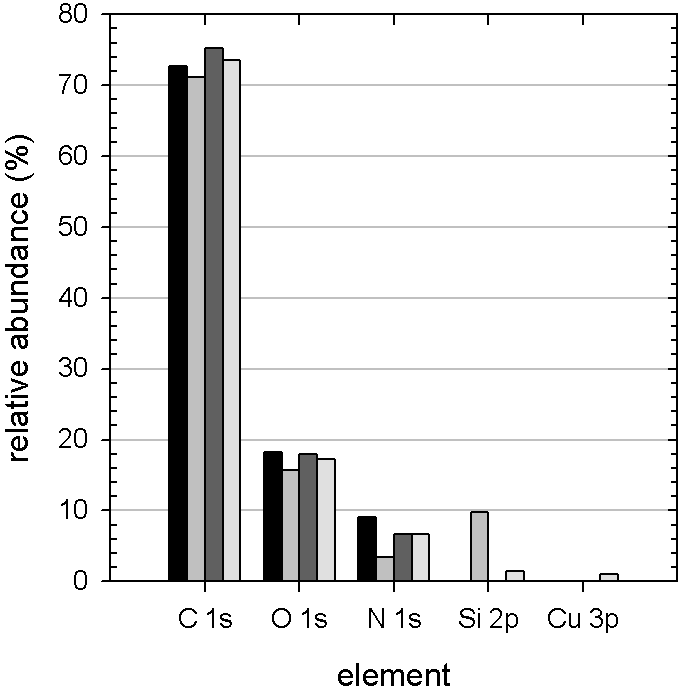}
	\caption[XPS composition of dopamine-melanins]{Atomic composition derived from XPS of dopamine-melanin deposits obtained by method A (grey, 32 immersions of 15 min), B with phosphate buffer instead of Tris (dark grey, 48 h reaction) or C (light grey, 24 h reaction) and theoretical composition of pure dopamine (black).}
	\label{fig:xpscomposition}
\end{figure}
\label{sec:resxpscomposition}
X-ray photoelectron spectroscopy (XPS) is used to determine the chemical composition of melanin deposits. Figure \ref{fig:xps32x15min} shows an exemplary XPS survey spectrum of a dopamine-melanin deposit obtained by method A and figure \ref{fig:xpscomposition} summarizes the atomic composition of different dopamine-melanin deposits compared to the theoretical composition of pure dopamine. The composition of the different samples is close to the one of dopamine with the following differences: The sample prepared by method A presents a notable silicon signal. This is probably due to the fact that this deposit is thinner than the other ones. Its thickness measured by ellipsometry at ambient conditions is ($24 \pm 3$) nm instead of ($68 \pm 8$) nm (method B) or ($39 \pm 2$) nm (method C). Dehydration in ultra-high vacuum might decrease its thickness below the probing depth of XPS (about 4.5 nm for the employed conditions, Equation \ref{eqn:xpsprobingdepth}) making the underlying silicon support visible. The sample prepared by method C contains 1 \% of copper confirming the incorporation of this element in the dopamine-melanin deposit. On the contrary, there is no detectable incorporation of phosphorus from the phosphate buffer employed in method B. The atom ratio carbon/nitrogen (\chem{C/N}) is 8 in dopamine and 4 in Tris. Thus the dopamine-melanin deposits should present a \chem{C/N} ratio between 4 and 8. Though in the experiments all samples have \chem{C/N > 8} probably due to organic contanimation making it impossible to conclude if dopamine-melanin films prepared in presence of Tris incorporate this buffering agent.

\begin{figure}
	\centering
		\includegraphics[width=85mm]{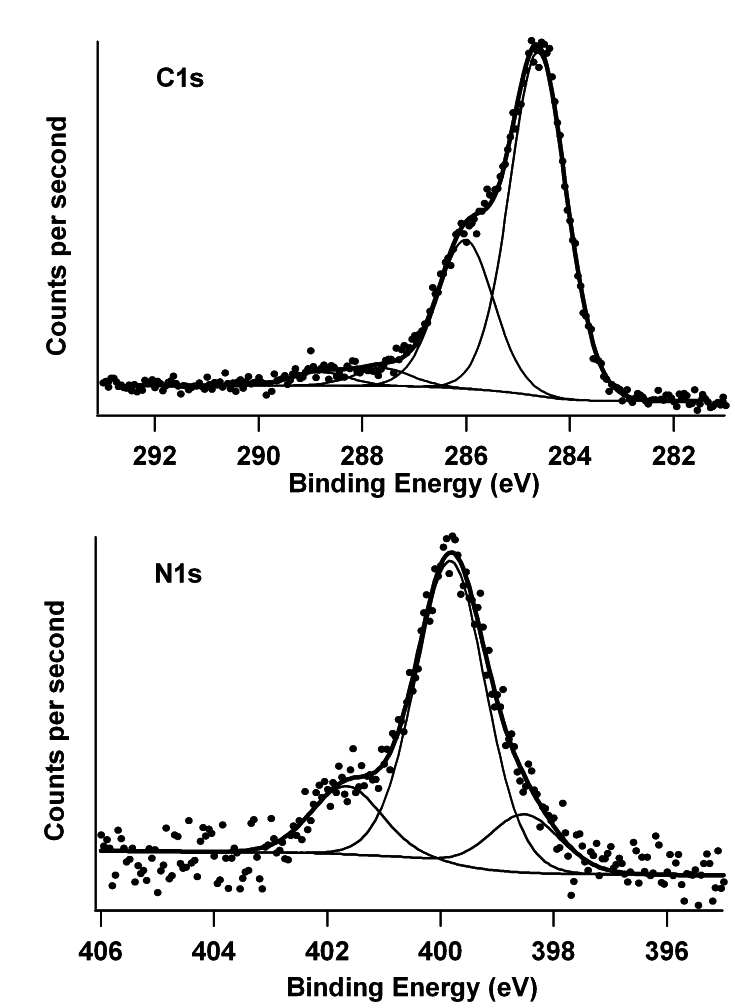}
	\caption[High resolution XPS of dopamine-melanin]{High resolution XPS of a dopamine-melanin deposit obtained by dipping of the support for 32 x 15 min in dopamine solutions at the \chem{C}1s peak (top) and the \chem{N}1s peak (bottom). The points correspond to the experimental data, the thin lines to the components used for fitting and the thick line to the sum of the fitting curves.}
	\label{fig:highresxps}
\end{figure}
\begin{figure}
	\centering
		\includegraphics[width=\textwidth]{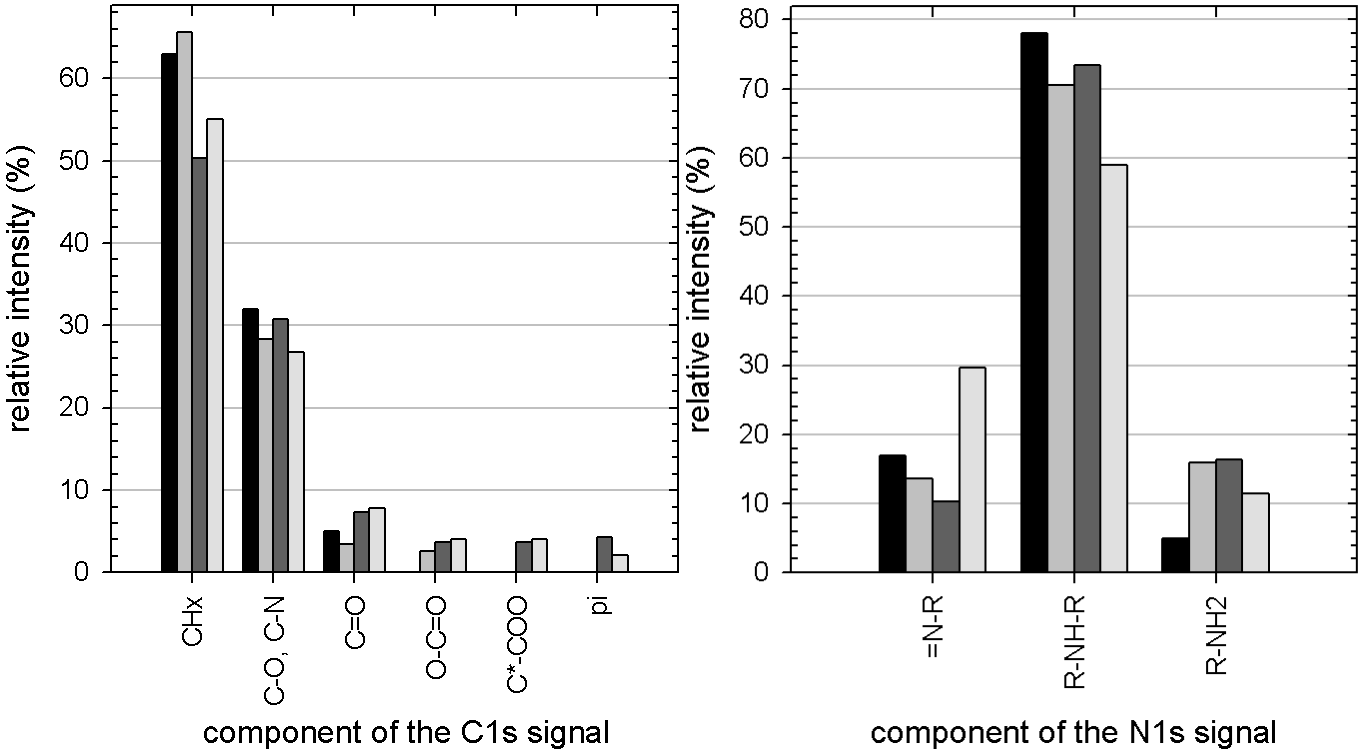}
	\caption[Deconvolution of XPS \chem{C}1s and \chem{N}1s peaks]{Relative intensities of the components of the \chem{C}1s (left) and \chem{N}1s (right) peaks in high-resolution XPS spectra of different melanin deposits. Black: from N-methyl-5,6-dihydroxyindole \cite{clark:1990}, grey: method A (32 immersions of 15 min), dark grey: method B with phosphate buffer instead of Tris (48 h reaction), light grey: method C (24 h reaction)}
	\label{fig:xpsc1sn1s}
\end{figure}

XPS also provides information about the chemical environment of the elements detected in the dopamine-melanin deposit. Figure \ref{fig:highresxps} shows exemplary high-resolution spectra of dopamine-melanin deposits at the signals of \chem{C}1s and \chem{N}1s and the deconvolution of these peaks. In the \chem{C}1s peak up to six components can be identified corresponding to aromatic carbons (\chem{CH_x} at a photoelectron energy of 284.6 e\/V), oxygen- or nitrogen-substituted carbons (\chem{C-O,C-N} at 286.0 e\/V), carbonyl groups (\chem{C=O} at 287.6 e\/V), carboxyl groups (\chem{O-C=O} at 288.9 e\/V), carbons next to a carboxyl group (\chem{C^*-COO} at 291.0 e\/V) and $\pi - \pi^*$ transitions at 291.0 e\/V. The \chem{N}1s signals contain three components corresponding to primary amine groups (\chem{R-NH_2} at 401.7 e\/V), secondary amine groups (\chem{R-NH-R} at 400.0 e\/V) and imino groups (\chem{=N-R} at 398.6 e\/V). The relative intensities of these components are shown in figure \ref{fig:xpsc1sn1s} for dopamine-melanin deposits obtained by methods A, B or C in comparison to the intensities found by Clark and others for melanin obtained from N-methyl-5,6-dihydroxyindole \cite{clark:1990}. The intensities are comparable for the different melanins and in accord with a molecular structure made of 5,6-dihydroxyindole and its carboxylated form \cite{clark:1990}. Notably excitations of delocated $\pi$ electrons are only detected in dopamine-melanin prepared by methods B and C. Maybe the intensity of the \chem{C}1s signal is not strong enough in the other cases to detect the $\pi - \pi^*$ transitions. It has not been possible to distinguish between aromatic carbons and aliphatic ones. This confirms the fully delocalized structure of melanin films, which is supposed to be responsible for its electrical properties \cite{meredith:2006.2}.

\begin{figure}
	\centering
		\includegraphics[width=85mm]{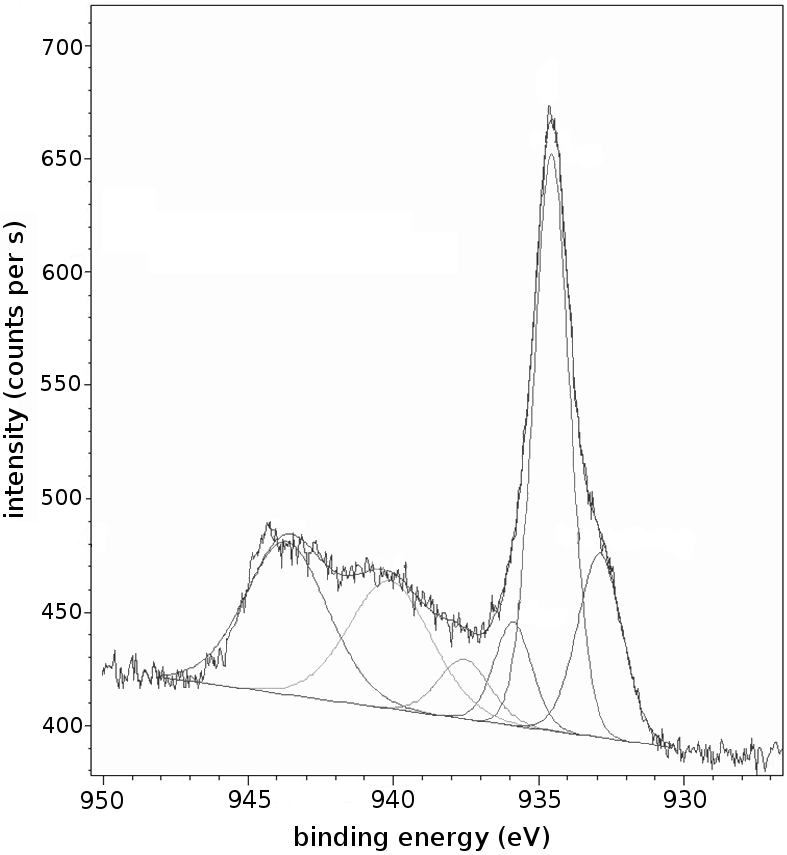}
	\caption[High resolution \chem{Cu}2p XPS in dopamine-melanin]{High resolution \chem{Cu}2p XPS of a dopamine-melanin deposit obtained by 23.5~h immersion in 2 g/L dopamine with 50 mmol/L Tris and 30 mmol/L \chem{CuSO_4}. The components used for fitting are displayed in addition to the measured spectrum.}
	\label{fig:xpscu2p}
\end{figure}
High resolution XPS was also employed to deconvolute the \chem{Cu}2p peak of dopamine-melanin samples prepared in the presence of \chem{CuSO_4} (Method C; Figure \ref{fig:xpscu2p}). Most of the copper incorporated in the film is in the oxidation state \chem{Cu}(II) (63 \%, at 934.6 e\/V). \chem{Cu}(I) and \chem{Cu}(0) (25 \%, at 932.9 e\/V) as well as \chem{CuSO_4} (12 \%, at 935.8 e\/V) cause the remaining intensity of the \chem{Cu}2p peak. The peaks at 940 e\/V and 944 e\/V are satellite peaks of the principal \chem{Cu}(II) peak. It is not surprising to find most of the copper in its oxidation state II, since others have shown that \chem{Cu}(II) can be tightly bound by phenolic groups of melanin at the pH of the melanin deposition environment \cite{hong:2007}.
  
According to the UV--visible absorbance and XPS results, the product deposited on the solid substrates is identified as dopamine-melanin like the reaction product in solution (Section \ref{sec:nmrresults}).

\subsection{Surface characteristics: Contact angles and morphology}
Since the wettability of a surface strongly influences protein adsorption \cite{elwing:1987}, a subject that will be treated in section \ref{sec:resultsproteinonmelanin}, static contact angles of water droplets on dopamine-melanin deposits are measured. For method A the contact angle does not depend on the number of immersions of the substrate in dopamine solutions and its mean value and standard deviation over 18 experiments of 1 to 14 immersions of 5 min are ($49 \pm 4$)\textdegree. These values agree with the ones reported by Xi \cite{xi:2009} (40\textdegree to 45\textdegree) and Lee \cite{lee:2007.2} (54\textdegree) for dopamine-melanin deposits on poly(ethylene) and on silicon, respectively. Copper-containing dopamine-melanin deposits from method C display a slightly higher contact angle of ($68 \pm 6$)\textdegree (3 samples, 24 h immersion time), but it is not sure if the difference between methods A and C is significant, because the contact angle on unmodified silicon slides was ten degrees higher in the measurements for method C (($63 \pm 1$)\textdegree) than in the measurements for method A (($53 \pm 4$)\textdegree).

\begin{figure}
	\centering
		\includegraphics[width=\textwidth]{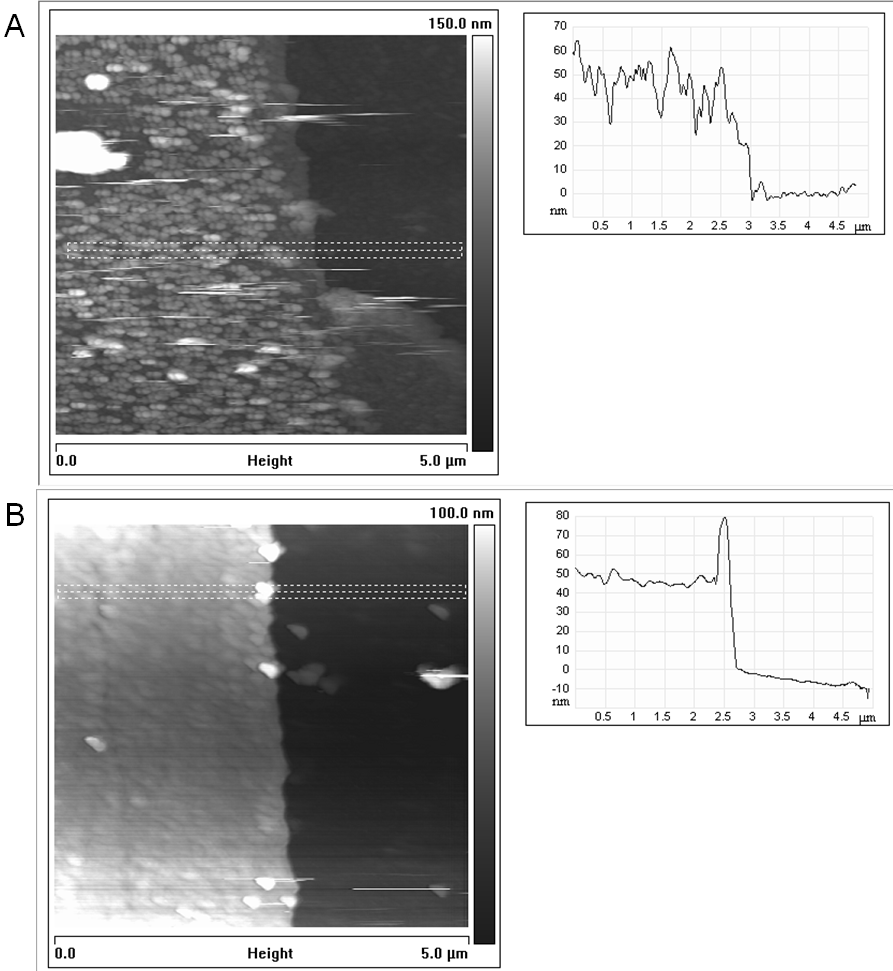}
	\caption[SFM height profiles of melanin deposits A and B]{SFM height images of partially removed melanin formed by method A on a \chem{SiO_2}-covered QCM crystal (85 x 5 min = 425 min total contact time) or by method B on a \chem{SiO_2}-covered silicon slide (22 h contact time). Height profiles on the right correspond to the mean height values in the dashed areas on the left.}
	\label{fig:compmelaninscratch1}
\end{figure}

\begin{figure}
	\centering
		\includegraphics[width=\textwidth]{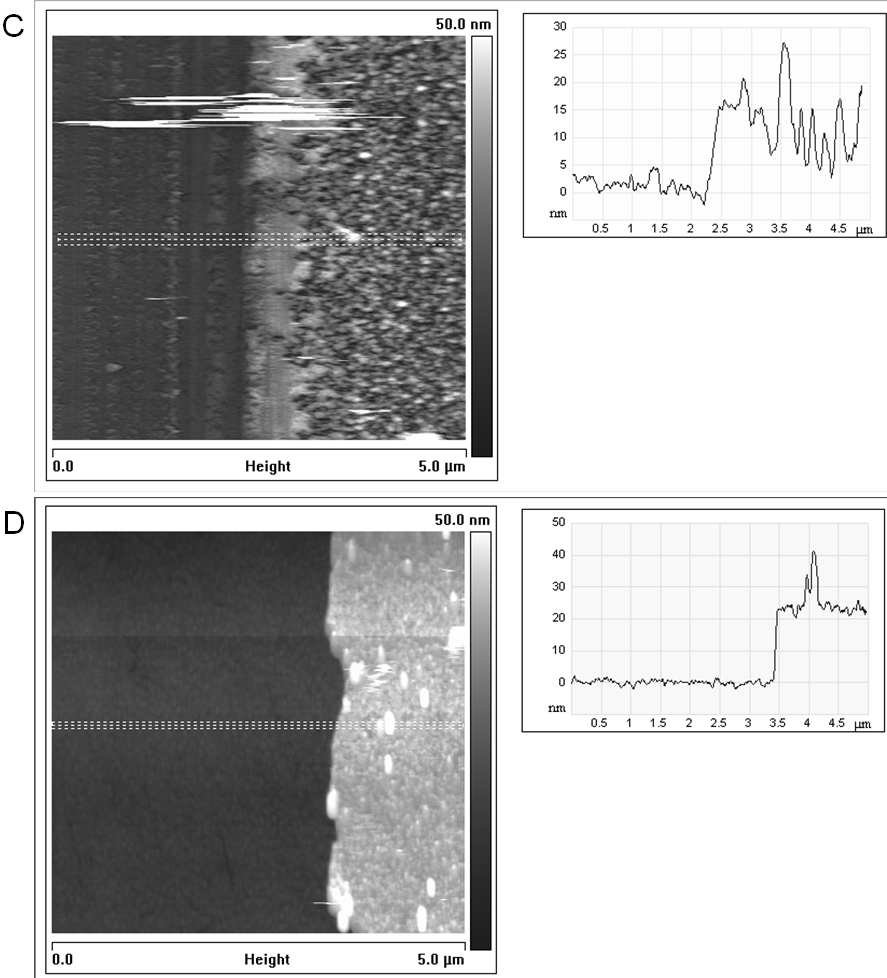}
	\caption[SFM height profiles of melanin deposits C and D]{SFM height images of partially removed melanin formed by method C on a \chem{SiO_2}-covered silicon slide (23.5 h contact time) or by method D on a \chem{Au}-covered QCM crystal (50 voltamperometry cycles). Height profiles on the right correspond to the mean height values in the dashed areas on the left.}
	\label{fig:compmelaninscratch2}
\end{figure}

\label{sec:ressfmcomp}
\begin{figure}
	\centering
		\includegraphics[width=\textwidth]{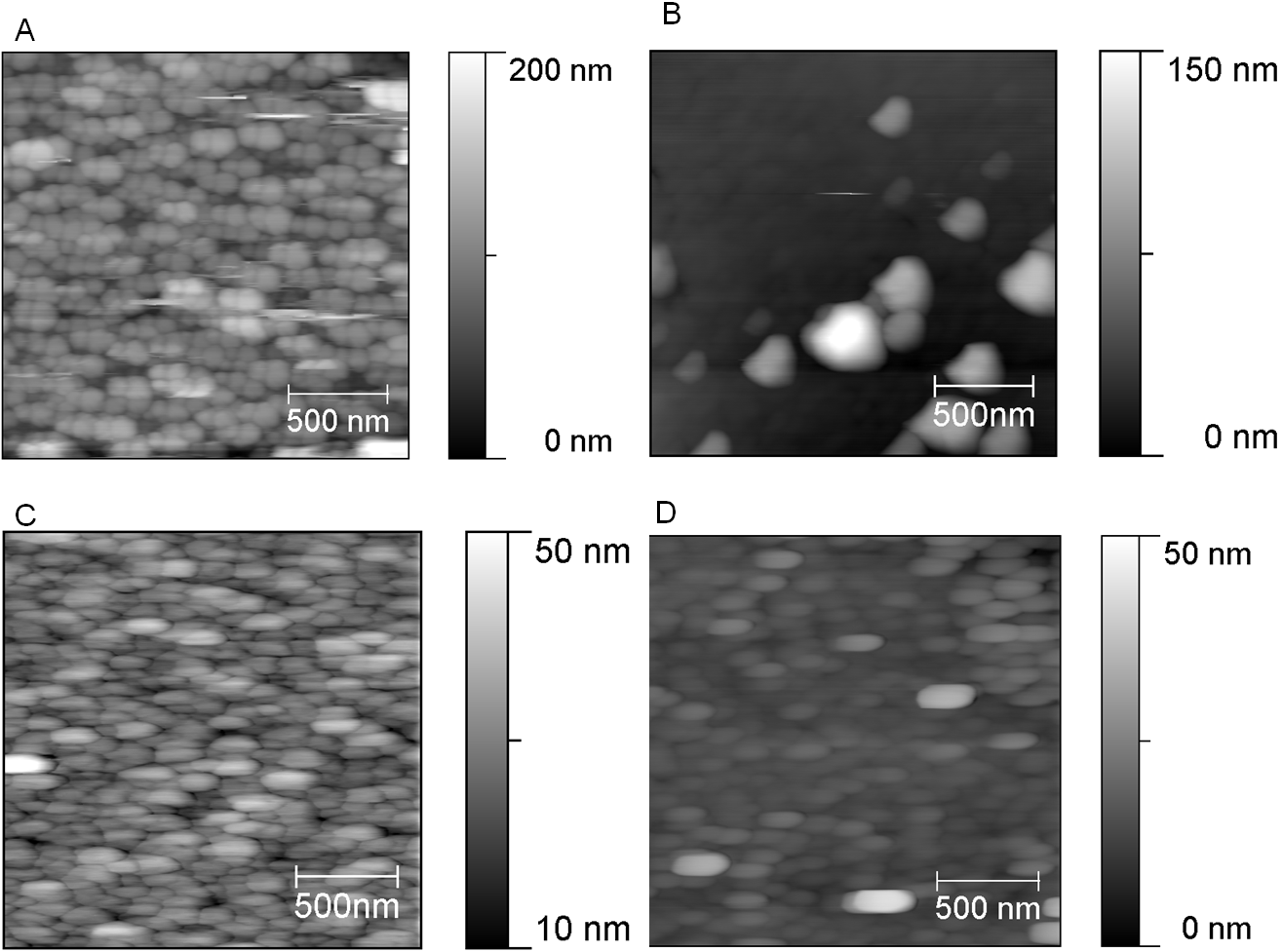}
	\caption[SFM images of melanin from different deposition methods]{SFM height images of dopamine-melanin formed by method A on a \chem{SiO_2}-covered QCM crystal (85 immersions of 5 min), by method B on a \chem{SiO_2}-covered silicon slide (22 h contact time), by method C on a \chem{SiO_2}-covered silicon slide (23.5 h contact time) or by method D on a \chem{Au}-covered QCM crystal (50 voltamperometry cycles).}
	\label{fig:compmelaninsfm}
\end{figure}
Figure \ref{fig:compmelaninsfm} shows scanning force microscopy (SFM) images in contact mode in air of dopamine-melanin films produced by the different methods. The surface morphology of the samples is very similar. In each case the entire sample surface is covered by aggregates about 200 nm to 300 nm long and about 100 nm large reminiscent of the aggregates found in melanin from glycera jaws \cite{moses:2006} or from sepia officinalis \cite{clancy:2001} \cite{nofsinger:2000} and in \chemr{(PDADMAC-melanin)_{10}} multilayers (Figure \ref{fig:sfmpdadmacmelanin}). Again the lateral extension of the aggregates is higher than the thickness of the deposits (Figures \ref{fig:compmelaninscratch1}, \ref{fig:compmelaninscratch2}) indicating a platelet-like shape of the aggregates. These observations suggest that the formation of such platelets is an intrinsic property of melanin.

In figures \ref{fig:compmelaninscratch1} and \ref{fig:compmelaninscratch2} part of the dopamine-melanin deposit is removed from the samples by scratching with a syringe needle to reveal the melanin thickness, which agrees well with the values calculated from ellipsometry or quartz crystal microbalance data for the same samples (Table \ref{tab:compmelaninsfm}). SFM does not detect any holes reaching down to the support in the dopamine-melanin deposits. Therefore dopamine-melanin forms \emph{continuous films} at the resolution of the SFM tip of about 20 nm. 
\begin{table}
	\centering
		\begin{tabular}{l|c|c|c} 
		Preparation method & SFM & QCM & Ellipsometry \\ \hline
		A: 85 x 5 min immersion & $(45 \pm 5)$ nm & $(38 \pm 2)$ nm & \\
		B: 22 h immersion & $(45 \pm 5)$ nm & & $(34 \pm 3)$ nm\\
		B with phosphate buffer: 25 h & $(45 \pm 5)$ nm & & $(44 \pm 7)$ nm \\
		C: 23.5 h immersion & $(15 \pm 5)$ nm & & $(22 \pm 1)$ nm\\
		D: 50 voltamperometry cycles & $(28 \pm 3)$ nm & $(28 \pm 2)$ nm & 
		\end{tabular}
		\caption[Thickness of dopamine-melanin deposits]{Thickness of dopamine-melanin deposits prepared by different methods measured by different techniques}
		\label{tab:compmelaninsfm}
\end{table}

\subsection{Permeability to electrochemical probes}
\begin{figure}
	\centering
		\includegraphics[width=85mm]{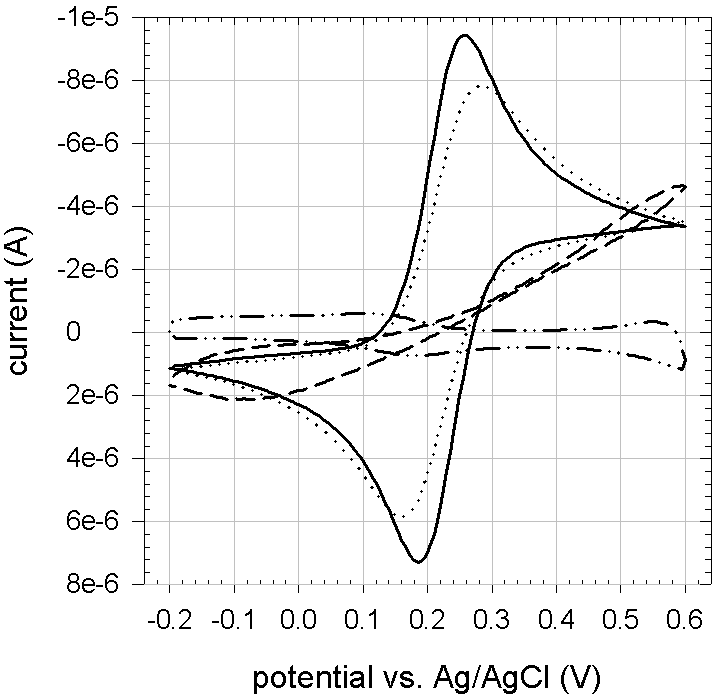}
	\caption[CV of \chem{Fe(CN)_6^{4-}} on melanin-modified electrodes]{Cyclic voltammograms corrected for capacitive currents, recorded on amorphous carbon working electrodes at a scan rate of 0.05 V/s in the presence of 1 mmol/L \chem{Fe(CN)_6^{4-}}, 10 mmol/L Tris and 0.15 mol/L \chem{NaNO_3} at pH 7.5. Dopamine-melanin was deposited on the electrode by method B for 0 min (full line), 10 min (dotted), 17 min (dashed) or 45 min (dash-dotted).}
	\label{fig:voltaaeratedmelanin}
\end{figure}
\begin{figure}
	\centering
		\includegraphics[width=85mm]{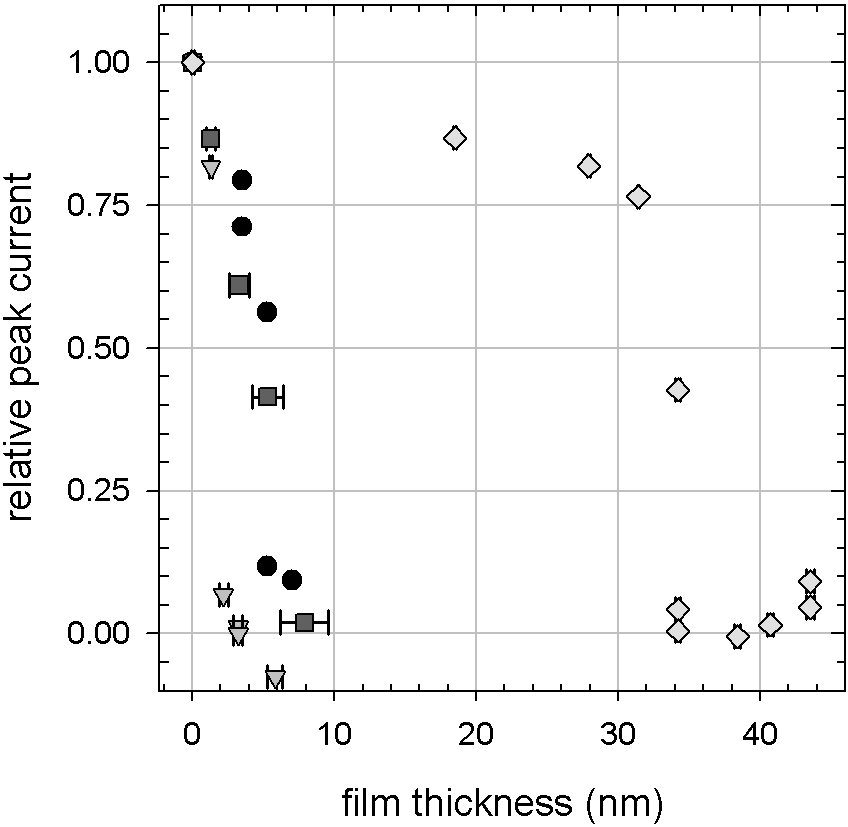}
	\caption[\chem{Fe(CN)_6^{4-}} permeability of melanin deposits from different methods]{CV peak currents of \chem{Fe(CN)_6^{4-}} versus thickness of melanin deposits on the working electrode obtained by method A (circles), B (triangles), C (squares) or D (diamonds). To get the reported values the absolute values of the oxidation and reduction peak currents are added. Then this sum is divided by the corresponding sum for the same electrode measured before melanin deposition to obtain the relative peak current. All currents are corrected for capacitive currents.}
	\label{fig:compmelaninvolta}
\end{figure}
\label{sec:compmelaninvolta}
The permeability of the dopamine-melanin films to electrochemical probes is tested using melanin deposits on the vitrous carbon working electrode of an electrochemical cell. Cyclic voltamperometry (CV) is done in a potassium hexacyanoferrate(II) solution (1 mmol/L \chem{K_4Fe(CN)_6} in 10 mmol/L Tris and 150 mmol/L \chem{NaNO_3}, pH 7.5) and the oxidation and reduction peak currents of hexacyanoferrate are monitored as a function of the melanin thickness inferred from ellipsometry or QCM data (Figure \ref{fig:compmelaninvolta}). Figure \ref{fig:voltaaeratedmelanin} shows exemplary cyclic voltammograms of dopamine-melanin prepared by method B. For all deposition methods the peak currents decrease with increasing melanin thickness and vanish at a finite thickness leading to two conclusions: First, the charges necessary to oxidise or reduce hexacyanoferrate are not conducted through the melanin deposits. Second, there are no more pits in the deposit allowing for hexacyanoferrate molecules to reach the working electrode beyond a given melanin thickness. For methods A and C the formation of an insulating melanin film occurs at a thickness of 8 nm corresponding to 12 deposition steps of 5 minutes (A) or to one immersion of nearly six hours in copper containing solutions (C). Obviously the small amounts of copper incorporated in the deposits from method C (Figure \ref{fig:xpscomposition}) do not make them electrically conducting. Using method B the oxidation and reduction currents vanish at a smaller thickness of 4 nm at a melanin deposition time of half an hour. A much thicker film (34 nm) is necessary to prevent charge transfer in the case of method D. Interestingly melanin deposition by cyclic voltamperometry (method D) slows down at the same melanin thickness (Figure \ref{fig:compmelaningrowth}) supporting the hypothesis that the thickness is limited, because the melanin deposits form insulating films. The difference in the thickness necessary to prevent charge transfer between method D and the others might be due to structural differences at a length scale below the resolution of the SFM images in section \ref{sec:ressfmcomp}, which do not show any difference.

\begin{figure}
	\centering
		\includegraphics[width=85mm]{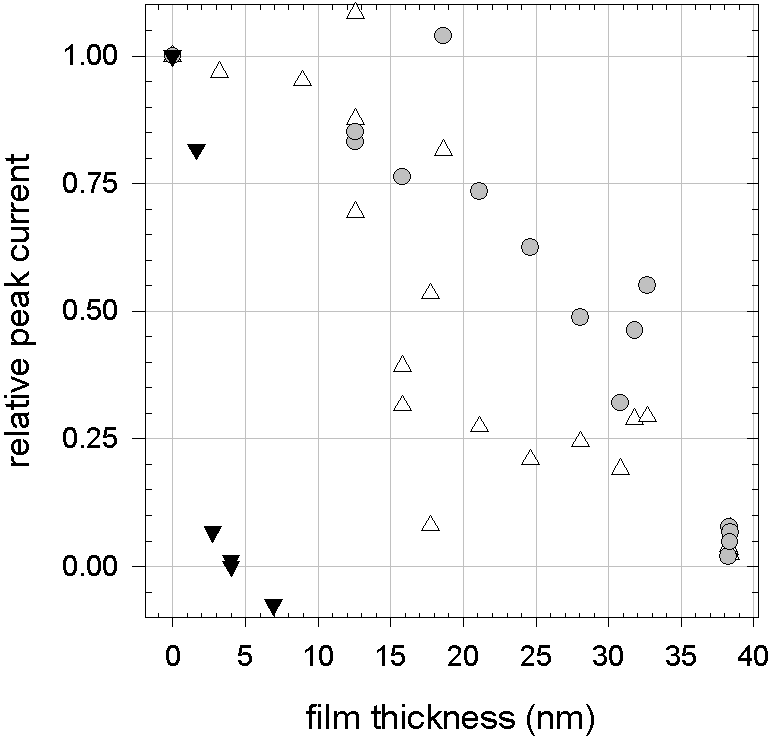}
	\caption[Permeability of melanin deposits to different electrochemical probes]{CV peak currents of \chem{Ru(NH_2)_6^{2+}} (white triangles), \chem{C_{11}H_{12}FeO} (circles) or  \chem{Fe(CN)_6^{4-}} (black triangles) on working electrodes covered with dopamine-melanin films obtained by method B versus melanin thickness. See caption of figure \ref{fig:compmelaninvolta} for calculation of the relative peak current.}
	\label{fig:voltaprobeairedmelanin}
\end{figure}
Li and others have observed that deposits obtained by electrochemical oxidation of dopamine are selectively permeable to cationic probes compared to anionic probes \cite{li:2006.1}. Thus a dopamine-melanin covered electrode can be used for electrochemical detection of dopamine, which is positively charged under physiological conditions, while anionic species like ascorbic acid that would disturb the measurement on an unprotected electrode are excluded. The feasibility of this approach, which opens perspectives for important biomedical applications, was shown by Rubianes and others \cite{rubianes:2001}.

To check if the dopamine-melanin deposits presented in this work also display electric charge dependent permselectivity, the permeability of deposits from aerated dopamine solutions (method B) to cationic hexaamineruthenium (\chem{Ru(NH_2)_6^{2+}}), neutral ferrocenemethanol (\chem{C_{11}H_{12}FeO}) and anionic hexacyanoferrate (\chem{Fe(CN)_6^{4-}}) is measured. The experiments with hexaamineruthenium and ferrocenemethanol show a large scattering in figure \ref{fig:voltaprobeairedmelanin}. Nevertheless it is clear that the film thickness of nearly 40 nm required to exclude both probes is about ten times as high as the corresponding thickness for hexacyanoferrate. As the electrochemical probes are of comparable size, the examined dopamine-melanin deposits are selectively impermeable to anionic species as reported by others for similar systems \cite{li:2006.1} \cite{rubianes:2001}. This behaviour indicates that the deposits bear a negative charge like the melanin grains examined in section \ref{sec:resultsmelaningrains}. The negative charge of the deposits will be confirmed by $\zeta$-potential measurements in section \ref{sec:resultszeta}.

\subsection{Influence of the buffer agent}
Up to now all experiments were performed in aqueous solutions buffered with tris\-(hydroxy\-methyl)\-amino\-methane, a molecule containing an amine group. Since melanin is able to bind amines (Section \ref{sec:resultsaminebinding}), the Tris molecules might influence the dopamine-melanin deposition. There is few information published about the role of buffer agents in melanin formation. For example Herlinger and others noted that the spontaneous oxidation of dopamine is inhibited by sodium tetraborate (borax) buffer \cite{herlinger:1995}. In this section dopamine-melanins deposited from aerated solutions (method B) in the presence of 50 mmol/L Tris or di-potassium hydrogenphosphate (\chem{K_2HPO_4}) at pH 8.5 are compared.

\begin{figure}
	\centering
		\includegraphics[width=85mm]{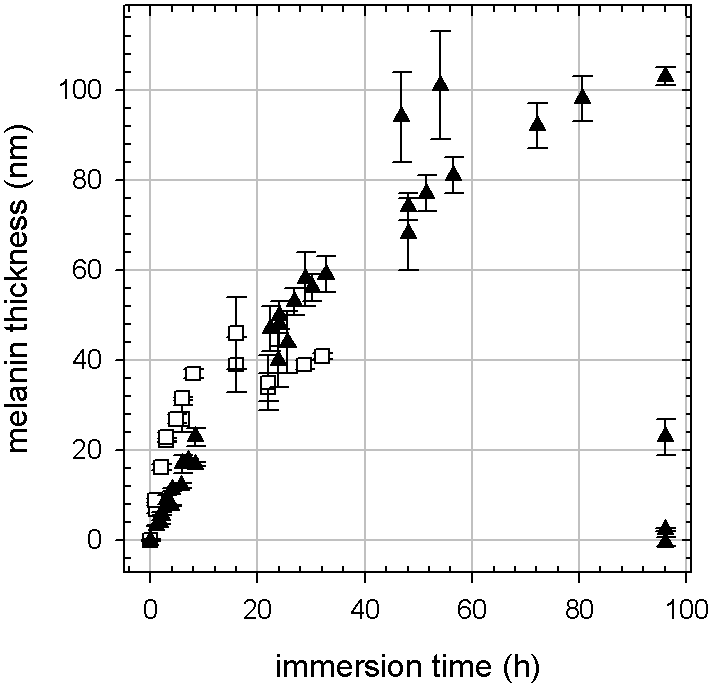}
	\caption[Influence of buffer on melanin growth]{Ellipsometric thickness of dopamine-melanin on silicon slides versus immersion time in aerated dopamine solutions buffered at pH 8.5 with 50 mmol/L Tris (squares) or \chem{K_2HPO_4} (triangles). A refractive index of $N=1.8 + 0.14 i$ is used for thickness calculation.}
	\label{fig:ellipsophosphatemelanin}
\end{figure}
Indeed there is a marked influence of the buffering agent on the growth of dopamine-melanin films (Figure \ref{fig:ellipsophosphatemelanin}). Using phosphate buffer the thickness of the deposit starts to grow linearly with the reaction time as in the case of Tris buffer but the initial growth rate is reduced by a factor of three to 2.5 nm/h. Instead of levelling off at 40 nm the thickness continues to grow up to 100 nm after 60 h to 100 h of immersion time in \emph{one single} dopamine solution containing phosphate. At this stage the deposits become unstable and dissolve. Probably their cohesion is too weak to withstand the shear stress generated by air bubbling of the solutions. There is no visible influence of intermediate rinsing and drying steps on the growth of dopamine-melanin.

\begin{figure}
	\centering
		\includegraphics[width=\textwidth]{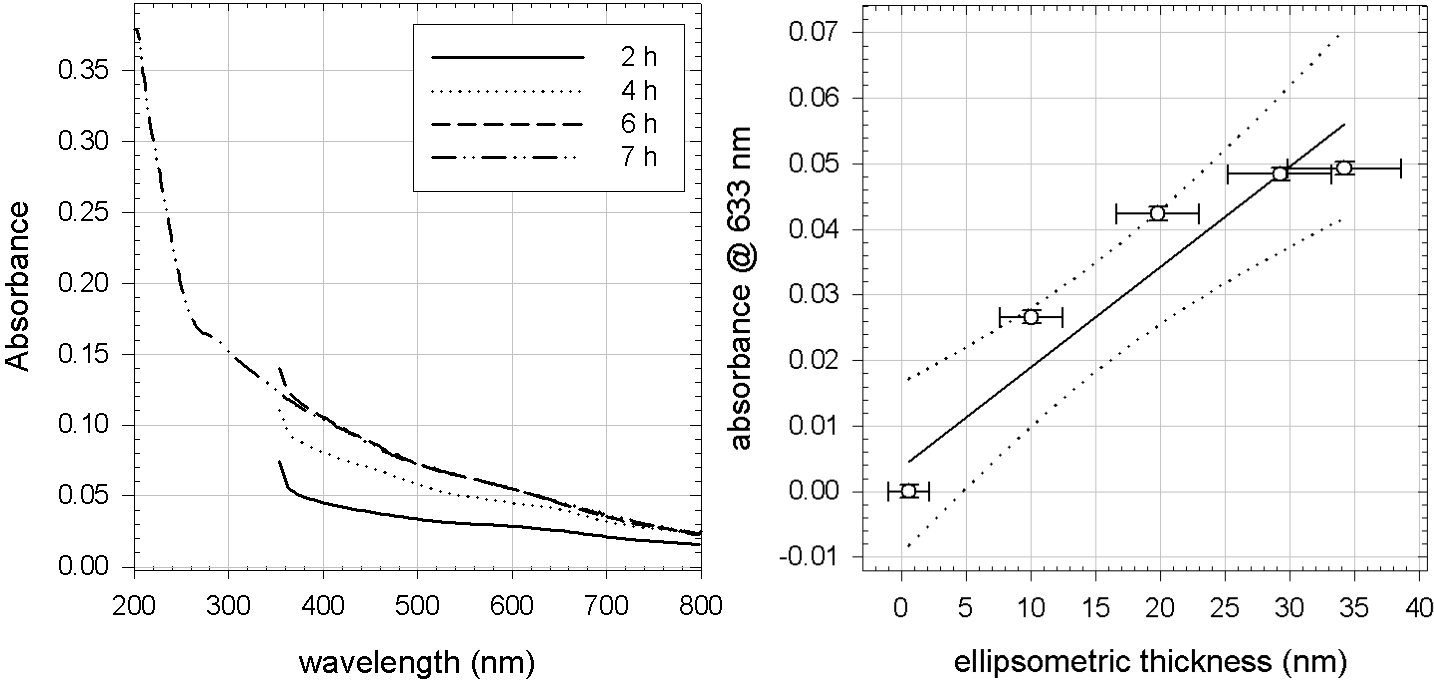}
	\caption[UV--visible spectroscopy of melanin from phosphate buffer]{Left: UV--visible absorbance spectra of melanin deposits obtained by immersion of a quartz slide for different times indicated in the figure in dopamine solutions (2 g/L in 50 mmol/L \chem{K_2HPO_4}, pH 8.5). Right: absorbance at 633 nm of the same deposits versus thickness. The full line represents a linear regression to the data, dotted lines delimit the 95 \% confidence interval.}
	\label{fig:uvvisphosphatemelanin}
\end{figure}
\begin{figure}
	\centering
		\includegraphics[width=85mm]{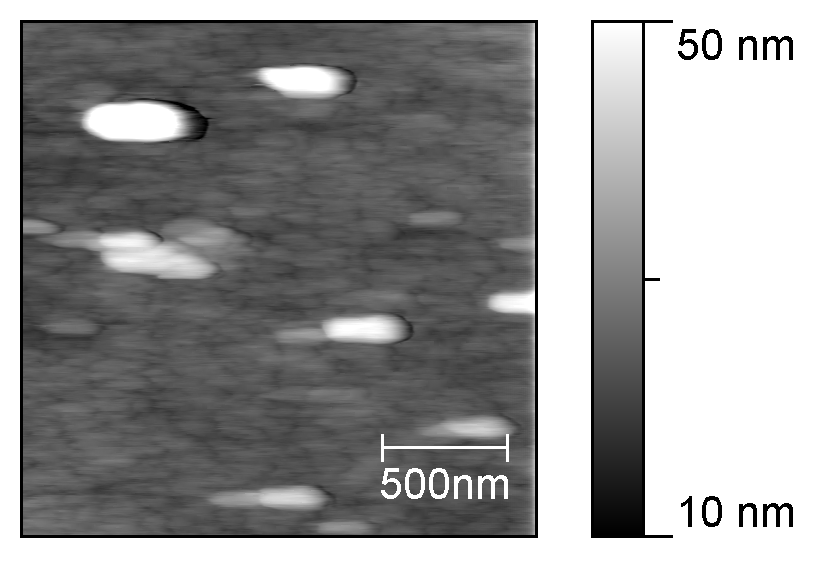}
	\caption[SFM images of melanin from phosphate buffered solutions]{SFM height image of melanin deposited on \chem{SiO_2}-covered silicon by 25 h contact with a dopamine solution (2 g/L in 50 mmol/L \chem{K_2HPO_4}, pH 8.5).}
	\label{fig:sfmphosphatemelanin}
\end{figure}
\begin{figure}
	\centering
		\includegraphics[width=\textwidth]{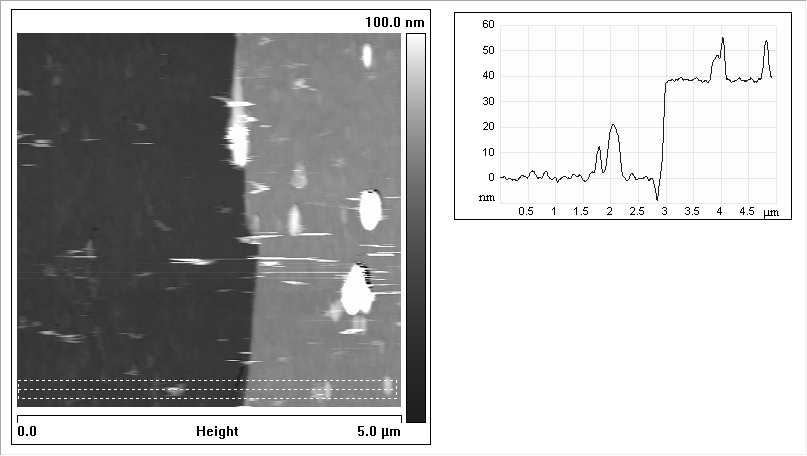}
	\caption[SFM height profiles of melanin from phosphate buffered solutions]{SFM height images of the same dopamine-melanin deposit as in figure \ref{fig:sfmphosphatemelanin} partially removed by syringe scratching. The height profile on the right corresponds to the mean height values in the dashed area on the left.}
	\label{fig:scratchphosphatemelanin}
\end{figure}
Nevertheless concerning the physical properties of dopamine-melanin, deposits obtained from phosphate buffered solutions are very similar to their counterparts from Tris buffered solutions:

The UV--vis absorbance spectra are monotonous (Figure \ref{fig:uvvisphosphatemelanin}) and the extinction coefficient at a wavelength of 633 nm is close to the ones of dopamine-melanin from other methods (Table \ref{tab:compmelaninabs}).

XPS does not detect any influence of the buffer agent on the chemical composition of dopamine-melanin (Figures \ref{fig:xpscomposition} \ref{fig:xpsc1sn1s})

The SFM images \ref{fig:sfmphosphatemelanin} and \ref{fig:scratchphosphatemelanin} closely resemble there previously presented counterparts \ref{fig:compmelaninsfm} and \ref{fig:compmelaninscratch1}. The melanin height measured by SFM matches the one calculated from ellipsometry data (Table \ref{tab:compmelaninsfm}).

The static water contact angle of ($59 \pm 6$)\textdegree (3 samples of 25 h to 54 h immersion time in phosphate buffered solutions) is only slightly higher than for samples prepared in Tris buffer (($49 \pm 4$)\textdegree).

Cyclic voltamperometry shows that dopamine-melanin films from phosphate buf\-fered solutions become impermeable to hexacyanoferrate at a thickness of 4 nm just like their Tris counterparts (Figure \ref{fig:compmelaninvolta}).

\subsection{Conclusion}
In conclusion all presented methods lead to the deposition of melanin films of controlled thickness from aqueous dopamine solutions. The deposits are hydrophilic, a useful property for their potential use as platform for biomolecule grafting. In SFM images all samples show a similar granular morphology that seems to be typical of melanin in different environments. Beyond a given thickness the deposits become impermeable to hexacyanoferrate, and for one deposition method it is confirmed that dopamine-melanin selectively excludes anionic electrochemical probes opening a promising route to applications in electrochemical dopamine sensing.

Concerning their differences the methods have the following advantages and disadvantages:
\begin{description}
	\item[Method A] is applicable to virtually any kind of substrate \cite{lee:2007.2}, easy to realise and leads to a linear and ``unlimited'' film growth, but this multistep process is tedious and implies a high dopamine consumption.
	\item[Method B] is also applicable to virtually any kind of substrate \cite{lee:2007.2}. It is a one step process for a thickness up to 40 nm and repeatable to obtain a higher thickness, but this method is difficult to realise in small measurement cells like in a QCM.
	\item[Method C] can be used in an acidic, anaerobic environment, and it is a one step process, but the supplementary reactant copper sulphate is needed.
	\item[Method D] is another one step process and there is no uncontrolled melanin formation in solution, but it needs an electrically conducting substrate, and the melanin thickness is limited to 45 nm.
\end{description}

Changing the buffer agent from Tris to phosphate leads to a modified growth of dopamine-melanin but the physical properties of the obtained deposits stay the same.

At the present state of our investigation the following questions remain to be answered: What exactly happens during the drying of melanin from aerated dopamine solutions to allow for further melanin deposition steps? How can copper sulphate initiate the formation of dopamine-melanin in an acidic environment? What causes the absorbance peaks in the UV range of dopamine-melanin prepared in the presence of copper sulphate? By which mechanism does the choice of a buffer agent influence the dopamine-melanin growth?

%% file: resultsnx5min.tex
\section{Melanin deposition by immersion in multiple dopamine solutions}
\label{sec:resnx15min}
This section contains a more detailed investigation of the properties of melanin deposits made by successive immersion of a silicon oxide substrate in multiple freshly prepared dopamine hydrochloride solutions (2 g/L, 50 mmol/L Tris, pH 8.5). This preparation method described as ``Method A'' in section \ref{sec:resdepostitionmethods} is chosen, because it is easily applied to all planned methods of analysis.

\subsection{Growth mechanism}
\begin{figure}
	\centering
		\includegraphics[width=85mm]{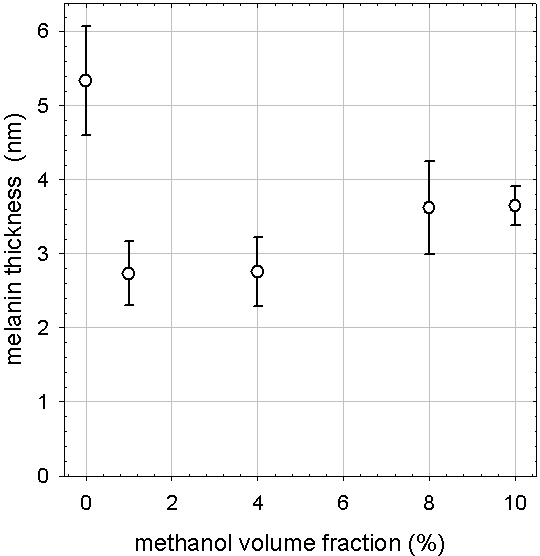}
	\caption[Influence of methanol on dopamine-melanin deposition]{Thickness of melanin deposits on silicon (4 x 15 min immersions in 2 g/L dopamine hydrochloride, 50 mM Tris, pH 8.5) measured by ellipsometry assuming a refractive index of 1.465 versus volume fraction of methanol in the dopamine solution.}
	\label{fig:ellipsomethanol}
\end{figure}
Section \ref{sec:resfirstobservations} implied that the deposition of dopamine-melanin is a process initiated by the interaction of monomeric species or small oligomers with the support surface. This suggests that the growth process may follow a mechanism similar to oxidative aniline polymerisation at interfaces \cite{sapurina:2001}. In this case a radical cation obtained from the oxidation of aniline adsorbs at the solution-solid interface and polymerisation occurs from these active centres. Similarly oxidation of dopamine can lead to the dopamine semiquinone radical \cite{barreto:2001} \cite{szpoganicz:2002}. To demonstrate the role of radical species in the deposition of dopamine-melanin, deposition experiments are performed from Tris buffer containing methanol, a well-known radical scavenger \cite{herlinger:1995}. In these experiments a silicon slide is immersed successively in four freshly prepared dopamine solutions  each immersion lasting for 15~min as in figure \ref{fig:ellipsomelanin}. By adding methanol to the dopamine solutions the thickness of the melanin deposit measured by ellipsometry is reduced although the deposition is not suppressed (Figure \ref{fig:ellipsomethanol}). Interestingly the strongest effect of methanol on the film thickness is reached at a methanol volume fraction as small as 1~\%. Herlinger and others observed that methanol also slows down the polymerisation of dopamine in solution without of totally suppressing it \cite{herlinger:1995}. These experiments suggest a radicalar process probably involving dopamine semiquinone, but do not constitute a definitive proof. To obtain such a proof, more sophisticated techniques like electron paramagnetic resonance (EPR) spectroscopy \cite{szpoganicz:2002} would have to be used.

\begin{figure}
	\centering
		\includegraphics[width=\textwidth]{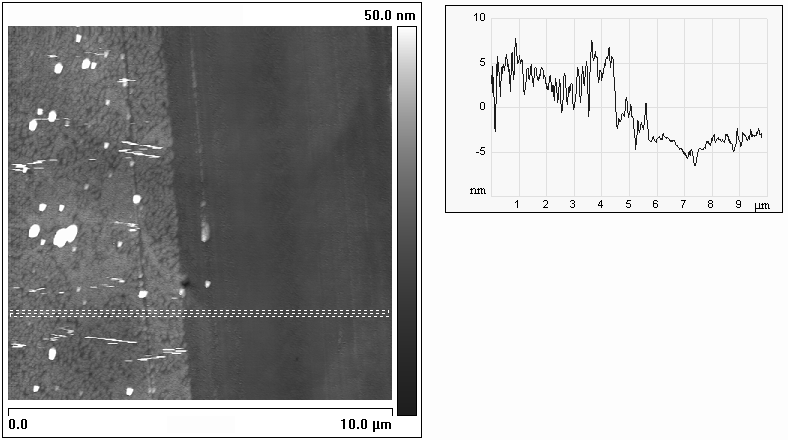}
	\caption[SFM of dopamine-melanin from method A]{Left: SFM height image at ambient humidity in contact mode of melanin deposited on \chem{SiO_2}-covered silicon by 8 x 15 min immersions in dopamine solutions. In the right part of the image the deposit is removed using a syringe needle. Right: Height profile of the dashed rectangles in the image on the left.}
	\label{fig:sfm8x15mindopamine}
\end{figure}
As mentioned in section \ref{sec:resxpscomposition}, a notable signal of the silicon support is visible in x-ray photoelectron spectra (XPS) (Figure \ref{fig:xps32x15min}) of melanin deposits from 32 immersions of 15~min in dopamine solutions. At ambient conditions these deposits have an ellipsometric thickness of ($24 \pm 3$) nm (Figure \ref{fig:ellipsomelanin}) and should be continuous films according to scanning force microscopy (SFM) images (Figure \ref{fig:sfm8x15mindopamine}) and electrochemical experiments (Section \ref{sec:compmelaninvolta}). Assuming the formation of a continuous film, the presence of a silicon signal can be explained if the dopamine-melanin deposit is altered by dehydration in the ultra-high vacuum of the XPS analysis chamber. Dehydration might reduce the thickness of the deposit below the XPS probing depth of 9 nm (for a take-off angle of 90\textdegree) or create cracks in the deposit making the silicon support visible.

\begin{figure}
	\centering
		\includegraphics[width=85mm]{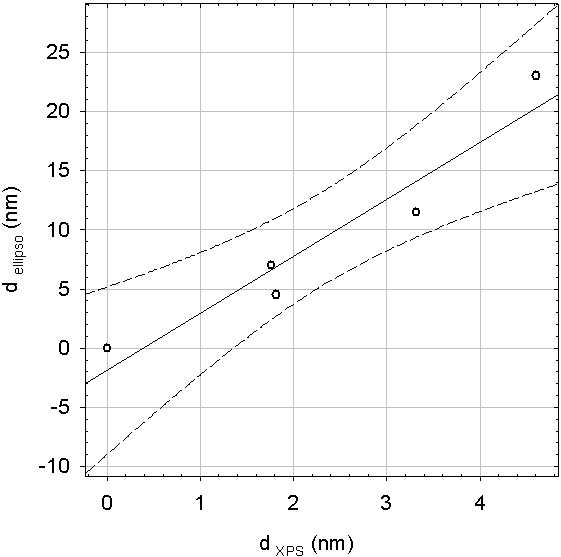}
	\caption[Ellipsometric thickness of dopamine-melanin deposits versus XPS thickness]{Thickness of dopamine-melanin deposits measured by ellipsometry (\chem{d _{ellipso}}) versus thickness of the same samples obtained from XPS measurements (\chem{d _{XPS}}). The full line is a linear regression and the dashed lines represent the limits of the 95 \% confidence interval}
	\label{fig:thicknessellipsoxps}
\end{figure}

For the following analysis, the first assumption is supposed to be the main reason for the appearance of a silicon signal. Thus the mean thickness of the dopamine-melanin deposits in ultra-high vacuum can be calculated using equation \ref{eqn:xpsthickness}. For each sample the thickness determined by XPS is lower than the one determined by ellipsometry. The correlation between the results from the two techniques is close to linear (Figure \ref{fig:thicknessellipsoxps}) with a factor of about four between corresponding values. Thus complete dehydration of the melanin deposits in ultra-high vacuum induces an important shrinking of the dopamine-melanin deposit. This observation is in line with the findings by other groups that dehydration strongly modifies the mass of melanin deposits \cite{subianto:2005} as well as other properties of melanin like its electrical conductivity \cite{jastrzebska:1995}.

\subsection{pH-dependent stability}
\begin{figure}
	\centering
		\includegraphics[width=85mm]{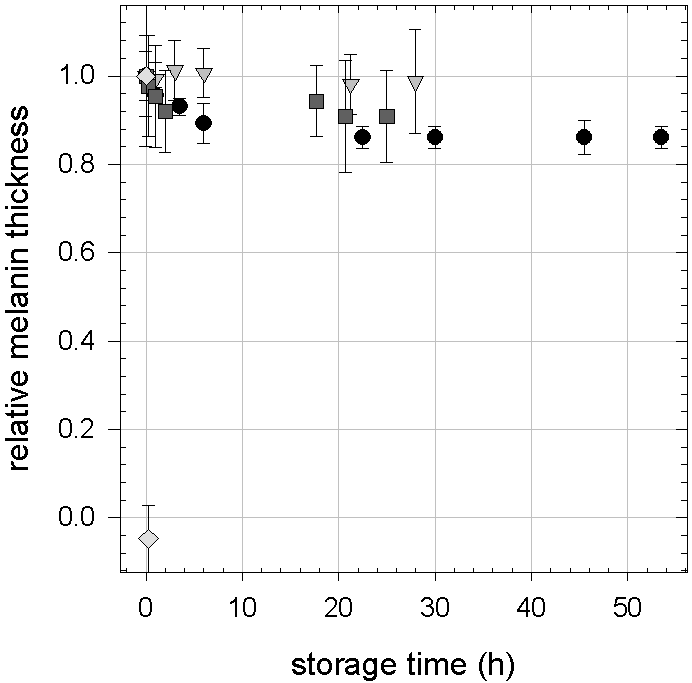}
	\caption[pH-dependent stability of dopamine-melanin deposits]{Evolution of the ellipsometric thickness of melanin deposits in solutions at pH 1 (circles), 3 (triangles), 11 (squares) or 13 (diamonds) relative to their thickness at the end of the melanin deposition by dipping a silicon support for 16 x 15 min in dopamine solutions.}
	\label{fig:melaninph}
\end{figure}
\label{sec:stabilitymelaninph}
The stability of melanin deposits is investigated in acidic environment in presence of hydrochloric acid at pH 3 and pH 1 as well as in basic environment in presence of sodium hydroxide at pH 11 and pH 13, because melanin is known to be soluble only in strongly alkaline solutions \cite{bothma:2008}. The thickness of the deposit measured by ellipsometry in the dry state decreases by no more than $14 \%$ after 54~h at pH 1 (Figure \ref{fig:melaninph}). No measurable changes in thickness are found at pH 3 and pH 11 for more than 24~h. On the contrary, exposition to pH 13 for only 15 min completely removes the dopamine-melanin deposit either due to the dissolution of the melanin deposit or to a partial dissolution of the silica support. Qualitative observations show that exposition to pH 13 can also remove dopamine-melanin from various other supports (polymers and stainless steel) except gold. Thus the dopamine-melanin deposit and not (only) the support is dissolved at pH 13. This finding has important practical significance because it shows that the surfaces covered with dopamine-melanin can be easily cleaned in strongly alkaline solutions.

\begin{figure}
	\centering
		\includegraphics[width=85mm]{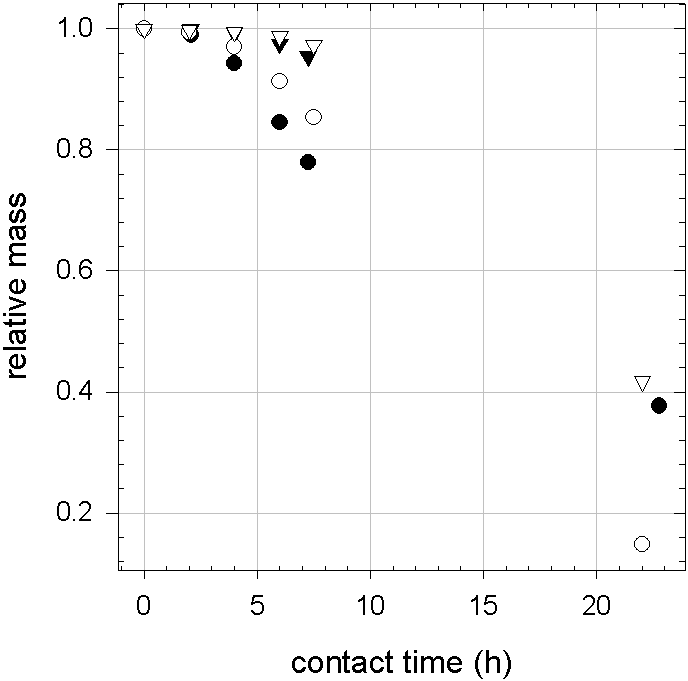}
	\caption[Corrosion of melanin-coated zinc plates]{Relative mass of zinc plates versus contact time with 0.1 mol/L \chem{HCl}. Black and white symbols represent two independent experiments with one melanin-coated (circles) and one uncoated (triangles) zinc plate each. Melanin coating prepared by 12 x 15 min immersions in dopamine solutions.}
	\label{fig:corrosionzn}
\end{figure}
Ou and others showed that silicon coated with 3-aminopropyl triethoxysilane presents a corrosion current density that decreases upon deposition of increasing amounts of dopamine-melanin on its surface \cite{ou:2009}. This observation, the stability of dopamine-melanin deposits in aqueous solutions and their adherence to metal surfaces, make them candidates for anti-corrosive coatings. To test this possible application, pristine zinc plates cleaned with ethanol and plates immersed twelve times for 15 min in dopamine solutions are given in dilute hydrochloric acid solution at pH 1 and their dry mass is regularly measured. Figure \ref{fig:corrosionzn} shows that in two independent experiments the mass relative to the initial mass of the melanin-coated plates decreases in a similar or even faster way than the one of the uncoated samples. Thus with the employed protocol dopamine-melanin cannot be used as an anti-corrosive coating.

The following reasons could be responsible for the absence of a protective effect: \chem{H^+} ions might diffuse through the dopamine-melanin coating due its cationic permselectivity (Section \ref{sec:compmelaninvolta}) or the coating is not thick enough or it was damaged by sample handling within the experiment. At the time of the corrosion tests, only melanin deposition method A was available, and the other methods to easily obtain thicker dopamine-melanin deposits were not established yet. Therefore this investigation was not continued. 

\subsection{Zeta-potential}
\label{sec:resultszeta}
\begin{figure}
	\centering
		\includegraphics[width=85mm]{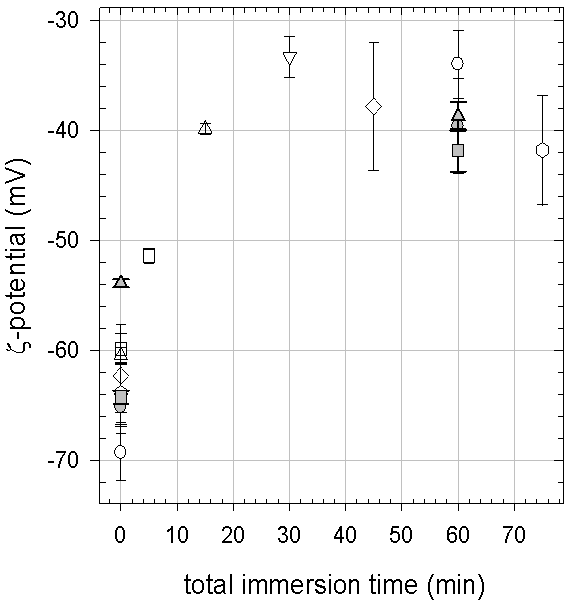}
	\caption[$\zeta$-potential of melanin on glass slides]{$\zeta$-potential measured in presence of 5 mmol/L Tris at pH 8.5 of dopamine-melanin deposits on glass made by successive immersions of 5 min in dopamine solutions. Each symbol represents the $\zeta$-potential measured before (immersion time 0) and after melanin deposition for a given substrate in an independent experiment.}
	\label{fig:zetamelanin}
\end{figure}
Figure \ref{fig:zetamelanin} shows the  $\zeta$-potential of dopamine-melanin deposits for increasing numbers of immersions lasting 5 min of the sample in dopamine solutions. The $\zeta$-potential increases from the value of the pristine glass slide (between -70 mV and -60 mV at pH 8.5) to stabilize around -40 mV after three immersion steps. Thus at the length scale probed by the $\zeta$-potential measurements, the substrate appears completely covered with dopamine-melanin after three steps (15 min total immersion time) instead of twelve steps (60 min total immersion time) in the electrochemical experiments described in section \ref{sec:compmelaninvolta}. This difference can be explained by the differences between the two measurement methods: With the streaming potential technique one measures the mobility of the counterions in the double layer above the film, whereas counterions present in small channels in the film are not affected by the liquid flow. On the contrary, these channels allow for hexacyanoferrate ions to reach the working electrode in electrochemical experiments. When the number of deposition steps increases from three to twelve, the channels are progressively filled or occluded with melanin leading to the observation of a continuous film by both methods.

After twelve immersion steps in dopamine solutions the $\zeta$-potential is $(-39 \pm 3)$ mV at pH 8.5. Application of the Grahame equation \ref{eq:grahame} with a supporting electrolyte concentration of $C_0 = 1.42$ mol m$^{-3}$ (only 28 \% of the Tris molecules are protonated at pH 8.5)\footnote{The dissociation constant $K_a$ for an acid $HB^+$ to form its conjugate base $B$ and a proton $H^+$ is defined by the equilibrium concentrations: $K_a = [B][H^+]/[HB^+]$. The relative concentration $C_r$ of protonated molecules at $pH=8.5$ with $pK_a=8.1$ (value for Tris according to the supplier) is
$$
C_r=\frac{[HB^+]}{[B]+[HB^+]}=\frac{1}{\frac{[B]}{[HB^+]}+1}=\frac{1}{\frac{[K_a]}{H^+}+1}=\frac{1}{10^{(pH-pK_a)}+1} \approx 0.28 .
$$}
yields a surface charge density of $-3.7 \cdot 10^{-3}$ C m$^{-2}$.

\begin{figure}
	\centering
		\includegraphics[width=0.9\textwidth]{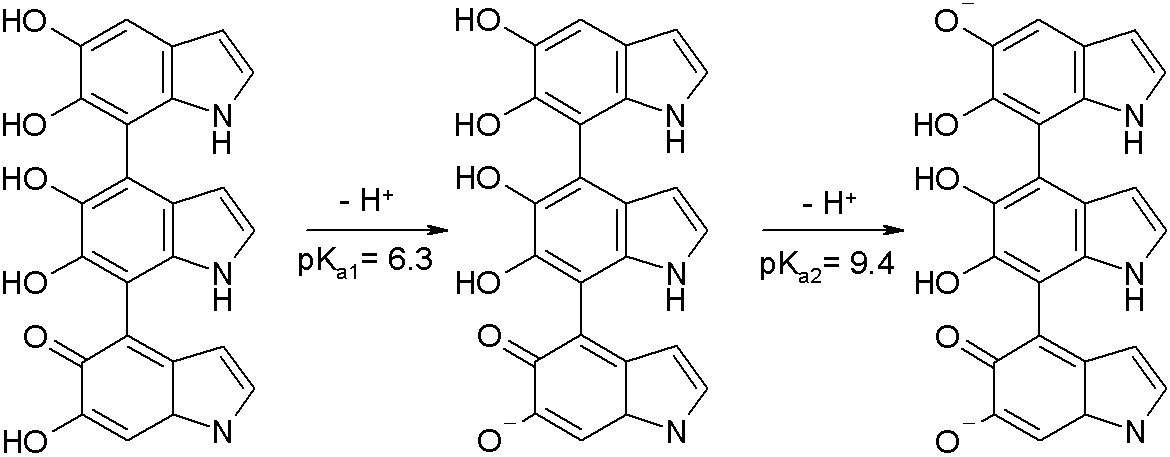}
	\caption[Deprotonation scheme of DHI-melanin]{Average building block of DHI-melanin and its first deprotonation steps as proposed by Szpoganicz et al. \cite{szpoganicz:2002}}
	\label{fig:szpoganicz}
\end{figure}
The negative surface charge at pH 8.5 may originate from the dissociation of quinone imine and catechol groups. Szpoganicz and others have performed potentiometric titrations of dihydroxyindole-melanin (DHI-melanin) in aqueous solution to obtain the dissociation constants ($K_a$) of acidic functionalities present on melanin \cite{szpoganicz:2002}. They numerically reproduced the behaviour of DHI-melanin using the average building block represented in figure \ref{fig:szpoganicz} to find the first two $pK_a$ ($pK_a =-\log(K_a)$) values as $pK_{a1} = 6.3$ and $pK_{a2} = 9.4$ for deprotonation of the quinone imine group and one catechol group.

\begin{figure}
	\centering
		\includegraphics[width=85mm]{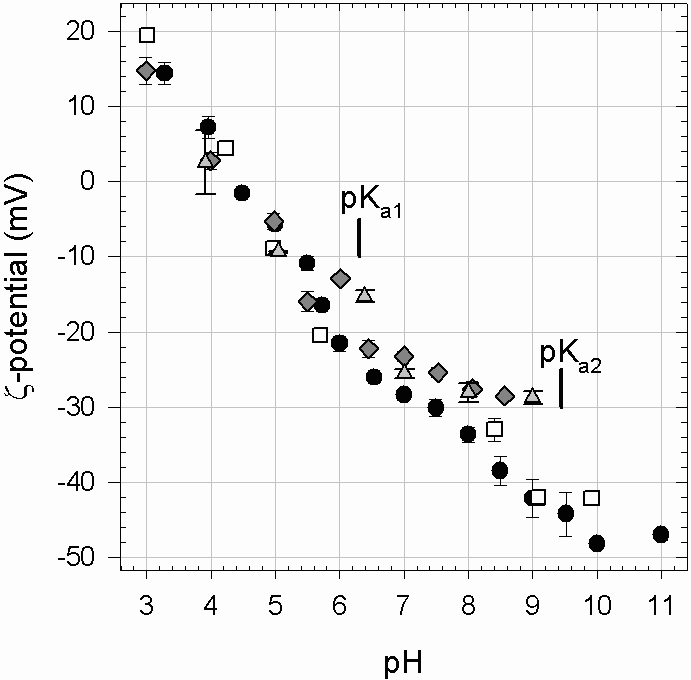}
	\caption[$\zeta$-potential of melanin versus measurement pH]{$\zeta$-potential of melanin deposits from 12 x 5 min immersions in dopamine solutions versus pH of the measurement solution containing 5 mmol/L Tris. Measurements are performed in the order of descending pH. Different symbols represent independent experiments. Please refer to the main text for the meaning of the $pK_a$ values indicated by vertical lines.}
	\label{fig:zetatitration}
\end{figure}
The $\zeta$-potential of dopamine-melanin films is monitored for \emph{descending pH} of the measurement solution (Figure \ref{fig:zetatitration}). Due to technical problems it was not possible to follow the $\zeta$-potential with ascending pH in a reproducible manner. As expected the absolute value of the $\zeta$-potential decreases with decreasing pH but it is not possible to discern individual protonation steps at the pH values corresponding to the values $pK_{a1}$ and $pK_{a2}$ observed by Szpoganicz and others \cite{szpoganicz:2002}. This might be caused by experimental scattering between different measurements or by \emph{chemical disorder} \cite{cheng:1994} \cite{dischia:2009} \cite{meredith:2006.1} \cite{tran:2006} within melanin. In a chemically disordered material chemical groups reside in a multitude of differing chemical environments leading to a large ensemble of $pK_a$ values.

%\begin{figure}
%	\centering
%		\includegraphics[width=85mm]{quinoneprotonation.png}
%	\caption[Sketch of quinone protonation]{Quinone protonation possibly responsible for the positive dopamine-melanin charge at a pH below 4.5.}
%	\label{fig:quinoneprotonation}
%\end{figure}
%The $\zeta$-potential reaches positive values when the solution pH decreases below 4.5. The positive charge can be caused by the protonation of amine groups within dopamine-melanin. Indeed the program Marvin Sketch (version 5.3.1, ChemAxon Ltd. , Budapest, Hungary, available free of charge at \href{http://www.chemaxon.com/marvin/sketch}{www.chemaxon.com/marvin/sketch}) predicts a $pK_a$ of 4.0 for the deprotonation of the secondary amine in the quinone structure represented in figure \ref{fig:quinoneprotonation}. This value is quite close to the point of zero charge of melanin deposits (Figure \ref{fig:zetatitration}).

\subsection{Protein adsorption}
\label{sec:resultsproteinonmelanin}
To investigate the adsorption of proteins on dopamine-melanin films by quartz crystal microbalance with dissipation monitoring (QCM-D), lysozyme, myoglobin and $\alpha$-lactalbumin are chosen as model proteins, because they have comparable molar masses but different isoelectric points (pI, Table \ref{tab:proteinmasses}). Melanin films are prepared by 12 injections of dopamine hydrochloride solutions (2 g/L in 50 mmol/L Tris at pH 8.5) lasting 5~min each. At the end of the dopamine deposition, which should lead to the formation of a continuous film (Section \ref{sec:compmelaninvolta}), the QCM measurement cell is rinsed with pure buffer solution and then a protein solution is injected.
%In addition the adsorption of alkaline phosphatase was studied because its enzymatic activity can be easily observed.
\begin{table}
	\centering
		\begin{tabular}{l|r|r|r|r}
		Protein & $M$ (kg/mol) & pI & $\Gamma$ (ng/cm$^2$) & $\Gamma_{SDS}$ (ng/cm$^2$)\\ \hline
		Lysozyme & 14.3  & 11.4 & $640 \pm 01$ & $291 \pm 01$ \\
		Myoglobin & 17.0 & 7.2 & $242 \pm 33$ & $227 \pm 57$  \\
		$\alpha$-Lactalbumine & 14.2  & 4.5 & $153\pm 80$& $130\pm 56$
		%Alkaline Phosphatase & 	162 \cite{brun:2007} & 4.4 \cite{brun:2007}\\ \hline
		\end{tabular}
	\caption[Protein properties and adsorption behaviour]{Molar mass $M$, isoelectric point pI and adsorption behaviour on dopamine-melanin of the examined proteins. $\Gamma$: Adsorbed mass per surface area on melanin films measured by QCM using the Sauerbrey approximation after buffer rinsing at the end of 20 min of protein adsorption from solutions containing 1 g/L protein in 50 mmol/L Tris buffer at pH 8.5. $\Gamma_{SDS}$: Remaining mass after 10 min to 15 min of further rinsing with SDS at 10 mmol/L and another buffer rinsing step. Values are reported as mean value $\pm$ one standard deviation over two independent experiments.}
	\label{tab:proteinmasses}
\end{table}

\begin{figure}
	\centering
		\includegraphics[width=85mm]{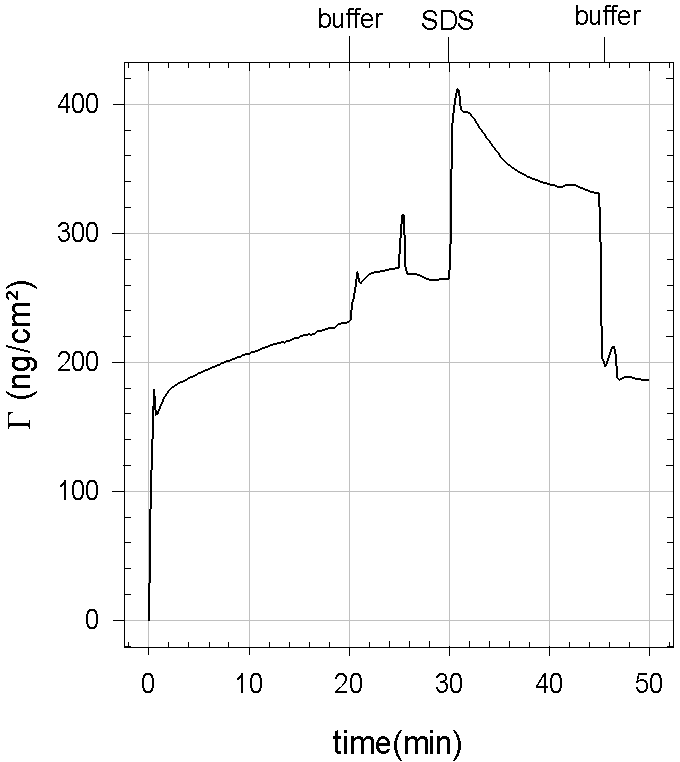}
	\caption[Myoglobin adsorption on dopamine-melanin followed by QCM]{Adsorbed mass per surface area ($\Gamma$) of myoglobin on a dopamine-melanin film calculated from the frequency shift at the third harmonic measured by QCM. During the first 20 min the quartz crystal is in contact with a protein solution (1 g/L in 50 mmol/L Tris buffer at pH 8.5) that is subsequently replaced by pure buffer solution, a SDS solution at 10 mmol/L and another buffer solution at the times indicated in the figure. The apparent increase in mass at the beginning of the first buffer rinsing step is an artefact due to the different viscosity of the pure buffer solution and the one containing the protein.}
	\label{fig:qcmmyoglobin}
\end{figure}

For the three proteins the frequency shifts at all overtones of the quartz crystal overlap within 5 \% during the adsorption experiments, and the dissipation is small ($<10^{-5}$), justifying the use of the Sauerbrey equation (Equation \ref{eqn:sauerbreynumerical}) to calculate the adsorbed mass. Since the frequency shift reaches a plateau within less than 20 min (Figure \ref{fig:qcmmyoglobin}), adsorption times are limited to this duration. Figure \ref{fig:qcmmyoglobin} shows a typical adsorption curve of myoglobin and its partial desorption during rinsing with an aqueous solution of 10 mmol/L sodium dodecylsulphate (SDS). The adsorbed mass per unit area increases upon injection of SDS as observed by others \cite{wahlgren:1990}. This reflects strong binding between the anionic surfactant and positively charged groups of the protein \cite{nelson:1971}. Protein desorption starts after this transient binding regime. SDS is not expected to bind to dopamine-melanin films owing to their hydrophilic nature and negative surface charge. Hence protein elution will occur through surfactant binding and solubilization of protein-surfactant aggregates.

Table \ref{tab:proteinmasses} summarizes the adsorbed masses of the different proteins before and after rinsing with SDS solution showing that lysozyme adsorbs in higher amounts than the other proteins. In the solutions employed for the adsorption experiments the pH is adjusted to 8.5 leading to a positive net charge on lysozyme. Thus a positive electric charge seems to be favourable for adsorption on the dopamine-melanin film, which is negatively charged under the employed conditions according to the  $\zeta$-potential measurements (Figure \ref{fig:zetatitration}). These observations point to an important role of electrostatic interactions probably between protonated amine groups of the proteins and negatively charged groups of dopamine-melanin. This is further supported by the observation that the  $\zeta$-potential of a dopamine-melanin film increases from $(-40 \pm 4)$~mV to $(-18 \pm 2)$~mV upon adsorption of lysozyme (20 min from a 1 g/L solution in 50 mmol/L Tris at pH 8.5). Nevertheless the adsorption of lysozyme does not induce a charge overcompensation at the surface. Since the adsorption has reached a saturation level, the absence of surface charge inversion is not due to an incomplete surface coverage. Probably lysozyme molecules, which have a highly anisotropic charge distribution \cite{haggerty:1993}, orient their positive charge to the melanin film exposing most of the negatively charged groups to the buffer solution.

\begin{figure}
	\centering
		\includegraphics[width=85mm]{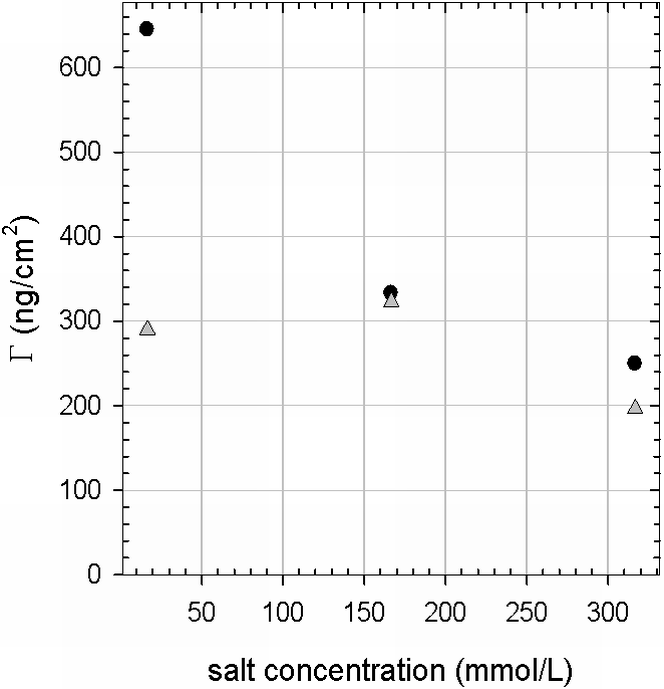}
	\caption[Adsorption of lysozyme on melanin versus salt concentration]{Adsorbed mass per surface area $\Gamma$ of lysozyme on dopamine-melanin films measured by QCM, quantified using the Sauerbrey equation and plotted versus the total salt concentration of the lysozyme solution (1 g/L in buffer containing 50 mmol/L Tris and 0 to 300 mmol/L NaCl at pH 8.5). Circles: Mass after buffer rinsing at the end of 20 min of protein adsorption. Triangles: Remaining mass after 10 to 15 min of further rinsing with SDS at 10 mmol/L and another buffer rinsing step.}
	\label{fig:qcmproteins}
\end{figure}
To confirm the importance of electrostatic interactions in protein adsorption, lysozyme (1 g/L) is adsorbed at different concentrations of sodium chloride (\chem{NaCl}, in presence of 50 mmol/L Tris at pH 8.5). The surface concentration of the positively charged protein on the negatively charged dopamine-melanin film decreases with increasing salt concentration, and hence with increasing screening of the electrostatic interactions (Figure \ref{fig:qcmproteins}). Furthermore the amount of lysozyme adsorbed in the presence of 150~mmol/L or 300~mmol/L \chem{NaCl} is close to the amount remaining after elution with 10 mmol/L SDS, when lysozyme adsorption was done in absence of \chem{NaCl}. Additionally no SDS-induced desorption is observed, when the adsorption was performed in presence of 150 mmol/L \chem{NaCl}. This suggests that at higher ionic strength the contribution of electrostatic interactions to protein adsorption is suppressed and a possibly covalent mechanism is predominant.

Myoglobin and $\alpha$-lactalbumin adsorb despite overall repulsive electrostatic interactions, which is a well known phenomenon for the adsorption of soft proteins on hydrophilic surfaces \cite{norde:2008}. Table \ref{tab:proteinmasses} shows that only part of the adsorbed lysozyme and very small quantities of myoglobin and  $\alpha$-lactalbumin desorb upon injection of SDS when the adsorption was performed in presence of 50 mmol/L of Tris. This indicates that the proteins are strongly bound to the melanin-modified surface. These strong interactions could originate from the formation of covalent bonds, hence the occurrence of \emph{chemisorption}. Indeed covalent bonds between amine groups of the proteins and catechol groups of the deposit could be involved as it was suggested by Lee \cite{lee:2009} and Merrit \cite{merrit:1996} for comparable systems. Furthermore the occurrence of strong, probably covalent bonds between amine groups and dopamine-melanin aggregates is confirmed in section \ref{sec:resultsaminebinding}. Interestingly the amount of strongly bound protein is similar for all proteins compared to the experimental standard deviation (Table \ref{tab:proteinmasses}). This can be explained by the fact that all three proteins having similar molecular masses expose a similar surface area upon their initial contact with the substrate. This initial contact area would then constitute the binding sites for covalent binding.

\subsection{Conclusion}
In this section melanin deposits obtained by successive immersions of a silica support in multiple dopamine solutions (2 g/L dopamine hydrochloride in 50 mmol/L Tris, pH 8.5) were further examined. Experiments in presence of methanol suggest that the initial formation of dopamine-melanin at the solution--support interface involves a radical like dopamine semiquinone. The attenuation of the support signal in XPS indicates that the thickness of dopamine-melanin deposits is strongly reduced upon dehydration in ultra-high vacuum. This shrinkage is in line with observations by other groups that physical properties of melanin strongly depend on its hydration state. The obtained dopamine-melanin films are stable in a pH range from 1 to 11, but they dissolve quickly at pH 13.

For the first time the $\zeta$-potential of melanin deposits was measured, revealing a value of about -40 mV at pH 8.5 for deposits made by at least three immersions for 5 min of the support in dopamine solutions. The $\zeta$-potential increases with decreasing pH and reaches positive values at a pH below 4.5. The varying charge of dopamine-melanin can be explained by the successive protonation of catechol, quinone imine and quinone groups with descending pH. These groups probably reside in an environment characterized by chemical disorder preventing the appearance of individual ``steps'' in the plots of the $\zeta$-potential as a function of pH.

The negative surface charge of dopamine-melanin deposits can partially explain the adsorption of the model proteins lysozyme, myglobin and $\alpha$-lactalbumin on melanin-coated QCM crystals by electrostatic interactions. Nevertheless elution experiments with sodium dodecylsulphate and adsorption experiments at increased salt concentration suggest, that a second interaction exists. Based on the quantification of amine binding sites on melanin particles and on a reaction scheme proposed in the literature, covalent bonds are supposed to form between amine groups of proteins and catechol groups of melanin. This binding mechanism makes melanin films a possible platform for controlled immobilization of various biomolecules, for example enzymes or antibodies for sensing applications.

%% file: resultsdopamineinpllha.tex
\section{Dopamine-melanin in polyelectrolyte films}
\label{sec:resdopaminepllha}
Polyelectrolyte films of poly(L-lysine) (PLL) and hyaluronate (HA) are an intensively studied system because of the superlinear growth of their thickness with the number of deposition steps and the biocompatibility of PLL and HA \cite{picart:2001} \cite{picart:2002}. One major drawback of \chemr{(PLL-HA)_n} films is their poor mechanical stability: They behave like a viscous liquid \cite{francius:2006}, and upon exposure to pure water they dissolve quickly. This section presents a new biomimetic approach to strengthen \chemr{(PLL-HA)_n} films using melanin obtained by oxidation of dopamine as presented in the previous sections.

Since the initial rate of dopamine-melanin formation in solution does not depend on the presence of HA or PLL (figure \ref{fig:dopaminepolymerisationrate}), melanin can probably be formed by oxidation of dopamine in \chemr{(PLL-HA)_{n}} films.

\subsection{Build-up of poly(L-lysine)-hyaluronate films}
\begin{figure}
\centering
\includegraphics[width=\textwidth]{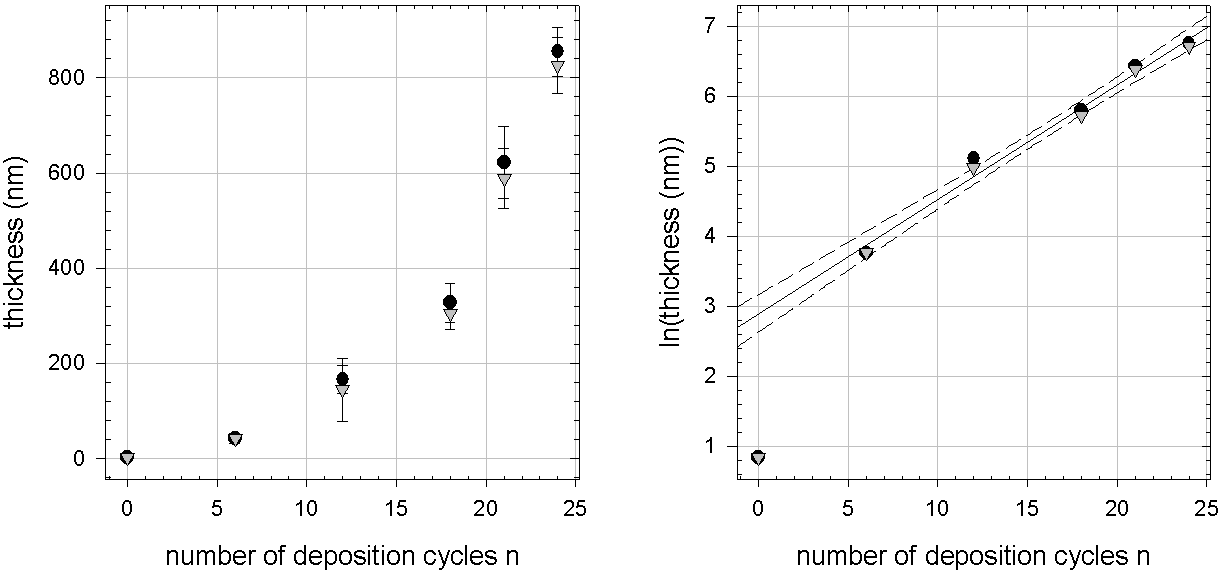}
\caption[Ellipsometric thickness of \chemr{(PLL-HA)_n}]{Thickness of two \chemr{(PLL-HA)_n} films (circles and triangles) during build-up on silicon measured by ellipsometry using a refractive index of 1.465. The straight line in the right plot is a linear regression for $n\geq 6$. The dashed lines delimit the 95 \% confidence interval. Error bars represent one standard deviation of 5 or 10 measurements on the same sample.}
\label{fig:ellipsopllha24}
\end{figure}
Before modifying \chemr{(PLL-HA)_n} films with dopamine-melanin, it is checked whether they can be built in the same buffer solution that will be used for dopamine-melanin formation (1 g/L of polyelectrolyte in 50 mmol/L Tris, pH 8.5).
 
Figure \ref{fig:ellipsopllha24} shows the thickness of a \chemr{(PLL-HA)_n} film during its build-up on silicon measured by ellipsometry supposing a refractive index of 1.465 close to the one reported for \chemr{(PLL-HA)_n} (1.42 to 1.43) \cite{picart:2002}. The logarithmic representation reveals an exponential growth with the number of deposition cycles for 6 to 24 cycles. In the end a dry thickness of about 800 nm is reached, a value close to the ones observed by Porcel and others for \chemr{(PLL-HA)_n} deposition from solutions containing about 1 g/L of polyelectrolyte and 0.15 mol/L of sodium chloride at pH 6.5 \cite{porcel:2006}. These are the conditions most commonly used to build \chemr{(PLL-HA)_n} films \cite{picart:2001} \cite{picart:2002}. In contrast to our observations, the films examined in \cite{porcel:2006} show an exponential growth only between 4 and 12 deposition cycles and a linear growth elsewhere.

\begin{figure}
	\centering
		\includegraphics[width=\textwidth]{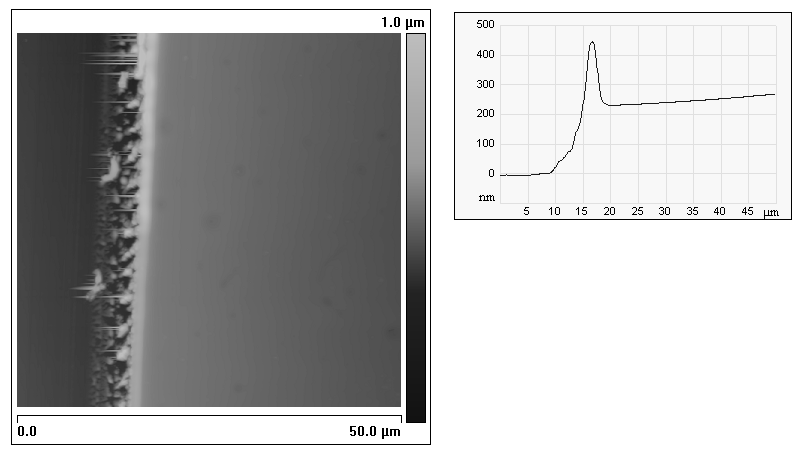}
	\caption[SFM height image of a dried \chemr{(PLL-HA)_{20}} film]{SFM height image in contact mode of a dried \chemr{(PLL-HA)_{20}} film. The plot at the right shows the mean height over the 512 lines of the image at the left.}
	\label{fig:pllha24_1_scratch}
\end{figure}

Single-wavelength ellipsometry leads to an ambiguity in the calculated thickness due to the periodicity of the measured polarisation angles (Equation \ref{eq:periodicity}). If $d_f$ is a solution, all \mbox{$d_f + m\cdot 281.6  \: \mathrm{nm} \; (m \in \mathbb{Z})$} are also possible solutions for the employed wavelength (632.8~nm) and refractive index (1.465). The correct value is chosen by comparing the final thickness to the thickness of  \chemr{(PLL-HA)_{20}} films built on cover glasses measured by scanning force microscopy. Therefore the films are dried and scratched with a syringe needle before imaging them in contact mode (Figure \ref{fig:pllha24_1_scratch}). This way a thickness between 200 nm and 700 nm is measured for two samples. These values are lower than the smallest possible values deduced from the ellipsometry measurements. By consequence the smallest possible values are regarded as the measured thickness in the ellipsometry experiments. To obtain these smallest values, it is only assumed that the film thickness increases with each polyelectrolyte deposition. The remaining discrepancy can be explained by the lower number of deposition cycles (20 instead of 24), the different deposition method (manual or automated dipping) and an intenser drying in the SFM experiments. 

\subsection{UV--visible spectroscopy}
First it is checked if dopamine is able to diffuse into \chemr{(PLL-HA)_n} films under conditions, where no melanin formation occurs, and if it is retained in the films upon rinsing with dopamine-free solutions. The aim of this experiment is to load polyelectrolyte films with dopamine and to trigger the formation of melanin by an external pH change. This would allow for separating the formation of melanin in the films from events occurring in solution.

\begin{figure}
	\centering
		\includegraphics[width=\textwidth]{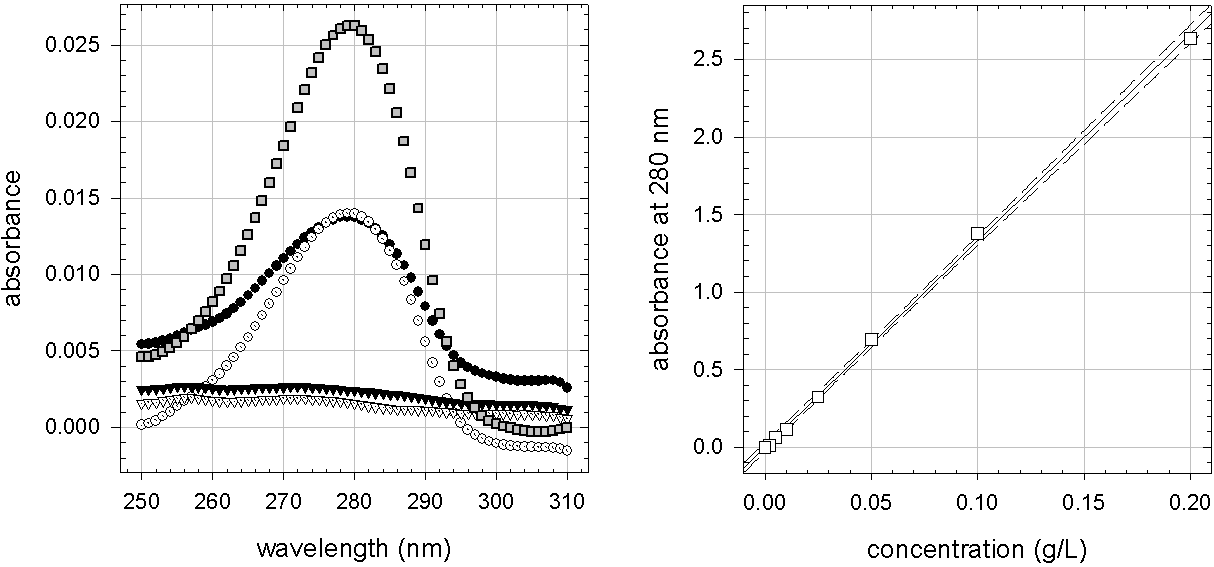}
	\caption[UV absorbance of dopamine]{Left: UV absorbance of a dopamine hydrochloride solution at 2 mg/L (squares) and of \chemr{(PLL-HA)_{12}} put in a dopamine hydrochloride solution (2 g/L) during 2 h and subsequently put in dopamine-free solution during 5 s (circles) or 4 min (triangles). Black and white symbols correspond to two samples. Right: Absorbance at 280 nm of dopamine hydrochloride solutions versus concentration. The full line is a linear regression with its 95 \% confidence intervals delimited by dashed lines. All solutions contain 0.15 mol/L \chem{NaCl} at pH 5.9}
	\label{fig:uvvispllhadopamine}
\end{figure}
\chemr{(PLL-HA)_{12}} films are prepared on quartz slides by manually dipping them into polyelectrolyte solutions. Afterwards the samples are put for two hours into an aqueous solution containing 2 g/L dopamine hydrochloride and 0.15 mol/L sodium chloride at pH 5.9. Then the samples are rinsed with dopamine-free solutions and their absorbance in the wet state around 280 nm is compared to the absorbance of a dilute dopamine solution. In accord with \cite{barreto:2001} an absorbance peak is present at this wavelength (Figure \ref{fig:uvvispllhadopamine}).

To estimate the concentration of dopamine in the polyelectrolyte films, the absorbance $A_s$ of dopamine hydrochloride solutions at a wavelength of 280 nm is measured as a function of the concentration $C_s$ (Figure \ref{fig:uvvispllhadopamine}). The linear regression $A_s/C_s=(13.3\pm 0.1)$~L/g and the optical path length $d_s=10^{-2}$ m are used to calculate the extinction coefficient $\epsilon$ with equation \ref{eq:extinctioncoefficient}:
\begin{equation}
\epsilon=\frac{A_s}{d_s \log(e) C_s}=\frac{1.33 \cdot 10^3}{\log(e)} \mathrm{\frac{L}{gm}}
\end{equation} 
After five seconds of rinsing the dopamine-loaded \chemr{(PLL-HA)_{12}} films have an absorbance at 280 nm of $A_f=0.014$. Their thickness is estimated by taking the dry thickness from figure \ref{fig:ellipsopllha24}, multiplying it by two to account for the two faces of the quartz slide and multiplying the result by four to account for swelling of the film upon hydration \cite{porcel:2006} to obtain $d_f \approx 1.1 \cdot 10^{-6}$m. These values lead to a dopamine hydrochloride concentration in the polyelectrolyte film of
\begin{equation}
C_f=\frac{A_f}{\epsilon d_f \log(e)}\approx 9 \ \mathrm{g/L}.
\end{equation}
This value is only a rough estimation, since the extinction coefficient was determined at concentrations below 0.2 g/L, and the condition of a dilute solution, necessary to define the extinction coefficient, might be violated at higher concentrations. Nevertheless this estimation shows that the dopamine concentration in polyelectrolyte films is higher than the concentration in solution.\footnote{Fortunately. Otherwise we would not have been able to detect it. Assuming a minimal detectable absorbance of 0.002 the minimal detectable concentration in the film would be about 1 g/L} Such a concentrating effect upon loading of polyelectrolyte films has also been reported for other systems, for example in \cite{srivastava:2008}.    

Figure \ref{fig:uvvispllhadopamine} shows furthermore that after having rinsed the \chemr{(PLL-HA)_{12}} films for four minutes, the peak at 280 nm has vanished indicating that at least 6/7 (considering a minimum detectable absorbance of 0.002) of the dopamine molecules have left the film. Hence dopamine is not retained in the polyelectrolyte film, and it has to be investigated if melanin can be formed \emph{simultaneously} in polyelectrolyte films and in solution.

\begin{figure}
	\centering
		\includegraphics[width=85mm]{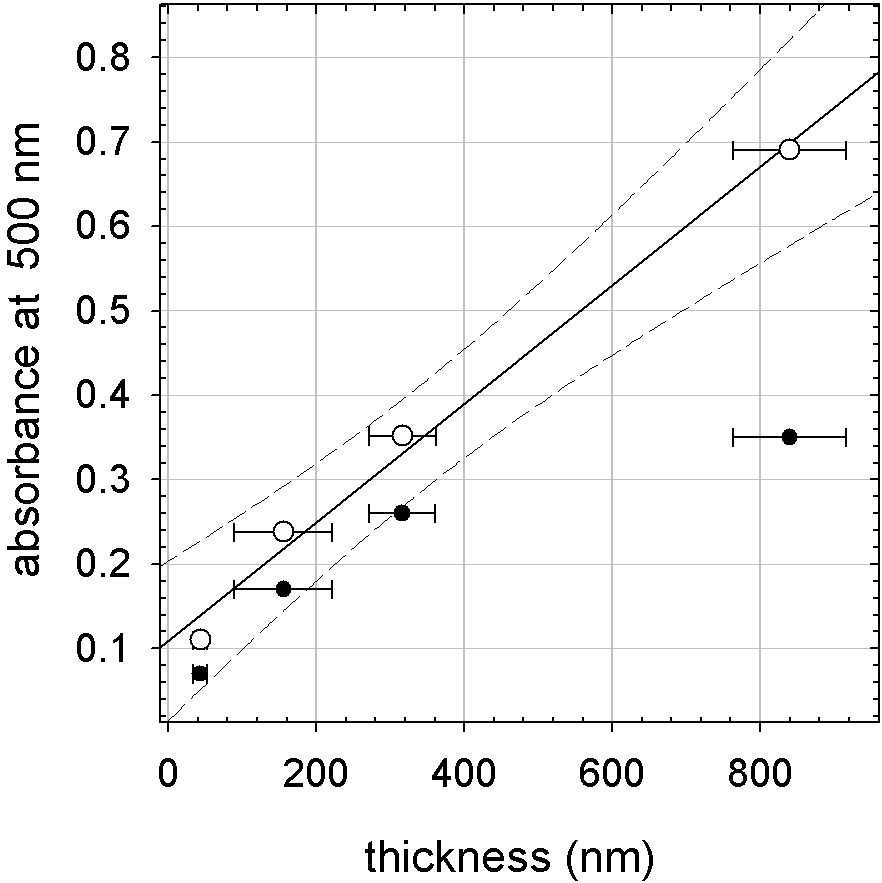}
	\caption[Absorbance of melanin in \chemr{(PLL-HA)_n}]{Absorbance at 500 nm of \chemr{(PLL-HA)_n} after contact with dopamine hydrochloride (2 g/L in 50 mmol/L Tris at pH 8.5) for 4 h (full circles) or 14 h (open circles) versus thickness of the films measured by ellipsometry in the dry state before contact with the dopamine solutions. The full line represents a linear regression with 95 \% confidence intervals delimited by dashed lines to the data for 14 h contact time.}
	\label{fig:dopamineincorporation}
\end{figure}

When \chemr{(PLL-HA)_n} films are put into a dopamine hydrochloride solution (2 g/L, 50 mmol/L Tris, pH 8.5), they develop a brown colour just like the solution itself indicating that melanin formation takes place within the film. UV--visible spectroscopy shows that after 14 h of contact between vertically held \chemr{(PLL-HA)_n} films on quartz slides and a dopamine solution, the absorbance of the films at 500 nm scales approximately in proportion to the film thickness (figure \ref{fig:dopamineincorporation}). Thus dopamine is able to diffuse into the \chemr{(PLL-HA)_n} films to form homogeneously distributed melanin within 14 h. If the reaction time is only 4 hours, a downward deviation occurs in the absorbance versus thickness curve. This demonstrates that for thicker films more time is needed to obtain an homogeneous filling with dopamine-melanin.

\subsection{Fourier-transform infrared spectroscopy}
\label{sec:resftir}
\begin{figure}
	\setcapindent*{0pt}
		\begin{captionbeside}[FTIR-ATR of melanin formation in \chemr{PEI-(PLL-HA)_{12}}]{A: Infrared absorbance spectra versus wavenumber (1/$\lambda$) of \chemr{PEI-(PLL-HA)_n}, with n = 6 (dashed black line), n = 9 (solid grey line) and n = 12 (black plus symbols). \\
B: Spectra of a dopamine hydrochloride solution at 2 g/L (in 50 mmol/L Tris at pH 6.6) (solid line) and of the \chem{ZnSe} crystal after contact with the same solution at pH 8.5 during 24 h (plus symbols). \\
C: \chemr{PEI-(PLL-HA)_{12}} spectra (thin grey line) and difference spectra of the same film after contact with dopamine solutions at pH 8.5 during 1 h (solid black line), 3 h (dashed black line) and 25 h (solid dark grey line).\\
Thick vertical bars mark the absorbance of 1) the amide I band of PLL, 2) \chem{COO^-} elongation in HA and 3) polysaccharide vibration of HA}
	\includegraphics[height=\textheight]{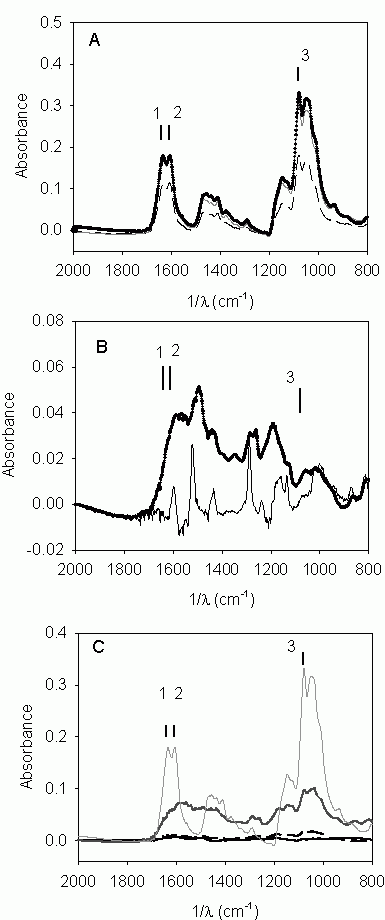}
	\end{captionbeside}
	\label{fig:ftirpllhadopamine}
\end{figure}
Fourier-transform infrared spectroscopy (FTIR) in the attenuated total reflection mode (ATR) is performed on \chemr{PEI-(PLL-HA)_n} films built on a zinc selenide (\chem{ZnSe}) crystal. Figure \ref{fig:ftirpllhadopamine} A shows that a \chemr{(PLL-HA)_{12}} film is thicker than the penetration depth of the evanescent wave, because there is no absorbance increase from \chemr{(PLL-HA)_9} to \chemr{(PLL-HA)_{12}}. When the latter film is put in contact with a solution containing 2 g/L dopamine hydrochloride in Tris buffer (50 mmol/L) at pH 8.5 an increase of the absorbance due to the formation of melanin is observed for contact times up to 25 hours (figure \ref{fig:ftirpllhadopamine} C). Thus melanin forms not only close to the surface but also in the lowest part of the film probed by the evanescent wave of the spectrometer. Furthermore figure \ref{fig:ftirpllhadopamine} C shows an increase of the absorbance caused by the \chemr{(PLL-HA)_{12}} film itself for long incubation times in dopamine solution. These bands are labelled 1, 2 and 3 in the figures. Their increase corresponds to an increase of the concentration of PLL and HA in the vicinity of the \chem{ZnSe} crystal, probably caused by shrinking of the film that can be explained by chemical crosslinking due to melanin formation.

\subsection{Scanning force microscopy}
\label{sec:ressfmpllha}
\begin{figure}
	\centering
		\includegraphics[width=\textwidth]{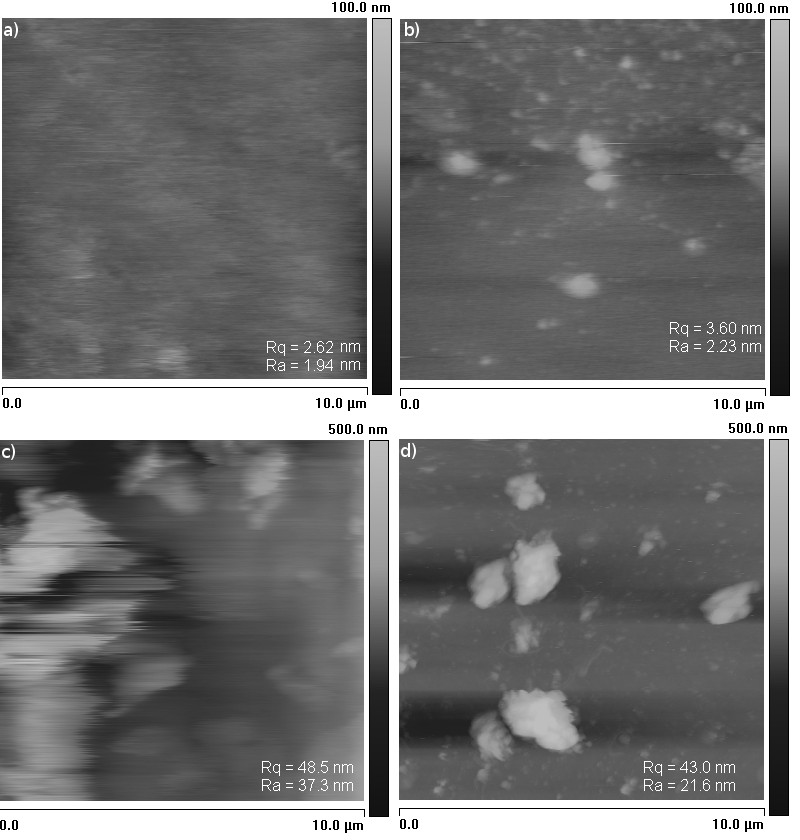}
	\caption[Surface topography of \chemr{(PLL-HA)_{30}} with melanin]{Surface topography of \chemr{(PLL-HA)_{30}} on cover glasses imaged by SFM in dynamic mode in 50 mmol/L Tris at pH 8.5. Rq and Ra are the roughness values of the whole images. The films were exposed for 0 min (a), 30 min (b), 1 h (c) or 9 h (d) to a solution of 2 g/L dopamine hydrochloride in Tris buffer.}
	\label{fig:roughness_pllha30_dopa}
\end{figure}
\chemr{(PLL-HA)_{30}} films on cover glasses are exposed to a solution of 2 g/L dopamine hydrochloride in Tris buffer (50 mmol/L, pH 8.5). Then the samples are rinsed with pure buffer solution and imaged by scanning force microscopy (SFM) in dynamic mode in the same buffer solution. For exposition times from 30 min to 9 h the surface roughness increases and large aggregates, probably made of melanin, appear (Figure \ref{fig:roughness_pllha30_dopa}). The strongest effect occurs between 30 min and 1 h of exposition time, but the surface roughness remains small compared to the film thickness (between 2 $\mu$m and 5 $\mu$m, Section \ref{sec:resclsm}). SFM imaging cannot answer the question whether the aggregates visible after longer exposition times are formed on top of or within the polyelectrolyte film.

\subsection{Confocal laser scanning microscopy}
\label{sec:resclsm}
Polyelectrolyte films of the type \chemr{(PLL-HA)_{30}-PLL_{FITC}} built on cover glasses and labelled with fluorescein isothiocyanate are exposed to dopamine solutions in Tris buffer (50 mmol/L) at pH 8.5. These films are imaged by confocal laser scanning microscopy (CLSM) in pure Tris buffer. 

\begin{figure}
\centering
\includegraphics[width=110mm]{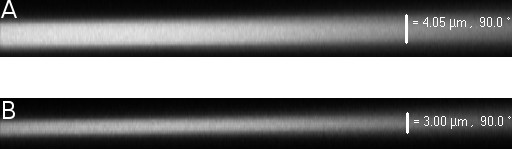}
\caption[Virtual z-sections of \chemr{(PLL-HA)_{30}-PLL_{FITC}} films]{\chemr{(PLL-HA)_{30}-PLL_{FITC}} films exposed (B) or not (A) for 1 h to a dopamine hydrochloride solution at 1 g/L (in 50 mmol/L Tris, pH 8.5) and imaged by CLSM. The length of the virtual z-sections in the sample plane is $230.3 \mu$m. The z-axis is stretched by a factor of three for clarity.}
\label{fig:confocalline}
\end{figure}
The CLSM virtual z-sections (Figure \ref{fig:confocalline}) confirm the observation from SFM (Section \ref{sec:ressfmpllha}), that the films remain flat upon contact with dopamine solutions. Furthermore the films appear homogeneous, and the fluorescence intensity progressively decreases for increasing contact times, because the formed melanin partially absorbs the exciting light as well as the light emitted by the fluorophore. Due to this inner filter effect, the contact time between polyelectrolyte films and dopamine solutions is limited to 1 h, because the film loses its transparency for longer contact times. According to the SFM results, the most important changes occur anyway within the first hour of contact with a dopamine solution.

The thickness of \chemr{(PLL-HA)_{30}-PLL_{FITC}} films measured in the virtual z-sections lies between $2 \mu$m and $5 \mu$m, and usually it decreases slightly during melanin formation. As mentioned in the FTIR-ATR results (Section \ref{sec:resftir}), the decrease in thickness might be due to chemical crosslinking of the polyelectrolyte film by melanin.

\begin{figure}
\centering
\includegraphics[width=85mm]{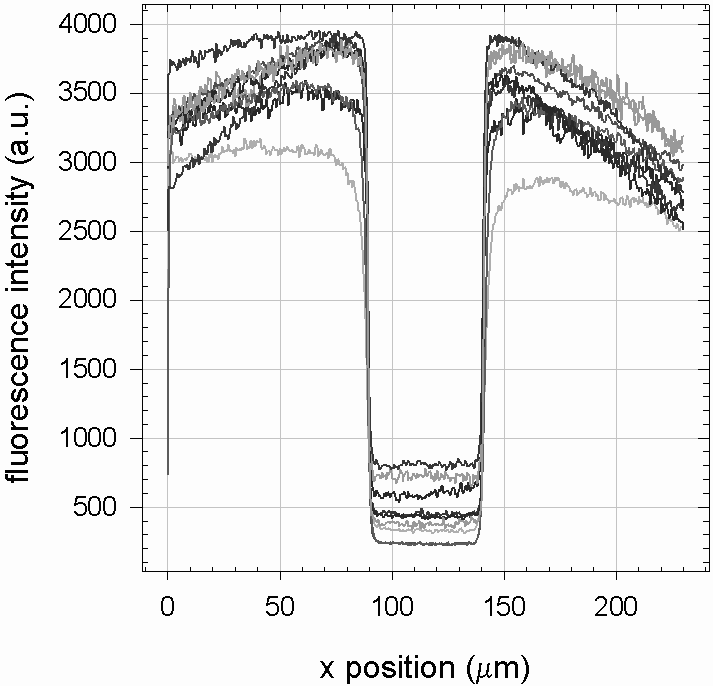}
\caption[Intensity profiles after photobleaching]{Intensity profiles measured right after photobleaching along a diameter of the bleached disk in \chemr{(PLL-HA)_{30}-PLL_{FITC}}. The fluorescence intensity values are mean values over 5 lines (total width: $2.2 \ \mu$m) in the acquired images. Different shades of grey correspond to independent experiments.}
\label{fig:frapprofiles}
\end{figure}

Once the homogeneity of a sample is checked by regarding z-sections and images in the sample plane at five spots, a circular area of the polyelectrolyte film is bleached. Immediately after photobleaching an image is taken in the sample plane, and intensity profiles are measured along a diameter of the bleached area. The exemplary profiles shown in figure \ref{fig:frapprofiles} confirm the uniformity of the residual fluorescence in the bleached area. Consequently the theoretical model developed in \cite{picart:2005} can be used to treat the data.

\begin{figure}
\centering
\includegraphics[width=85mm]{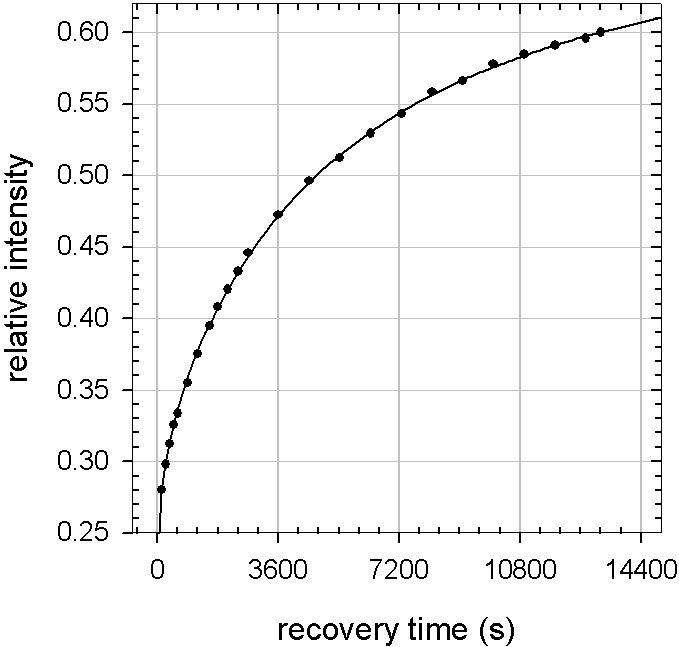}
\caption[Fluorescence recovery in a \chemr{(PLL-HA)_{30}-PLL_{FITC}} film]{Fluorescence recovery in a \chemr{(PLL-HA)_{30}-PLL_{FITC}} film exposed for one hour to a dopamine hydrochloride solution at 0.5 g/L in 50 mmol/L Tris at pH 8.5. The relative fluorescence intensity in the bleached area (circles) is fitted by the theoretical function (line) given by equation \ref{eq:fraptheory}}
\label{fig:frapfit}
\end{figure}

The recovery of the fluorescence is followed as a function of time and the resulting curves of relative intensity versus time are fitted by the theoretical function as described in section \ref{sec:expfrap}. The fitted curves excellently match the experimental points as shown by the example in figure \ref{fig:frapfit}.

\begin{figure}
\centering
\includegraphics[width=85mm]{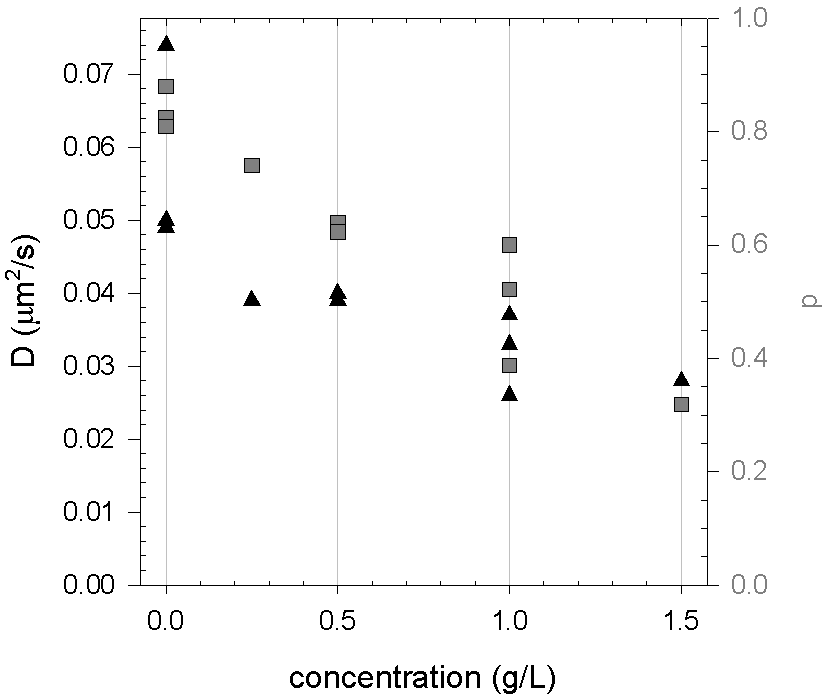}
\caption[Influence of melanin on the mobility of PLL chains]{Diffusion coefficient D (triangles) and fraction of mobile PLL chains p (squares) in \chemr{(PLL-HA)_{30}-PLL_{FITC}} films exposed for 1 h to dopamine hydrochloride solutions of different concentrations in 50 mmol/L Tris at pH 8.5}
\label{fig:pllmobility}
\end{figure}

The diffusion coefficient $D$ and the fraction of mobile PLL chains $p$ obtained as parameters by the fits are shown in figure \ref{fig:pllmobility} as a function of the dopamine hydrochloride concentration the polyelectrolyte films were exposed to. Both parameters decrease with increasing dopamine concentration, showing that melanin formed in the polyelectrolyte film impedes the diffusion of \chemr{PLL_{FITC}}-chains. The reduction of $p$ indicates that melanin does not act as a physical obstacle to the PLL-chains. Instead it chemically binds PLL-chains leading to a reticulation of the film. According to the literature \cite{lee:2009} \cite{merrit:1996} and to our results concerning the adsorption of proteins (section \ref{sec:resultsproteinonmelanin}) and 2-(2-pyridinedithiol)ethylamine (section \ref{sec:resultsaminebinding}) to dopamine-melanin, the primary amines present in PLL can covalently bind catechol groups of melanin. Together with the observed shrinkage of \chemr{(PLL-HA)_{n}} films during melanin formation, these results indicate that melanin formation induces a chemical reticulation of \chemr{(PLL-HA)_{n}} films.

It is also observed that the diffusion coefficient and the fraction of mobile PLL-chains depends on the age of the films. In the experiments mentioned above, the recovery of fluorescence is always followed the day after the build-up, fluorescence labelling and possible dopamine modification of the films. In one experiment the polyelectrolyte films are aged four days at room temperature before fluorescence labelling and dopamine modification. Consequently the FRAP experiments are performed on the fifth day after film build-up instead of the first day. Table \ref{tab:frapage} shows  that at both concentrations of dopamine hydrochloride the diffusion coefficient and the fraction of mobile molecules are reduced for the aged film.
\begin{table}
	\centering
		\begin{tabular}{l|l|l|l}
		$C$ (g/L)& age (days) & $D$ ($\mu$m$^2$/s) & $p$ \\ \hline
		0.0 & 1 & $0.058 \pm 0.012$ & $0.79 \pm 0.05$ \\
		0.0 & 5 & $0.014$ & $0.67$ \\ \hline
		0.5 & 1 & $0.040 \pm 0.001$ & $0.63 \pm 0.01$ \\
		0.5 & 5 & $0.028$ & $0.46$		
		\end{tabular}
	\caption[Age-dependent mobility of PLL chains]{Influence of the age of \chemr{(PLL-HA)_{30}-PLL_{FITC}} films on the diffusion coefficient $D$ and the fraction of mobile PLL-chains $p$. The films were exposed for one hour to a dopamine hydrochloride solution of concentration $C$ in Tris buffer (50 mmol/L, pH 8.5). Five days old films were measured only once, for the other films mean values $\pm$ one standard deviation over two or three samples are given.}
	\label{tab:frapage}
\end{table}

\subsection{Free-standing membranes}
\begin{figure}
	\centering
	\includegraphics[width=68mm]{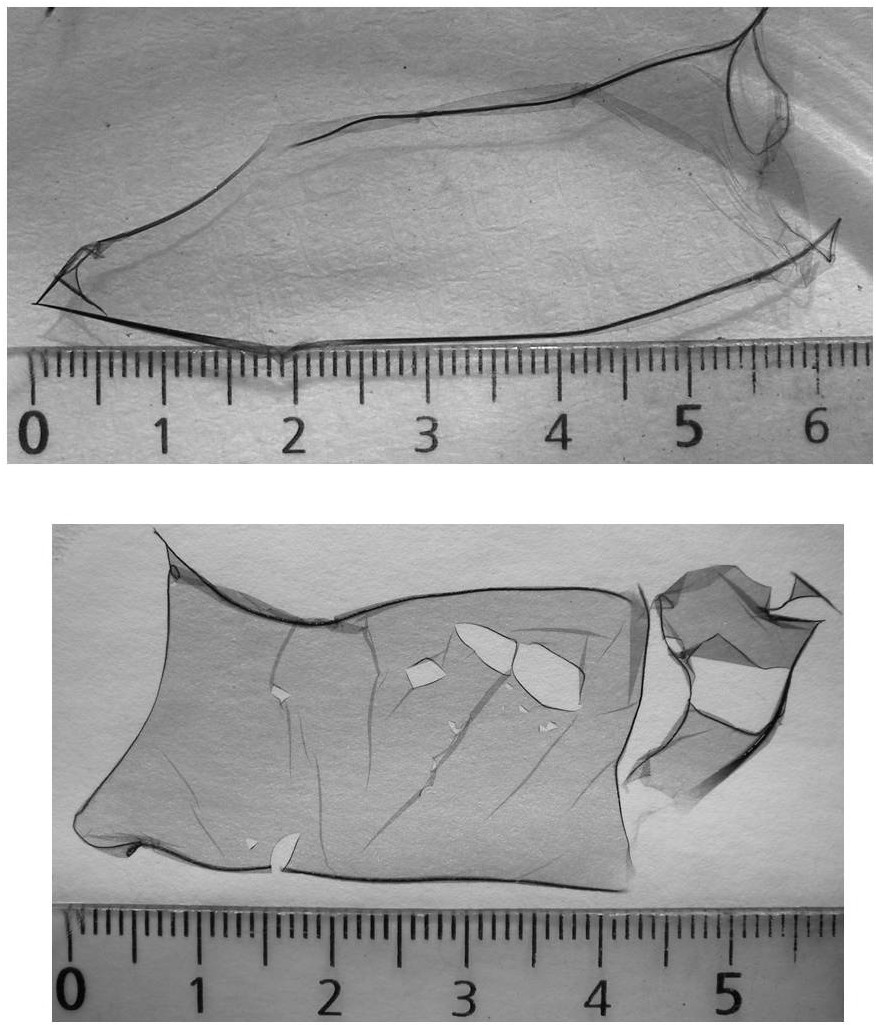}	
	\caption[\chemr{(PLL-HA)_{30}} + melanin free standing membrane]{Grayscale photographs of a free standing membrane obtained from a \chemr{(PLL-HA)_{30}} film exposed for 24 h to a dopamine hydrochloride solution (2 g/L in 50 mmol/L Tris at pH 8.5). Top: floating in 0.1 mol/L \chem{HCl} solution. Bottom: deposited on paper and dried. The scale is in cm.}
	\label{fig:membranepreparation}
\end{figure}
While unmodified \chemr{(PLL-HA)_{n}} films decompose spontaneously when rinsed with pure water, the same films put in contact with dopamine (in Tris buffer at pH 8.5) during at least 4 h remain unchanged and adherent to the support. After exposition to dopamine, the edges of the samples are cut with a razor blade. Then the samples are immersed in a hydrochloric acid (\chem{HCl}) solution at pH 1 leading to detachment of the films from their glass support. For contact times with dopamine up to 4 h the modified films detach in small pieces, while contact times of at least 10 h allow for detachment of membranes as large as the support (Figure \ref{fig:membranepreparation}). The separation of the membrane and the support occurs within some minutes \chem{HCl} solution at 0.1 mol/L, but no membrane detachment is observed when a sample is put in pure water for one day.

The obtained membranes withstand shear stress in solution and can be deposited on paper (Figure \ref{fig:membranepreparation}). Since no film detachment is obtained when a sample is put in \chem{HCl} solutions at $10^{-3}$ mol/L or $10^{-2}$ mol/L, detachment probably occurs through positive charging of the silanol groups of the glass slide, whose point of zero charge lies close to pH 2. The \chemr{(PLL-HA)_{n}} film is anchored to the glass slide with the positively charged PLL. Thus the membrane detachment can be explained by electrostatic repulsion when the pH of the solution is decreased to 1.

Section \ref{sec:stabilitymelaninph} showed that a dopamine-melanin deposit without polyelectrolytes stays adherent to a \chem{SiO_2}-covered silicon slide in a \chem{HCl} solution at 0.1 mol/L. This indicates that the membranes presented in this section are not pure dopamine-membranes prepared in a \chemr{(PLL-HA)_{n}} template, which is washed away in the water rinsing step. Instead, the membranes are a composite material of polyelectrolytes and dopamine-melanin, probably held together by covalent bonds between primary amines of PLL and catechol groups of dopamine-melanin. It is unknown whether there are any interactions between hyaluronate and melanin. If there is no binding between HA and melanin, HA might be rinsed from the composite membrane with 0.1 mol/L \chem{HCl}, because in strongly acidic solution it looses its negative charge ($pK_a=3$ \cite{burke:2003}), which is necessary for electrostatic binding to PLL.

\subsection{Comparison to PDADMA-PAA films}
\label{sec:respdadmapaadopamine}
To find out if the method of membrane preparation with dopamine can be generalised to other polyelectrolyte films, 30 bilayers of poly(dimethyldiallyl ammonium) (PDADMA) and poly(acrylic acid) (PAA) are built on a glass slides. 

After contact with a solution of dopamine hydrochloride (2 g/L in 50 mmol/L Tris at pH 8.5) for 24 hours the films present a brown colour. The absorbance at 500 nm is about 4 times as strong as the one of glass slides after immersion in an equivalent dopamine solution for the same time. The absorbance stays constant during immersion in water for 24 hours. Afterwards it is tried to detach the melanin-modified \chemr{(PDADMA-PAA)_{30}} films from their glass support using the protocol previously developed for \chemr{(PLL-HA)_n} films. Upon immersion of the samples in hydrochloric acid solution at 0.1 mol/L, membranes seem to detach from the support but they dissolve very quickly in solution.

These experiments show that dopamine-melanin can form in \chemr{(PDADMA-PAA)_{30}} and that it stays there during water rinsing, but \chemr{(PDADMA-PAA)_{30}} films are not enforced in a way comparable to \chemr{(PLL-HA)_{30}}. One important difference between the two polyelectrolyte films is, that a lysine unit contains a primary amine while a dimethyldiallylammonium unit contains a quaternary amine. The latter cannot bind to melanin by the mechanism proposed in \cite{lee:2009} and \cite{merrit:1996}. Thus the experiments with \chemr{(PDADMA-PAA)_{30}} films give another hint, that melanin can covalently bind PLL by the proposed mechanism.  

\subsection{Conclusion}
This section has presented a new biomimetic approach to strengthen polyelectrolyte films. Strengthening of biocompatible \chemr{(PLL-HA)_n} films is of particular interest, because it largely enhances cell adhesion and proliferation \cite{engler:2004} \cite{richert:2004}. UV--visible and infrared spectroscopy confirmed that dopamine-melanin can form in \chemr{(PLL-HA)_n} polyelectrolyte films. Scanning force microscopy and confocal laser scanning microscopy showed that the polyelectrolyte films remain homogeneous and relatively flat upon incorporation of dopamine-melanin. The lateral mobility of PLL chains within the polyelectrolyte films decreases progressively, when these films are exposed to increasing concentrations of dopamine (in 50 mmol/L Tris, pH 8.5) as evidenced by fluorescence recovery after photobleaching experiments.

The most convincing illustration that dopamine-melanin strengthens polyelectrolyte films is the formation of free-standing composite membranes for long contact times ($\geq 10$~h) between \chemr{(PLL-HA)_{30}} films and dopamine solutions. Furthermore these membranes of macroscopic lateral size and some $\mu$m in thickness are obtained under mild conditions (in 0.1 mol/L \chem{HCl}). For biomedical applications, for example tissue engineering, this is an important advantage compared to other methods to build polyelectrolyte membranes involving the dissolution of a sacrificial support layer with acetone \cite{mamedov:2000} or hydrofluoric acid \cite{mamedov:2002} \cite{podsiadlo:2007} \cite{tang:2003}. Up to now there are only few reports of free-standing polyelectrolyte membranes obtained in aqueous basic \cite{lavalle:2005} or even neutral \cite{ono:2006} solution.

This work does not provide a direct proof of chemical crosslinking between PLL chains and dopamine-melanin. Nevertheless the following hints lead to the conclusion, that the polyelectrolyte films are strengthened by covalent bonds between catechol groups of dopamine-melanin and primary amines of PLL: Dopamine-melanin is capable of covalently binding 2-(2-pyridine\-di\-thiol)\-ethyl\-amine via catechol--amine bonds (Section \ref{sec:resultsaminebinding}), which are probably involved in protein binding (\cite{lee:2009} \cite{merrit:1996}, Section \ref{sec:resultsproteinonmelanin}). Infrared spectroscopy and confocal microscopy indicate that the thickness of \chemr{(PLL-HA)_{n}} films decreases during dopamine-melanin incorporation. The number of free PLL chains within the film decreases, a behaviour difficult to explain if melanin acted only as a physical obstacle slowing down the diffusion of PLL. And finally it is not possible to obtain free-standing membranes using a polyelectrolyte film without primary amines. To definitely proof the existence of chemical crosslinking, further experiments, for example nuclear magnetic resonance spectroscopy, would have to be done.

%% file: conclusion.tex
\chapter{Conclusion}
\section{Summary}
Based on the work of Lee and others \cite{lee:2007.2}, the spontaneous oxidation of dopamine in slightly alkaline solutions was investigated and the reaction product was identified as dopamine-melanin. The ability of melanin to covalently bind amine functional groups was confirmed by quantification of the corresponding binding sites on dopamine-melanin aggregates. Furthermore it is possible to redissolve dopamine-melanin aggregates in strongly alkaline solutions. The obtained small melanin grains were used to build layer-by-layer deposits with poly(diallyldimethylammonium) (PDADMA) implying a preferential adsorption of larger dopamine-melanin aggregates to the layer-by-layer film.

After the investigation of dopamine-melanin formation in solution, different methods of dopamine oxidation to grow melanin films at solid--liquid interfaces were developed. All examined methods lead to continuous dopamine-melanin films with very similar surface morphologies. Interestingly these morphologies, made of highly anisotropic platelet-shaped aggregates, are also found on the \chemr{(PDADMA-melanin)_n} films presented before as well as on natural and synthetic melanin samples investigated by other groups. Therefore the formation of platelet-shaped aggregates seems to be an intrinsic property of melanin. Dopamine-melanin films become impermeable to electrochemical probes at a thickness in the 10 nm range. In this context a higher permeability for positively charged and neutral probes than for negatively charged ones was confirmed for one preparation method.

The adsorption of proteins on dopamine-melanin coatings was explained as combination of electrostatic and strong, most probably covalent, interactions. To obtain this explanation, the $\zeta$-potential of dopamine-melanin deposits has been measured as a function of pH.

Finally the formation of melanin by dopamine oxidation in layer-by-layer films of poly(L-lysine) (PLL) and hyaluronate (HA) was studied. It was observed that melanin is able to homogeneously fill \chemr{(PLL-HA)_n} films inducing only small changes in film morphology. On the contrary, the mobility of PLL chains in the films is strongly reduced in presence of dopamine-melanin. Furthermore the obtained polyelectrolyte--melanin composites can be detached from their substrate as free-standing membranes prepared by a biomimetic method in relatively mild conditions. Numerous indications lead to the conclusion that the observed strengthening of \chemr{(PLL-HA)_n} films is caused by chemical crosslinking of PLL chains with dopamine-melanin.

\section{Open questions}
The most important questions evoked in this work that should be answered by further studies are:

What is the mechanism making electrochemically prepared dopamine-melanin films more permeable to hexacyanoferrate ions than films prepared by the other methods? No differences in morphology were detected by scanning force microscopy. Thus the answer might be found using techniques sensing structural parameters at the nanometer scale, for example small angle x-ray scattering.

Why is copper sulphate able to induce melanin formation in acidic medium contrary to the other examined oxidants? And why do dopamine-melanin samples containing very low amounts of copper present intense UV absorbance peaks that are not seen in absence of copper? Probably the peaks are not directly related to copper but to structural changes in dopamine-melanin induced by copper or by the acidic environment during melanin formation.

How does the choice of a particular buffering agent influence the growth of dopamine-melanin deposits, although the buffering agents are not detected in the deposits? An answer to this question might be given by analysis of the dopamine-melanin molecular structure using x-ray diffraction or nuclear magnetic resonance spectroscopy.

Is it possible to ``freeze'' the formation of melanin to examine intermediate products? A route to answer this question might be the modification of dopamine binding sites with sugars as proposed in \cite{pezzella:2009}.

\section{Outlook}
The work initiated with this thesis might be continued for example by the following projects:

Polyelectrolyte--melanin composite membranes might be functionalised with biomolecules to obtain cell culture sheets or bioactive patches.

Since the examined dopamine-melanin films can be easily deposited on various supports and present chemically active sites, they might be used as a versatile platform for further functionalisation. The reduction potential of melanin films can be used for example to build silver particles for bactericidal coatings.

The iridescent colours of peacock feathers are caused by diffraction at periodic nanostructures rich in melanin \cite{kinoshita:2005} \cite{yoshioka:2002} \cite{zi:2003}. Photonic crystals made by periodic arrangement of melanin grains and a material of different refractive index, for example silica beads in a colloidal crystal \cite{norris:2004}, might mimic the diffraction behaviour of peacock feathers. In addition to their beauty, the obtained crystals might find applications in optical data transmission and data treatment.

Embedding dopamine-melanin grains in a material having a high coefficient of thermal expansion would lead to a composite with a temperature-dependent refractive index. This might be useful to build thermometers or bolometers due to the broadband absorption of melanin.

%% file: appendix.tex
\begin{appendix}
%\chapter{Un nouveau chapitre de la thèse}
%``Le nouveau chapitre de la th\`ese'' is a project of the Association Bernard Grégory aiming to better communicate the value of a Ph. D. research project as a professional experience. Therefore a group of Ph. D. students from different faculties meets a human resources manager (Michael Rendler in the present case) to discuss how to describe their research as a professional project considering the following subjects:
%\begin{itemize}
%\item Presentation of the project and its environment in non-academic language
%\item Estimation of the project's expenses and identification of the resources 
%\item Description of the course of the project
%\item Identification of scientific, technical and personal skills developed during the project
%\item Development of a career perspective
%\end{itemize}
%These points are used to prepare an oral presentation and the written report hereafter.

%\begin{otherlanguage*}{french}
%\input{ncttext}
%\end{otherlanguage*}

\chapter{ImageJ Macros}
\subsection*{frap.txt}
\label{sec:fraptxt}
\begin{verbatim}
// Measure FRAP images (512 x 512) with a bleached circle in the centre,
// reference region between a square and a circle.
// all lengths in pixels

b=32; // border width
d=250; // circle diameter
smalld=111; // diameter of bleached circle
xshift=0; // shift of the bleached circle from the centre
yshift=0;

// set measurements to measure mean grey value
run("Set Measurements...", "  mean redirect=None decimal=0");

// measure reference region:
makeRectangle(b+xshift,b+yshift,512-2*b,512-2*b);
setKeyDown("alt");
makeOval((512-d)/2+xshift, (512-d)/2+yshift, d, d);
run("Measure");

// measure bleached region:
makeOval((512-smalld)/2+xshift, (512-smalld)/2+yshift, smalld, smalld);
run("Measure");
\end{verbatim}

\subsection*{hprofile.txt}
\begin{verbatim}
// Create horizontal intensity profile 
// in the middle of a (512 x 512) image:
// all lengths in pixels

d=5; // height of the profile in the image

// Set profile options: output as formatted graph and list:
run("Profile Plot Options...", "width=512 height=200 minimum=0 
-> maximum=4096 fixed list");

// Select region for profile and make it:
makeRectangle(0,256-d/2,512,d);
run("Plot Profile");
\end{verbatim}

\subsection*{vprofile.txt}
\begin{verbatim}
// Create vertical intensity profile
// in the middle of a (512 x 512) image:
// all lengths in pixels

d=5; // width of the profile in the image

// Set profile options: vertical profile,
-> output as formatted graph and list:
run("Profile Plot Options...", "width=512 height=200 minimum=0
-> maximum=4096 fixed vertical list");
 
// Select region for profile and make it:
makeRectangle(256-d/2,0,d,512);
run("Plot Profile");
\end{verbatim}

\chapter{Publications of the author during thesis preparation}
\section*{Melanin-containing films: Growth from dopamine solutions versus layer-by-layer deposition}
Falk Bernsmann, Ovidu Ersen, Jean-Claude Voegel, Edward Jan, Nicholas A. Kotov, Vincent Ball, submitted to \emph{ChemPhysChem}
\begin{quotation}
Films made by oxidation of dopamine become an interesting new functionalisation method for solid-liquid interfaces owing to their versatility. There is nevertheless an urgent need to be able to modulate the properties, for example the permeability to ionic species and the absorption coefficient, of such films. Indeed melanin films produced by oxidation of dopamine absorb strongly over the whole UV--visible part of the electromagnetic spectrum and are impermeable to anions even for a film thickness as low as some nm.

Herein we combine oxidation of dopamine to produce a solution containing melanin particles and their alternated deposition with poly\-(di\-allyl\-dimethyl\-ammonium chloride) to produce films having nearly the same morphology as pure dopamine-melanin films but being less compact, more transparent and more permeable to ferrocyanide anions.
\end{quotation}

\section*{Permeability of Thin Melanin Films Made by Dopamine Oxidation}
Falk Bernsmann, Jean-Claude Voegel and Vincent Ball, oral presentation in \emph{6th International Conference on Diffusion in Solids and Liquids}, 2010, Paris, France.
\begin{quotation}
Melanins are a class of biomolecules with interesting properties like efficient conversion of electromagnetic radiation into heat and hydration-de\-pen\-dent electrical conductivity. Furthermore they are involved in the innate immune system and in neurodegenerative diseases. Thin melanin films, which can be deposited on various kinds of substrates, might find applications as photoabsorbers, as a platform for biomolecule immobilisation or as a model to study the physiological role of melanin.

In previous work we have investigated the physicochemical properties of and protein adsorption on melanin depositions made by dipping of a silica support in multiple alkaline dopamine solutions in contact with ambient oxygen. Herein we compare new methods to obtain thin melanin films with particular interest in their permeability to different redox probes. In one method melanin deposition is enhanced by aerating the dopamine solution. In the other methods the oxygen necessary to start dopamine polymerisation by its oxidation is replaced by other oxidising species, namely a carbon electrode connected to a potentiostat or copper(II) ions in solution.
\end{quotation}

\section*{Protein adsorption on dopamine-melanin films: Role of electrostatic interactions inferred from zeta-potential measurements versus chemisorption}
Falk Bernsmann, Benoît Frisch, Christian Ringwald and Vincent Ball. \emph{Journal of Colloid and Interface Science}, 344: 54-60, 2010, doi: \href{http://dx.doi.org/10.1016/j.jcis.2009.12.052}{10.1016/j.jcis.2009.12.052}. 
\begin{quotation}
We recently showed the possibility to build dopamine–melanin films of controlled thickness by successive immersions of a substrate in alkaline solutions of dopamine [F. Bernsmann, A. Ponche, C. Ringwald, J. Hemmerlé, J. Raya, B. Bechinger, J.-C. Voegel, P. Schaaf, V. Ball, \emph{J. Phys. Chem. C} 113: 8234–8242, 2009]. In this work the structure and properties of such films are further explored.

The $\zeta$-potential of dopamine–melanin films is measured as a function of the total immersion time to build the film. It appears that the film bears a constant $\zeta$-potential of $(-39 \pm 3)$ mV after 12 immersion steps. These data are used to calculate the surface density of charged groups of the dopamine–melanin films at pH 8.5 that are mostly catechol or quinone imine chemical groups. Furthermore the $\zeta$-potential is used to explain the adsorption of three model proteins (lysozyme, myoglobin, $\alpha$-lactalbumin), which is monitored by quartz crystal microbalance. We come to the conclusion that protein adsorption on dopamine–melanin is not only determined by possible covalent binding between amino groups of the proteins and catechol groups of dopamine–melanin but that electrostatic interactions contribute to protein binding. Part of the adsorbed proteins can be desorbed by sodium dodecylsulfate solutions at the critical micellar concentration. The fraction of weakly bound proteins decreases with their isoelectric point. Additionally the number of available sites for covalent binding of amino groups on melanin grains is quantified.
\end{quotation}

\section*{Polyoxometalates in Polyelectrolyte Multilayer Films: Direct Loading of \chem{[H_7P_8W_{48}O_{184}]^{33-}} vs. Diffusion into the Film}
Vincent Ball, Falk Bernsmann, Sandra Werner, Jean-Claude Voegel, Luis Fernando Piedra-Garza and Ulrich Kortz. \emph{European Journal of Inorganic Chemistry}, 2009(34): 5115-5124, 2009, doi: \href{http://dx.doi.org/10.1002/ejic.200900603}{10.1002/ejic.200900603}.
\begin{quotation}
The immobilization of polyoxometalates (POMs) into thin organized films is of major importance for the development of functional devices like electrochromic films and catalysts. We investigate herein two methods to incorporate the wheel-shaped tungstophosphate \chem{[H_7P_8W_{48}O_{184}]^{33-}} (P8W48) polyoxometalates (POM) into polyelectrolyte multilayer (PEM) films, which offer the advantage of controlled thickness and ease of processability.

The first method, whose feasibility has been widely demonstrated, consists in the alternate deposition of a polycation and the negatively charged POM. The second method has never been used to load POMs into thin films: an exponentially grown polyelectrolyte multilayer film, made from sodium hyaluronate (HA) and poly-L-lysine (PLL), is brought in contact with the POM solution to allow its passive diffusion into the PEM film. Both methods of incorporation are compared in terms of effective POM concentration in the film. In addition it is shown that the second method leads to a maximum in the amount of incorporated POM as a function of its concentration in solution.
\end{quotation}

\section*{Characterization of Dopamine-Melanin Growth on Silicon Oxide}
Falk Bernsmann, Arnaud Ponche, Christian Ringwald, Joseph Hemmerlé, Jesus Raya, Burkhard Bechinger, Jean-Claude Voegel, Pierre Schaaf and Vincent Ball. \emph{Journal of Physical Chemistry C}, 113: 8234-8242, 2009, doi: \href{http://dx.doi.org/10.1021/jp901188h}{10.1021/jp901188h}.
\begin{quotation}
It has recently been demonstrated that dopamine solutions put in contact with a variety of solid substrates allow the production of thin coatings probably made of melanin (Science 2007, 318, 426). In this article, we show that the thickness of these coatings can be controlled to allow a growth regime that is proportional to the reaction time if fresh dopamine is regularly provided. 

We propose that the growth is initiated by the adsorption of a radical compound. When dopamine polymerization or aggregation has reached a steady state in solution, the produced species do not adhere anymore to the substrate, emphasizing the role played by unoxidized dopamine. X-ray photoelectron spectroscopy showed that the thickness of the deposit increases linearly with the number of immersion steps, but the thickness measured in ultravacuum is about 4 times smaller than the thickness measured by ellipsometry in conditions of ambient humidity. This suggests that the drying of the deposit has a considerable influence on its properties. The Si2p signal characteristic of the silicon substrate decreases progressively when the number of deposition steps increases but does vanish even after 32 deposition steps. This observation will be discussed with respect to the formation of a continuous film. Cyclic voltammetry experiments showed that a deposit impermeable to ferrocyanide is obtained after the immersion in nine freshly prepared dop\-amine solutions, demonstrating the formation of a film. The atomic composition of the film determined by X-ray photoelectron spectroscopy is compatible with that of melanin. Finally, we show that the deposit can be quantitatively removed from the substrate when put in a strongly alkaline solution.
\end{quotation}

\section*{Polyelectrolyte Multilayer Films Built from Poly(l-lysine) and a Two-Component Anionic Polysaccharide Blend}
Vincent Ball, Falk Bernsmann, Cosette Betscha, Clarisse Maechling, Serge Kauffmann, Bernard Senger, Jean-Claude Voegel, Pierre Schaaf and Nadia Benkirane-Jessel. \emph{Langmuir}, 25(6): 3593-3600, 2009, doi: \href{http://dx.doi.org/10.1021/la803730j}{10.1021/la803730j}.
\begin{quotation}
The buildup of polyelectrolyte multilayer films made from poly(L-lysine) (PLL) as a polycation and from a blend of two anionic polysaccharides, namely, $\beta$-1,3 glycan sulfate (GlyS) and alginate (Alg), was investigated as a function of the mass fraction, x, of GlyS in the blend, at a constant total weight concentration in polyanions. 

We find that the film thickness, after the deposition of a given number of layer pairs, reaches a minimum for x values lower than 0.1 (the position of this minimum could not be more precisely localized) and that the film thickness at intermediate values of x is the same as that of films built at the same concentration of GlyS in the absence of Alg (pure GlyS solution). Infrared spectroscopy in the attenuated total reflection mode shows that the weight fraction of GlyS in the multilayer films is much higher than its weight fraction, x, in the blend used to build the film. This preferential incorporation of GlyS over Alg is related to preferential interactions of GlyS as compared to Alg with PLL in solution, as measured by means of isothermal titration calorimetry. We also demonstrate that GlyS is able to displace Alg almost quantitatively from \chemr{(PLL/Alg)_n} films but that in contrast Alg is not able to exchange GlyS from \chemr{(PLL/GlyS)_n} films.

These results, which combine adsorption from blended polyanion solutions, exchange of one polyanion already present in the film by the other in solution, and thermodynamic measurements, suggest that sulfated polymers are able to interact with polycations preferentially over polymers carrying carboxylated charged groups. These results give a first structural basis to the mechanism of preferential incorporation of a given polyanion with respect to another.
\end{quotation}

\section*{Use of dopamine polymerisation to produce free-standing membranes from \chem{(PLL-HA)_n} exponentially growing multilayer films}
Falk Bernsmann, Ludovic Richert, Bernard Senger, Philippe Lavalle, Jean-Claude Voe\-gel, Pierre Schaaf and Vincent Ball. \emph{Soft Matter}, 4: 1621-1624, 2008,
\\ doi: \href{http://dx.doi.org/10.1039/b806649c}{10.1039/b806649c}.
\begin{quotation}
In this communication, we demonstrate that dopamine is able to undergo a polymerisation process in \chemr{(PLL-HA)_n} polyelectrolyte multilayer films, and that this polymerisation is of the same nature as in solution at pH 8.5. This polymerisation changes the chemical composition and decreases the mobility of the PLL chains in the film, and ultimately allows the easy detachment of the film as free-standing membranes with 0.1 M HCl solutions.
\end{quotation}

%\section*{Protein films adsorbed on experimental dental materials: ToF-SIMS with multivariate data analysis}
%Falk Bernsmann, Nicole Lawrence, Matthias Hannig, Hubert Gnaser and Christiane Ziegler. \emph{Analytical and Bioanalytical Chemistry}, 391: 545-554, 2008.
%\begin{quotation}
%The proteins lysozyme, amylase, and bovine serum albumin (BSA) were adsorbed on two experimental dental materials, made of fluoroapatite particles embedded in polymer matrices, and on silicon wafers. The protein films were prepared as single-component layers, as binary mixtures, and as double layers. These systems were investigated by time-of-flight secondary ion mass spectrometry (ToF-SIMS) and the multivariate data analysis technique of discriminant principal-component analysis (DPCA).

%During adsorption of a single protein film on to the solid surfaces, the three proteins could be clearly distinguished by the scores of their mass spectra after selection of amino acid-related peaks and DPCA. Furthermore, very similar results were obtained on the two different fluoroapatite substrates. For samples coated with binary layers of two proteins adsorbed simultaneously, it was found for both substrate types that BSA shows the strongest ability to adsorb followed by lysozyme, while amylase has the smallest ability. By contrast, the consecutive adsorption of two protein layers showed a strong influence of substrate type on the adsorption ability of the proteins.
%\end{quotation}

\chapter{Curriculum vitae}
\subsection*{General}
\begin{tabular}{lll}
Contact:& Falk Bernsmann& \\
	&13 rue du Maréchal Joffre&Phone: +33 6 73 14 35 18 \\ 
	&67000 Strasbourg, France&Email: bernsmann@web.de \\
Date of birth:& 12 October 1982&\\
Nationality:& German& \\
\end{tabular}

\subsection*{University Studies}
\begin{tabular}{ll}
2007 -- 2010:& Ecole Doctorale de Physique et Chimie Physique, Strasbourg\\
2004 -- 2007:& Integrated French-German Course in Physics, Universit\'e Louis \\
	& Pasteur, Strasbourg and Technische Universit\"at Kaiserslautern\\
2002 -- 2004:& Vordiplomstudium Physik, Westf\"alische Wilhelms-Universit\"at,\\
&M\"unster, Germany
\end{tabular}

\subsection*{Diplomas}
\begin{tabular}{ll}
2007: &Diplom Physik (sehr gut), Technische Universit\"at Kaiserslautern, Germany \\
&\emph{Award of the Kreissparkassenstiftung Kaiserslautern}\\
2007: &Master Matériaux (mention très bien),\\
&Spécialisation Polymères, Universit\'e Louis Pasteur\\
2006: &Ma\^{\i}trise Physique (mention tr\`es bien), Universit\'e Louis Pasteur\\
2005: &Licence Physique(mention bien), Universit\'e Louis Pasteur\\
2002: &Abitur (735/840 points), Johann-Conrad-Schlaun Gymnasium, Münster
\end{tabular}

\subsection*{Working Experience}
\begin{tabular}{ll}
2007& \emph{Experimental PhD studies} on melanin made by oxidation of dopamine;\\ 
-- 2010& \emph{techniques:} confocal laser scanning and scanning force microscopy,\\
	&ellipsometry, quartz crystal microbalance, cyclic voltammetry;\\
	&Institut National de la Sant\'e et de le Recherche M\'edicale, Unit\'e 977:\\
	&Biomat\'eriaux et Ing\'enierie Tissulaire, Strasbourg\\
2008& \emph{European School on Nanoscience and Nanotechnology,} Grenoble, France \\
2006& \emph{Experimental master studies} on dental implant materials and adsorbed \\
-- 2007&protein films; \emph{techniques:} time-of-flight secondary ion mass spectroscopy,\\
	&discriminant principal component analysis; Labor Grenzfl\"achen, \\
	&Nanomaterialien und Biophysik, Technische Universit\"at Kaiserslautern
\end{tabular}

\end{appendix}

%% file: thesis.bbl
\begin{thebibliography}{100}

\bibitem{abbas:2009}
M.~Abbas, F.~D'Amico, et~al.
\newblock {Structural, electrical, electronic and optical properties of melanin
  films}.
\newblock {\em The European Physical Journal E}, 28:285--291, 2009.
\newblock \href{http://dx.doi.org/10.1140/epje/i2008-10437-9}{doi:
  10.1140/epje/i2008-10437-9}.

\bibitem{adhyaru:2003}
B.~B. Adhyaru, N.~G. Akhmedov, et~al.
\newblock {Solid-state cross-polarization magic angle spinning 13C and 15N NMR
  characterization of Sepia melanin, Sepia melanin free acid and Human hair
  melanin in comparison with several model compounds}.
\newblock {\em Magnetic Resonance in Chemistry}, 41:466--474, 2003.
\newblock \href{http://dx.doi.org/10.1002/mrc.1193}{doi: 10.1002/mrc.1193}.

\bibitem{albuquerque:2006}
J.~E. Albuquerque, C.~Giacomantonio, et~al.
\newblock {Study of optical properties of electropolymerized melanin films by
  photopyroelecric scepctroscopy}.
\newblock {\em European Biophysics Journal}, 35:190--195, 2006.
\newblock \href{http://dx.doi.org/10.1007/s00249-005-0020-z}{doi:
  10.1007/s00249-005-0020-z}.

\bibitem{bard:2001}
A.~J. Bard and L.~R. Faulkner.
\newblock {\em {Electrochemical Methods: Fundamentals and Applications}}.
\newblock John Wiley and Sons, second edition, 2001.

\bibitem{barreto:2001}
W.~J. Barreto, S.~R.~G. Barreto, et~al.
\newblock {Interruption of the MnO2 oxidative process on dopamine and L-dopa by
  the action of S2O3}.
\newblock {\em Journal of Inorganic Biochemistry}, 84:89--96, 2001.
\newblock \href{http://dx.doi.org/10.1016/S0162-0134(00)00207-5}{doi:
  10.1016/S0162-0134(00)00207-5}.

\bibitem{binnig:1986}
G.~Binnig, C.~F. Quate, and C.~Gerber.
\newblock {Atomic force microscope}.
\newblock {\em Physical Review Letters}, 56(9):930--933, 1986.
\newblock \href{http://dx.doi.org/10.1103/PhysRevLett.56.930}{doi:
  10.1103/PhysRevLett.56.930}.

\bibitem{boman:1987}
H.~G. Boman and D.~Hultmark.
\newblock {Cell-Free Immunity in Insects}.
\newblock {\em Annual Review of Microbiology}, 41:103--126, 1987.
\newblock \href{http://dx.doi.org/10.1146/annurev.mi.41.100187.000535}{doi:
  10.1146/annurev.mi.41.100187.000535}.

\bibitem{bothma:2008}
J.~P. Bothma, J.~de~Boor, et~al.
\newblock {Device-Quality Electrically Conducting Melanin Thin Films}.
\newblock {\em Advanced Materials}, 20:3539--3542, 2008.
\newblock \href{http://dx.doi.org/10.1002/adma.200703141}{doi:
  10.1002/adma.200703141}.

\bibitem{burke:2003}
S.~E. Burke and C.~J. Barret.
\newblock {pH-Responsive Properties of Multilayered Poly(L-lysine)/Hyaluronic
  Acid Surfaces}.
\newblock {\em Biomacromolecules}, 4:1773--1783, 2003.
\newblock \href{http://dx.doi.org/10.1021/bm034184w}{doi: 10.1021/bm034184w}.

\bibitem{butt:2006}
H.-J. Butt, K.~Graf, and M.~Kappl.
\newblock {\em {Physics and Chemistry of Interfaces}}.
\newblock WILEY-VCH Verlag, Weinheim, second edition, 2006.

\bibitem{carlsson:1978}
J.~Carlsson, H.~Drevin, and R.~Axén.
\newblock {Protein Thiolation and Reversible Protein-Protein Conjugation}.
\newblock {\em Biochemical Journal}, 173:723--737, 1978.

\bibitem{cheng:1994}
J.~Cheng, S.~C. Moss, et~al.
\newblock {X-Ray Characterization of Melanins - II}.
\newblock {\em Pigment Cell Research}, 7:263--273, 1994.
\newblock \href{http://dx.doi.org/10.1111/j.1600-0749.1994.tb00061.x}{doi:
  10.1111/j.1600-0749.1994.tb00061.x}.

\bibitem{clancy:2001}
C.~M.~R. Clancy and J.~D. Simon.
\newblock {Ultrastructural Organization of Eumelanin from Sepia officinalis
  Measured by Atomic Force Microscopy}.
\newblock {\em Biochemistry}, 40:13353--13360, 2001.
\newblock \href{http://dx.doi.org/10.1021/bi010786t}{doi: 10.1021/bi010786t}.

\bibitem{clark:1990}
M.~B. Clark, J.~A. Gardella, et~al.
\newblock {Solid-state analysis of eumelanin biopolymers by electron
  spectroscopy for chemical analysis}.
\newblock {\em Analytical Chemistry}, 62(9):949--956, 1990.
\newblock \href{http://dx.doi.org/10.1021/ac00208a011}{doi:
  10.1021/ac00208a011}.

\bibitem{collins:1993}
R.~W. Collins and K.~Vedam.
\newblock {Ellipsometers}.
\newblock In G.~L. Trigg, editor, {\em {Encyclopedia of Applied Physics}},
  volume~6, pages 191--205. VCH Publishers Inc., New York, USA, 1993.

\bibitem{dadachova:2007}
E.~Dadachova, R.~A. Bryan, et~al.
\newblock {The radioprotective proterties of fungal melanin are a function of
  its chemical composition, stable radical presence and spatial arrangement}.
\newblock {\em Pigment Cell and Melanoma Research}, 21:192--199, 2007.
\newblock \href{http://dx.doi.org/10.1111/j.1755-148X.2007.00430.x}{doi:
  10.1111/j.1755-148X.2007.00430.x}.

\bibitem{diaz:2005}
P.~Daz, Y.~Gimeno, et~al.
\newblock {Electrochemical Self-Assembly of Melanin Films on Gold}.
\newblock {\em Langmuir}, 21(13):5924--5930, 2005.
\newblock \href{http://dx.doi.org/10.1021/la0469755}{doi: 10.1021/la0469755}.

\bibitem{decher:1997}
G.~Decher.
\newblock {Fuzzy Nanoassemblies: Toward Layerd Polymeric Multicomposites}.
\newblock {\em Science}, 277(5330):1232--1237, 1997.
\newblock \href{http://dx.doi.org/10.1126/science.277.5330.1232 }{doi:
  10.1126/science.277.5330.1232 }.

\bibitem{discher:2005}
D.~E. Discher, P.~Janmey, and Y.-L. Wang.
\newblock {Tissue Cells Feel and Respond to the Stiffness of Their Substrate}.
\newblock {\em Science}, 310:1139--1143, 2005.
\newblock \href{http://dx.doi.org/10.1126/science.1116995}{doi:
  10.1126/science.1116995}.

\bibitem{dischia:2009}
M.~d'Ischia, A.~Napolitano, et~al.
\newblock {Chemical and Stuctural Diversity in Eumelanins: Unexplored
  Bio-Optoelectronic Materials}.
\newblock {\em Angewandte Chemie International Edition English}, 48:3914--3921,
  2009.
\newblock \href{http://dx.doi.org/10.1002/anie.200803786}{doi:
  10.1002/anie.200803786}.

\bibitem{drude:1887}
P.~Drude.
\newblock {Ueber die Gesetze der Reflexion und Brechung des Lichtes an der
  Grenze absorbirender Krystalle}.
\newblock {\em Annalen der Physik}, 268(12):584--625, 1887.
\newblock \href{http://dx.doi.org/10.1002/andp.18872681205}{doi:
  10.1002/andp.18872681205}.

\bibitem{elwing:1987}
H.~Elwing, S.~Welin, et~al.
\newblock {A wettability gradient method for studies of macromolecular
  interactions at the liquid/solid interface}.
\newblock {\em Journal of Colloid and Interface Science}, 119(1):203--210,
  1987.
\newblock \href{http://dx.doi.org/10.1016/0021-9797(87)90260-8 }{doi:
  10.1016/0021-9797(87)90260-8 }.

\bibitem{engler:2004}
A.~J. Engler, L.~Richert, et~al.
\newblock {Surface probe measurements of the elasticity of sectioned tissue,
  thin gels and polyelectrolyte multilayer films: Correlations between
  substrate stiffness and cell adhesion}.
\newblock {\em Surface Science}, 570:142--154, 2004.
\newblock \href{http://dx.doi.org/10.1016/j.susc.2004.06.179}{doi:
  10.1016/j.susc.2004.06.179}.

\bibitem{fei:2008}
B.~Fei, B.~Qian, et~al.
\newblock {Coating carbon nanotubes by spontaneous oxidative polymerization of
  dopamine}.
\newblock {\em Carbon}, 46:1795--1797, 2008.
\newblock \href{http://dx.doi.org/10.1016/j.carbon.2008.06.049}{doi:
  10.1016/j.carbon.2008.06.049}.

\bibitem{forest:1998}
S.~E. Forest and J.~D. Simon.
\newblock {Wavelength-dependent Photoacoustic Calorimetry Study of Melanin}.
\newblock {\em Photochemistry and Photobiology}, 68(3):296--298, 1998.
\newblock \href{http://dx.doi.org/10.1111/j.1751-1097.1998.tb09684.x}{doi:
  10.1111/j.1751-1097.1998.tb09684.x}.

\bibitem{francius:2006}
G.~Francius, J.~Hemmerlé, et~al.
\newblock {Effect of Crosslinking on the Elasticity of Polyelectrolyte
  Multilayer Films Measured by Colloidal Probe AFM}.
\newblock {\em Microscopy Research and Technique}, 69:84--92, 2006.
\newblock \href{http://dx.doi.org/10.1002/jemt.20275}{doi: 10.1002/jemt.20275}.

\bibitem{francius:2007}
G.~Francius, J.~Hemmerlé, et~al.
\newblock {Stiffening of Soft Polyelectrolyte Architectures by Multilayer
  Capping Evidenced by Viscoelastic Analysis of AFM Indention Measurements}.
\newblock {\em Journal of Physical Chemistry C}, 111:8299--8306, 2007.
\newblock \href{http://dx.doi.org/10.1021/jp070435+}{doi: 10.1021/jp070435+}.

\bibitem{gallas:1999}
J.~M. Gallas, K.~C. Littrell, et~al.
\newblock {Solution Structure of Copper Ion-Induced Molecular Aggregates of
  Tyrosine Melanin}.
\newblock {\em Biophysical Journal}, 77:1135--1142, 1999.
\newblock \href{http://dx.doi.org/10.1016/S0006-3495(99)76964-X}{doi:
  10.1016/S0006-3495(99)76964-X}.

\bibitem{goncalves:2006}
P.~J. Gonçalves, O.~Baffa Filho, and C.~F.~O. Graeff.
\newblock {Effects of hydrogen on the electronic properties of synthetic
  melanin}.
\newblock {\em Journal of Applied Physics}, 99:104701--104705, 2006.
\newblock \href{http://dx.doi.org/10.1063/1.2201691}{doi: 10.1063/1.2201691}.

\bibitem{haggerty:1993}
L.~Haggerty and A.~M. Lenhoff.
\newblock {Analysis of ordered arrays of adsorbed lysozyme by scanning
  tunneling microscopy}.
\newblock {\em Biophysical Journal}, 64(3):886--895, 1993.
\newblock \href{http://dx.doi.org/10.1016/S0006-3495(93)81448-6}{doi:
  10.1016/S0006-3495(93)81448-6}.

\bibitem{hawley:1967}
M.~D. Hawley, S.~V. Tatawawadi, et~al.
\newblock {Electrochemical Studies of the Oxidation Pathways of
  Catecholamines}.
\newblock {\em Journal of the American Chemical Society}, 89(2):447--450, 1967.
\newblock \href{http://dx.doi.org/10.1021/ja00978a051}{doi:
  10.1021/ja00978a051}.

\bibitem{he:2005}
H.~He, Q.~Xie, and S.~Yao.
\newblock {An electrochemical quartz crystal impedance study on anti-human
  immunoglobulin G immobilization in the polymer grown during dopamine
  oxidation at an Au electrode}.
\newblock {\em Journal of Colloid and Interface Science}, 289:446--454, 2005.
\newblock \href{http://dx.doi.org/10.1016/j.jcis.2005.03.085}{doi:
  10.1016/j.jcis.2005.03.085}.

\bibitem{herlinger:1995}
E.~Herlinger, R.~F. Jameson, and W.~Linert.
\newblock {Spontaneous Autoxidation of Dopamine}.
\newblock {\em Journal of the Chemical Society, Perkin Transactions 2},
  (2):259--263, 1995.
\newblock \href{http://dx.doi.org/10.1039/P29950000259 }{doi:
  10.1039/P29950000259 }.

\bibitem{herve:1994}
M.~Hervé, J.~Hirschinger, et~al.
\newblock {A 13C solid-state NMR study of the structure and auto-oxidation
  process of natural and synthetic melanins}.
\newblock {\em Biochimica et Biophysica Acta - Protein Structure and Molecular
  Enzymology}, 1204:19--27, 1994.
\newblock \href{http://dx.doi.org/10.1016/0167-4838(94)90027-2}{doi:
  10.1016/0167-4838(94)90027-2}.

\bibitem{hong:2007}
L.~Hong and J.~D. Simon.
\newblock {Current Understanding of the Binding Sites, Capacity, Affinity, and
  Biological Significance of Metals in Melanin}.
\newblock {\em Journal of Physical Chemistry B}, 111:7938--7947, 2007.
\newblock \href{http://dx.doi.org/10.1021/jp071439h}{doi: 10.1021/jp071439h}.

\bibitem{howell:2008}
R.~C. Howell, A.~D. Schweitzer, et~al.
\newblock {Chemosorption of radiometals of interest to nuclear medicine by
  synthetic melanins}.
\newblock {\em Nuclear Medicine and Biology}, 35:353--357, 2008.
\newblock \href{http://dx.doi.org/10.1016/j.nucmedbio.2007.12.006}{doi:
  10.1016/j.nucmedbio.2007.12.006}.

\bibitem{hunter:1988}
R.~J. Hunter.
\newblock {\em {Zeta Potential in Colloid Sciences. Principles and
  Applications}}.
\newblock Academic Press, London, 3rd edition, 1988.

\bibitem{ito:1986}
S.~Ito.
\newblock {Reexamination of the structure of eumelanin}.
\newblock {\em Biochimica et Biophysica Acta - General Subjects}, 883:155--161,
  1986.
\newblock \href{http://dx.doi.org/10.1016/0304-4165(86)90146-7}{doi:
  10.1016/0304-4165(86)90146-7}.

\bibitem{ito:2003}
S.~Ito.
\newblock {A Chemist's View of Melanogenesis}.
\newblock {\em Pigment Cell Research}, 16:230--236, 2003.
\newblock \href{http://dx.doi.org/10.1034/j.1600-0749.2003.00037.x}{doi:
  10.1034/j.1600-0749.2003.00037.x}.

\bibitem{izquierdo:2005}
A.~Izquierdo, S.~S. Ono, et~al.
\newblock {Dipping versus Spraying: Exploring the Deposition Conditions for
  Speeding Up Layer-by-Layer Assembly}.
\newblock {\em Langmuir}, 21:7558--7567, 2005.
\newblock \href{http://dx.doi.org/10.1021/la047407s}{doi: 10.1021/la047407s}.

\bibitem{jaber:2010}
M.~Jaber and J.-F. Lambert.
\newblock {A New Nanocomposite: L-DOPA/Laponite}.
\newblock {\em The Journal of Physical Chemistry Letters}, 1:85--88, 2010.
\newblock \href{http://dx.doi.org/10.1021/jz900020m}{doi: 10.1021/jz900020m}.

\bibitem{jastrzebska:1995}
M.~M. Jastrzebska, J.~Isotalo, and J.~Paloheimo.
\newblock {Electrical conductivity of synthetic DOPA-melanin polymer for
  different hydration states and temperatures}.
\newblock {\em Journal of Biomaterials Science Polymer Edition}, 7:577--586,
  1995.
\newblock \href{http://dx.doi.org/10.1163/156856295X00490}{doi:
  10.1163/156856295X00490}.

\bibitem{kaxiras:2006}
E.~Kaxiras, A.~Tsolakidis, et~al.
\newblock {Structural Model of Eumelanin}.
\newblock {\em Physical Review Letters}, 97(21):218102, 2006.
\newblock \href{http://dx.doi.org/10.1103/PhysRevLett.97.218102}{doi:
  10.1103/PhysRevLett.97.218102}.

\bibitem{kienzl:1999}
E.~Kienzl, K.~Jellinger, et~al.
\newblock {Iron as Catalyst for Oxidative Stress in the Pathogenesis of
  Parkinson's Desease?}
\newblock {\em Life Scences}, 65(18/19):1973--1976, 1999.
\newblock \href{http://dx.doi.org/10.1016/S0024-3205(99)00458-0}{doi:
  10.1016/S0024-3205(99)00458-0}.

\bibitem{kinoshita:2005}
S.~Kinoshita and S.~Yoshioka.
\newblock {Structural Colors in Nature: The Role of Regularity and Irregularity
  in the Structure}.
\newblock {\em ChemPhysChem}, 6:1442--1459, 2005.
\newblock \href{http://dx.doi.org/10.1002/cphc.200500007}{doi:
  10.1002/cphc.200500007}.

\bibitem{ladam:2000}
G.~Ladam, P.~Schaad, et~al.
\newblock {In Situ Determination of the Structural Properties of Initially
  Deposited Polyelectrolyte Multilayers}.
\newblock {\em Langmuir}, 16(3):1249--1255, 2000.
\newblock \href{http://dx.doi.org/10.1021/la990650k}{doi: 10.1021/la990650k}.

\bibitem{lavalle:2005}
P.~Lavalle, F.~Boulmedais, et~al.
\newblock {Free standing membranes made of biocompatible polyelectrolytes using
  the layer by layer method}.
\newblock {\em Journal of Membrane Science}, 253:49--56, 2005.
\newblock \href{http://dx.doi.org/10.1016/j.memsci.2004.12.033}{doi:
  10.1016/j.memsci.2004.12.033}.

\bibitem{lee:2007.2}
H.~Lee, S.~M. Dellatore, et~al.
\newblock {Mussel-Inspired Surface Chemistry for Multifunctional Coatings}.
\newblock {\em Science}, 318:426--430, 2007.
\newblock \href{http://dx.doi.org/10.1126/science.1147241}{doi:
  10.1126/science.1147241}.

\bibitem{lee:2007.1}
H.~Lee, B.~P. Lee, and P.~B. Messersmith.
\newblock {A reversible wet/dry adhesive inspired by mussels and geckos}.
\newblock {\em Nature}, 448:338--341, 2007.
\newblock \href{http://dx.doi.org/10.1038/nature05968}{doi:
  10.1038/nature05968}.

\bibitem{lee:2008}
H.~Lee, Y.~Lee, et~al.
\newblock {Substrate-Independent Layer-by-Layer Assembly by Using
  Mussel-Adhesive-Inspired Polymers}.
\newblock {\em Advanced Materials}, 20(9):1619--1623, 2008.
\newblock \href{http://dx.doi.org/10.1002/adma.200702378}{doi:
  10.1002/adma.200702378}.

\bibitem{lee:2009}
H.~Lee, J.~Rho, and P.~B. Messersmith.
\newblock {Facile Conjugation of Biomolecules onto Surfaces via Mussel Adhesive
  Protein Inspired Coatings}.
\newblock {\em Advanced Materials}, 21:431--434, 2009.
\newblock \href{http://dx.doi.org/10.1002/adma.200801222}{doi:
  10.1002/adma.200801222}.

\bibitem{lee:2006}
H.~Lee, N.~F. Scherer, and P.~B. Messersmith.
\newblock {Single-molecule mechanics of mussel adhesion}.
\newblock {\em Proceedings of the National Academy of Sciences USA},
  103(35):12999--13003, 2006.
\newblock \href{http://dx.doi.org/10.1073/pnas.0605552103}{doi:
  10.1073/pnas.0605552103}.

\bibitem{li:2006.1}
Y.~Li, M.~Liu, et~al.
\newblock {Electrochemical quartz crystal microbalance study on growth and
  property of the polymer deposit at gold electrodes during oxidation of
  dopamine in aqueous solutions}.
\newblock {\em Thin Solid Films}, 497:270--278, 2006.
\newblock \href{http://dx.doi.org/10.1016/j.tsf.2005.10.048}{doi:
  10.1016/j.tsf.2005.10.048}.

\bibitem{ligonzo:2009}
T.~Ligonzo, M.~Ambrico, et~al.
\newblock {Electrical and optical properties of natural and synthetic melanin
  biopolymer}.
\newblock {\em Journal of Non-Crystalline Solids}, 355:1221--1226, 2009.
\newblock \href{http://dx.doi.org/10.1016/j.jnoncrysol.2009.05.014}{doi:
  10.1016/j.jnoncrysol.2009.05.014}.

\bibitem{linert:2000}
W.~Linert and G.~N.~L. Jameson.
\newblock {Redox reactions of neurotransmitters possibly involved in the
  progression of Parkinson’s Disease}.
\newblock {\em Journal of Inorganic Biochemistry}, 79:319--326, 2000.
\newblock \href{http://dx.doi.org/10.1016/S0162-0134(99)00238-X}{doi:
  10.1016/S0162-0134(99)00238-X}.

\bibitem{lutkenhaus:2005}
J.~L. Lutkenhaus, K.~D. Hrabak, et~al.
\newblock {Elastomeric Flexible Free-Standing Hydrogen-Bonded Nanoscale
  Assemblies}.
\newblock {\em Journal of the American Chemical Society}, 127:17228--17234,
  2005.
\newblock \href{http://dx.doi.org/10.1021/ja053472s}{doi: 10.1021/ja053472s}.

\bibitem{mackintosh:2001}
J.~A. Mackintosh.
\newblock {The Antimicrobial Properties of Melanocytes, Melanosomes and Melanin
  and the Evolution of Black Skin}.
\newblock {\em Journal of Theoretical Biology}, 211:101--113, 2001.
\newblock \href{http://dx.doi.org/10.1006/jtbi.2001.2331}{doi:
  10.1006/jtbi.2001.2331}.

\bibitem{mamedov:2000}
A.~A. Mamedov and N.~A. Kotov.
\newblock {Free-Standing Layer-by-Layer Assembled Films of Magnetite
  Nanoparticles}.
\newblock {\em Langmuir}, 16:5530--5533, 2000.
\newblock \href{http://dx.doi.org/10.1021/la000560b}{doi: 10.1021/la000560b}.

\bibitem{mamedov:2002}
A.~A. Mamedov, N.~A. Kotov, et~al.
\newblock {Molecular design of strong single-wall carbon
  nanotube/polyelectrolyte multilayer composites}.
\newblock {\em Nature Materials}, 1:190--194, 2002.
\newblock \href{http://dx.doi.org/10.1038/nmat747}{doi: 10.1038/nmat747}.

\bibitem{meredith:2006.1}
P.~Meredith, B.~J. Powell, et~al.
\newblock {Towards structure-property-function relationships for eumelanin}.
\newblock {\em Soft Matter}, 2:37--44, 2006.
\newblock \href{http://dx.doi.org/10.1039/b511922g}{doi: 10.1039/b511922g}.

\bibitem{meredith:2006.2}
P.~Meredith and T.~Sarna.
\newblock {The physical and chemical properties of eumelanin}.
\newblock {\em Pigment Cell Research}, 19:572--594, 2006.
\newblock \href{http://dx.doi.org/10.1111/j.1600-0749.2006.00345.x}{doi:
  10.1111/j.1600-0749.2006.00345.x}.

\bibitem{merrit:1996}
M.~E. Merrit, A.~M. Christensen, et~al.
\newblock {Detection of Intercatechol Cross-Links in Insect Cuticle by
  Solid-State Carbon-13 and Nitrogen-15 NMR}.
\newblock {\em Journal of the American Chemical Society}, 118:11278--11282,
  1996.
\newblock \href{http://dx.doi.org/10.1021/ja961621o}{doi: 10.1021/ja961621o}.

\bibitem{mondon:2002}
M.~Mondon.
\newblock {\em {Untersuchungen zur Proteinadsorbtion auf medizinisch relevanten
  Oberflächen mit Rasterkraftspektroskopie und dynamischer
  Kontaktwinkelanalyse}}.
\newblock {PhD thesis}, Universität Kaiserslautern, 2002.

\bibitem{moses:2006}
D.~N. Moses, M.~A. Mattoni, et~al.
\newblock {Role of melanin in mechanical properties of Glycera jaws}.
\newblock {\em Acta Biomaterialia}, 2:521--530, 2006.
\newblock \href{http://dx.doi.org/10.1016/j.actbio.2006.05.002}{doi:
  10.1016/j.actbio.2006.05.002}.

\bibitem{nelson:1971}
C.~A. Nelson.
\newblock {The Binding of Detergents to Proteins}.
\newblock {\em The Journal of Biological Chemistry}, 246:3895--3901, 1971.

\bibitem{nofsinger:2000}
B.~J. Nofsinger, S.~E. Forest, et~al.
\newblock {Probing the Building Blocks of Eumelanins Using Scanning Electron
  Microscopy}.
\newblock {\em Pigment Cell Research}, 13:179--184, 2000.
\newblock \href{http://dx.doi.org/10.1034/j.1600-0749.2000.130310.x}{doi:
  10.1034/j.1600-0749.2000.130310.x}.

\bibitem{norde:2008}
W.~Norde.
\newblock {My voyage of discovery to proteins in flatland... and beyond}.
\newblock {\em Colloids and Surfaces B: Biointerfaces}, 61:1--9, 2008.
\newblock \href{http://dx.doi.org/10.1016/j.colsurfb.2007.09.029}{doi:
  10.1016/j.colsurfb.2007.09.029}.

\bibitem{norris:2004}
D.~J. Norris, E.~G. Arlinghaus, et~al.
\newblock {Opaline Photonic Crystals: How Does Self-Assembly Work?}
\newblock {\em Advanced Materials}, 16(16):1393--1399, 2004.
\newblock \href{http://dx.doi.org/10.1002/adma.200400455}{doi:
  10.1002/adma.200400455}.

\bibitem{odh:1994}
G.~Odh, R.~Carstam, et~al.
\newblock {Neoromelanin of the Human Substantia Nigra: A Mixed-Type Melanin}.
\newblock {\em Journal of Neurochemistry}, 62(5):2030--2036, 1994.
\newblock \href{http://dx.doi.org/10.1111/j.1471-4159.1994.tb11160.x}{doi:
  10.1111/j.1471-4159.1994.tb11160.x}.

\bibitem{ono:2006}
S.~S. Ono and G.~Decher.
\newblock {Preparation of Ultrathin Self-Standing Polyelectrolyte Multilayer
  Membranes at Physiological Conditions Using pH-Responsive Film Segments as
  Sacrificial Layers}.
\newblock {\em Nano Letters}, 6(4):592--598, 2006.
\newblock \href{http://dx.doi.org/10.1021/nl0515504}{doi: 10.1021/nl0515504}.

\bibitem{ou:2009}
J.~Ou, J.~Wang, et~al.
\newblock {Microtribological and electrochemical corrosion behaviors of
  polydopamine coating on APTS-SAM modified Si substrate}.
\newblock {\em Applied Surface Science}, 256:894--899, 2009.
\newblock \href{http://dx.doi.org/10.1016/j.apsusc.2009.08.081}{doi:
  10.1016/j.apsusc.2009.08.081}.

\bibitem{patrick:2003}
G.~L. Patrick.
\newblock {\em {Chimie Pharmaceutique}}.
\newblock de Boeck, Paris, France, 2003.

\bibitem{pelham:1997}
R.~J.~J. Pelham and Y.-L. Wang.
\newblock {Cell locomotion and focal adhesions are regulated by substrate
  flexibility}.
\newblock {\em Proceedings of the National Academy of Sciences USA},
  94(25):13661--13665, 1997.

\bibitem{peter:1989}
M.~G. Peter and H.~Förster.
\newblock {On the Structure of Eumelanins: Identification of Constitutional
  Patterns by Solid-state NMR Spectroscopy}.
\newblock {\em Angewandte Chemie International Edition English},
  28(6):741--743, 1989.
\newblock \href{http://dx.doi.org/10.1002/anie.198907411}{doi:
  10.1002/anie.198907411}.

\bibitem{pezzella:2009}
A.~Pezzella, A.~Iadonisi, et~al.
\newblock {Disentangling Eumelanin “Black Chromophore”: Visible Absorption
  Changes As Signatures of Oxidation State- and Aggregation-Dependent Dynamic
  Interactions in a Model Water-Soluble 5,6-Dihydroxyindole Polymer}.
\newblock {\em Journal of the American Chemical Society}, 131:15270--15275,
  2009.
\newblock \href{http://dx.doi.org/10.1021/ja905162s}{doi: 10.1021/ja905162s}.

\bibitem{pezzella:1997}
A.~Pezzella, A.~Napolitano, et~al.
\newblock {Identification of Partially Degraded Oligomers of
  5,6-Dihydroxyindole-2-carboxylic Acid in Sepia Melanin by Matrix-assisted
  Laser Desorption/Ionization Mass Spectrometry}.
\newblock {\em Rapid Commucications in Mass Spectrometry}, 11:368--372, 1997.
\newblock
  \href{http://dx.doi.org/10.1002/(SICI)1097-0231(19970228)11:4<368::AID-RCM859>3.0.CO;2-E}{doi:
  10.1002/(SICI)1097-0231(19970228)11:4<368::AID-RCM859>3.0.CO;2-E}.

\bibitem{pezzella:2007}
A.~Pezzella, L.~Panzella, et~al.
\newblock {5,6-Dihydroxyindole Tetramers with “Anomalous” Interunit Bonding
  Patterns by Oxidative Coupling of 5,5',6,6'-Tetrahydroxy-2,7'-biindolyl:
  Emerging Complexities on the Way toward an Improved Model of Eumelanin
  Buildup}.
\newblock {\em Journal of Organic Chemistry}, 72:9225--9230, 2007.
\newblock \href{http://dx.doi.org/10.1021/jo701652y}{doi: 10.1021/jo701652y}.

\bibitem{picart:2001}
C.~Picart, P.~Lavalle, et~al.
\newblock {Buildup Mechanism for Poly(L-lysine)/Hyaluronic Acid Films onto a
  Solid Surface}.
\newblock {\em Langmuir}, 17:7412--7424, 2001.
\newblock \href{http://dx.doi.org/10.1021/la010848g}{doi: 10.1021/la010848g}.

\bibitem{picart:2002}
C.~Picart, J.~Mutterer, et~al.
\newblock {Molecular basis for the explanation of the exponential growth of
  polyelectrolyte multilayers}.
\newblock {\em Proceedings of the National Academy of Sciences USA},
  99:12531--12535, 2002.
\newblock \href{http://dx.doi.org/10.1073/pnas.202486099}{doi:
  10.1073/pnas.202486099}.

\bibitem{picart:2005}
C.~Picart, J.~Mutterer, et~al.
\newblock {Application of Fluorescence Recovery After Photobleaching to
  Diffusion of a Polyelectrolyte in a Multilayer Film}.
\newblock {\em Microscopy Research and Technique}, 66:43--57, 2005.
\newblock \href{http://dx.doi.org/10.1002/jemt.20142}{doi: 10.1002/jemt.20142}.

\bibitem{podsiadlo:2007}
P.~Podsiadlo, Z.~Liu, et~al.
\newblock {Fusion of Seashell Nacre and Marine Bioadhesive Analogs:
  High-Strength Nanocomposite by Layer-by-Layer Assembly of Clay and
  L-3,4-Dihydroxyphenylalanine Polymer}.
\newblock {\em Advanced Materials}, 19:949--955, 2007.
\newblock \href{http://dx.doi.org/10.1002/adma.200602706}{doi:
  10.1002/adma.200602706}.

\bibitem{porcel:2006}
C.~Porcel, P.~Lavalle, et~al.
\newblock {From Exponential to Linear Growth in Polyelectrolyte Multilayers}.
\newblock {\em Langmuir}, 22:4376--4383, 2006.
\newblock \href{http://dx.doi.org/10.1021/la053218d}{doi: 10.1021/la053218d}.

\bibitem{postma:2009}
A.~Postma, Y.~Yan, et~al.
\newblock {Self-Polymerisation of Dopamine as a Versatile and Robust Technique
  to Prepare Polymer Capsules}.
\newblock {\em Chemistry of Materials}, 21(14):3042--3044, 2009.
\newblock \href{http://dx.doi.org/10.1021/cm901293}{doi: 10.1021/cm901293}.

\bibitem{rawlins:1992}
D.~J. Rawlins.
\newblock {\em {Light Microscopy}}.
\newblock Introduction to Biotechniques. BIOS Scientific Publishers Ltd.,
  Oxford, United Kingdom, 1992.

\bibitem{reinheimer:1999}
P.~Reinheimer, J.~Hirschinger, et~al.
\newblock {Cross-polarization/magic-angle-spinning nuclear magnetic resonance
  in selectively 13C-labeled synthetic eumelanins}.
\newblock {\em Biochimica et Biophysica Acta - General Subjects},
  1472:240--249, 1999.
\newblock \href{http://dx.doi.org/10.1016/S0304-4165(99)00127-0}{doi:
  10.1016/S0304-4165(99)00127-0}.

\bibitem{richert:2004}
L.~Richert, F.~Boulmedais, et~al.
\newblock {Improvement of Stability and Cell Adhesion Properties of
  Polyelectrolyte Multilayer Films by Chemical Cross-Linking}.
\newblock {\em Biomacromolecules}, 5(2):284--294, 2004.
\newblock \href{http://dx.doi.org/10.1021/bm0342281}{doi: 10.1021/bm0342281}.

\bibitem{riesz:2006}
J.~Riesz, J.~Gilmore, and P.~Meredith.
\newblock {Quantitative Scattering of Melanin Soutions}.
\newblock {\em Biophysical Journal}, 90:4137--4144, 2006.
\newblock \href{http://dx.doi.org/10.1529/biophysj.105.075713}{doi:
  10.1529/biophysj.105.075713}.

\bibitem{rubianes:2001}
M.~D. Rubianes and G.~A. Rivas.
\newblock {Highly selective dopamine quantification using a glassy carbon
  electrode modified with a melanin-type polymer}.
\newblock {\em Analytica Chimica Acta}, 440:99--108, 2001.
\newblock \href{http://dx.doi.org/10.1016/S0003-2670(01)01059-5}{doi:
  10.1016/S0003-2670(01)01059-5}.

\bibitem{sapurina:2001}
I.~Sapurina, A.~Riede, and J.~Stejskal.
\newblock {In-situ polymerized polyaniline films 3. Film formation}.
\newblock {\em Synthetic Metals}, 123:503--507, 2001.
\newblock \href{http://dx.doi.org/10.1016/S0379-6779(01)00349-6}{doi:
  10.1016/S0379-6779(01)00349-6}.

\bibitem{sauerbrey:1959}
G.~Sauerbrey.
\newblock {Verwendung von Schwingquarzen zur Wägung dünner Schichten und zur
  Mikrowägung}.
\newblock {\em Zeitschrift für Physik}, 155:206--222, 1959.
\newblock \href{http://dx.doi.org/10.1007/BF01337937}{doi: 10.1007/BF01337937}.

\bibitem{sawyer:1995}
D.~T. Sawyer, A.~Sabkawiak, and J.~L. Roberts.
\newblock {\em {Electrochemistry for Chemists}}.
\newblock J. Wiley and Sons, New York, USA, second edition, 1995.

\bibitem{schmitt:2006}
F.~Schmitt.
\newblock {\em {Cantilevermodifizierungen zur Untersuchung der Biofilmbildung
  auf Implantatmaterialien und der Bestimmung der Adhäsion von Botrytis Cinerea
  Konidien auf Modelloberflächen mit Hilfe der Rasterkraftmikroskopie}}.
\newblock {Diploma Thesis}, TU Kaiserslautern, 2006.

\bibitem{seo:2008}
J.~Seo, J.~L. Lutkenhaus, et~al.
\newblock {Effect of the Layer-by-Layer (LbL) Deposition Method on the Surface
  Morphology and Wetting Behavior of Hydrophobically Modified PEO and PAA LbL
  Films}.
\newblock {\em Langmuir}, 24:7995--8000, 2008.
\newblock \href{http://dx.doi.org/10.1021/la800906x}{doi: 10.1021/la800906x}.

\bibitem{simon:2000}
J.~D. Simon.
\newblock {Spectroscopic and Dynamic Studies of the Epidermal Chromophores
  trans-Urocanic Acid and Eumelanin}.
\newblock {\em Accounts of Chemical Research}, 33(5):307--313, 2000.
\newblock \href{http://dx.doi.org/10.1021/ar970250t}{doi: 10.1021/ar970250t}.

\bibitem{srivastava:2008}
S.~Srivastava, V.~Ball, et~al.
\newblock {Reversible loading and unloading of nanoparticles in "Exponentially"
  growing polyelectrolyte LBL films}.
\newblock {\em Journal of the American Chemical Society}, 130:3748--3749, 2008.
\newblock \href{http://dx.doi.org/10.1021/ja7110288}{doi: 10.1021/ja7110288}.

\bibitem{stark:2003}
K.~B. Stark, J.~M. Gallas, et~al.
\newblock {Spectroscopic Study and Simulation from Recent Structural Models for
  Eumelanin: II. Oligomers}.
\newblock {\em Journal of Physical Chemistry B}, 107:11558--11562, 2003.
\newblock \href{http://dx.doi.org/10.1021/jp034965r}{doi: 10.1021/jp034965r}.

\bibitem{stark:2005}
K.~B. Stark, J.~M. Gallas, et~al.
\newblock {Effect of Stacking and Redox State on Optical Absorption Spectra of
  Melanins - Comparison of Theoretical and Experimental Results}.
\newblock {\em Journal of Physical Chemistry B}, 109:1970--1977, 2005.
\newblock \href{http://dx.doi.org/10.1021/jp046710z}{doi: 10.1021/jp046710z}.

\bibitem{statz:2005}
A.~R. Statz, R.~J. Meagher, et~al.
\newblock {New Peptidomimetic Polymers for Antifouling Surfaces}.
\newblock {\em Journal of the American Chemical Society}, 127:7972--7973, 2005.
\newblock \href{http://dx.doi.org/10.1021/ja0522534}{doi: 10.1021/ja0522534}.

\bibitem{subianto:2005}
S.~Subianto, G.~Will, and P.~Meredith.
\newblock {Electrochemical synthesis of melanin free-standing films}.
\newblock {\em polymer}, 46:11505--11509, 2005.
\newblock \href{http://dx.doi.org/10.1016/j.polymer.2005.10.068}{doi:
  10.1016/j.polymer.2005.10.068}.

\bibitem{szpoganicz:2002}
B.~Szpoganicz, S.~Gidanian, et~al.
\newblock {Metal binding by melanins: studies of colloidal
  dihydroxyindole-melanin and its complexation by Cu(II) and Zn(II) ions}.
\newblock {\em Journal of Inorganic Biochemistry}, 89:45--53, 2002.
\newblock \href{http://dx.doi.org/10.1016/S0162-0134(01)00406-8 }{doi:
  10.1016/S0162-0134(01)00406-8 }.

\bibitem{tang:2003}
Z.~Tang, N.~A. Kotov, et~al.
\newblock {Nanostructured artificial nacre}.
\newblock {\em Nature Materials}, 2:413--418, 2003.
\newblock \href{http://dx.doi.org/10.1038/nmat906}{doi: 10.1038/nmat906}.

\bibitem{tang:2006.1}
Z.~Tang, Y.~Wang, et~al.
\newblock {Biomedical Applications of Layer-by-Layer Assembly: From Biomimetics
  to Tissue Engineering}.
\newblock {\em Advanced Materials}, 18:3203--3224, 2006.
\newblock \href{http://dx.doi.org/10.1002/adma.200600113}{doi:
  10.1002/adma.200600113}.

\bibitem{tanuma:1993}
S.~Tanuma, C.~J. Powell, and D.~R. Penn.
\newblock {Calculations of electron inelastic mean free paths. V. Data for 14
  organic compounds over the 50-2000 eV range}.
\newblock {\em Surface and Interface Analysis}, 21:165--176, 1993.
\newblock \href{http://dx.doi.org/10.1002/sia.740210302}{doi:
  10.1002/sia.740210302}.

\bibitem{tran:2006}
M.~L. Tran, B.~J. Powell, and P.~Meredith.
\newblock {Chemical and Structural Disorder in Eumelanins: A Possible
  Explanation for Broadband Absorbance}.
\newblock {\em Biophysical Journal}, 90(3):734--752, 2006.
\newblock \href{http://dx.doi.org/10.1529/biophysj.105.069096}{doi:
  10.1529/biophysj.105.069096}.

\bibitem{voegel:2003}
J.-C. Voegel, G.~Decher, and P.~Schaaf.
\newblock {Multicouches de polyélectrolytes dans le domaine des
  biotechnologies}.
\newblock {\em L'actualité chimique}, 269:30--38, 2003.

\bibitem{voinova:1999}
M.~V. Voinova, M.~Rodahl, et~al.
\newblock {Viscoelastic Acoustic Response of Layered Polymer Films at
  Fluid-Solid Interfaces: Continuum Mechanics Approach}.
\newblock {\em Physica Scripta}, 59:391--396, 1999.
\newblock \href{http://dx.doi.org/10.1238/Physica.Regular.059a00391}{doi:
  10.1238/Physica.Regular.059a00391}.

\bibitem{wahlgren:1990}
M.~C. Wahlgren and T.~Arnebrant.
\newblock {Interaction of cetyltrimethylammonium bromide and sodium dodecyl
  sulfate with beta-lactoglobulin and lysozyme at solid surfaces}.
\newblock {\em Journal of Colloid and Interface Science}, 142(2):503--511,
  1990.
\newblock \href{http://dx.doi.org/10.1016/0021-9797(91)90080-R}{doi:
  10.1016/0021-9797(91)90080-R}.

\bibitem{waite:2001}
J.~H. Waite and X.~Qin.
\newblock {Polyphosphoprotein from the Adhesive Pads of Mytilus edulis}.
\newblock {\em Biochemistry}, 40:2887--2893, 2001.
\newblock \href{http://dx.doi.org/10.1021/bi002718x}{doi: 10.1021/bi002718x}.

\bibitem{wang:2009}
J.~Wang, L.~Xiao, et~al.
\newblock {A facile surface modification of Nafion membrane by the formation of
  self-polymerized dopamine nano-layer to enhance the methanol barrier
  property}.
\newblock {\em Journal of Power Sources}, 192:336--343, 2009.
\newblock \href{http://dx.doi.org/10.1016/j.jpowsour.2009.03.014}{doi:
  10.1016/j.jpowsour.2009.03.014}.

\bibitem{watt:2009}
A.~A.~R. Watt, J.~P. Bothma, and P.~Meredith.
\newblock {The supramolecular structure of melanin}.
\newblock {\em Soft Matter}, 5:3754--3760, 2009.
\newblock \href{http://dx.doi.org/10.1039/b902507c}{doi: 10.1039/b902507c}.

\bibitem{wilhelm}
S.~Wilhelm, B.~Gröbler, et~al.
\newblock {\em {Confocal Laser Scanning Microscopy}}.
\newblock Microscopy from Carl Zeiss - Principles. Carl Zeiss, Jena, Germany.

\bibitem{xi:2009}
Z.-Y. Xi, Y.-Y. Xu, et~al.
\newblock {A facile method of surface modification for hydrophobic polymer
  membranes based on the adhesive behavior of poly(DOPA) and poly(dopamine)}.
\newblock {\em Journal of Membrane Science}, 327:244--253, 2009.
\newblock \href{http://dx.doi.org/10.1016/j.memsci.2008.11.037}{doi:
  10.1016/j.memsci.2008.11.037}.

\bibitem{yoshioka:2002}
S.~Yoshioka and S.~Kinoshita.
\newblock {Effect of Macroscopic Structure in Iridescent Color of the Peacock
  Feathers}.
\newblock {\em Forma}, 17:169--181, 2002.

\bibitem{yu:2009}
B.~Yu, D.~A. Wang, et~al.
\newblock {Robust polydopamine nano/microcapsules and their loading and release
  behavior}.
\newblock {\em Chemical Communications}, 2009:6789--6791, 2009.
\newblock \href{http://dx.doi.org/10.1039/b910679k}{doi: 10.1039/b910679k}.

\bibitem{zajac:1994}
G.~W. Zajac, J.~M. Gallas, et~al.
\newblock {The fundamental unit of synthetic melanin: a verification by
  tunneling microscopy of X-ray scattering results}.
\newblock {\em Biochimica et Biophysica Acta - General Subjects},
  1199(3):271--278, 1994.
\newblock \href{http://dx.doi.org/10.1016/0304-4165(94)90006-X}{doi:
  10.1016/0304-4165(94)90006-X}.

\bibitem{zhang:2009}
Y.~Zhang, H.~Wang, et~al.
\newblock {Mussel-inspired fabrication of encoded polymer films for
  electrochemical identification}.
\newblock {\em Electrochemistry Communications}, 11:1936--1939, 2009.
\newblock \href{http://dx.doi.org/10.1016/j.elecom.2009.08.024}{doi:
  10.1016/j.elecom.2009.08.024}.

\bibitem{zhou:2009}
W.-H. Zhou, L.~Chun-Hua, et~al.
\newblock {Mussel-inspired molecularly imprinted polymer coating
  superparamagnetic nanoparticles for protein recognition}.
\newblock {\em Journal of Materials Chemistry}, 20(5):880--883, 2009.
\newblock \href{http://dx.doi.org/10.1039/b916619j }{doi: 10.1039/b916619j }.

\bibitem{zi:2003}
J.~Zi, X.~Yu, et~al.
\newblock {Coloration strategies in peacock feathers}.
\newblock {\em Proceedings of the National Academy of Sciences of the United
  States of America}, 100(22):12576--12578, 2003.
\newblock \href{http://dx.doi.org/10.1073/pnas.2133313100}{doi:
  10.1073/pnas.2133313100}.

\end{thebibliography}
